\documentclass[12pt,a4paper,oneside]{article}

\usepackage{graphicx}
\usepackage{titlepic}
\usepackage[left=1.1in,right=1.1in, top=1in, bottom= 1in]{geometry}
\usepackage{amsfonts}
\usepackage{amssymb}
\usepackage{amsmath}
\usepackage{fancyhdr}
\usepackage{hyperref}
\usepackage{etoolbox}
\usepackage[nottoc]{tocbibind}
\usepackage{appendix}
\usepackage{multicol}
\usepackage{leftidx}
\graphicspath{{figures/}}
\usepackage{ragged2e}
\usepackage{mathtools}
\usepackage{units}
\usepackage{float}
\usepackage{subcaption}
\usepackage{commath}
\usepackage{amsthm}
\usepackage[none]{hyphenat} % Avoids to go out of margin

\usepackage{subfiles}

\usepackage{tikz}
\usepackage[compat=1.1.0]{tikz-feynman}
\usepackage{relsize}
\tikzset{invclip/.style={clip,insert path={{[reset cm]
      (-16383.99999pt,-16383.99999pt) rectangle (16383.99999pt,16383.99999pt)
    }}}}
\usetikzlibrary{calc}
\definecolor{allOrderBlue}{rgb}{0.4,0.5,1}
\definecolor{patternBlue}{rgb}{0,0,1}
\definecolor{photonRed}{rgb}{1,0.2,0.2}

\usepackage{cancel}

\def\bf{\mathbf}

\def\ie{\textit{i.e.\;}}
\def\eg{\textit{e.g.\;}}

\def\eq#1{Eq.~({\ref{#1}})}
\def\app#1{App.~({\ref{#1}})}
\def\Refe#1{Ref.~\cite{#1}}
\def\sect#1{Sec.~\ref{#1}}
\def\fig#1{Fig.~(\ref{#1})}

\def\a{\alpha}
\def\b{\beta}
\def\g{\gamma}
\def\d{\delta}
\def\e{\epsilon}
\def\ve{\varepsilon}
\def\k{\kappa}
\def\l{\lambda}
\def\tl{\tilde\lambda}
\def\x{\chi}
\def\tx{\tilde\chi}
\def\m{\mu}
\def\n{\nu}
\def\r{\rho}
\def\s{\sigma}
\def\t{\tau}
\def\z{\zeta}

\def\identity{{\rlap{1} \hskip 1.6pt \hbox{1}}}
\def\iden{\identity}

\def\tD{\tilde{\mathcal{D}}}
\def\IC{\mathbb{C}}
\def\IN{\mathbb{N}}

\def\la{\langle}
\def\ra{\rangle}
\def\lb{[}
\def\rb{]}
\newcommand{\qla}[1]{\la #1 |}
\newcommand{\qlb}[1]{\lb #1 |}
\newcommand{\qra}[1]{| #1 \ra}
\newcommand{\qrb}[1]{| #1 \rb}
\newcommand{\wa}[2]{\la #1 #2 \ra}
\newcommand{\wb}[2]{\lb #1 #2 \rb}
\newcommand{\wab}[3]{\la #1 #2 #3 \rb}
\newcommand{\wba}[3]{\lb #1 #2 #3 \ra}

\newcommand{\wbb}[4]{\lb #1 #2 #3 #4 \rb}

\newcommand{\bra}[1]{\langle #1|}
\newcommand{\ket}[1]{|#1\rangle}
\newcommand{\braket}[2]{\langle #1 |#2 \rangle}
\def\Lexp{\biggl\langle\!\!\!\biggl\langle}
\def\Rexp{\biggr\rangle\!\!\!\biggr\rangle}
\newcommand{\de}[1]{\frac{\partial}{\partial #1}}
\newcommand{\dee}[2]{\frac{\partial #1 }{\partial #2}}
\newcommand{\Res}[1]{\mathop{\mathrm{Res}}_{#1}}

\def\nn{\nonumber}

\def\AC{\mathcal{A}}
\def\MC{\mathcal{M}}
\def\NC{\mathcal{N}}
\def\kb{\bar{k}}
\def\tS{\tilde{S}}

\usepackage{caption}

\title{Scattering waveforms for Kerr black holes from the soft expansion}	             
\author{Damiano Barcaro}				     
\date{\today}								

\makeatletter
\let\thetitle\@title
\let\theauthor\@author
\let\thedate\@date
\makeatother

\begin{document}
\numberwithin{equation}{section}

%%%%%%%%%%%%%%%%%%%%%%%%%%%%%%%%%%%%%%%%%%%%%%%%
%----------------------------------------------%
%%%%%%%%%%%%%%%%%%%%%%%%%%%%%%%%%%%%%%%%%%%%%%%%

\begin{titlepage}
	\centering
    \vspace*{0.5 cm}
    \includegraphics[scale = 0.75]{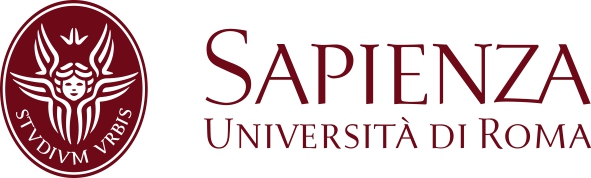}
    \\[1.0 cm]	

    \vspace*{-0.4cm}
    \textsc{\large Department of Physics}
    \\[2.0 cm]	
    \vspace*{1cm}

    { \fontsize{20.74pt}{18.5pt}\selectfont\bfseries \thetitle \par } % Title

    \vspace*{0.25cm}
    \textsc{\Large Theoretical physics}\\[0.5 cm] % Course Name

    \vspace*{3.6cm}
	\begin{minipage}{0.4\textwidth} % 0.4
		\begin{flushleft} \large
			\textbf{Professors:}\\
			Paolo Pani\\
            Vittorio Del Duca\\
            Riccardo Gonzo\\
		\end{flushleft}
	\end{minipage}~
	\begin{minipage}{0.3\textwidth} %0.4
		\begin{flushright} \large
		\begin{minipage}{1\textwidth}
		\begin{flushleft} \large
			\textbf{Students:} \\
			\theauthor
        \end{flushleft}
        \end{minipage}
		\end{flushright}
	\end{minipage}\\[2 cm]

    \vspace{4cm}
    \rule{\linewidth}{0.2 mm} \\[0.3 cm]
    \vspace*{-0.2cm}
    Academic Year 2023/2024
\end{titlepage}

%%%%%%%%%%%%%%%%%%%%%%%%%%%%%%%%%%%%%%%%%%%%%%%%
%----------------------------------------------%
%%%%%%%%%%%%%%%%%%%%%%%%%%%%%%%%%%%%%%%%%%%%%%%%

%
{\color{white} .}
\vspace{6cm}

\begin{abstract}
\noindent In this thesis, I will study the classical scattering problem of two Kerr black holes in general relativity with novel quantum field theory techniques in the Post-Minkowskian (PM) expansion, generalizing the subleading soft theorem to the case of spinning particles. 
The leading order term in the soft expansion is uniquely determined by the universal Weinberg pole, but the subleading one depends on the angular momentum of the external particles and receives a new spin corrections for classically spinning black holes. 
Using this approach, I will compute the gravitational Compton amplitude, both in the quantum and classical theory.
I will then compute the tree-level five-point amplitude of two spinning point particles emitting a graviton, using the spinor-helicity formalism combined with the soft expansion. 
Finally, I will use the KMOC formalism to derive the analytic time-domain waveform for the scattering of two Kerr black holes at leading order in the PM expansion. 
\end{abstract}

%%%%%%%%%%%%%%%%%%%%%%%%%%%%%%%%%%%%%%%%%%%%%%%%
%----------------------------------------------%
%%%%%%%%%%%%%%%%%%%%%%%%%%%%%%%%%%%%%%%%%%%%%%%%

\clearpage
\newpage
\tableofcontents
\newpage
%

%%%%%%%%%%%%%%%%%%%%%%%%%%%%%%%%%%%%%%%%%%%%%%%%
%----------------------------------------------%
%%%%%%%%%%%%%%%%%%%%%%%%%%%%%%%%%%%%%%%%%%%%%%%%

\section{Introduction}\label{sec:introduction}

The advent of gravitational-wave (GW) science~\cite{Abbott:2016blz, TheLIGOScientific:2017qsa,LIGOScientific:2021qlt} has already revolutionized multiple domains of astronomy, cosmology, and particle physics. However, this is merely a glimpse of the vast potential yet to be unlocked~\cite{LISA:2017pwj,Kalogera:2021bya}. \\
Future developments of GW science require a robust theoretical framework to support precision calculations of GW signals. Over the next decade, both space- and ground-based observatories are expected to detect and characterize millions of merger events annually, with sensitivity far surpassing that of current LIGO/Virgo facilities (see, e.g.,~\cite{Favata:2013rwa,Samajdar:2018dcx,Purrer:2019jcp,Huang:2020pba,Gamba:2020wgg}). 
In particular, these progresses will address the need for high precision in upcoming LIGO-Virgo-KAGRA (LVK) runs, in space-based detectors such as LISA~\cite{LISA:2017pwj}, and in future ground-based detectors such as LIGO-India~\cite{Saleem:2021iwi}, Cosmic Explorer~\cite{Reitze:2019iox} and Einstein Telescope~\cite{Punturo:2010zz}. To fully exploit the discovery potential of these increasingly sensitive detectors, the development of high-precision waveform models will be essential~\cite{LISA:2017pwj,Sathyaprakash:2019yqt,Maggiore:2019uih,Kalogera:2021bya,Berti:2022wzk}. 

One of the key challenges will be advancing the theoretical modeling of compact binary coalescences to produce accurate gravitational waveforms.
\begin{figure}[H]
\begin{center}
\includegraphics[scale=.21]{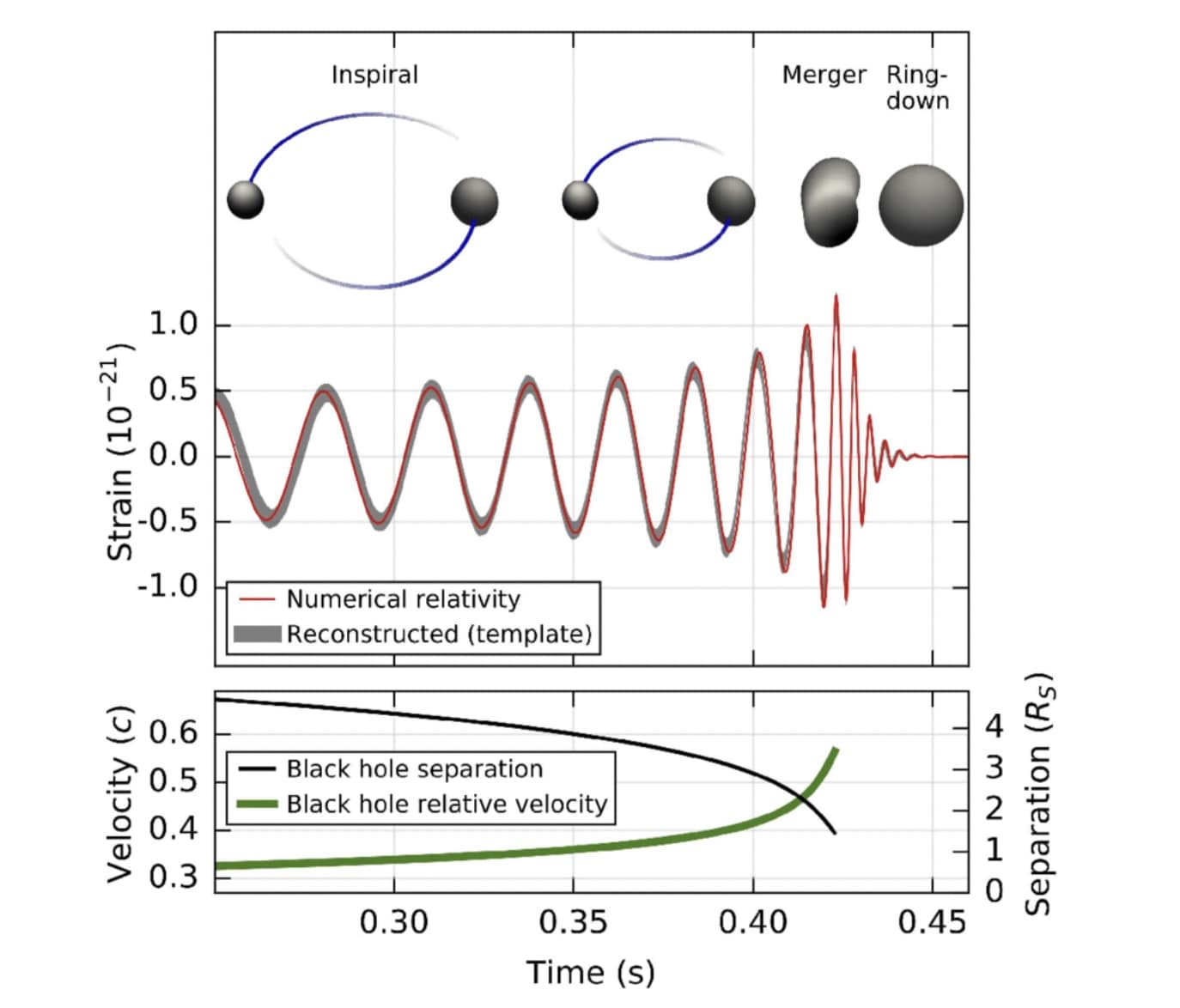}
\end{center}
\vspace{-0.5cm}
\caption{The GW150914 signal observed at the Hanford observatory. The three stages of the
coalescing process are indicated. The lower plot shows the velocity of the components as function of their
spatial separation. Figure reproduced from \Refe{Abbott:2016blz}.}
\label{fig:ffff}
\end{figure}
The theoretical modeling of gravitational-wave (GW) sources poses significant challenges due to the presence of multiple physical scales, which are intricately coupled through the nonlinear framework of general relativity. Interestingly, these very \\complications —nonlinearity and multi-scale dynamics— are analogous to the challenges that spurred groundbreaking advancements in quantum field theory (QFT) over the past few decades. \\
The modern scattering amplitudes program, born from these efforts, has uncovered deep mathematical structures in both gauge theory and gravity. This has not only provided new physical insights but also led to highly efficient computational methods and inspired the development of more ambitious theoretical frameworks.\\
For recent overviews, see~\cite{Blanchet:2013haa, Porto:2016pyg, Schafer:2018kuf, Barack:2018yvs, Levi:2018nxp}. Particularly noteworthy is the seminal work~\cite{Goldberger:2004jt}, which introduced nonrelativistic effective field theory (EFT) concepts from particle physics into the worldline approach for binary dynamics. This innovation has led to several landmark results, with recent developments thoroughly documented in the Snowmass White Paper on NRGR~\cite{NRGRWhitePaper2022}.

Scattering amplitudes are gauge-invariant, universal objects that compactly and analytically encode the perturbative dynamics of point particles. They offer numerous advantages: often expressed through concise analytic formulas, they provide deep physical insights and benefit from a flexible, scalable formalism. This adaptability makes it straightforward to incorporate subleading contributions or new physics, such as spin, tidal effects, or extensions beyond general relativity. \\
However, there are notable challenges. To connect the mathematical formalism of scattering amplitudes with the physics of bound black hole systems, a scatter-to-bound map is required. Moreover, a resummation scheme is necessary, as the perturbative results derived from amplitude methods must be resummed to accurately reflect the non-perturbative nature of bound systems.

The Parke-Taylor formula~\cite{Parke:1986gb} dramatically simplified gluon scattering calculations, reducing pages of Feynman diagrams to a concise half-line expression. This landmark result showcased the profound importance of uncovering the theoretical structures underlying scattering amplitudes. In recent decades, this field has experienced major advances due to two parallel developments.\\
The first is the emergence of novel methods that reformulate quantum field theory (QFT) without relying on explicit quantum fields, instead focusing directly on physical observables. These "on-shell methods," revitalized by twistor string theory ideas~\cite{Witten:2003nn, Roiban:2004yf, Gukov:2004ei, Cachazo:2004kj}, have become highly efficient tools for both tree-level~\cite{Britto:2005fq} and loop-level~\cite{Bern:1994zx, Fusing, TripleCuteeJets, BCFUnitarity} calculations in gauge and gravity theories. For comprehensive overviews, see~\cite{BDKUniarityReview, ElvangHuangReview, Elvang:2015rqa, JJHenrikReview, BernHuangReview}.\\
The second development is a radically new insight into gravity: gravitational scattering amplitudes, \(\mathcal{M}_{\rm gravity}\), can be expressed as a "double copy" of gauge theory amplitudes, \(\mathcal{M}_{\rm gauge}\)~\cite{BCJ, BCJLoop}.
\begin{equation}
{\cal M}_{ \rm gauge} \times {\cal M}_{ \rm gauge}  \sim {\cal M}_{\rm gravity} \,.   
\end{equation}
This correspondence has provided a powerful bridge between gauge and gravity theories, revealing deep structural connections between the two.

\begin{figure}[t]
\begin{center}
\includegraphics[scale=.25]{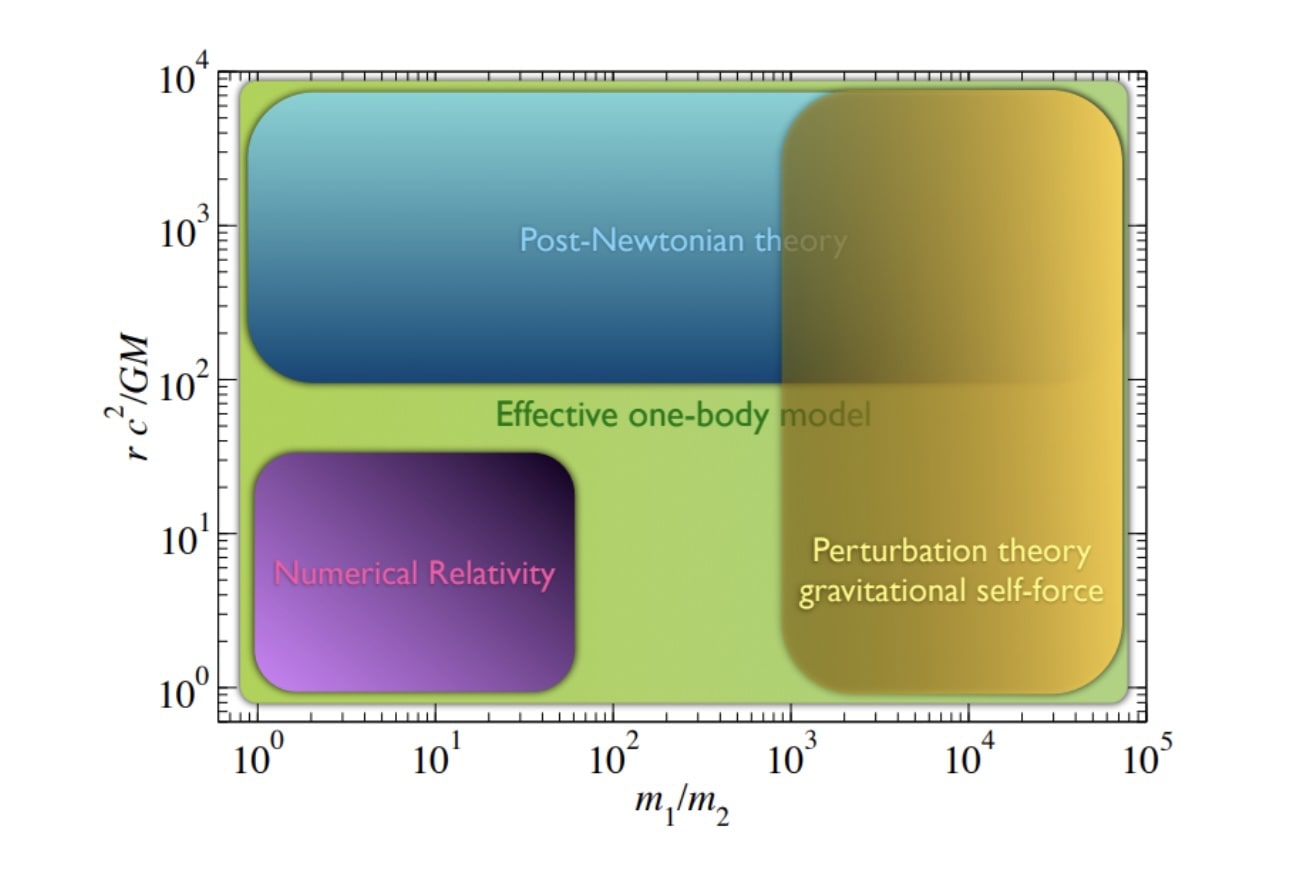 }
\end{center}
\vspace{-0.5cm}
\caption{To model accurately the entire parameter space of the two-body dynamics, various communities need to work together: Gravitational self-force (GSF), Post-Minkowskian (PM), Post-Newtonian (PN) and numerical relativity (NR).
In the above figure the validity of the different methods producing gravitational wave templates in a typical BBH
system is shown. Figure reproduced from \Refe{buonanno2015sourcesgravitationalwavestheory}.}
\label{fig:fffffa}
\end{figure}

\bigskip
The new approach, leveraging tools from theoretical high-energy physics, aims to complement, and has significantly benefited from, decades of successful efforts using traditional methods to solve the relativistic two-body. Some of the most remarkable approaches are the post-Newtonian (PN) approximation~\cite{Einstein:1938yz, Einstein:1940mt, Ohta:1973je, Jaranowski:1997ky, Damour:1999cr, Blanchet:2000nv, Damour:2001bu, Damour:2014jta, Jaranowski:2015lha}, the gravitational self-force (GSF) formalism~\cite{Mino:1996nk, Quinn:1996am}, the effective-one-body (EOB) framework~\cite{Buonanno:1998gg, Buonanno:2000ef}, and the nonrelativistic general relativity (NRGR) approach~\cite{Goldberger:2004jt}. Other important frameworks include the post-Minkowskian (PM) approximation~\cite{Bertotti:1956pxu, Kerr:1959zlt, Bertotti:1960wuq, Portilla:1979xx, Westpfahl:1979gu, Portilla:1980uz, Bel:1981be, Westpfahl:1985tsl, Damour:2016gwp, Damour:2017zjx}, and numerical relativity (NR) methods~\cite{Pretorius:2005gq, Campanelli:2005dd, Baker:2005vv}. \\
It's important to remark that amplitudes demonstrate strong synergies with PN and GSF approaches, enhancing their utility across various frameworks, see \fig{fig:fffffa}. \\

\begin{figure}[H]
\begin{center}
\includegraphics[scale=.3]{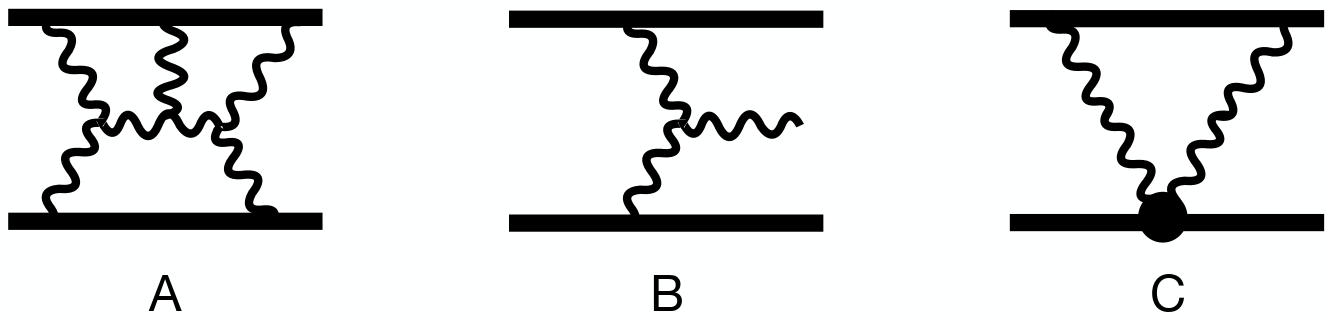}
\\[10pt]
\includegraphics[scale=.4]{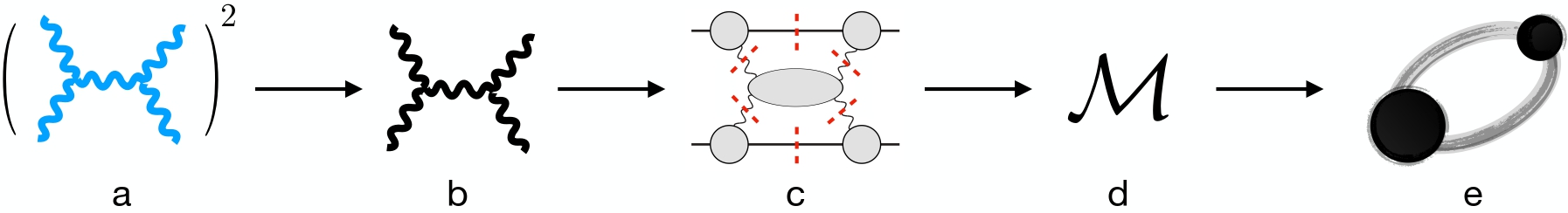}
\end{center}
\vspace{-0.7cm}
\caption{\textbf{(top)} Classical binary dynamics is encoded in scattering amplitudes of massive particles (thick lines) interacting through gravitons (wavy lines):  (A) four-point scattering encodes higher-order corrections to conservative binary dynamics, (B) five-point scattering encodes radiative effects due to graviton emission, and (C) higher-dimension operators (solid circle) encode tidal deformation of neutron stars. Spinning black holes can be described by higher-spin representations in QFT. \hspace{0.15cm} \textbf{(bottom)} A sample calculational pipeline using the tools of theoretical high-energy physics: Starting from tree-level gauge theory amplitudes (a), corresponding gravitational amplitudes (b) are obtained using the double-copy. These are then fused into loop amplitudes (c) using generalized unitarity. The integrated amplitude (d) is obtained using advanced multiloop integration methods developed in high-energy physics, in combination with EFT. The amplitude can then be mapped, using a variety of methods, to the EOB Hamiltonian used for producing waveforms (e). The classical limit is applied at every stage, leading to vast simplifications. Figure reproduced from \Refe{buonanno2022snowmasswhitepapergravitational}.}
\label{fig:method}
\end{figure}

Leveraging newly developed tools from theoretical particle physics offers significant advantages for gravitational wave physics. First, the structure of perturbation theory is greatly streamlined by using special relativity, on-shell methods, and the double-copy framework, resulting in compact expressions that expose the underlying theoretical structures. Second, advanced techniques and a well-established knowledge base for loop integration in quantum field theory —refined over decades for collider physics—can be directly applied. This includes integration-by-parts systems~\cite{Chetyrkin:1981qh, Laporta:2000dsw, Smirnov:2008iw} and the use of differential equations~\cite{Kotikov:1990kg,Bern:1993kr,Remiddi:1997ny,Gehrmann:1999as,Henn:2013pwa,Henn:2013nsa,Parra-Martinez:2020dzs}, which are now highly developed for efficient loop calculations.
Finally, effective field theory (EFT) methods are well-suited for systematically addressing different contributions in the classical limit, allowing for targeted and precise predictions across a range of processes. Figure~\ref{fig:method} illustrates the application of these tools in GW physics.

\begin{figure}[H]
\begin{center}
\includegraphics[scale=.4]{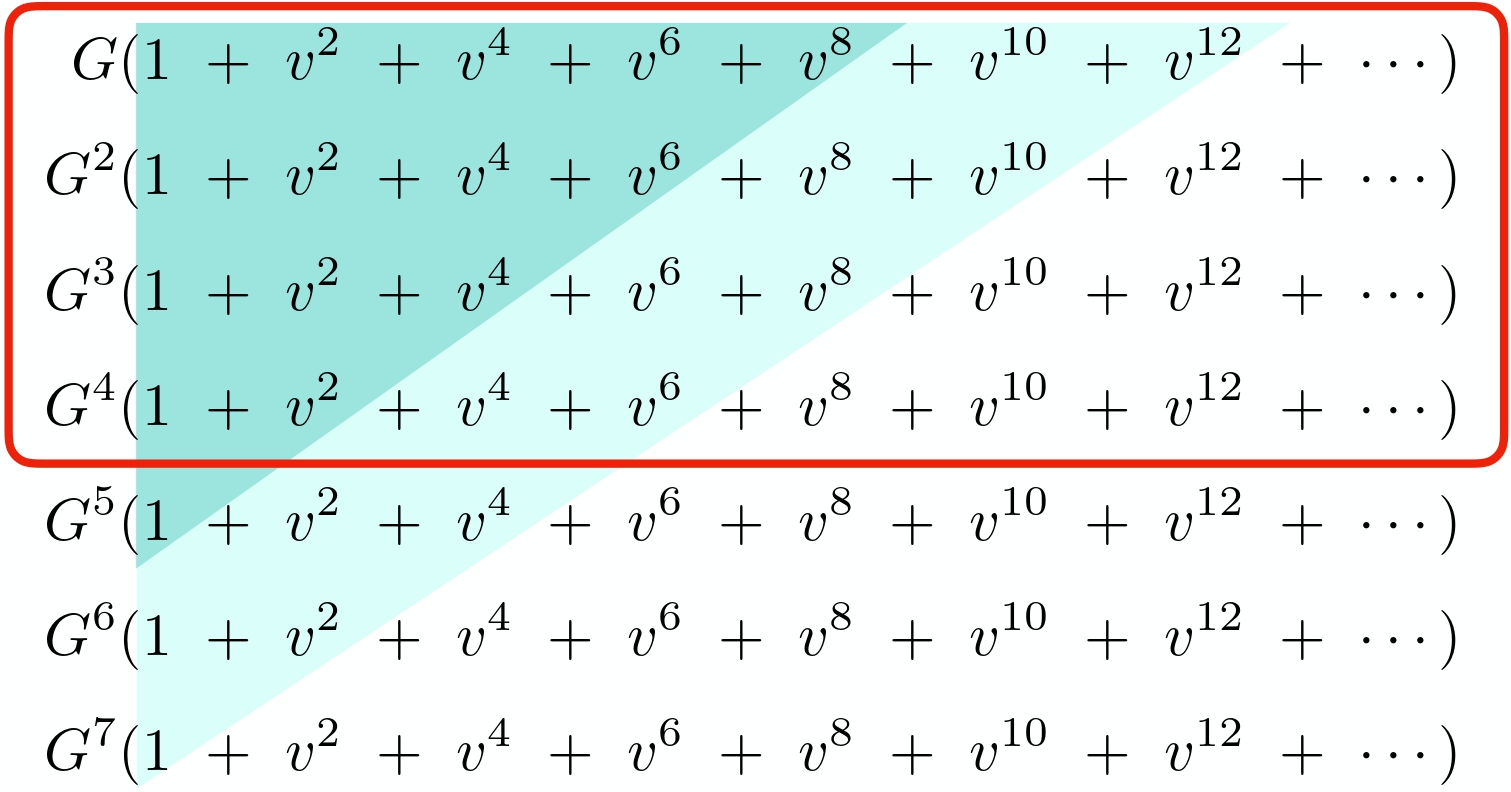}
\end{center}
\vspace{-0.5cm}
\caption{Map of perturbative corrections to Newton's potential, where $G$ is Newton's constant and $v$ is the relative velocity of the binary constituents. New results through $O(G^4)$ were recently obtained using QFT tools (red box). They are valid to all orders in velocity, and overlap with the state-of-the-art from the PN expansion (dark triangle) and the contributions required by future detectors (light triangle) (see, e.g., ~\cite{Favata:2013rwa, Samajdar:2018dcx,Purrer:2019jcp,Huang:2020pba,Gamba:2020wgg}). Figure reproduced from \Refe{buonanno2022snowmasswhitepapergravitational}.}
\label{fig:mapPT}
\end{figure}

In addition to providing state-of-the-art predictions, the new approach using particle physics tools seeks to uncover theoretical structures that emerge in the classical limit of scattering amplitudes. Some of these structures may be well-known in particle physics but remain hidden in traditional treatments of the two-body problem in general relativity. Others may originate in the classical regime and have yet to be fully explored from the perspective of quantum field theory. 
Key examples include universality in the high-energy limit, the interaction between conservative and dissipative effects, nonperturbative connections to classical solutions, and the development of perturbation theory in curved backgrounds. \\
Ultimately, scattering amplitudes serve as powerful tools for both understanding and precisely modeling gravitational wave sources, in much the same way as they are used to describe interactions between fundamental particles.

\paragraph{Classical Limit}\label{classical}
The correspondence principle asserts that classical physics arises as the macroscopic limit of quantum theory, where conserved quantities (charges) such as mass, electric charge, spin, and orbital angular momentum become large. Perturbations around such macroscopic configurations, which are subleading relative to the large charges, can be systematically included and naturally interpreted as quantum corrections.\\
In the context of gravitational waves, the classical limit of scattering amplitudes is characterized by two key properties that distinguish compact binaries from their quantum counterparts:
\begin{itemize}
\item \textbf{Large angular momentum}: 
Bound compact objects, such as black holes or neutron stars, exhibit large angular momentum \( J \gg \hbar \), in contrast to quantum bound states where \( J \sim \hbar \equiv 1 \).
\item \textbf{Large gravitational charges}: 
Compact objects possess significantly large gravitational charges, for example, black holes or neutron stars have \( M_{\odot}/M_{\rm Planck} \sim 10^{38} \), compared to the much smaller ratio of \( e/Q_{\rm Planck} \sim 10^{-1} \) for the electric charges of elementary particles. Here, \( M_{\odot} \) represents the solar mass and \( e \) the electron charge, while \( M_{\rm Planck} \) and \( Q_{\rm Planck} \) are the Planck mass and Planck charge, respectively.
\end{itemize}

In other words, for a system of two gravitationally interacting spinless bodies with masses \( m_1 \) and \( m_2 \), the classical regime emerges when the angular momentum \( J \gg \hbar \) and the masses \( m_1, m_2 \gg M_{\rm Planck} \). This regime corresponds to the condition where the de Broglie wavelength \( \lambda \) of the particles is much smaller than the separation between the particles \( |\mathbf{b}| \), which is conjugate to the momentum transfer \( \mathbf{q} \). Thus, the classical physics regime is characterized by momentum transfer being much smaller than the incoming momenta, \( |\mathbf{p}| \gg |\mathbf{q}| \).\\
Other charges that may characterize classical particles follow similar scaling behavior in the classical limit, \eg:
the spin \( S \) and finite size \( R \) scale as \( |\mathbf{q}|R \sim \mathcal{O}(1) \) and \( |\mathbf{q}||\mathbf{S}| \sim m_{1,2} \), respectively.\\
These scalings are consistent with the classical form of Newton's potential, where \( |\mathbf{b}| \gg Gm \), and they impose constraints on the general structure of four-point scattering amplitudes and the generating functions for observable quantities.

Another perspective on the classical limit is presented in~\cite{Kosower:2018adc}, which is commonly referred to as the KMOC formalism. The details of this formalism will be elaborated in \sect{sec:cap2}.

A significant implication of the classical limit is that loop amplitudes involving massive particles can contain classical contributions. Importantly, this classical limit can be applied at the initial stages of calculations, leading to substantial simplifications before the integration process. This approach allows for the computation of amplitudes that would otherwise be impractical to evaluate if quantum effects were included from the outset.

\bigskip
Recent insights have highlighted a remarkable connection between the soft limit in quantum theories and memory effects in classical dynamics~\cite{Strominger:2014pwa}. This relationship can be explored within the KMOC formalism by examining the radiated momentum. In the long-wavelength limit, the scattering amplitude associated with radiation simplifies to a soft factor multiplied by a lower-point amplitude, effectively recovering the impulse~\cite{bautista2019}. \\
Classically, this impulse represents the "memory" of a step-change in the field, characterized by its very low-frequency Fourier components. More broadly, the KMOC formalism elucidates a rich interplay between classical physics, soft or low-frequency radiation, and scattering amplitudes~\cite{Laddha:2018rle,Laddha:2018myi,Sahoo:2018lxl,Manu:2020zxl,Bautista:2021llr}.

\bigskip
A comprehensive overview of the latest state-of-the-art predictions can be found in \Refe{buonanno2022snowmasswhitepapergravitational}.
\bigskip

This thesis is organized as follows:
in the first chapter \sect{sec:cap1} I will describe the spinor-helicity formalism, that I will use throughout the entirety of this work.
Thanks to this formalism many breakthroughs in particle physics were made in the previous decades, furthermore the application of this formalism to gravity scattering amplitudes led to many elegant results. \\

In the second chapter \sect{sec:cap2} I will introduce the aforementioned KMOC formalism, describing its intricacies and providing a general method to compute observables. \\

In the third chapter (\sect{sec:cap3}), I will begin computing amplitudes, starting with the simplest case: three-point amplitudes. I will first calculate the massless and equal-mass three-point amplitudes in a quantum framework, and then derive their classical counterparts. I will also provide a straightforward method for computing the classical (infinite-spin) limit.\\

In the fourth chapter \sect{sec:cap4} I will discuss a general method to obtain higher-point amplitudes, introducing the BCFW recursion relations. 
Furthermore, I will introduce the soft recursion relations to compute $(n+1)$-amplitudes with one additional external graviton starting from the corresponding $n$-amplitudes.\\

In the fifth chapter \sect{sec:cap5} I will apply the soft recursion relations of \sect{sec:cap3} to the three-point amplitudes obtained in \sect{sec:cap4} to compute the gravitational Compton amplitude, both in the spinless and spinning black holes cases.
I will then evaluate their classical counterpart.
At the end of this chapter I will also introduce a classical version of the soft recursion relations to obtain higher-point classical amplitudes from lower-point ones in a purely classical setting.\\

In the sixth chapter \sect{sec:cap6} I will first use the BCFW recursion relations introduced in \sect{sec:cap3} to compute the 2-to-2 black holes scattering amplitudes without the emission of radiation. 
I will then use the soft recursions to compute the five-point amplitude, and then take its classical limit.\\

In the seventh chapter \sect{sec:cap7} I will use the formalism discussed in \sect{sec:cap2} to compute the waveform at leading order for the black holes scattering with emission of radiation both in the spinning and spinless cases.

%%%%%%%%%%%%%%%%%%%%%%%%%%%%%%%%%%%%%%%%%%%%%%%%
%----------------------------------------------%
%%%%%%%%%%%%%%%%%%%%%%%%%%%%%%%%%%%%%%%%%%%%%%%%

\clearpage

\section{Scattering amplitudes}\label{sec:cap1}

Scattering amplitudes are central to particle physics: their modulus squared is the most important ingredient in the computation of cross sections. The traditional approach to the computation of an un-polarized cross section is to square the amplitude and sum over the polarization of the external states, averaging over the initial ones. The outcome is an expression in terms of Mandelstam invariants and masses. \\
The traditional workflow is based on the following steps: 
we choose the irreducible representations of the Poincaré group, whom are subsequently associated to the fields of the particle present in the theory of interest; following Gell-Mann totalitarian principle we 
now construct the most general Lagrangian. From this Lagrangian we extract the Feynman rules and use them to obtain the Feynman diagrams. We now have all ingredients necessary to evaluate any scattering amplitude, which is finally squared and integrated in order to get the cross section which can be compared with the experimental data.
The bottleneck in this procedure happens when we square the amplitude, as if we are presented with \textit{n} Feynman diagrams, $n^2$ terms will appear in the cross section. The computation quickly becomes intractable as the number of external particles grows, even with state-of-the-art computers. \\
Fixing the polarizations of the external particles, \ie for massless particles the helicities, improves a lot the computation:
for a given helicity configuration the amplitude is just a complex number; furthermore, different helicity configurations do not interfere with each other. Therefore if the amplitude has \textit{n} helicity configurations, the cross section is simply the sum of the \textit{n} squared contributions at fixed helicity. As a final bonus, different helicity configurations are often related by charge conjugation or parity, greatly reducing the number or configurations needed to be computed.\\
Finally, while Feynman diagrams are local in space-time, they are off-shell and not gauge invariant (individually). The ultimate advantage of amplitudes is that they are on-shell and gauge invariant objects.
In fact, one might want to re-think QFTs starting from the fundamental pillars quantum mechanics and special relativity.
%-----------
\subsection{The Little group} 
Amplitudes scatter particles. Let us see then how we can characterize quantum one-particle states. I will mainly follow the work presented in \Refe{Weinberg_libro, Wigner}.\\
Physical states are represented by rays $\mathcal{R}$ in Hilbert space, \ie by normalized vectors $\ket{\psi}$ with $\braket{\psi}{\psi}=1 $, where we identify states $\ket{\psi}$ up to an arbitrary overall phase $e^{i \gamma}$, with $\gamma \in \mathbb{R}$, \ie 
$\ket{\psi}$ and $\ket{\psi'}=e^{i \gamma} \ket{\psi}$ belong to the same ray $\mathcal{R}$. \\
Physical observables \textsl{A} are represented by Hermitian operators. A state $\ket{\psi}$ has definite value $\alpha$ for the observable \textsl{A} if $\ket{\psi}$ is an eigenvector of \textsl{A},
with $A\ket{\psi}=\alpha\ket{\psi}, \, \alpha \in \mathbb{R}$. \\
A transformation between two different observers \textsl{O} and \textsl{O'} who observe the same system is a symmetry transformation.
In such a case, if \textsl{O} observes the states $\ket{\psi_i} \in \mathcal{R}_i$, and \textsl{O'} the states $\ket{\psi'_i} \in \mathcal{R}'_i$, 
then the transition probabilities, 
$\mathcal{P}\left(\mathcal{R}_1 \rightarrow \mathcal{R}_2\right) \equiv \left|\braket{\psi_1}{\psi_2}\right|^2$, 
are conserved, \ie $\mathcal{P}\left(\mathcal{R}_1\rightarrow\mathcal{R}_2\right) = \mathcal{P}\left(\mathcal{R}'_1\rightarrow\mathcal{R}'_2\right) $. \\
Wigner showed that the operator $\mathcal{U}$ which implements the symmetry transformation must be unitary and linear (or anti-unitary and anti-linear). In particular, transformations which are continuously connected to the identity are represented by a linear unitary operator $\mathcal{U}$. \\
The symmetry transformations $\Lambda$ form a group, and the corresponding operators $\mathcal{U}$ acting on the rays mimic the group structure. In fact, one can show that 
\begin{equation}
\mathcal{U} (\Lambda_1 ) \mathcal{U} (\Lambda_2 )= e^{i \phi\left(\Lambda_1,\Lambda_2\right)}\mathcal{U} (\Lambda_1, \Lambda_2 ).   
\end{equation}
If $\phi=0$, then the $\mathcal{U}(\Lambda)$ provide a representation of $\Lambda$. For $\phi\neq 0$, they provide a projective representation.

The tenets of any QFT, even on-shell, are quantum mechanics and special relativity, therefore any transformation under the Poincaré group P must be symmetry transformations. The Poincaré group itself is the minimal subgroup of the affine group which includes all translations and Lorentz transformations. More precisely, it is a semidirect product of the spacetime translations group and the Lorentz group. In particular under the action of a Poincaré transformation, a vector $x^\mu$ transforms as $x^\mu \xrightarrow{P} \Lambda^\mu_{~\nu} x^\nu+b^\mu$, with $\left(det\Lambda\right)^2=1$.

One-particle states (1PS) are classified by the irreducible representations of the Poincaré group, which, Wigner showed, can be classified by the irreducible representations of the little group, \ie the subgroup of transformations which leave the momentum $k$ invariant: $W^\mu_{~\nu}k^\nu=k^\mu$.\\
We want to classify our 1PS by a maximal set of commuting operators, whose eigenvalues will be our quantum numbers, in order to identify the states univocally. We recall that commuting operators can be simultaneously diagonalized.
The components of the momentum commute with each other, so their eigenvalues can be chosen to be our first quantum number; furthermore, the momentum operator $\text{P}^\mu$ commutes with the operator $\text{P}^2$, which will be associated to the second quantum number. The discrete degrees of freedom (\eg the helicity, \dots) will be labeled with $\sigma$.\\
So the 1PS can be written as $\ket{p;\sigma}$, with $\mathrm{P}^\mu\ket{p;\sigma}=p^\mu\ket{p;\sigma}$.

Under a general Lorentz transformation $\mathrm{P}^\mu$ transforms as a vector:
\begin{equation}
\mathrm{P}^\mu \rightarrow \mathrm{P'}^\mu \equiv \mathcal{U}(\Lambda)\mathrm{P}^\mu\mathcal{U}(\Lambda)=\left(\Lambda^{-1}\right)^\mu_{~\nu}\mathrm{P}^\nu = \Lambda^{~\mu}_{\nu}\mathrm{P}^\nu.
\end{equation}
The action of $\mathcal{U}(\Lambda)$ on the state $\ket{p;\sigma}$ is to produce an eigenvector of $\mathrm{P}^\mu$ with eigenvalue $\Lambda_p$.
We now might act with $\mathrm{P}^\mu$ on the state $\mathcal{U}(\Lambda)\ket{p;\sigma}$:
\begin{equation}
\mathrm{P}^\mu \mathcal{U}(\Lambda)\ket{p;\sigma}=\mathcal{U}(\Lambda)
\left(\mathcal{U}^{-1}(\Lambda)\mathrm{P}^\mu \mathcal{U}(\Lambda)\right)\ket{p;\sigma}= 
\mathcal{U}(\Lambda) (\Lambda^{-1})_\nu^{~\mu}\mathrm{P}^\nu \ket{p;\sigma}=\Lambda^\mu_{~\nu}p^\nu \mathcal{U}(\Lambda)\ket{p;\sigma}.
\end{equation}
Therefore $\mathcal{U}(\Lambda)\ket{p;\sigma}$ must be a linear combination of Lorentz-transformed 1PS,
\begin{equation}
\label{eq:1_coeff_little_group}
\mathcal{U}(\Lambda)\ket{p;\sigma}=C_{\sigma \sigma'}(\Lambda,p) \ket{\Lambda p;\sigma'}.
\end{equation}
The crucial point is then to express the coefficients $C_{\sigma \sigma'}(\Lambda,p)$ in terms of irreducible representations of the Poincaré group.

The only functions of $p^\mu$ which are left invariant by (proper orthochronous) Lorentz transformations are $p^2=m^2$ and, if $m^2>0$, the sign of $p^0$. For each $p^2$, we can choose a reference momentum $k^\mu$ such that $p^\mu=L^\mu_{~\nu}(p;k)k^\nu$ and we can define the 1PS as $\ket{p; \sigma}=\mathcal{U}\big(L(p;k)\big)\ket{k;\sigma}$.
Under a general Lorentz transformation, $\ket{p;\sigma}$ transforms in the following way:
\begin{equation}
\mathcal{U}(\Lambda)\ket{p;\sigma}=
\mathcal{U}\big(\Lambda L(p;k)\big)\ket{k;\sigma}=
\mathcal{U}\big(L(\Lambda p;k)\big)\mathcal{U}\big(L^{-1}(\Lambda p;k)\Lambda L(p;k)\big)\ket{k;\sigma}
\end{equation}
We now define the transformation $W=L^{-1}(\Lambda p;k)\Lambda L(p;k)$. This transformation maps $k$ to $p=L(p;k)K$, then to $\Lambda p$, then back to $k$, hence it leaves $k^\mu$ invariant, \ie $W^\mu_{~\nu}k^\nu=k^\mu$.

As previously mentioned, the subgroup of the Lorentz group composed of the Lorentz maps that leave $k^\mu$ invariant is the little group.
The action of $W$ on the reference state $\ket{k;\sigma}$ yields a linear combination of reference states, 
\begin{equation}
\mathcal{U}\big(W(\Lambda,p;k)\big)\ket{k;\sigma}=
D_{\sigma \sigma'}\big(W(\Lambda,p;k)\big)\ket{k;\sigma'},
\end{equation}
where the coefficients $D_{\sigma \sigma'}$ provide a representation of the little group.

In light of everything shown above, we come back to the original problem in \eq{eq:1_coeff_little_group}:
\begin{flalign}
\mathcal{U}(\Lambda)\ket{p;\sigma} & = 
\mathcal{U}\big(L(\Lambda p;k)\big)\mathcal{U}\big(W(\Lambda,p;k)\big)\ket{k;\sigma}= \nn \\
& = D_{\sigma \sigma'}\big(W(\Lambda,p;k)\big)\mathcal{U}\big(L(\Lambda p;k)\big)\ket{k;\sigma'}=
D_{\sigma \sigma'}\big(W(\Lambda,p;k)\big)\ket{\Lambda p;\sigma'} &
\end{flalign}
Thus the issue of determining the coefficients $C_{\sigma \sigma'}(\Lambda, p)$ has been reduced to finding the representations $D_{\sigma \sigma'}\big(W(\Lambda,p;k)\big)$ of the little group. 

To sum up, a particle is labeled by its momentum and transforms under representations of the little group. Thus an n-point scattering amplitude $\MC_n$ is labeled by $\ket{p_i;\sigma_i}$, with $i=1\dots n$. Furthermore, Poincaré invariance implies that
\begin{align}
& \MC(p_i;\sigma_i)=\delta^D(p_1^\mu +\dots+ p_n^\mu)M_n(p_i;\sigma_i), \\
& M^\Lambda(p_i;\sigma_i)=\prod_{i=1}^n D_{\sigma_i \sigma'_i}(W)M\big((\Lambda p)_i;\sigma_i\big).    
\end{align}
In $D$ spacetime dimensions, the little group for massive particles is $SO(D{-}1)$. For massless particles the little group is the the group of Euclidean symmetries in $(D{-}2)$ dimensions. Finite-dimensional representations require choosing all states to have vanishing eigenvalues under these translations, and hence the little group is just $SO(D{-}2)$.
Beyond particles of spin zero, fields are manifestly ``off-shell", and transform as Lorentz tensors (or spinors), while particle states transform instead under the little group. The objects we compute directly with Feynman diagrams in quantum field theory, which are Lorentz tensors, have the wrong transformation properties to be called ``amplitudes". 
This infinite redundancy is hard-wired into the usual field-theoretic description of scattering amplitudes for gauge bosons and gravitons.

The modern on-shell approach to scattering amplitudes departs from the conventional approach to field theory by directly working with objects that transform properly under the little group.
In $D=4$ spacetime dimensions, the little groups are $SO(2) = U(1)$ for massless particles, and $SO(3) = SU(2)$ for massive particles.

In four dimensions, we label massless particles by their helicity $h$. Massive particles transform as some spin $S$ representation of $SU(2)$. 
We will label states of spin $S$ as a symmetric tensor of $SU(2)$ with rank $2S$. 

An important property of the Little group is that it is defined for each individual momenta separately. In other words, only the spinor variables of a given leg can carry its Little group index.

%-----------
\subsection{Spinor-helicity formalism}
It is a well known fact that the Lorentz group $SO^+(1,3)$ is equivalent to $SL(2,\mathbb{C})/\mathbb{Z}_2$, indeed $SL(2,\mathbb{C})$ is the universal covering of the Lorentz group.

We recall that the algebra of the Lorentz group is the algebra of $su(2)_\mathbb{C}\oplus su(2)_\mathbb{C}$ (consistent with $SL(2, \mathbb{C})$ being parametrized by 3 complex parameters or 6 real ones).
Note that, since the (finite dimensional) representations of $SU(2)_\mathbb{C}$ are equivalent to the (finite dimensional) representations of $SL(2, \mathbb{C})$,
we may express the finite dimensional representations of the Lorentz group by either $SU(2)_\mathbb{C}\otimes SU(2)_\mathbb{C}$ or $SL(2, \mathbb{C})\otimes SL(2,\mathbb{C})$.\\
Then the representations of the Lorentz group are labelled by a pair of indices $(i,j)$ taking half-integer values, and have dimensionality $(2i+1)\times(2j+1)$.

The spinor representations are $(\frac{1}{2},0)$ and $(0,\frac{1}{2})$, where the former labels the negative-\\chirality spinor representation (holomorphic spinors), and the latter the positive chirality spinor representation (anti-holomorphic spinors). They transform separately under the respective $SL(2,\mathbb{C})$ factors.\\

We work with the conventions of \app{Conventions}. 
Any real 4-vector $p^\mu$ can be traded for a 2x2 hermitian matrix. We introduce the Pauli matrices
\begin{equation}
  \sigma^1 = 
   \left( 
     \begin{array}{cc}
        0 & 1 \\
        1 & 0 
      \end{array}
   \right) \, ,~~~~~
  \sigma^2 = 
   \left( 
     \begin{array}{cc}
        0 & -i \\
        i & 0 
      \end{array}
   \right) \, ,~~~~~
  \sigma^3 = 
   \left( 
     \begin{array}{cc}
        1 & 0 \\
        0 & -1 
      \end{array}
   \right) 
   \, ,
   \label{pauliM}
\end{equation}
that we can package in a 4-vector 
$\sigma^\mu = (1,\vec\sigma)$, $\bar{\sigma}^\mu = (1,-\vec\sigma) $. Then
\begin{equation}
p\s \equiv p^\mu\sigma_\mu = \left( \begin{array}{cc}
     p^0-p^3  & -p^1+ip^2\\
     -p^1-ip^2& p^0+p^3 
\end{array}
\right),
\qquad \text{with}\quad \mathrm{det}(p^\mu \sigma_\mu)=p^2
\end{equation}
Therefore, any real 4-vector is bijective to a 2x2 hermitian matrix.  Hermiticity is preserved under the $GL(2,\mathbb{C})$ mapping $p^\mu\sigma_\mu\rightarrow\tau p^\mu\sigma_\mu\tau^+$, with $\tau$ an arbitrary 2x2 matrix, furthermore, $\mathrm{det}(p^\mu \sigma_\mu)=p^2$ is preserved if $|\text{det}\tau|=1$. These transformations are simply the ones of the little group in this new language of 2x2 matrices.

Since $p^2=\text{det}(p\cdot\sigma)=\text{det}(p\cdot\bar\sigma)$, 
for $p^2\neq0$, $(p\cdot\sigma)$ and $(p\cdot\bar\sigma)$ have rank 2, while for
$p^2=0$, $(p\cdot\sigma)$ and $(p\cdot\bar\sigma)$ have rank 1.\\

A Lorentz vector, such as the momenta, can be written as a bi-fundamental tensor under $SL(2,\mathbb{C})$:
\begin{equation}
p^\mu \rightarrow p^\mu\sigma_{\mu,\alpha\dot{\alpha}}\equiv   p_{\alpha\dot{\alpha}}
\end{equation}
 where $\alpha, \dot{\alpha}=1,2$. The usual Lorentz invariant inner products are then mapped to the contraction of these tensors with the $2\times 2$ Levi-Civita tensor:
\begin{equation}
p_i^\mu p_{j\nu}=\frac{1}{2}\epsilon^{\alpha\beta}\epsilon^{\dot{\alpha}\dot{\beta}}\,p_{i\alpha\dot{\alpha}}\,p_{j\beta\dot{\beta}}\,.  
\end{equation}
\paragraph{Massless momenta} For massless momenta, the $2\times 2$ tensor $p^{\alpha\dot{\alpha}}$ is of rank 1 so we can write it as the outer product of 2 generic vectors:
\begin{equation}
p_{\alpha\dot{\alpha}}=\lambda_\alpha \tilde{\lambda}_{\dot{\alpha}}\,.  
\end{equation}
This relation is invariant under the following transformation:
\begin{equation}
\lambda\rightarrow e^{-i\frac{\theta}{2}} \lambda,\quad \tilde\lambda\rightarrow e^{i\frac{\theta}{2}} \tilde\lambda    
\end{equation}
Note that this is precisely the definition of the Little group! Thus we identify the spinors $\lambda, \tilde{\lambda}$ as having $(-\frac{1}{2}, +\frac{1}{2})$ Little group weight respectively. Using these bosonic spinors it is then convenient to define the following Lorentz invariant, Little group covariant building blocks:
\begin{equation}
\langle i j \rangle \equiv\lambda_i^\alpha\lambda_j^\beta\epsilon_{\alpha\beta},\quad [ij]\equiv\tilde\lambda_{i\dot\alpha}\tilde\lambda_{j\dot\beta}\epsilon^{\dot\alpha\dot\beta}\,.   
\end{equation}
In terms of these blocks, the usual Mandelstam variables are given as $2p_i\cdot p_j=\langle ij\rangle[ji]$. 

\paragraph{Massive momenta} For massive momenta, $p_{\alpha\dot{\alpha}}$ has rank 2 and we have
\begin{equation}
p_{\alpha\dot{\alpha}}=\lambda^{I}_{\alpha}\tilde{\lambda}_{I\dot{\alpha}}\,,
\end{equation}
where $I=1,2$. The index $I$ indicate that they form a doublet under the $SU(2)$ massive Little group. Indeed the momentum is invariant under the following transformations:
\begin{equation}
\lambda^{I\alpha}\rightarrow U^I\,_J\,\lambda^{J\alpha},\quad \tilde\lambda^{I\dot\alpha}\rightarrow U^I\,_J\,\lambda^{J\alpha}\,,
\end{equation}
where $U$ is an element of $SU(2)$. One can convert between the two spinors via the Dirac equation
\begin{equation}\label{eq:SHvarDiracEq}
p_{\alpha\dot\alpha}\tilde{\lambda}^{I\dot{\alpha}}=m\lambda^I_{\alpha},\quad p_{\alpha\dot\alpha}\lambda^{I\alpha}=-m\tilde\lambda^I_{\dot\alpha}\,, 
\end{equation}
where work with the chiral representation of the gamma matrices
\begin{equation}
\gamma^\mu = \left( \begin{array}{cc}
     0  & \sigma^\mu\\
     \bar{\sigma}^\mu & 0 
\end{array}
\right), \qquad
\gamma^5 = \left( \begin{array}{cc}
     -1  & 0\\
     0 & 1 
\end{array}
\right).
\end{equation}
In the following I will use the labels $\x$ and $\l$ interchangeably for the massive spinors.

\paragraph{Polarizations}
When we consider gauge bosons, we must take into account also their polarizations. We start by considering photons and gluons (graviton's polarization tensors will be simply obtained by joining gluon's ones).\\
The physical polarization of a gluon or photon with momentum $p$ and helicity $h=\pm$ is given by a 4-vector $\ve^\mu_h(p,k)$, with respect to arbitrary reference null vector $k$, $k^2=0$, non-collinear to $p$, \ie $p\cdot k \neq 0$, and has the following properties:
\begin{align}
& \left(\ve^\m_\pm(p,k)\right)^*=\ve^\m_\mp(p,k), 
\quad p_\m\ve^\m_\pm(p,k)= k_\m\ve^\m_\pm(p,k)=0, 
\quad \ve_h(p,k)\cdot \ve_{h'}^*(p,k)=-\d_{h h'} \nn \\
& \hspace{3.5cm} \sum_h \ve_h^\m(p,k)\ve_h^{\n*}(p,k)=-\eta^{\m\n}+\frac{p^\m k^\n+p^\n k^\m}{p\cdot k}
\end{align}
Polarization vectors satisfying these properties are:
\begin{equation}\label{eq:tensori_polarizzazione_spin1}
\ve_+^{\m}(p,k)=\frac{\wab{k}{\s^\m}{p}}{\sqrt{2}\wa{k}{p}}, \qquad
\ve_-^{\m}(p,k)=-\frac{\wba{k}{\bar\s^\m}{p}}{\sqrt{2}\wb{k}{p}}.
\end{equation}
It is easy to check that the choice of the reference vector $k$, does not affect the final amplitude, as the different choices are equivalent up to a factor that cancels due to the Ward (or Slavnov-Taylor) identities. This result is to be expected, as $k$ parametrises the gauge redundancy.

The polarisation tensors for higher-spin particles are constructed as symmetric products of \eq{eq:tensori_polarizzazione_spin1}

%%%%%%%%%%%%%%%%%%%%%%%%%%%%%%%%%%%%%%%%%%%%%%%%
%----------------------------------------------%
%%%%%%%%%%%%%%%%%%%%%%%%%%%%%%%%%%%%%%%%%%%%%%%%

\clearpage

\section{Waveforms from Amplitudes}\label{sec:cap2}
In this chapter I present a formalism, the KMOC formalism, for computing classically measurable quantities directly from on-shell quantum scattering amplitudes. I will mainly follow the works presented in \Refe{KMOC, waveforms_1,waveforms_2}.

Our goal is to understand how to systematically extract the classical result using on-shell quantum- mechanical scattering amplitudes in order to take full advantage of amplitude methods in the gravitational-wave problem.

\paragraph{Restoring $\hbar$}
As a first step we restore all factors $\hbar$. A pragmatic way to do it is by dimensional analysis. We denote the dimensions of mass and length by $[M]$ and $[L]$, respectively. We will use relativistically natural units, with
$c=1$. \\
We keep the dimensions of an $n$-point scattering amplitude (in four dimensions) to be $[M]^{4-n}$ even when $\hbar \neq 1$. This is consistent with choosing the dimensions of creation and annihilation operators so that,
\begin{equation}
[a_p, a^\dagger_{p'}] = (2\pi)^3 \delta^{(3)}(\vec{p} - \vec{p}')\,,
\end{equation}
We define single-particle states by,
\begin{equation}
\ket{p} = \sqrt{2 E_p} \, a^\dagger_p \ket{0}\,.
\end{equation}
The dimension of $\ket{p}$ is thus $[M]^{-1}$ (The vacuum state is taken to be dimensionless). We further define $n$-particle asymptotic states as tensor products of these normalised single particle states.\\
The scattering matrix $S$ and the transition matrix $T$ are both dimensionless. \\
In the course of restoring powers of $\hbar$ by dimensional analysis, we first treat the momenta of all particles as genuine momenta. We also treat any mass as a mass rather than the associated Compton wavelength.

Let us now imagine restoring the $\hbar$s in a given amplitude. When $\hbar \neq 1$, the dimensions of the momenta and masses in the amplitude are unchanged. Similarly there is no change to the dimensions of polarisation vectors or tensors of any Yang--Mills theories. However, a factor of $1/\sqrt{\hbar}$ appears as the appropriate coupling, \eg for gravity, $\kappa = \sqrt{32 \pi G/ \hbar}$. 

In putting the factors of $\hbar$ back into the couplings, we have however not yet made manifest all of the physically relevant factors of $\hbar$, as certain momenta also scale with $\hbar$: for massless particles it is convenient to distinguish between the momentum $p^\mu$ of a particle and its wavenumber $\bar p$:
\begin{align}
\bar p \equiv p / \hbar.
\label{eq:notationWavenumber_2}
\end{align}
Finally, in order for our classical amplitudes to be sensible, we must ensure our system to be in a ``classical" regime, see \sect{sec:Goldilock_1}

%----------------
\subsection{The Incoming State}
\def\in{\text{in}}
\def\out{\text{out}}
We examine scattering events in which two widely separated particles are prepared at $t \rightarrow -\infty$, and then shot at each other with transverse impact parameter $b^\mu$. 
At the quantum level, the particles are described by wavefunctions. For massive particles we will use the point-particle description, for massless particles a sensible treatment relies on coherent states, see \sect{sec:Goldilock_1}.

As we prepare the particles in the far past, the appropriate incoming states are $\ket{\psi}_\in$. We describe the incoming particles by wavefunctions $\phi_i(p_i)$, which are taken to have reasonably well-defined positions and momenta. 
The initial state is then, 
\begin{equation}\label{InitialState_0}
\ket{\psi}_\in = \int \! d^4 p_1 d^4 p_2 \, \hat\d^{(+)}(p_1^2 - m_1^2) 
\hat\d^{(+)}(p_2^2 - m_2^2) \, \phi_1(p_1) \phi_2(p_2) \, e^{i b \cdot p_1/\hbar} \ket{p_1 p_2}_\in \,,
\end{equation}
where we use for convenience the notation:
\begin{align}
& \hat\d^{(n)}(p)\equiv (2\pi)^n \d^{(n)}(p), \quad 
d^n p \equiv \frac{d^n p}{(2 \pi)^n}\,, \quad
\hat\d^{(+)}(p^2-m^2) \equiv 2\pi\Theta(p^0)\d(p^2-m^2), \nn \\
& \quad d\Phi(p_i) \equiv d^4 p_i \, \hat\d^{(+)}(p_i^2-m_i^2)\,, \quad
\hat\d_\Phi(p_1-p_1') \equiv  2 E_{p'_1} \hat\d^{(3)} (\vec{p}_1-\vec{p}_1\,\!\!')\equiv \braket{p'}{p} \,. 
\end{align}
Thus
\begin{equation}\label{eq:initial_state}
\ket{\psi}_\in = \int \! d\Phi(p_1) d\Phi(p_2)\;
  \phi_1(p_1) \phi_2(p_2) \, e^{i b \cdot p_1/\hbar} \ket{p_1 p_2}_\in \,.
\end{equation}
It is now easy to check that the normalisation condition ${}_\in\braket{\psi}{\psi}_\in=1$ is realized if we require both wavefunctions $\phi_i$ to be normalized to unity, \ie
\begin{equation}
\int \! d\Phi(p_i)\; |\phi_i(p_i)|^2 = 1, \quad \text{with} \,\, i=1,2.
\label{WavefunctionNormalization_0}
\end{equation}
In the following, we will often omit the subscript ``in'', any unlabeled state being understood to be an ``in'' state.

\def\KK{\mathbb{K}}
\subsubsection{The Momentum Radiated during a collision}
To understand the subtleties of the approach outlined above, it is useful to compute the expectation value of the four-momentum radiated during a scattering process, as it is a classical observable \footnote{As in this work I'm not interested in global observables, I will just outline the main passages. The full computation can be found in chapter 3 of \Refe{waveforms_1}}.\\ 
To define the observable, let us imagine to surround the collision with detectors which will cover the full solid angle $4\pi$ around the interaction point. 
We will call the radiated particles `messengers'. 

Let $\KK^\mu$ be the momentum operator for whatever field is radiated. The expectation of the radiated momentum is
\begin{equation}
\langle k^\mu \rangle = {}_\out \bra{\psi} \KK^\mu \ket{\psi}_\out = {}_\in\bra{\psi} \, U(\infty, -\infty)^\dagger \KK^\mu U(\infty, -\infty) \, \ket{\psi}_\in,
\end{equation}
where $U(\infty, -\infty)$ is the time evolution operator, which is simply the S-matrix.\\
After rewriting $S = 1 + i T$, and observing that $\KK^\mu \ket{\psi}_\in = 0$ (there are no quanta of radiation in the incoming state), this expression becomes:
\begin{align}
R^\mu \equiv \langle k^\mu \rangle &= {}_\in\bra{\psi} \, S^\dagger \KK^\mu S \, \ket{\psi}_\in= {}_\in\bra{\psi} \, T^\dagger \KK^\mu T \, \ket{\psi}_\in,
\end{align}
We can now use the completeness of our Hilbert space to insert complete set of states $\ket{X\, k\, r_1\, r_2}$, containing at least one radiated messenger of momentum $k$. 
We assume the presence of at least 2 particles, with momenta $r_1$, $r_2$ in the intermediate states, as this is the first non-zero classical contribution to the emitted radiation. Then
\begin{align}
R^\m & = \sum_X \int \! d\Phi(k)d\Phi(r_1)d\Phi(r_2)\; \bra{\psi}\, T^\dagger \,\ket{X\, k\, r_1\, r_2}k_X^\mu \bra{X\, k\, r_1\, r_2}\, T \,\ket{\psi} =\nn \\
& = \sum_X \int \! d\Phi(k)d\Phi(r_1)d\Phi(r_2)\; k_X^\mu \left|\bra{X\, k\, r_1\, r_2}\, T \,\ket{\psi}\right|^2
\end{align}
We now expand the initial state. Diagrammatically, we find:
\begin{equation}
\usetikzlibrary{decorations.markings}
\usetikzlibrary{positioning}
\begin{aligned}
R^\mu & = \sum_X \int d\Phi(k) \prod_{i = 1, 2} d\Phi(r_i) d\Phi(p_i) d\Phi(p'_i)\; k_X^\mu  \, e^{i b \cdot (p_1 - p'_1)/\hbar} \hat\d^{(4)}(p_1 +p_2 - r_1 - r_2 - k - r_X) \\
& \hspace{2cm}\times \hat\d^{(4)}(p'_1 + p'_2 - r_1 - r_2 - k - r_X) \nn
\begin{tikzpicture}[scale=1.0, baseline={([yshift=-\the\dimexpr\fontdimen22\textfont2\relax] current bounding box.center)},] 
\begin{feynman}
\begin{scope}
	\vertex (ip1) ;
	\vertex [right=2 of ip1] (ip2);
    \node [] (X) at ($ (ip1)!.5!(ip2) $) {};
	\begin{scope}[even odd rule]
    \begin{pgfinterruptboundingbox} % useful to avoid the rectangle in the bounding box
	\path[invclip] ($  (X) - (4pt, 30pt) $) rectangle ($ (X) + (4pt,30pt) $) ;
	\end{pgfinterruptboundingbox} 

	\vertex [above left=0.66 and 0.33 of ip1] (q1) {$ \phi_1(p_1)$};
	\vertex [above right=0.66 and 0.33 of ip2] (qp1) {$ \phi^*_1(p'_1)$};
	\vertex [below left=0.66 and 0.33 of ip1] (q2) {$ \phi_2(p_2)$};
	\vertex [below right=0.66 and 0.33 of ip2] (qp2) {$ \phi^*_2(p'_2)$};
	
	\diagram* {
		(ip1) -- [photon, out=30, in=150, photonRed]  (ip2);
		(ip1) -- [photon, out=330, in=210]  (ip2)};

	\begin{scope}[decoration={markings, mark=at position 0.4 with {\arrow{Stealth}}}] 
		\draw[postaction={decorate}] (q1) -- (ip1);
		\draw[postaction={decorate}] (q2) -- (ip1);
	\end{scope}
	\begin{scope}[decoration={markings, mark=at position 0.7 with {\arrow{Stealth}}}] 
		\draw[postaction={decorate}] (ip2) -- (qp1);
		\draw[postaction={decorate}] (ip2) -- (qp2);
	\end{scope}
	\begin{scope}[decoration={markings, mark=at position 0.38 with {\arrow{Stealth}}, mark=at position 0.74 with {\arrow{Stealth}}}] 
		\draw[postaction={decorate}] (ip1) to [out=90, in=90,looseness=1.7] node[above left] {{$ r_1$}} (ip2);
		\draw[postaction={decorate}] (ip1) to [out=270, in=270,looseness=1.7]node[below left] {${r_2}$} (ip2);
	\end{scope}

	\node [] (Y) at ($(X) + (0,1.5)$) {};
	\node [] (Z) at ($(X) - (0,1.5)$) {};
	\node [] (k) at ($ (X) - (0.35,-0.55) $) {$k$};
	\node [] (x) at ($ (X) - (0.35,0.55) $) {$r_X$};

	\filldraw [color=white] ($ (ip1)$) circle [radius=8pt];
	\filldraw  [fill=allOrderBlue] ($ (ip1) $) circle [radius=8pt];
		
	\filldraw [color=white] ($ (ip2) $) circle [radius=8pt];
	\filldraw  [fill=allOrderBlue] ($ (ip2) $) circle [radius=8pt];
	
\end{scope} 
\end{scope}
	  \draw [dashed] (Y) to (Z);
\end{feynman}
\end{tikzpicture},
\end{aligned}
\end{equation}
We may now introduce the momentum transfers, $q_i=p'_i-p_i$, and trade the integrals over $p'_i$ for integrals over the $q_i$. One of the four-fold $\d$ functions will become $\hat\d^{(4)}(q_1+q_2)$, and we can
use it to perform the $q_2$ integrations.  
Finally we relabel $q_1\rightarrow q$ and define the scattering
momentum transfers $w_i = r_i-p_i$. The final result is
\begin{align}\label{eq:radiazione_emessa}
R^\m &= \sum_X \int d\Phi(k) \prod_{i=1,2} d\Phi(p_i)d^4 w_i d^4 q \; \hat\d(2p_i\cdot w_i +w_i^2) \Theta(p_i^0+w_i^0) \,\times \nn \\
& \quad \times \, \hat\d(2p_1\cdot q +q^2)\hat\d(2p_2\cdot q +q^2) \Theta(p_1^0+w_i^0)\Theta(p_2^0+w_i^0) \,\times  \\
& \quad  \times \, \phi_1(p_1) \phi_2(p_2) \phi_1^*(p_1+q) \phi_2^*(p_2-q) \, k_X^\mu \, e^{-i b \cdot q/\hbar}  \hat\d^{(4)}(w_1+w_2+ k+ r_X)\, \times \nn \\
& \quad \times \AC(p_1\,,p_2 \rightarrow p_1+w_1\,,p_2+w_2\,,k\,,r_X)
\AC(p_1+q\,,p_2-q \rightarrow p_1+w_1\,,p_2+w_2\,,k\,,r_X) \nn
\end{align}

\subsection{Classical Point Particles}\label{sec:Goldilock_1}
\def\pcl{\breve{p}}
\def\ucl{{u}}
\def\spread{\sigma^2}
We now discuss the issue of suitable wavefunctions.
We expect the point-particle description to be valid when the separation of the two scattering particles is very large compared to their (reduced) Compton wavelengths $\ell_c^{(i)}\equiv \hbar/m_i$. Furthermore, the wavefunctions have another intrinsic scale, given by the spread of the wavepackets, $\ell_w$.
Let us formalize this intuition.

Heuristically, the wavefunctions for the scattered particles must satisfy
two separate conditions: we will take these to be wavepackets, characterized by a spread in momenta not too large, so that the interaction with the other particle cannot peer into the details of the wavepacket, at the same time, the details of the wavepacket should not be sensitive to quantum effects. \\
The dimensionless parameter controlling the approach to the classical
limit in momentum space is the square of the ratio of the Compton wavelength $\ell_c$
to the intrinsic spread~$\ell_w$,
\begin{equation}
\xi \equiv \left(\frac{\ell_c}{\ell_w}\right)^2\,.
\end{equation}
The classical result is achieved in the limit $\xi\rightarrow 0$. Note that in this limit the wavefunctions are sharply peaked around the classical
value for the momenta $\pcl_i = m_i \ucl_i$, with the classical four-velocities
$\ucl_i$ normalized to $\ucl_i^2 = 1$. \\
Because of the phase-space integrals over the initial-state momenta, which enforce the on-shell conditions $p_i^2=m_i^2$, the only Lorentz invariant non-constant parameters are the dimensionless combination $p\cdot \ucl/m$ and $\xi$.
The simplest combination we can build out of these parameters will be the combination $p\cdot\ucl/(m\xi)$. This function is very useful, as large deviations from $p=m \,\ucl$ will be suppressed in a classical quantity.

As one nears the classical limit, the wavefunction and its conjugate should both represent the particle, hence $\phi^*(p+q) \sim \phi^*(p)$.
Note that we will integrate the momentum mismatch $q$ over all possible values, so it is somewhat a matter of taste how we normalize it. Nonetheless, if we take $q_0$ to be a ‘characteristic’ value of $q$, this requirement is equivalent to 
\begin{equation}
\frac{q_0\cdot\ucl_i}{m\xi} \ll 1\,.
\label{qConstraint}
\end{equation}
Replacing the momentum by a wavenumber, this constraint takes the form:
\begin{equation}
\label{eq:qbConstraint}
\bar{q}_0\cdot\ucl_i\,\ell_w \ll \sqrt{\xi}.
\end{equation}
We next examine the delta functions in $q$ arising from the on-shell constraints on the conjugate momenta $p_i$: 
\begin{equation}
\hat{\d}(2p_i\cdot q+q^2) = \frac1{\hbar m_i}\hat{\d}(2\bar{q}\cdot u_i+\ell_c \bar{q}^2)\,.
\label{universalDeltaFunction}
\end{equation}
In addition to its dependence on $\xi$, this function depends on two additional dimensionless ratios,
\begin{equation}
\ell_c \sqrt{-\bar{q}^2} \qquad \textrm{and}\qquad
\frac{\bar{q}\cdot u_i}{\sqrt{-\bar{q}^2}}\,.
\end{equation}
Let us call $1/\sqrt{-\bar{q}^2}$ a `scattering length' $\ell_s$. We expect this quantity to be somewhat similar to the impact parameter, \ie $\ell_s \sim \sqrt{-b^2}$. \\
As suggested by the nonrelativistic limit, the smallest reasonable power of $\xi$ we can imagine emerging as a constraint from the later $\bar{q}$ integration is one-half, therefore
\begin{equation}
\frac{\ell_c}{\ell_s} \lesssim \sqrt{\xi}\,,\qquad {\bar{q}\cdot u_i}\,\ell_s \lesssim \sqrt{\xi}\,.
\label{eq:deltaConstraints}
\end{equation}
If we take a higher power of $\xi$, the constraints would grow stronger.

Combining the second constraint of \eq{eq:deltaConstraints} with  \eq{eq:qbConstraint}, we obtain the constraint $\ell_w \ll \ell_s$.
The first constraint is weaker, $\ell_w \lesssim \ell_s$, but on physical grounds we should expect the stronger one. \\
Combining the stronger constraint with $\xi \ll 1$, we obtain the ‘Goldilocks’ inequalities,
\begin{equation}\label{eq:Goldilock}
\ell_c \ll \ell_w \ll \ell_s \sim \sqrt{-b^2}
\end{equation}

In computing the classical observable, we cannot simply set $\xi = 0$. Indeed, we don’t even want to fully take the $\xi \rightarrow 0$ limit. Rather, we want to extact the leading term in that limit.

What about loop integrations?  Unitarity considerations suggest that we should
choose the loop momentum to be that of a massless line in the loop, if there is
one, and replace them by wavenumbers.

To sum up, the quantum-mechanical expectation values of observables are \\well-approximated by the corresponding classical ones, when the packet spreads are in the `Goldilocks' zone, $\ell_c \ll \ell_w\ll \sqrt{-b^2}$.

Let us finally introduce a convenient notation to allow us to manipulate integrands under the eventual approach to the $\hbar\rightarrow0$ limit. 
We will use large angle brackets for the purpose,
\begin{equation}
\Lexp f(p_1,p_2, \dots) \Rexp \equiv 
\int d\Phi(p_1)d\Phi(p_2)\;|\phi_1(p_1)|^2\,|\phi_2(p_2)|^2\,
  f(p_1,p_2,\dots) \,,
\label{eq:angleBrackets}
\end{equation}
where the integration over both $p_1$ and $p_2$ is implicit. \\
Within the angle brackets, we have approximated $\phi(p+q)\simeq \phi(p)$, and 
when evaluating the integrals, we will also set $p_i\simeq m_i \ucl_i$, along with the other simplifications discussed above.

\subsubsection{Classical Radiation}
\def\cl{\text{cl}}
\def\RadKer{\mathcal{R}}
To fully understand the methodology described above, let us compute the classical limit of the radiated momentum in \eq{eq:radiazione_emessa}
\begin{align}\label{eq:ExpectedMomentum2recast}
R^\m_\cl & = \sum_X \Lexp \int d\Phi(k) \prod_{i=1,2} d^4 w_i d^4 q \; \hat\d(2p_i\cdot w_i +w_i^2) \Theta(p_i^0+w_i^0) \,\times  \\
& \quad \times \, \hat\d(2p_1\cdot q +q^2)\hat\d(2p_2\cdot q +q^2) \Theta(p_1^0+w_i^0)\Theta(p_2^0+w_i^0) k_X^\mu \, e^{-i b \cdot q/\hbar}  \hat\d^{(4)}(w_1+w_2+ k+ r_X)\, \times \nn \\
& \quad \times \AC(p_1\,,p_2 \rightarrow p_1+w_1\,,p_2+w_2\,,k\,,r_X)
\AC(p_1+q\,,p_2-q \rightarrow p_1+w_1\,,p_2+w_2\,,k\,,r_X)\Rexp \nn
\end{align}
We will focus on the leading contribution, with $X=\emptyset$. \\
We now rescale $q \rightarrow \hbar\bar{q}$, and drop the $q^2$ inside the on-shell delta functions. Next we extract the overall factor of $g^6$, and the respective $\hbar$s, from the two amplitudes.
We define $\bar\AC^{(L)}$ the reduced L-loop amplitude, \ie the L-loop amplitude with a factor of $g/\sqrt{\hbar}$ removed for every interaction.\\
In addition, we rescale the momentum transfers $w\rightarrow \hbar\bar{w}$ and the radiation momenta, $k\rightarrow\hbar \bar{k}$. 
Finally, at leading order, we may drop the $w_i^2$ inside the on-shell delta functions. 
\begin{align}\label{eq:ExpectedMomentum2classicalLO}
R^{\m, (0)}_\cl & = g^6 \Lexp \hbar^4 \int d\Phi(\bar{k}) \prod_{i=1,2} d^4 \bar{w}_i d^4 \bar{q} \; \hat\d(2p_i\cdot \bar{w}_i) \hat\d(2p_1\cdot \bar{q})\hat\d(2p_2\cdot \bar{q}) 
\bar{k}^\mu \, e^{-i b \cdot \bar{q}}  \hat\d^{(4)}(\bar{w}_1+\bar{w}_2+ \bar{k})\, \times \nn \\
&  \hspace{-0.5 cm} \times \bar{\AC}^{(0)}(p_1\,,p_2 \rightarrow p_1+\hbar\bar{w}_1\,,p_2+\hbar\bar{w}_2\,,\hbar\bar{k})
\bar{\AC}^{(0)}(p_1+\hbar\bar{q}\,,p_2-\hbar\bar{q} \rightarrow p_1+\hbar\bar{w}_1\,,p_2+\hbar\bar{w}_2\,,\hbar\bar{k})\Rexp 
\end{align}
It is often convenient to rewrite this result as the integral over a perfect square, see \Refe{waveforms_1}:
\begin{align}\label{eq:radiatedMomentumClassical}
R^\m_\cl = \sum_X \hbar^{-3}\Lexp \int \! d\Phi(k) \, k_X^\mu
\left |\RadKer(k, r_X) \right|^2 \Rexp ,
\end{align}
where we have defined the radiation kernel
\begin{align}\label{eq:defOfR}
\RadKer(k, r_X) = \hbar^{3/2}\prod_{i = 1, 2} \int \! & d^4 w_i \; 
     \hat\d(2 p_i \cdot w_i + w_i^2) \Theta(p_i^0+w_i^0)
     \hat\d^{(4)}( w_1+  w_2- k- r_X)\times \nn \\
& \times  e^{i b \cdot  w_1/\hbar}  \,   
  \AC( p_1 +  w_1,  p_2 +  w_2 \rightarrow p_1\,,  p_2\,, k\,, r_X)   
\end{align}
At leading-order, we simply get
\begin{align}\label{eq:radiatedMomentumClassicalLO}
R^{\mu, (0)}_\cl &= g^6 \Lexp \int \! d\Phi(\bar k) \, \bar k^\mu
\left | \RadKer^{(0)}(\bar k) \right|^2 \Rexp   
\end{align}

%--------------
\subsection{Classical Limit for Massless Particles}
\label{MasslessClassicalLimitSection}
We now want to include initial-state massless classical waves in the aforementioned formalism.
However a particle's Compton wavelength diverges when its mass goes to zero, making it impossible to satisfy the required conditions \eq{eq:Goldilock}. 
Nonetheless, it does not make sense to treat messengers (photons or gravitons) as point-like particles.
Indeed, Newton and Wigner \Refe{N_W} and Wightman \Refe{Wightman} proved rigorously long ago that a strict localization of known massless particles in position space is impossible. 
A proper treatment instead relies on coherent states.  

\paragraph{Coherent States of the Electromagnetic Field}
\def\opA{\mathbb{A}}
\def\opC{\mathbb{C}}
\def\opN{\mathbb{N}}
\def\opK{\mathbb{K}}
\def\opF{\mathbb{F}}
We can write the electromagnetic field operator as,
\begin{equation}
\opA_\mu(x) = \frac 1{\sqrt{\hbar}}\sum_\eta \int d\Phi(k) \, 
\bigl[ a_{(\eta)}(k) \ve^{(\eta)*}_\mu(k)\, e^{-i k\cdot x/\hbar}+
a^\dagger_{(\eta)}(k) \ve^{(\eta)}_\mu(k)\, e^{+i k\cdot x/\hbar}\bigr] \,,
\label{eq:aField}
\end{equation}
where $\eta = \pm$ labels the helicity, and the polarization vectors satisfy,
\begin{equation}
\bigl[\ve^{(\eta)}_\mu(k) \bigr]^* = 
\ve^{(-\eta)}_\mu(k)\,.
\end{equation}
The commutation relations are
\begin{equation}
\bigl[ a_{\eta}(k), a^\dagger_{\eta'}(k') \bigr] = 
  \d_{\eta, \eta'} \hat\d_\Phi(k-k') \,.
\end{equation}
Using the form of the electromagnetic field in \eq{eq:aField}, the electromagnetic field strength operator is,
\begin{equation}\label{eq:fieldstrengthdn}
\opF_{\m\n}(x)=-\frac{2i}{\hbar^{3/2}}\sum_\eta \int d\Phi(k) \, 
\bigl[ a_{(\eta)}(k) k_{[\m}\ve^{(\eta)*}_{\n]}(k)\, e^{-i k\cdot x/\hbar}-
a^\dagger_{(\eta)}(k) k_{[\m}\ve^{(\eta)}_{\n]}(k)\, e^{+i k\cdot x/\hbar}\bigr] \,,
\end{equation}
where as usual the subscripted square brackets denote antisymmetrization.

We now introduce the coherent-state operator,
\begin{equation}
\opC_{\alpha,(\eta)} \equiv 
\NC_{\alpha} \exp \biggl[ \int d\Phi(k) \, \a(k) 
a^\dagger_{(\eta)}(k) \biggr]\,,
\end{equation}
with normalization $\NC_{\alpha}$. 
We can build coherent states of the electromagnetic field using this operator, \eg
\begin{equation}\label{CoherentStateDefinition}
\ket{\alpha^{+}} =  \opC_{\alpha,(+)} \ket{0} \,.
\end{equation}
More generally, we could consider coherent states containing both helicities; nonetheless, coherent-state operators for different helicities commute and every polarization vector can be decomposed in the helicity basis, so there is no loss of generality in making a specific helicity choice for the coherent states we consider. \\
Finally, notice that coherent state operators are unitary,
\begin{equation}
(\opC_{\alpha,(\eta)})^{\dagger} = 
(\opC_{\alpha,(\eta)})^{-1} \,.
\end{equation}
The normalization factor $\NC_{\alpha}$ is fixed by the condition
$\langle \alpha^{+}|\alpha^{+}\rangle = 1$. Using Baker--\\Campbell--Hausdorff formula we find
\begin{equation}
\NC_{\alpha} = \exp\left[ -\frac{1}{2} \int d\Phi(k) \, |\alpha(k)|^2 \right] \,.
\end{equation}
The coherent-state creation operator acting on the vacuum can be rewritten as a displacement operator \Refe{Ilderton}, yielding 
\begin{align}\label{CoherentStateDefinitionII}
&\hspace{2cm}\opC_{\alpha,(\eta)} \ket{0} =\exp\biggl[\int d\Phi(k) \left(\alpha(k)a^\dagger_{\eta}(k) - \alpha^*(k)a_{\eta}(k)\right) \biggr] \ket{0} \,, \\
&\opC_{\alpha,(\eta)}^{\dagger} a_{\rho}(k)\opC_{\alpha,(\eta)}=a_{\rho}(k)+\d_{\eta \rho}\, \alpha(k)\,, \qquad
\opC_{\alpha,(\eta)}^{\dagger}a^\dagger_{\rho}(k)\opC_{\alpha,(\eta)}= a^\dagger_{\rho}(k) +\d_{\eta \r} \, \a^*(k)\,. \nn
\end{align}

To interpret the state, let us compute $\bra{\alpha^+}\opA^\mu(x) \ket{\alpha^+}$.\\ 
Notice that \eg 
$a_{(+)}(k)\ket{\alpha^+}= \alpha(k) \ket{\alpha^+}$,
which incidentally imply that the dimension of $\alpha(k)$ is the same as the dimension of the annihilation operator, $[M]^{-1}$.
\begin{align}\label{eq:Aclassico_1}
A_{\cl, \m}(x) \equiv \bra{\alpha^+} \opA_\mu(x) \ket{\alpha^+} &=
\frac{1}{\sqrt{\hbar}} \int d\Phi(k) \, 
\bigl[ \alpha(k) \ve^{(+)*}_\mu(k) e^{-i k\cdot x/\hbar} 
+ \alpha^*(k) \ve^{(+)}_\mu(k) e^{+ik \cdot x/\hbar} \bigr] = \nn \\
&\hspace{-1.5cm}= \int d\Phi(\bar{k}) \, 
\bigl[ \bar\a(\bar{k}) \ve^{(+)*}_\mu(\bar{k}) e^{-i \bar{k}\cdot x} 
 + \bar\a^*(\bar{k}) \ve^{(+)}_\mu(\bar{k}) e^{+i \bar{k} \cdot x} \bigr],
\end{align}
where we have defined $\bar\a(\bar{k} ) \equiv \hbar^{\frac32}\alpha(k)$.

Finally, the most general solution of the classical Maxwell equation in vacuum, in Fourier space, is:
\begin{equation}\label{eq:Aclassico_2}
\sum_\eta A_{\cl, \m}^{(\eta)}(x) = \sum_\eta \int \! d\Phi(\bar{k})  
\bigl[ \widetilde{A}_\eta(\bar{k}) \ve^{(\eta)*}_\mu(\bar{k}) e^{-i \bar{k} \cdot x} +
\widetilde{A}^*_\eta(\bar{k}) \ve^{(\eta)}_\mu(\bar{k}) e^{+i \bar{k} \cdot x} \bigr].
\end{equation}
Comparing \eq{eq:Aclassico_1} with \eq{eq:Aclassico_2}, we can naturally identify $\bar\a(\bar{k})$ with the Fourier coefficients $\widetilde{A}_\eta(\bar{k})$.

\paragraph{Classical Coherent States}
The coherence of a state does not suffice for it to behave classically: we must also require factorization of expectation values, \ie
\begin{equation}\label{eq:classicalFactorization}
\bra{\alpha^+} \opA^\mu(x) \opA^\nu(y) \ket{\alpha^+} \simeq \bra{\alpha^+} \opA^\mu(x) \ket{\alpha^+}  \bra{\alpha^+} \opA^\nu(y)\ket{\alpha^+} \,.
\end{equation}
To this end it is useful to define an operator that measures the number of photons:
\begin{equation}
\opN_\gamma = \sum_\eta \int d\Phi(k) \, a^\dagger_{\eta}(k) \a_{\eta}(k) \,.    
\end{equation}
The expectation number $N_\g$  of photons in our coherent state is,
\begin{equation}\label{eq:nphotons}
N_\g = \bra{\alpha^+}\opN_\gamma\ket{\alpha^+}= \int d\Phi(k) |\alpha(k)|^2
= \frac1{\hbar}\int d\Phi(\bar{k}) |\bar\a(\bar{k})|^2\,.
\end{equation}
The classical limit $\hbar\rightarrow 0$ corresponds to the limit of a large number of photons \footnote{we fix $\a$ in such a way that the integral in
the last line of \eq{eq:nphotons} is not parametrically small as $\hbar \rightarrow 0$. A simple way to do so is to chose
$\bar\alpha$ independent of $\hbar$}. The desired factorization property \eq{eq:classicalFactorization} will thus hold when, $N_\g \gg 1$.

\paragraph{Localized Beams of Light}
When considering the scattering of light from a point-like object, to be able to use the standard QFT technology, the incoming wave must be spatially separated from the incoming particle in the far past. 
Consequently, we need to understand how to describe a localized incoming beam of light.

To localize the wave, we ``broaden'' the delta function:
\begin{equation}
\d_\s(\bar{k}) \equiv \frac{1}{\s \sqrt\pi}\exp \left[ - \frac{\bar{k}^2}{\sigma^2}\right].
\end{equation}
We have two measures of beam spread, $\sigma_\parallel$ and $\sigma_\perp$, along and transverse to the wave direction respectively.
Let us now consider $\sigma_\parallel$ to be very small compared to the other two scales, $\sigma_\perp$ and $\omega=\bar{k}_\odot^t$, this way the beam can be considered almost monochromatic, \ie $\d_{\s\parallel}\sim \d$. 
Nonetheless, the broadened distribution $\delta_{\sigma_\perp}$ does allow components of momentum in the perpendicular beam directions. These components should be subdominant, therefore we require ${\lambda^{-1}} \gg  \sigma_\perp$. \\
We may define, respectively, a transverse size of the beam and a `pulse length':
\begin{equation}
\ell_\perp = {\s_\perp^{-1}} \,, \qquad \ell_\parallel = {\s_\parallel^{-1}}.
\end{equation}

The requirement $\lambda^{-1} \gg \s_\perp$ becomes: $\lambda \ll \ell_\perp$; this condition is in some respects analogous to the first part of the `Goldilocks' condition \eq{eq:Goldilock}, notice however it is just a sufficient condition, not necessary to have a classical beam of light.

%--------------
\subsection{Point-like Observables}\label{ObserversSection}
\def\bhn{\hat{\mathbf{n}}}
\def\Gret{G_{\mathrm ret}}
\def\Gadv{G_{\mathrm adv}}
\def\Jtilde{{\widetilde J}}
In the previous sections, we analyzed what we may call \textit{global\/} observables. The measurement of such quantities require an array of detectors covering the celestial sphere at infinity.
In the gravitational context, where we would be looking to detect emission from scattering of distant black holes, such a measurement would be hopelessly impractical. \\
We now turn our attention to what we may call
\textit{local\/} observables, which can be measured with a localized detector, albeit still sitting somewhere on the celestial sphere, say at $x$.
The archetype for such a measurement is that of the waveform of radiation emitted during a scattering event $W(t,\bhn;x)$, where $\bhn$ is the versor pointing from the event at the coordinate origin.\\
For simplicity, we will present the formalism in the context of electromagnetic radiation, nevertheless it will be the same also for the gravitational case.
\paragraph{General Structure of Local Observables}
It is convenient to work with the Fourier transform of the waveform in the frequency domain. We will refer to this as the spectral waveform $f(\omega,\bhn; x)$:
\begin{equation}\label{SpectralFunction}
f(\omega,\bhn;x) = \int_{-\infty}^{+\infty} dt\; W(t,\bhn;x)\, e^{i\omega t}\,.
\end{equation}
We are interested in radiation produced by long-range forces, thus the idealized waveforms for the scattering processes we will consider stretch infinitely far back and forward in time.\\
Let us fix the point of closest approach during the scattering event at the coordinate origin, $(t,{\bf x})=(0,{\bf 0})$. 
We can treat the scattering as occurring in a box of temporal length $\Delta t_s$, and of spatial size $\Delta x_s$.  
Radiation is emitted inside the box during the scattering event, and then spreads out. We will measure the radiation in some direction $\bhn$, very far away (both spatially and temporally).\\
The details of the scattering, (\eg the particles' interactions, \dots), will determine the radiation emitted inside the box, however, those details will have no effect on the propagation of the radiation out to the distant measuring apparatus. 
We thus expect the form of the result to be a (retarded) Green's
function convoluted with a source, which we can expand in the large-distance limit.

The details of the scattering inside the box give rise to a field-strength current, in wavenumber-space, $\Jtilde_{\vec\m}(\bar{k})$,
where $\vec\m$ is a generic placeholder for the indices of any specific current (\eg for gravity, we have $\Jtilde_{\vec\m}=\Jtilde_{\m\n\r\s}$). \\
We can go from this wavenumber-space current to the position-space current and viceversa by taking a Fourier transform
\begin{equation}
J_{\vec\mu}(x) = \int d^4 \bar{k}\; \Jtilde_{\vec\mu}(\bar{k})\,e^{-i\bar{k}\cdot x}\,, \quad 
\Jtilde_{\vec\mu}(\bar{k}) = \int d^4 x\; J_{\vec\mu}(x)\,e^{i \bar{k}\cdot x}\,.
\end{equation}

As I will show in detail in the next paragraph, the radiation observable takes the the general form,
\begin{equation}\label{eq:OriginalRadiation}
R_{\vec\mu} = i\int d\Phi(\bar{k})\;\bigl[\Jtilde_{\vec\mu}(\bar{k})\, e^{-i\bar{k}\cdot x}-\Jtilde_{\vec\mu}^\dagger(\bar{k})\, e^{+i\bar{k}\cdot x}\bigr]\,,
\end{equation}
that is, as an integral of the source $\Jtilde_{\vec\mu}(\bar{k})$
over the on-shell massless phase space for the radiated messenger. 
Furthermore, the reality condition of our currents in position space leads to the relation,
\begin{equation}\label{eq:realityCondition}
\Jtilde_{\vec\mu}(-\bar{k}) = \Jtilde_{\vec\mu}^\dagger(\bar{k}) \,.    
\end{equation}
We may now express the observable of \eq{eq:OriginalRadiation} in terms of the spatial current $J_{\vec\mu}(x)$, yielding
\begin{equation}\label{eq:RadiationII}
R_{\vec\mu} = i\int d\Phi(\bar{k})\, d^4 y\; J_{\vec\mu}(y)\bigl[e^{-i\bar{k}\cdot (x-y)}- e^{+i\bar{k}\cdot (x-y)}\bigr]\,.
\end{equation}
We can interpret this expression as the difference of retarded and advanced Green's functions, 
\begin{align}\label{eq:RadiationIII}
R_{\vec\mu}(x) & = \int d^4 y J_{\vec\m}(y) \left(\frac{i}{(2\pi)^3} \int d^4\bar{k} \,\Theta(\bar{k}^0)\d(\bar{k}^2)\left[e^{-i\bar{k}\cdot (x-y)}- e^{+i\bar{k}\cdot (x-y)}\right] \right) \,= \nn \\
&=\int d^4 y\; J_{\vec\mu}(y)\bigl[G_{\mathrm ret}(x-y)- G_{\mathrm adv}(x-y)\bigr]\,.
\end{align}
It is important to point out that in the far future, where the observer measures the wavetrain emitted from the scattering event, $\Gadv$ will vanish.  \\
Finally, we explicit $\Gret$, and switch back to the wavenumber-space current in order to make the complete dependence of the integrand on $x$ and $y$ manifest. The result is,
\begin{align}\label{eq:RadiationIV}
R_{\vec\mu}(x) & = \int d\omega d^3 \bar{\bf{k}} \, d^4y\;  \Jtilde_{\vec\mu}(\bar{k})\,e^{-i\bar{k}\cdot y}\,
\frac{\d(x^0-y^0-|\bf{x}-\bf{y}|)}{4\pi |\bf{x}-\bf{y}|} \, = \nn \\
 & = \int d\omega d^3 \bf{\bar{k}}\, d^3\bf{y}\;  \Jtilde_{\vec\mu}(\bar{k})\,
\frac{e^{-i\omega x^0}\,e^{+i\omega|\bf{x}-\bf{y}|}\,e^{+i\bf{\bar k}\cdot\bf{y}}}{4\pi |\bf{x}-\bf{y}|} \,
\end{align}
From the earlier discussion, we know that $J_{\vec\mu}(y)$ is concentrated around $y\simeq 0$, whereas $x$ is far away ($x\gg y$).  Accordingly we can expand the integrand there, using,
\begin{equation}\label{eq:Expansion}
|\bf{x}-\bf{y}| \sim |\bf{x}| \Bigl(1-\frac{\hat{\bf n}\cdot\bf{y}}{|\bf{x}| } \Bigr)\,.
\end{equation}
Substituting the expansion \eq{eq:Expansion} into \eq{eq:RadiationIV} and performing the $\bf{y}$ and $\bf{k}$ integrals, we obtain,
\begin{align}
R_{\vec\mu}(x) &= \int d\omega d^3 \bf{\bar{k}}\, d^3\bf{y}\;  \Jtilde_{\vec\mu}(\bar{k})\,
\frac{e^{-i\omega x^0}\,e^{+i\omega|\bf{x}|}e^{-i\omega \hat{\bf n}\cdot\bf{y}}\,e^{+i\bf{\bar{k}}\cdot\bf{y}}}
{4\pi |\bf{x}|}\, = \nn \\
&= \frac1{4\pi |\bf{x}|}\int d\omega \;  \Jtilde_{\vec\mu}(\omega, \omega\hat{\bf n})\,
e^{-i\omega (x^0-|\bf{x}|)} \,.
\end{align}
We can identify the waveform with the coefficient of the leading-power term $|\bf{x}|^{-1}$, 
\begin{equation}
\begin{aligned}
W_{\vec{\mu}} (t,\hat{\bf n};x) &= \frac1{4\pi}\int d\omega \;  \Jtilde_{\vec\mu}(\omega, \omega\hat{\bf n})\,
e^{-i\omega (x^0-|\bf{x}|)}
\end{aligned}\,.
\label{WaveformI}
\end{equation} 
Choosing $t = x^0-|\bf{x}|$ for the observer's clock time, the corresponding spectral waveform (for positive frequencies) is:
\begin{equation}\label{eq:spectral_waveform}
f_{\vec{\mu}}(\omega, \hat{\bf n}) = \frac1{4\pi} \Jtilde_{\vec\mu}(\omega, \omega\hat{\bf n})    
\end{equation}

\paragraph{Deriving the Spectral Waveforms}
As discussed in the above paragraph, once we know the current $\Jtilde_{\vec{\m}}(\bar k)$, we can immediately write down the spectral waveform.
To obtain this current, we need to choose a specific local radiation observable using its definition \eq{eq:OriginalRadiation}.

Let us begin by considering the field-strength tensor in electrodynamics, \eq{eq:fieldstrengthdn}. The observable is 
\begin{equation}\label{eq:radiation_observable}
\la F_{\m\n}^\out(x) \ra \equiv {}_\out\bra{\psi}\opF_{\m\n}(x)\ket{\psi}_\out =
{}_\in \bra{\psi} S^\dagger \opF_{\m\n}(x) S \ket{\psi}_\in
\end{equation}
Using \eq{eq:fieldstrengthdn}, and converting to integrals over wavenumbers, we find,
\begin{align}
\la F_{\m\n}^\out(x) \ra = 
-2i \hbar^{3/2} \sum_\eta \int d\Phi(\bar{k}) \, 
\bigl[ &\bra{\psi} S^\dagger a_{(\eta)}(k) S \ket{\psi} \,\bar{k}_{[\mu} \ve^{(\eta)*}_{\nu]}(\bar{k})\, e^{-i \bar{k}\cdot x} \\
& - \bra{\psi} S^\dagger a^\dagger_{(\eta)}(k) S \ket{\psi} \,\bar{k}_{[\mu} \ve^{(\eta)}_{\nu]}(\bar{k})\, e^{+i\bar{k} \cdot x} \bigr] \,,
\end{align}
We can now read off the current $\Jtilde_{\vec\m}(\bar{k})$ as
\begin{equation}
\Jtilde_{\mu\nu}(\bar{k}) = -2 \hbar^{3/2} \sum_\eta \bra{\psi} S^\dagger a_{(\eta)}(k) S \ket{\psi} \,\bar{k}_{[\mu} \ve^{(\eta)*}_{\nu]}(\bar{k}) \,.    
\end{equation}
From \eq{eq:spectral_waveform} the corresponding spectral waveform (for positive frequency) is,
\begin{equation}\label{eq:allOrdersSpectralEM}
f_{\mu\nu}(\omega,\hat{\bf n}) = -\frac1{2\pi} \hbar^{3/2} \sum_\eta \bra{\psi} S^\dagger a_{(\eta)}(k) S \ket{\psi} \,\bar{k}_{[\mu} \ve^{(\eta)*}_{\nu]}(\bar{k}) \Big|_{\bar{k} = (\omega, \omega \hat{\bf n})}.
\end{equation}
Notice that this is a non-perturbative result.

It is straightforward to extend this result to gravity. We work in Einstein gravity, and assume that the spacetime is asymptotically Minkowskian.
In this case our observer is the local spacetime curvature $\la R^\out_{\m\n\r\s}(x) \ra$. The corresponding spectral waveform (for positive frequency) is nothing 
but the double copy of \eq{eq:allOrdersSpectralEM},
\begin{equation}\label{eq:fRiemann}
f_{\m\n\r\s}(\omega,\hat{\bf n}) = \frac{i \kappa}{2\pi} \hbar^{3/2} \sum_\eta \bra{\psi} S^\dagger a_{(\eta)}(k) S \ket{\psi}\,
\bar{k}_{[\m} \ve^{(\eta)*}_{\n]}(\bar{k}) 
\;\bar{k}_{[\r} \ve^{(\eta)*}_{\s]}(\bar{k}) 
\Big|_{\bar{k} = (\omega, \omega \hat{\bf n})} \,.    
\end{equation}

\paragraph{Computing the Observables}
Assuming temporarily that the observable is measured at finite distance and there is no gravitational radiation in the infinite past, the waveform is given by \eq{eq:OriginalRadiation},
\begin{align}\label{eq:gravitational_Radiation}
\la R^\out_{\m\n\r\s}(x) \ra  & = i\int d\Phi(\bar{k})\;\bigl[\Jtilde_{\m\n\r\s}(\bar{k})\, e^{-i\bar{k}\cdot x}-\Jtilde_{\m\n\r\s}^*(\bar{k})\, e^{+i\bar{k}\cdot x}\bigr]\,= \nn \\
&= \frac1{4\pi |\bf{x}|}\int d\omega \;  \left[ \Jtilde_{\vec\mu}(\omega, \omega\hat{\bf n})\, e^{-i\omega (x^0-|\bf{x}|)} 
\Jtilde^\dagger_{\vec\mu}(\omega, \omega\hat{\bf n})\, e^{+i\omega (x^0-|\bf{x}|)}\right]\,.
\end{align}
The waveform $W_{\m\n\r\s}(t, \hat{\bf n})$ for the curvature tensor and the spectral waveform, $f_{\m\n\r\s}(\omega, \hat{\bf n})$, are given by
\begin{subequations}
\begin{align}
& \la R^\out_{\m\n\r\s}(x) \ra  \Big|_{|\bf{x}| \rightarrow \infty}=\frac1{|\bf{x}|}W(t,\hat{\bf n};x) = \frac1{|\bf{x}|} \int_{-\infty}^{\infty} d\omega \,f_{\m\n\r\s}(w,\hat{\bf n};x)e^{-i \omega t} \\
& \quad f_{\m\n\r\s}(w,\hat{\bf n};x)= \frac1{4\pi}\left[\Theta({\omega})\Jtilde_{\m\n\r\s}(\omega, \omega \hat{\bf n})+ \Theta({-\omega})\Jtilde^\dagger_{\m\n\r\s}(|\omega|, |\omega| \hat{\bf n})  \right]
\end{align}  
\end{subequations}
A convenient presentation of the curvature tensor (and consequently of the gravitational waveform) is in terms of Newman-Penrose scalars \Refe{Newman}. They are constructed as projections of the curvature tensor on a complex basis of null vectors. Following \Refe{waveforms_1} we choose these vectors to be
\begin{equation}\label{eq:Pen_New_basis}
L^\mu = \bar{k}^\mu / \omega = (1, \hat{\bf n})^\mu, \quad N^\mu = \zeta^\mu, 
\quad M^\mu = \ve^{(+)\mu}, \quad M^{*\mu} = \ve^{(-)\m} \,.
\end{equation}
The null vector $\zeta$ is simply a gauge choice, satisfying $\zeta \cdot \ve^{(\pm)} = 0$ and $L \cdot N = L \cdot \zeta = 1$. Furthermore note that $M\cdot M^* = -1$.
The independence of $L$ on the frequency of the outgoing graviton makes \eq{eq:Pen_New_basis} a suitable basis both for the waveform and the spectral waveform. \\
The Newman-Penrose scalars are defined by the independent contractions of the Weyl tensor with the vectors in \eq{eq:Pen_New_basis}.
The leading radiating scalar, typically denoted by $\Psi_4$, describes the transverse radiation propagating along $L$
\begin{equation}
\Psi_4(x) = - N^\mu M^{*\nu} N^\rho M^{*\sigma} \langle R^\out_{\mu\nu\rho\sigma}(x) \rangle \,=    
\frac{1}{|\bf x|} \Psi_4^0  + \cdots \,.
\end{equation}
Using the transversality and null property of $M^\m$ and \eq{eq:gravitational_Radiation}, we can write the spectral representation of 
${\Psi}_4^0$ as 
\begin{align}
\tilde \Psi_4^0(\omega, \hat{\bf n}) = - \frac{\kappa}{4\pi} \hbar^{3/2} \Bigl[ & \Theta(\omega)  (-i)\omega^2 \bra{\psi} S^\dagger a_{(--)}(\omega, \omega \hat{\bf n}) S \ket{\psi} \nn \\
& +\Theta(-\omega) (+i)\omega^2 \bra{\psi} S^\dagger a^\dagger_{(++)}(|\omega|, |\omega| \hat{\bf n}) S \ket{\psi} \Bigr]
\end{align}
For an asymptotically flat spacetime, outgoing radiation at large distances is described by linearized general relativity in the transverse traceless gauge. Using that $k\cdot N = \omega$, the Newman-Penrose scalar $\Psi_4$ takes the form
\begin{equation}
\Psi_4=\frac{\kappa}{8 \pi |\bf{x}|}(\ddot{h}_+^\infty +i\ddot{h}_\times^\infty)\, ,
\end{equation}
where the subscript $\times$ and $+$ denote the graviton polarizations, defined with respect to the vector $L^\m$ pointing along the graviton momentum.\\
The metric perturbation $h_{\m\n}$ is in transverse traceless gauge and is normalized in such a way that, at spatial infinity, it falls off as
\begin{equation}
g_{\m\n}\Big|_{|\bf{x}|\rightarrow \infty}= \eta_{\m\n}+\frac{\k}{8 \pi |\bf{x}|}h^\infty_{\m\n}
\end{equation}
Therefore, we may directly identify the strain at future null infinity in terms of the frequency-space Newman-Penrose scalar as:
\begin{align}
h(x)&=\frac{\k}{8 \pi |\bf{x}|}h_{\m\n}^\infty=\frac{\k}{8 \pi |\bf{x}|}(h_{+\, \m\n}^\infty+h_{\times \, \m\n}^\infty) = \nn \\
&= \frac{\k}{8 \pi |\bf{x}|} \int_{-\infty}^{\infty} d\omega \, e^{-i \omega t}
\Bigl[\Theta(\omega)  (-i) \bra{\psi} S^\dagger a_{(--)}(\omega, \omega \hat{\bf n}) S \ket{\psi} \\
& \hspace{4cm}+\Theta(-\omega) (+i) \bra{\psi} S^\dagger a^\dagger_{(++)}(|\omega|, |\omega| \hat{\bf n}) S \ket{\psi}\Bigr] \nn
\end{align}
Let us emphasize once again that these results are non-perturbative.

\subsection{The Detected Wave at Leading Order}
We will now study the expectation of the radiative field-strength tensor in perturbation theory. For simplicity, we will start by studying the electro-magnetic case.\\
Consider \eq{eq:radiation_observable}, in a perturbative fashion we rewrite the scattering matrix in terms of the transition matrix, \ie $S=1+i T$, yielding:
\begin{align}\label{FexpectationII}
\la F^\out_{\m\n}(x) \ra &= \nn
\bra{\psi}(1-iT^\dagger)\opF_{\m\n}(x)(1+i T)\ket{\psi}\,=\\ &= 
\bra{\psi}|\opF_{\m\n}(x)\ket{\psi}
+ 2 Re \,i\bra{\psi}\opF_{\m\n}(x) T\ket{\psi}
+\bra{\psi}T^\dagger\opF_{\m\n}(x) T\ket{\psi}\,.
\end{align}
At leading order in perturbation theory, only the second term in \eq{FexpectationII} contributes.
If we now explicit the field strength operator and the initial state using \eq{eq:fieldstrengthdn} and \eq{eq:initial_state}, we obtain (at leading order)
\begin{align}\label{FFirstTerm}
\langle F^{\mu\nu}(x)\rangle_1 &= 
\frac{4}{\hbar^{3/2}} Re\sum_{\eta} \int d\Phi(p_1) d\Phi(p_2) d\Phi(p'_1) d\Phi(p'_2) d\Phi(k)\, \times  \\
& \quad \times e^{-i b\cdot (p_1'-p_1)/\hbar}\phi(p_1)\phi^{*}(p'_1) \phi(p_2)\phi^{*}(p'_2)\, k^{[\mu}\ve^{(\eta)\nu]*} e^{-ik\cdot x/\hbar} \langle p'_1\, p'_2| a_{(\eta)}(k) \,T|p_1\,p_2\rangle \nn = \\
& = \frac{4}{\hbar^{3/2}} Re\sum_{\eta} \int d\Phi(p_1) d\Phi(p_2) d\Phi(p'_1) d\Phi(p'_2) d\Phi(k)\, \times \nn \\
& \qquad \times e^{-i b\cdot (p_1'-p_1)/\hbar}\phi(p_1)\phi^{*}(p'_1) \phi(p_2)\phi^{*}(p'_2)\, k^{[\mu}\ve^{(\eta)\nu]*} e^{-ik\cdot x/\hbar} \langle p'_1\, p'_2\,k^\eta|\,T|p_1\,p_2\rangle  \nn
\end{align}
We can identify the matrix element as a five-point amplitude,
\begin{equation}\label{FirstTermAmplitudes}
\langle p'_1\, p'_2\,k^{\eta}| T|p_1\,p_2\rangle = \AC(p_1,p_2\rightarrow p'_1,p'_2,k^{\eta}) \hat\d^4(p_1+p_2-p'_1-p'_2-k)\,.
\end{equation}
At leading order, we replace the amplitude by its LO contribution, given
by a tree-level expression. 

We now want to extract the classical contribution from this quantum amplitude.
Following \Refe{waveforms_2},  we define the momentum mismatches,
\begin{equation}\label{MomentumMismatchesWaveform}
q_1 = p'_1-p_1\,, \\\qquad
q_2 = p'_2-p_2\,.
\end{equation}
We now take \eq{FFirstTerm} and trade the integrals over the $p_i'$ for integrals over the $q_i$
\begin{equation}\label{FatLO}  
\begin{aligned}
\langle F^{\mu\nu}&(x)\rangle_1 = \\
&\hspace*{-6mm} =\frac{4}{\hbar^{3/2}} Re\sum_{\eta}\int d\Phi(p_1) d\Phi(p_2) d^4 q_1 d^4 q_2\, d\Phi(k)\; \hat\d(2p_1\cdot q_1+q_1^2)\hat\d(2p_2\cdot q_2+q_2^2)
\\[-2mm] &\hspace*{12mm}\times e^{-i b\cdot q_1/\hbar} \Theta(p_1^0+q_1^0) \Theta(p_2^0+q_2^0)  \phi(p_1)\phi^{*}(p_1+q_1) \phi(p_2)\phi^{*}(p_2+q_2)
\\ &\hspace*{12mm} \times   k^{[\mu}\ve^{(\eta)\nu]*} e^{-ik\cdot x/\hbar}
\AC(p_1,p_2\rightarrow p_1+q_1,p_2+q_2,k^{\eta}) \hat\d^4(q_1+q_2+k) \,. 
\end{aligned}
\end{equation}
We now take the classical limit,
\begin{align}\label{FatLOcl}
\langle F^{\mu\nu}&(x)\rangle_{1,\cl} =  {g^3} \Lexp \hbar^{2} Re\sum_\eta\int d\Phi(\bar{k}) \bar{k}^{[\mu}\ve^{(\eta)\nu]*} e^{-i\bar{k}\cdot x}\\ 
& \hspace*{1cm}\times
\prod_{i=1,2}\int d^4 \bar{q}_i\;\hat\d(p_i\cdot \bar{q}_i)\; e^{-i b\cdot \bar{q}_1} \hat\d^4(\bar{q}_1+\bar{q}_2+\bar{k}) \bar{\AC}(p_1,p_2\rightarrow p_1+\hbar \bar{q}_1,p_2+\hbar \bar{q}_2,\hbar \bar{k}^{\eta}) \Rexp\,.  \nn 
\end{align}
We recognize the second line as the radiation kernel, \eq{eq:defOfR}. At LO we can finally write
\begin{equation}\label{FatLOcl2}
\langle F^{\mu\nu}(x)\rangle_{1,\cl} = 
  g^3 \Lexp Re\sum_{\eta}\int d\Phi(\bar{k})
  \bar{k}^{[\mu}\ve^{(\eta)\nu]*} e^{-i\bar{k}\cdot x}
      \RadKer^{(0)}(\bar{k}^{\eta};b)
\Rexp\,.
\end{equation}
We now identify the expectation of $F^{\mu\nu}(x)$ as the spatial current $J_{\vec\m}(x)$, we can then obtain the spectral waveform and compute the desired observable.\\

Once again, it is straightforward to extend this result to gravity:
we identify the expectation of the Riemann tensor $R^{\mu\nu\r\s}(x)$ as the spatial current $J_{\vec\m}(x)$, and with the same procedure we are able to obtain the spectral waveform.\\

It is important to remark that the spectral waveform for gravitational radiation is nothing but the double copy of the electro-magnetic one, after subtracting the contribute from the dilaton, for a more detailed discussion see \Refe{Luna_2018} and \Refe{Goldberger_2018}.

%%%%%%%%%%%%%%%%%%%%%%%%%%%%%%%%%%%%%%%%%%%%%%%%
%----------------------------------------------%
%%%%%%%%%%%%%%%%%%%%%%%%%%%%%%%%%%%%%%%%%%%%%%%%

\clearpage

\section{General structure of the three-point amplitude}\label{sec:cap3}

In this section we introduce a formalism to describe four-dimensional (three-point) scattering amplitudes for particles of any mass and spin; an understanding of amplitudes for general mass and spin removes the distinction between ``on-shell" observables like scattering amplitudes and ``off-shell" observables like correlation functions.
We will use the spinor-helicity formalism described in \sect{sec:cap1}.\\
I will start by discussing the case of amplitudes involving only massless particles, and later generalize to amplitudes involving also massive ones. 

Spinor-helicity variables transform nicely under both the Lorentz and Little groups; thus amplitudes for massless particles are simply functions of these. 
The relation to the little group is clearly suggested by the fact that it is impossible to uniquely associate a pair $\lambda_{\alpha}, \tilde{\lambda}_{\dot{\alpha}}$ with some $p_{\alpha \dot{\alpha}}$, since we can always rescale $\lambda_\alpha \to w^{-1} \lambda_\alpha, \tilde{\lambda}_{\dot{\alpha}} \to w \tilde{\lambda}_{\dot{\alpha}}$ keeping $p_{\alpha \dot{\alpha}}$ invariant.
An important property of the Little group is that it is defined for each individual momenta separately, thus
\begin{equation}
M(w^{-1} \lambda, w \tilde{\lambda}) = w^{2 h} M(\lambda,\tilde{\lambda})
\end{equation}
Similarly, even for massive particles $\lambda^{I}, \tilde{\lambda}_I$ can't uniquely be associated with a given $p$: we can perform an $SL(2)$ transformation $\lambda^I \to W^I_J \lambda^J, \tilde{\lambda}_I \to (W^{-1})_I^J \tilde{\lambda}_J$,
where $W$ is a generic transformation belonging to the little group
$SU(2)$.
So amplitudes for massive particles are Lorentz-invariant functions for $\lambda^I, \tilde{\lambda}_I$ which are symmetric rank $2S$ tensors $\{I_1, \cdots, I_{2S}\}$ for spin $S$ particles.
As the Little group is defined for each individual momenta separately, only the spinor variables of a given leg can carry its Little group index, thus
\begin{equation}
M^{\{I_1 \cdots I_{2S}\}} = \lambda_{\alpha_1}^{I_1} \cdots \lambda_{\alpha_{2S}}^{I_{2S}} M^{\{\alpha_1 \cdots \alpha_{2S}\}}
\end{equation}
where $M^{\{\alpha_1 \cdots \alpha_{2S}\}}$ is totally symmetric in the $SL(2,\IC)$ indices.

\subsection{Massless Three-Particle Amplitudes}\label{Massless34}

\subsubsection{Complex Momenta}\label{complex_momenta}
The discussion of three-point amplitudes in the on-shell formalism is quite delicate. Consider a three-point scattering amplitude involving only massless particle. For simplicity we will consider all particles to be incoming.
By kinematic considerations, this configuration should be impossible:
\begin{equation}
p_1^\m+p_2^\m+p_3^\m=0, \quad p_1^2=p_2^2=p_3^2=0
\Rightarrow s_{12}=(p_1+p_2)^2=p_3^2=0, \, s_{13}=0,\, s_{23}=0
\end{equation}
Accordingly, $\wa{i}{j}=\wb{i}{j}=0$ with $i,j=1,2,3$. 
Therefore any possible three-point amplitude is seemingly null.\\
This results follows from the assumption that momenta are real.
If we allow momenta to be complex there is a way out: for complex momenta, $\tilde\l_{\dot \a}$ is not the hermitian conjugate of $\l_\a$, accordingly, $\wb{p}{k}$ is not the complex conjugate of $\wa{k}{p}$, although $s_{pk}=2p\cdot k=\wa{p}{k}\wb{k}{p}$ still holds. Let us formalize this intuition. 

For a massless three-particle scattering, momentum conservation reads $p_1^\m+p_2^\m+p_3^\m=0$, \ie 
\begin{equation}\label{eq:conservazione_impulso}
\qra{1}\qlb{1}+\qra{2}\qlb{2}+\qra{3}\qlb{3}=0\,.
\end{equation}
With complex momenta, two chirally conjugate solutions exist, obtained by multiplying \eq{eq:conservazione_impulso} by either $\qla{1}$ or $\qla{2}$:
\begin{equation}
\wa{1}{2}\qlb{2}+\wa{1}{3}\qlb{3}=0\, , \qquad
\wa{2}{1}\qlb{1}+\wa{2}{3}\qlb{3}=0\, ;
\end{equation}
then for $\wa{i}{j}\neq 0$ we obtain
\begin{equation}
\qlb{2}=-\frac{\wa{1}{3}}{\wa{1}{2}}\qlb{3}\,, \qquad \qlb{3}=-\frac{\wa{2}{1}}{\wa{2}{3}}\qlb{1}\,,
\end{equation}
hence $\qlb{1}\propto \qlb{2}\propto \qlb{3}$.
This result implies that $\wb{1}{2}=\wb{1}{3}=\wb{2}{3}=0$.\\
To recap, if we work with complex momenta and assume $\wa{i}{j}\neq0$ for $i,j=1,2,3$, $\qlb{1}\propto \qlb{2}\propto \qlb{3}$,
thus $\wb{i}{j}=0$ and $s_{ij}=0$. Similarly, if we assume $\wb{i}{j}\neq0$ for $i,j=1,2,3$, $\qla{1}\propto \qla{2}\propto \qla{3}$,
thus $\wa{i}{j}=0$ and $s_{ij}=0$.\\
Therefore, if we work with complex momenta, a solution exists but it must be constructed out of either only right-handed spinors or only left-handed ones, but not both.

\subsubsection{Massless Three-Particle Amplitudes}
We can now compute the desired three-point amplitudes for massless particles of any spin using just the little group scaling and the above results.
We recall that under little group scaling,
\begin{equation}
\begin{aligned}
\qra{p}\rightarrow w^{-1}\qra{p}\,, \quad \qla{p}\rightarrow w^{-1}\qla{p}\,, \qquad 
\qrb{p}\rightarrow w\qrb{p}\,, \quad \qlb{p}\rightarrow w\qlb{p}\,,
\end{aligned}
\end{equation}
with $w\in U(1)$; and the polarization vectors scale as
\begin{equation}\label{eq:scaling_vettori_polarizzazione}
\ve_+^\m(p,k)\equiv \frac{\wab{k}{\s^\m}{p}}{\sqrt{2}\wa{k}{p}}
\rightarrow w^2 \ve_+^\m(p,k)\,,\quad
\ve_-^\m(p,k)\equiv \frac{\wba{k}{\s^\m}{p}}{\sqrt{2}\wb{k}{p}}
\rightarrow w^{-2} \ve_+^\m(p,k).
\end{equation}
The polarization tensor of any massless particles is built out of the outer product of such polarization vectors, therefore \eq{eq:scaling_vettori_polarizzazione} can be generalized to
\begin{equation}
\ve_\pm^\m(p,k)\rightarrow w^{\pm2h}\ve_\pm^\m(p,k)\,,
\end{equation}
where $h$ is the helicity of the desired particle.

We now consider a massless three-point amplitude. After applying a little group transformation to all three particles, the amplitude must scale as
\begin{equation}\label{eq:scaling_1}
M(1,2,3)\rightarrow \Bigl(\prod_{i=1}^3 w^{2h_i}\Bigr)M(1,2,3)\,,
\end{equation}
where $h_i$ is the helicity of the i-{th} massless particle.\\
We recall that three-point amplitudes, in order not to vanish, must be functions of either only $\wa{i}{j}$ or $\wb{i}{j}$, so we can parametrize them as
\begin{equation}\label{eq:scaling_2}
M_3(1^{h_1},2^{h_2},3^{h_3})=M_{\la\cdot\ra}(1^{h_1},2^{h_2},3^{h_3})+M_{\lb\cdot\rb}(1^{h_1},2^{h_2},3^{h_3})\,,
\end{equation}
where $M_{\la\cdot\ra}$ and $M_{\lb\cdot\rb}$ are functions of the spinor products of the form:
\begin{equation}
\begin{cases}
M_{\la\cdot\ra}=A_{abc}\wa{1}{2}^{x_{12}}\wa{2}{3}^{x_{23}}\wa{1}{3}^{x_{13}} \\
M_{\lb\cdot\rb}=B_{abc}\wb{1}{2}^{y_{12}}\wb{2}{3}^{y_{23}}\wb{1}{3}^{y_{13}} \\
\end{cases}
\end{equation}
and where $A_{abc}$ and $B_{abc}$ are couplings, eventually dimensionful, in which we include internal degrees of freedom (\eg colour for QCD).\\
We now apply little group transformation individually for each particle and impose \eq{eq:scaling_1} and \eq{eq:scaling_2}.
We obtain the following system of equations:
\begin{equation}
\begin{cases}
y_{12}+y_{13}=2h_1 \\
y_{12}+y_{23}=2h_2 \\
y_{13}+y_{23}=2h_3 
\end{cases}
\Rightarrow
\begin{cases}
y_{12}=h_1+h_2-h_3 \\  
y_{13}=h_1-h_2+h_3 \\ 
y_{23}=-h_1+h_2+h_3 \\ 
\end{cases}
\end{equation}
Likewise,
\begin{equation}
\begin{cases}
x_{12}+x_{13}=-2h_1 \\
x_{12}+x_{23}=-2h_2 \\
x_{13}+x_{23}=-2h_3 
\end{cases}
\Rightarrow
\begin{cases}
x_{12}=-y_{12}=-h_1-h_2+h_3 \\  
x_{13}=-y_{13}=-h_1+h_2-h_3 \\ 
x_{23}=-y_{23}=h_1-h_2-h_3 \\ 
\end{cases}
\end{equation}
Thus, little group scaling fixes uniquely the dependence of $M_{\la\cdot\ra}$ and $M_{\lb\cdot\rb}$ on the spinor products:
\begin{equation}
\begin{cases}
M_{\la\cdot\ra}=A_{abc}\wa{1}{2}^{-h_1-h_2+h_3}\wa{2}{3}^{h_1-h_2-h_3}\wa{1}{3}^{-h_1+h_2-h_3} \\
M_{\lb\cdot\rb}=B_{abc}\wb{1}{2}^{h_1+h_2-h_3}\wb{2}{3}^{-h_1+h_2+h_3}\wb{1}{3}^{h_1-h_2+h_3} \\
\end{cases}
\end{equation}
Now, the three-point amplitude $M_3$ must have the correct physical behaviour for real momenta, \ie it must vanish when both $\wa{i}{j}$ and $\wb{i}{j}$ go to zero.

Since $x_{12}+x_{13}+x_{23}=-(y_{12}+y_{13}+y_{23})=-(h_1+h_2+h_3)$, for $h_1+h_2+h_3>0$, $A_{abc}$ must be zero in order to have a sensible result, otherwise $M_{\la\cdot\ra}$ would blow up. Conversely, for $h_1+h_2+h_3<0$, $B_{abc}$ must vanish, otherwise $M_{\lb\cdot\rb}$ would blow up. \\
So we can write in general:
\begin{equation}\label{Massless3}
M^{h_1 h_2 h_3} =
\begin{cases} B_{abc} [12]^{h_1 + h_2 - h_3} [23]^{h_2 + h_3 - h_1} [31]^{h_3 + h_1 - h_2} \qquad {\rm for} \, h_1 + h_2 + h_3 > 0 \\ A_{abc} \langle 1 2 \rangle^{h_3 - h_1 - h_2} \langle 2 3 \rangle^{h_1 - h_2 - h_3} \langle 3 1 \rangle^{h_2 - h_3 - h_1} \!\!\!\!\qquad {\rm for} \, h_1 + h_2 + h_3 < 0
\end{cases}
\end{equation}
Notice that this treatment is not suitable to describe the case $h_1+h_2+h_3=0$, which require a specific treatment depending on the individual process.

Finally, using Bose symmetry it is possible to show that for odd-(integer)-spin particles $A_{abc}$ and $B_{abc}$ must be completely antisymmetric, mimicking the role of the structure constants $f_{abc}$. For even-(integer)-spin particles we do not have such requirement.

If we now consider a theory of a single self-interacting particle of spin-s, the only possible three-point amplitudes are:
\begin{align}
& M_3(1_a^{-s},2_b^{-s},3_c^{+s})=A_{abc}\biggl( \frac{\wa{1}{2}^3}{\wa{2}{3}\wa{3}{1}}\biggr)^s\,, \quad
M_3(1_a^{+s},2_b^{+s},3_c^{-s})=B_{abc}\biggl( \frac{\wb{1}{2}^3}{\wb{2}{3}\wb{3}{1}}\biggr)^s\,, \\
& M_3(1_a^{-s},2_b^{-s},3_c^{-s})=A_{abc}\bigl(\wa{1}{2}\wa{2}{3}\wa{3}{1}\bigr)^s\,, \quad
M_3(1_a^{+s},2_b^{+s},3_c^{+s})=B_{abc}\bigl(\wb{1}{2}\wb{2}{3}\wb{3}{1}\bigr)^s\,. \nn
\end{align}
Furthermore, for parity invariant theories, such as gravity, parity symmetry requires $A_{abc}=B_{abc}$.

We finally focus on pure gravity amplitudes. In D=4 dimensions, $[M_n]=4-n$, the gravitational coupling is $(M_{PL})^{-1}= \sqrt{32 \pi G}= \k/2$, so it has mass dimensions $-1$.
Since the spinor products have mass dimension 1, then $[M_3(1^{\mp s},2^{\mp s},3^{\pm s})]=1$ implies that $[A_{abc}]=[B_{abc}]=-1$, which is precisely the dimension of the gravitational coupling.
On the other hand, $[M_3(1^{\pm s},2^{\pm s},3^{\pm s})]=1$ implies that $[A_{abc}]=[B_{abc}]=-5$. Even taking into account the dimensionful coupling, we need at least four momenta to justify this scaling. In a renormalization picture, this suggests that such amplitudes are not really ``elementary" but are rather the result of composite operators, hence we can neglect them on the basis of dimensional analysis.\\
To sum up the only relevant amplitudes in pure gravity are:
\begin{equation}\label{eq:ampiezze_3_gravitoni}
M_3(1^{-2},2^{-2},3^{+2})=\frac{\k}{2} \frac{\wa{1}{2}^6}{\wa{2}{3}^2\wa{3}{1}^2}\,, \quad
M_3(1^{+2},2^{+2},3^{-2})=\frac{\k}{2}\frac{\wb{1}{2}^6}{\wb{2}{3}^2\wb{3}{1}^2}\,,. \\
\end{equation}

\subsection{Three-Particle Amplitudes Involving Massive Particles}
I will now study the case of amplitudes involving massive particles. \\
As discussed in \sect{sec:cap3}, the amplitude will be labeled by the spin-S representation of the SU(2) little group for massive legs and helicities for the massless legs. \\
We recall that, as the Little group is defined for each individual momenta separately, only the spinor variables of a given leg can carry its Little group index. 
For amplitudes involving massive legs, it will be convenient to expand in terms of only $\l^I_\a$, since any dependence on $\tl^I_{\dot\a}$ can be converted using Dirac equation, \eq{eq:SHvarDiracEq}.
Thus we can pull out overall factors of $\lambda_i^I$ from the amplitude, 
\begin{equation}\label{ChiralDef}
M_n^{\cdots\{I_1,I_2,\cdots,I_{2s_i} \}\cdots}=\lambda^{I_1}_{i\,\alpha_1}\lambda^{I_2}_{i\,\alpha_2}\cdots \lambda^{I_{2s_i}}_{i\,\alpha_{2s_i}}M_n^{\cdots\{\alpha_1,\alpha_2,\cdots,\alpha_{2s_i} \}\cdots}\;.    
\end{equation}
leaving behind a function that is symmetric in $SL(2,\IC)$ indices instead.
We will refer to this representation as the chiral basis, reflecting the fact that we are using the un-dotted $SL(2, \IC)$
indices. Notice that one may equally use the anti-chiral basis.

For a three-point amplitude, we have three possible momenta, related by momentum conservation. As the amplitude is now just a tensor in the $SL(2,\IC)$ Lorentz indices, the problem reduces to finding two linear independent 2-component spinors that span this space, which
we will denote as $(v_\a, u_\a)$. 

The full categorization of three-point amplitude with arbitrary masses is discussed in detail in \Refe{arkanihamed2021}, however I will focus only on the case of one-massless two-massive particles, and in particular on the equal mass case. 

\subsubsection{Three Particle Amplitudes, one-massless two-massive}

For two massive legs, the three-point amplitude is labeled by $(h, S_1, S_2)$ 
\begin{equation}
    \vcenter{\hbox{\includegraphics[scale=0.5]{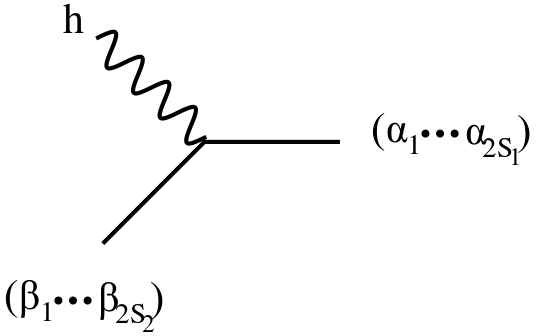}}} \quad M^{h}\,_{\{\alpha_1\alpha_2\cdots\alpha_{2S_1}\},\,\{ \beta_1\beta_2\cdots\beta_{2S_2}\}}
\end{equation}
where h is the helicity of the massless leg \footnote{Figure reproduced from \Refe{arkanihamed2021}.}. Now we have a $\{2s\} \otimes \{2s\} \,SL(2, \IC)$ tensor, and we are interested in the general structure of all possible couplings. As mentioned above, this entails the need of a basis to span the two-dimensional space. It is preferable to use the kinematic variables of
the problem to serve as a basis.\\

For unequal mass, one of the basis spinor can be $\lambda$ of the massless leg, while the remaining can be chosen to be $\tl$ contracted with one of the massive momentum. For example one can choose: 
\begin{equation}\label{uv}
\left(v_\alpha, u_\alpha\right)=\left(\l_{3\alpha}, \; \frac{p_{1\alpha\dot{\beta}}}{m_1}\tl_3^{\dot\beta}\right)\,,
\end{equation}
where $p_1$ is the momentum of one of the massive particles, and $p_3$ the momentum of the massless one.\\

From now on we will focus on the equal mass case.\\
If the masses are identical, then $u$ and $v$ are no longer independent. Indeed, using momentum conservation,
\begin{equation}
p_1+p_2+p_3=0 \quad \Rightarrow \quad p_1+p_3=-p_2  \quad \Rightarrow \quad p^2_1+p^2_3+2p_1 \cdot p_3=p^2_2 \,,
\end{equation}
and, as we are on-shell, $p^2_1=p^2_2=m^2,\, p^2_3=0$. Thus
\begin{equation}
0=\frac{2 p_3\cdot p_1}{m}=\frac{\langle 3|p_1|3]}{m}\equiv v^\a u_\a\,,
\end{equation}
\ie the spinor $\lambda_3^\alpha$ must be proportional to $\tilde{\lambda}_{3\dot{\alpha}}p_{1}^{\dot{\alpha}\a}$. \\
Taking into account this result, we introduce as the expansion basis
\begin{equation}
\lambda_{3\alpha}, \; \epsilon_{\alpha\beta}\,,
\end{equation}
where $\l_{3\a}$ is once again the chiral spinor of the massless leg, while $\e_{\a\b}$ is the Levi-Civita symbol.

Notice however that $\lambda_3$ carries helicity weight $-\frac{1}{2}$ of the massless leg. In order for amplitude to be little group invariant, one should also have a variable that carries positive weights. We call this variable ``$x$", and it is defined as the (dimensionless) proportionality constant between $\l_{3\a}$ and $p_{1 \a\dot\a}\tl_3^{\dot\a}$:
\begin{equation}\label{xDef}  
x\l_{3\a}=\frac{p_{1\a\dot\a}}{m}\tl_3^{\dot{\a}}\,,\quad x^{-1}\tl^{\dot{\a}}_3=\frac{p_1^{\dot{\a}\a}\l_{3\a}}{m}\,. 
\end{equation}
Note that $x$ carries $+1$ little group weight of the massless leg. Furthermore, $x$ cannot be expressed in a manifestly local way. Indeed contracting both sides of the above equation with a reference spinor $\zeta$ yields:
\begin{equation}
x=\frac{\langle \zeta |p_{1}|3]}{m\langle \zeta 3\rangle},  
\end{equation}
so while $x$ is independent of $\zeta$, any concrete expression for it has an apparent, spurious pole in $\zeta$. \\
The above shows that $x$ can be nicely written in terms of polarization vectors:
\begin{equation}
mx=\sqrt{2}\,\ve^+_\m \, p_1^\m=\frac1{\sqrt{2}}\,\ve^+_\m (p_1-p_2)^\m
\end{equation}
where the auxiliary spinor is identified with the reference spinor of the polarization vector.

Equipped with this new variable, we can write down the general structure of a three point amplitude for two spin $s$ and a helicity $h$ state: the only objects we have carrying $SL(2,\IC)$ indices are $\lambda_{3\a}$ and $\epsilon_{\alpha\beta}$, thus
\begin{equation}\label{eq:M3epsilon_lambda}
M_3^{h,\{\alpha_1,\cdots,\alpha_{2s}\},\{\beta_1,\cdots,\beta_{2s}\}}=
(m x)^{h}\left[\sum_{a=0}^{2s} g_a\epsilon^{2s-a}\biggl(x\frac{\lambda_3\lambda_3}{m}\biggr)^{\!a}\right]^{\{\alpha_1,\cdots,\alpha_{2s}\},\{\beta_1,\cdots,\beta_{2s}\}}\,,
\end{equation}
where the $\{2s\}\otimes \{2s\}$ separately symmetrized $SL(2,\IC)$ indices are distributed across the Levi-Civita tensors $\epsilon$ and the $\lambda_3$. Thus we see that there are in total $2s{+}1$ structures for spin $s$ states, and we have normalized the couplings such that the $g_i$ are dimensionless. 
 
It will be convenient to make connection with the amplitudes computed from the usual Feynman diagram approach. For this, we simply put back the $\lambda_i^I$ factors that was pulled out that defined the chiral basis in \eq{ChiralDef}. For example,
\begin{align}\label{eq:poshel3pt} 
M_3^{h,s,s} &= (mx)^h \left[ g_0 \frac{\wa{\bf 2}{\bf 1}^{2s}}{m^{2s-1}} + g_1 x \frac{\wa{\bf 2}{\bf 1}^{2s-1} 
 \wa{\bf 2}{3}\wa{3}{\bf 1}}{m^{2s}} + \dots + g_{2s} x^{2s} \frac{\wa{\bf 2}{3}^{2s} \wa{3}{\bf 1}^{2s}}{m^{4s-1}} \right]\,. 
\end{align}

\subsubsection{The simplest three-point Amplitude}
In the previous section, we have seen that for a massive spin-$s$ particle, whether it is fundamental or composite, the emission of a photon or graviton can in general be parameterized by \eq{eq:poshel3pt}. 
This parameterization is unique in the sense that the expansion basis is defined on kinematic grounds unambiguously. The expansion is organized in terms of powers of $1/m$ , with higher order terms hinting at potential problems in the UV region, \ie in the $m \rightarrow 0$ limit. 

Consider an amplitude with only the leading term in \eq{eq:poshel3pt}. The above discussion would indicate that not only is the amplitude simple in the number of terms involved, but is also simple in the sense of having the best UV behavior: at high energies our massive states behave, at leading order, as massless ones, and we can ask what amplitude in the UV does this pure $x$-piece matches to. 
The two massive spin-$s$ states will translate into a $+s$ and $-s$ helicity state separately at high energies. To avoid a singular piece we must have $\lambda_1\sim \lambda_2\sim \lambda_3$. Thus minimal coupling in the UV uniquely picks out 
\begin{equation}\label{MinimalCoupling}
M_{UV}^{h,s,s}=\frac{g_0}{m}\,(m x)^h\left(\frac{\langle \bf{1}\bf{2}\rangle}{m}\right)^{2s}
\end{equation}
For this reason it is natural to refer to the choice of setting all coupling constants except $g_0$ to zero as \emph{minimal coupling}. One can straightforwardly check that subleading terms in \eq{eq:poshel3pt} match to higher derivative couplings in the UV, see \Refe{Chung_2019}.

The minimal coupling three-point amplitude of a graviton with two massive spin-$s$ particles is given by:
\begin{equation}\label{eq:Minimal_Graviton_3pt}
M^{+2,s,s}_3 = x^2 \,\frac{\k}{2} \frac{\wa{\bf 1}{\bf 2}^{2s}}{m^{2s-2}}, \qquad M^{-2,s,s}_3 = x^{-2}\, \frac{\k}{2} \frac{\wb{\bf 1}{\bf 2}^{2s}}{m^{2s-2}}
\end{equation}

\subsubsection{Classical Spin-operators from Amplitudes}\label{sec:classical_spin}
Following \Refe{Chung_2019, Guevara_2019}, we see that it is very convenient to view the three-point amplitude as an operator acting on the Hilbert space of $SL(2,\IC)$ irreps; in particular this operator 
maps the spin-$s$ representation in the Hilbert space of particle 1 to that of particle 2, \ie
\begin{equation}
M^{+h,s,s}= \frac{1}{m^{2s-1}}\lambda^{I_1\alpha_1}_1\cdots \lambda^{I_{2s}\alpha_{2s}}_1  \mathcal{O}_{\{\alpha_1,\cdots,\alpha_{2s}\},\{\beta_1,\cdots,\beta_{2s}\}} \lambda^{J_1\beta_1}_2\cdots \lambda^{J_{2s}\beta_{2s}}_2   
\end{equation}
The operator $\mathcal{O}_{\{\alpha_1,\cdots,\alpha_{2s}\},\{\beta_1,\cdots,\beta_{2s}\}} $ is a linear combination of polynomials in $\epsilon_{\alpha\beta}$ and $\lambda_{3\alpha}\lambda_{3\beta}$, see \eq{eq:M3epsilon_lambda}. 
As we will now demonstrate, the former can be naturally identified as the identity operator, while the latter as the \emph{spin-operator}.

The spin vector $S^\mu$ is defined, in the Hilbert space of our spinors, as the Pauli-Lubanski pseudo-vector $S^\mu=-\frac{1}{2m}\epsilon^{\mu\nu\rho\sigma}p_{1\nu}J_{\rho\sigma}$. 
We now focus on the Lorentz generator $J_{\mu\nu}$. 
Using the conventions explained in \app{Conventions}, it is easy to check that
\begin{equation}\label{eq:SL2Celemdef}
(J^{\mu\nu})_\text{spinor} 
= \frac{i}{4} [\g^\m , \g^\n]  
= \frac{i}{2}
\begin{pmatrix}
(\s^{[\m} \bar{\s}^{\n]})_\a{}^\b & 0 
\\
0 & (\bar{\s}^{[\m} \s^{\n]})^{\dot{\a}}{}_{\dot{\b}} 
\end{pmatrix}
= 
\begin{pmatrix}
(J^{\m\n})_\a{}^\b & 0 
\\
0 & (J^{\m\n})^{\dot{\a}}{}_{\dot{\b}} 
\end{pmatrix}\,,
\end{equation}
forms a representation of the Lorentz algebra. 
We are now interested in its action on $SL(2,\IC)$ irreps. 
For spin-$s$, using the fact that the irreps are symmetric tensors of the 2s indices, it is possible to write:
\begin{equation}
(J_{\mu\nu})_{\alpha_1\alpha_2\cdots\alpha_{2s}}\,^{\beta_1\beta_2\cdots\beta_{2s}}=\sum_{i=1}^{2s}(J_{\mu\nu})_{\alpha_i}\,^{\beta_i}\,\bar{\iden}_i=2s(J_{\mu\nu})_{\alpha_1}\,^{\beta_1}\,\bar{\iden}_1,\quad (J_{\mu\nu})_{\alpha}\,^{\beta}=\frac{i}{2}\sigma_{[\mu}\bar{\sigma}_{\nu]}\,,   
\end{equation}
where $\bar{\iden}_i=\delta_{\alpha_1}^{\beta_1}\cdots \delta_{\alpha_{i{-}1}}^{\beta_{i{-}1}}\cancel{\d_{\a_i}^{\b_i}}\d_{\alpha_{i{+}1}}^{\beta_{i{+}1}}\cdots \delta_{\alpha_{2s}}^{\beta_{2s}}$, and I have used the symmetry of the 2s indices to pick, without loss of generality, 2s-copies of the tensor $(J_{\mu\nu})_{\alpha_1}\,^{\beta_1}$.
Using this, we find that
\begin{equation}\label{eq:spinopangsqdefs}
\begin{cases}
m \left( S_\m \right)_{\a}^{~\b} &= \frac{1}{4} \left[ \s_\m (p_1 \cdot \bar{\s}) - (p_1 \cdot \s) \bar{\s}_\m \right]_{\a}^{~\b}
\\ m \left( S_\m \right)^{\dot{\a}}_{~\dot{\b}} &= - \frac{1}{4} \left[ \bar{\s}_\m (p_1 \cdot \s) - (p_1 \cdot \bar{\s}) \s_\m \right]^{\dot{\a}}_{~\dot{\b}}  
\end{cases}\,.
\end{equation}
If we now multiply each side of these equations by the massless momentum $p_3$, we obtain:
\begin{equation}\label{PSDef} 
(p_3\cdot S)_{\a}^{~\b} = \frac{x}{2} \lambda_{3\a} \lambda_3^\b\equiv \frac{x}{2} |3\rangle\langle3|\,,
\qquad (p_3\cdot S)^{\dot\a}_{~\dot\b} =-\frac{\lambda_3^{\dot\a} \lambda_{3\dot\b}}{2x}\equiv-\frac{|3][3|}{2x} \,.    
\end{equation}
Therefore, the operator $\mathcal{O}_{\{\alpha_1,\cdots,\alpha_s\},\{\beta_1,\cdots,\beta_s\}}$ is comprised of identity operators and spin vector operators.

%------------------
\subsection{Multipole expansion of three-point amplitudes}\label{sec:multipole}
In this section, I will introduce the key steps in the multipole expansion of three-point amplitudes. This methodology becomes crucial in the analysis of Kerr black holes scattering, as it organizes our amplitudes in an extremely convenient way in angular momentum multipoles. \\
In this section, I will use the latin letters $a$, $b$ for the massive little group indices.

Prior to delving into the specifics,  we may construct the spin-$s$ external wavefunctions, using the relation
$p^\m=\frac12 \s^\m_{\a\dot \b} p^{\a\dot \b}=\frac12 \l^\a_a \s^\m_{\a\dot \b} \tl^{\dot \b}_a=\frac12
\qla{\bf{p}_{a}}\s^\m\qrb{\bf{p}^{a}}$, as
\begin{equation}\label{eq:polvectorsmassive}
    \ve_{p,\mu}^{ab}
    = \frac{\bra{\bf{p}^{(a}}\sigma_\mu|\bf{p}^{b)}]}{\sqrt{2}m}\,,
\end{equation}
where the round brackets indicate a symmetrization. With this definition, the following relations hold: 
\begin{equation}
p\cdot\ve_p^{ab}=0\,, \qquad 
\ve^{ab}_{p,\m}\ve_{p,\n \,ab}=\eta_{\m\n}-\frac{p_\m p_\n}{m^2}\,, \qquad \varepsilon_{p,11}\cdot\varepsilon_{p}^{11} 
    = \varepsilon_{p,22}\cdot\varepsilon_{p}^{22}
    = 2\varepsilon_{p,12}\cdot\varepsilon_{p}^{12} = 1\,.
\end{equation}

\subsubsection{Massive spin-1 Matter}
\label{sec:spin1matter}
To study the details of the multipole expansion it is instructive to consider the simple case of graviton emission by two massive vector fields.\\
Consider the massive spin-1 Lagrangian
\begin{equation}\label{eq:proca}
{\cal L} = -\frac{1}{4}F_{\mu\nu}F^{\mu\nu} + \dfrac{1}{2}m^2 A_\mu A^\mu ,
\end{equation}
where $F^{\mu\nu}$ is the field strength $F^{\mu\nu}=\partial^\mu A^\nu-\partial^\nu A^\mu$.\\
To compute the energy-momentum tensor in flat space, we can apply Noether's theorem for translations, which gives us the following result:
\begin{equation}\label{eq:tuvna}
T_N^{\mu\nu}=-F^{\mu\sigma} \partial^\nu A_\sigma - \eta^{\mu\nu} {\cal L} \qquad \Rightarrow \qquad\partial_\mu T_N^{\mu\nu} = 0 .
\end{equation}
However, its contraction with an on-shell graviton, \(\varepsilon_{\mu\nu} T_N^{\mu\nu}\), fails to produce the correct three-point amplitude (since \(\partial_\nu T_N^{\mu\nu} \neq 0\)); consequently, the (dual) orbital angular momentum
\begin{equation}
L^{\lambda \mu \nu} = x^\mu T_N^{\lambda \nu} - x^\nu T_N^{\lambda \mu}   
\end{equation}
is not conserved.
Notice however that it is possible to generalize $T_N^{\mu\nu}$ to a larger class of tensors that still satisfy \eq{eq:tuvna}:
\begin{equation}\label{eq:wqq_1}
T^{\mu\nu} = T_N^{\mu\nu} + \partial_\lambda B^{\lambda\mu\,\nu} , \qquad \quad B^{\lambda\mu\,\nu} = -B^{\mu\lambda\,\nu} \qquad \Rightarrow \qquad
\partial_\mu T^{\mu\nu} = 0 ,
\end{equation}
furthermore the Belinfante tensor $B^{\mu\nu\rho}$ may be adjusted to yield a symmetric energy- momentum tensor matching the gravitational one.\\
To ensure the conservation of total angular momentum, we can then apply Noether's theorem to Lorentz transformations, which gives:
\begin{equation}\label{eq:wqq_2}
T_N^{\mu\nu}-T_N^{\nu\mu} =-\partial_\lambda J^{\lambda\,\mu\nu} , \qquad \quad
J^{\lambda\,\mu\nu}
 = -i\frac{\partial{\cal L}}{\partial(\partial_\lambda A^\sigma)}
    \Sigma^{\mu\nu,\sigma}_{~~~~\tau} A^\tau
= i F^{\lambda\sigma} \Sigma^{\mu\nu}_{\sigma\tau} A^\tau \,,
\end{equation}
where $\Sigma_{\mu\nu}$ are the Lorentz generators
$\Sigma^{\mu\nu,\sigma}_{~~~~\tau}
=i[\eta^{\mu\sigma} \delta^\nu_\tau - \eta^{\nu\sigma} \delta^\mu_\tau]$.\\
We may now substitute \eq{eq:wqq_1} into \eq{eq:wqq_2}, and after imposing that \(T^{\mu\nu}\) is symmetric, we obtain the following expression:
\begin{equation}
\partial_{\lambda}B^{\lambda[\mu\,\nu]}
=\frac{1}{2}\partial_{\lambda}J^{\lambda\,\mu\nu} \,, 
\end{equation}
which is solved by 
\begin{equation}
B^{\lambda\mu\,\nu} = \frac{1}{2} \big[ J^{\lambda\,\mu\nu} + J^{\mu\,\nu\lambda} - J^{\nu\,\lambda\mu} \big] .
\end{equation}
Contracting the resulting energy-momentum tensor with a traceless symmetric graviton~$h_{\mu\nu}$ and integrating by parts, we obtain the gravitational interaction vertex
\begin{equation}
-h_{\mu\nu} T^{\mu\nu}
= h_{\mu\nu} F^{\mu\sigma} \partial^\nu A_\sigma
- i (\partial_\lambda h_{\mu\nu}) F^{\nu\sigma} \Sigma^{\lambda\mu}_{\sigma\tau}  A^\tau \,.
\end{equation}
The corresponding three-particle amplitude reads (for more details see \Refe{Guevara_2019}):
\begin{equation}
    {\cal M}_3(p_1,p_2,k)
     =-2(p\cdot\varepsilon)
       \big[ (p\cdot\varepsilon) (\varepsilon_1\!\cdot\varepsilon_2)
            - 2 k_\mu \varepsilon_\nu
              \varepsilon_1^{[\mu} \varepsilon_2^{\nu]}
       \big] , \qquad \text{with} \,\,
    p = \frac{1}{2} (p_1-p_2) .
\label{eq:spin1amp3pt}
\end{equation}
Naively, we may interpret the combination $\ve_1^{[\mu} \ve_2^{\nu]}$ as being proportional to the classical spin tensor.
However, the spin-1 amplitude contains up to quadratic spin interactions (see \Refe{Vaidya_2015}), whereas only the linear piece is apparent in \eq{eq:spin1amp3pt}.
To rewrite this contribution in terms of multipoles, we can use a redefined spin tensor (see \app{app:spin_tensor}):
\begin{equation}
   S^{\mu\nu}
    =-\frac{i}{\varepsilon_1\!\cdot \varepsilon_2} \bigg\{
      2\varepsilon_1^{[\mu} \varepsilon_2^{\nu]}
    - \frac{1}{m^2} p^{[\mu}
      \big( (k\cdot\varepsilon_2) \varepsilon_1
          + (k\cdot\varepsilon_1) \varepsilon_2 \big)^{\nu]} \bigg\}\, .
\label{eq:spintdef}   
\end{equation}
Inserting this spin tensor in \eq{eq:spin1amp3pt}, we rewrite the above amplitude as
\begin{align}\label{eq:spin1amp3pt2}
{\cal M}_3(p_1,p_2,k^\pm) &
      =-m^2 x_\pm^{2} (\varepsilon_1\!\cdot\varepsilon_2)
        \bigg[ 1 - \frac{i\sqrt{2}}{m x_\pm} k_\mu \varepsilon^\pm_\nu S^{\mu\nu}
             + \frac{ (k\cdot\varepsilon_1) (k\cdot\varepsilon_2) }
                    { m^2 (\varepsilon_1\!\cdot\varepsilon_2) }
        \bigg] \,,
\end{align}
where $x_\pm=\frac{\sqrt{2}}{m}p\cdot \ve^\pm$. 
It is important to note that, when the polarization of the graviton is negative, we have the so called $x_-$, related to the $x$ previously discussed and presented in \Refe{arkanihamed2021} (labeled $x_+$) via the relation $x_+ x_- =-1$ (whenever omitted $x\equiv x_+$).\\
We now want to express this amplitude, obtained from Feynman rules, in spinor-helicity variables to possibly rewrite it in a more convenient form. Choosing the polarization of the graviton to be negative, we have
\begin{subequations}\label{eq:vectors2spinors}
\begin{flalign}
\label{eq:vectors2spinors0}
& \hspace{1.5cm}\ve_1^{a_1 a_2}\cdot\ve_2^{b_1 b_2} 
    = -\frac{1}{m^2} \wa{\bf{1}^{(a_1}}{\bf{2}^{(b_1}}
      \bigg[ \wa{\bf{1}^{a_2)}}{\bf{2}^{b_2)}}
           - \frac{1}{m x_-} \wa{\bf{1}^{a_2)}}{k} \wa{k}{\bf{2}^{b_2)}}
      \bigg]\, , \\ 
\label{eq:vectors2spinors1}
& \hspace{-0.9cm}\big[(\ve_1\cdot\ve_2)
    k_\m \ve_\n^- S^{\m\n}\big]^{a_1 a_2 b_1 b_2} 
    = i\frac{\wa{\bf{1}^{(a_1}}{k}}{\sqrt{2}m^2} 
      \bigg[ \wa{\bf{1}^{a_2)}}{\bf{2}^{(b_1}}
           - \frac{1}{2m x_-} \wa{\bf{1}^{a_2)}}{k} \wa{k}{\bf{2}^{(b_1}}
      \bigg] \wa{k}{\bf{2}^{b_2)}} , \\
\label{eq:vectors2spinors2}
& \hspace{1.7cm} (k\cdot\ve_1^{a_1 a_2}) (k\cdot\ve_2^{b_1 b_2}) 
    =-\frac{1}{2m^2 x_-^2} \wa{\bf{1}^{(a_1}}{k} \wa{\bf{1}^{a_2)}}{k}
      \wa{k}{\bf{2}^{(b_1}} \wa{k}{\bf{2}^{b_2)}} ,
\end{flalign} 
\end{subequations}
where we have reduced all $[1^a|$ and $|2^b]$ to the chiral spinor basis of $\bra{1^a}$ and $\ket{2^b}$ using the identities:
\begin{align}
\wb{\bf 2}{\bf 1}= \wa{\bf 2}{\bf 1}-\frac{\wa{\bf 2}{3}\wa{3}{\bf 1}}{mx_-}
= \qla{\bf{2}} \left( \iden + \frac{\qla{3} \qra{3}}{mx_-} \right) \qra{\bf{1}} \label{eq:3ptkin1}
\\ 
\wb{\bf 2}{3} \wb{3}{\bf 1}=-\frac{\wa{\bf 2}{3}\wa{3}{\bf 1}}{x_-^2}
= \qla{\bf{2}} \left( - \frac{\qra{3} \qla{3}}{x_-^2} \right) \qra{\bf{1}}\,. \label{eq:3ptkin2}   
\end{align}
Written in this form, we can easily extract the factors $\bra{1^{(a_1}}\otimes\bra{1^{a_2)}}$ and $\ket{2^{(b_1}}\otimes\ket{2^{b_2)}}$, from each of these identities. Furthermore, we can interpret them as representations of the massive-particle states 1 and 2.
Introducing the symbol $\odot$ for the symmetrized tensor product,
we can rewrite \eq{eq:vectors2spinors0} as:
\begin{align}
\ve_1\cdot\ve_2 &= -\frac{1}{m^2} \qla{\bf 1}^{\odot 2}
\bigg[ \iden \odot \iden - \frac{1}{m x_-} \iden \odot \Big(\qra{k} \qla{k}\Big) \bigg] \qra{\bf 2}^{\odot 2}\, = \nn \\
&= -\frac{1}{m^2}
\bigg[ \wa{\bf 1}{\bf 2}^{\odot 2}- \frac{1}{m x_-} \wa{\bf 1}{\bf 2} \odot \Big(\wa{\bf 1}{k} \wa{k}{\bf 2}\Big) \bigg]\, .    
\end{align}
Combining all the terms in \eq{eq:vectors2spinors} into the amplitude, we obtain
\begin{align}
{\cal M}_3(p_1,p_2,k^-) & = x_-^2
\bigg[ \wa{\bf 1}{\bf 2}^{\odot 2}
- \frac{2}{m x_-} \wa{\bf 1}{\bf 2}\odot \Big(\wa{\bf 1}{k}\wa{k}{\bf 2}\Big)
+ \frac{1}{m^2 x_-^2}\wa{\bf 1}{k}^{\odot 2} \wa{k}{\bf 2}^{\odot 2}\bigg] .
\label{eq:spin1amp3pt3}
\end{align}

In \app{app:angularmomentum} we construct the differential form of the angular-momentum operator in momentum space. We can now act with the operator $k_\mu \ve_\nu J^{\mu\nu}$ on the product state
$\ket{p^a}^{\odot 2}=\ket{p^{a_1}}\otimes\ket{p^{a_2}}$.
For the negative helicity of the graviton, we have
\begin{equation}\label{eq:angular_2}
k_\mu \ve_\nu^- J^{\mu\nu}
= \frac{1}{4\sqrt{2}} \l^\alpha \l^\beta
  \e^{\dot{\a}\dot{\b}} J_{\a\dot{\a},\b\dot{\b}}
= -\frac{i}{\sqrt{2}} \wa{k}{p^a}
  \wa{k}{\frac{\partial~}{\partial \l_p^a}} , \qquad 
  \wa{k}{\frac{\partial~}{\partial \l_p^b}} \ket{p^a}
= \ket{k}\d^a_b \,.
\end{equation}
Thus,
\begin{align}
i\sqrt{2} (k_\mu \ve_\nu^- J^{\mu\nu}) \ket{p^a}^{\odot 2} &
 = \wa{k}{p^b} \bigg\{ 
  \bigg[ \wa{k}{\frac{\partial~}{\partial \l_p^b}} \ket{p^{a_1}}
  \bigg]\otimes\ket{p^{a_2}}
+ \ket{p^{a_1}}\otimes
  \bigg[ \wa{k}{\frac{\partial~}{\partial \l_p^b}} \ket{p^{a_2}}
  \bigg] \bigg\} \nn \\ &
= \ket{k} \wa{k}{p^{a_1}}\otimes\ket{p^{a_2}}
+ \ket{p^{a_1}}\otimes\ket{k}\wa{k}{p^{a_2}}
= 2 \ket{k} \wa{k}{p^a}\odot\ket{p^a} \,.
\end{align}
Furthermore, the action of the operator $(k_\m\ve_\n J^{\m\n})^j\,,\,\,j \in \IN$ on the state $\ket{p^a}^{\odot 2}$ is \footnote{To keep the notation simple we may omit the $SU(2)$ indices of the massive little group and the symmetrization symbol for the external massive momenta, nevertheless in more rigorous formulation their presence is understood}:
\begin{subequations}\label{eq:raising}
\begin{align}
\bigg(\frac{i k_\mu \varepsilon_\nu^- J^{\mu\nu}}{p\cdot\varepsilon^-}
\bigg) \ket{\bf p}^2 &
 = \frac{2}{m x_-} \ket{k} \wa{k}{\bf p} \ket{\bf p} ,
\label{eq:raising1} \\
\bigg(\frac{i k_\mu \varepsilon_\nu^- J^{\mu\nu}}{p\cdot\varepsilon^-}
\bigg)^{\!2} \ket{\bf p}^2 &
= \frac{2}{m^2 x_-^2} \ket{k}^2 \wa{k}{\bf p}^{2} ,
\label{eq:raising2} \\
\bigg(\frac{i k_\mu \varepsilon_\nu^- J^{\mu\nu}}{p\cdot\varepsilon^-}
\bigg)^{\!j} \ket{\bf p}^2 & = 0 ,
  \qquad \qquad \qquad \qquad \qquad j \geq 3 .
\label{eq:raising3}
\end{align} 
\end{subequations}
Notice that one of the possible algebraic realizations of $J^{\mu\nu}$ is the tensor-product version, $-(\s^{\m\n}\otimes\iden + \iden\otimes\s^{\m\n})$, of the standard $SL(2,\IC)$ chiral generator
$\sigma^{\mu\nu}=\frac{i}{2}\sigma^{[\mu}\bar{\sigma}^{\nu]}$.
Therefore, it is direct to check that
\begin{equation}\label{eq:linspi} 
\frac{i k_\mu \ve_\nu^- J^{\mu\nu}}{p\cdot\ve^-}
= \frac{\ket{k} \bra{k}}{m x_-} \otimes \iden
+ \iden \otimes \frac{\ket{k} \bra{k}}{m x_-} \,.
\end{equation}
In light of the previous observations, we may rewrite the last two terms in \eq{eq:spin1amp3pt3} as:
\begin{subequations}
\begin{align}
&-\frac{2}{mx_-} \wa{\bf1}{\bf2}\wa{\bf1}{k}\wa{k}{\bf2} = 
\bra{\bf 2}^2
\bigg( \frac{ik_\mu \ve_\nu^- J_1^{\mu\nu}}{p_1\cdot\ve^-} \bigg) \ket{\bf 1}^2 \,, \\
&\,\,\,\,\frac{1}{m^2 x_-^2} \wa{\bf1}{k}^2 \wa{k}{\bf2}^2 =
\frac{1}{2} \bra{\bf2}^2 \bigg( \frac{ik_\mu \varepsilon_\nu^- J_1^{\mu\nu}}{p_1\cdot\varepsilon^-}\bigg)^{\!2} \ket{\bf1}^2 \,.
\end{align}    
\end{subequations}
Therefore
\begin{equation}
{\cal M}_3(p_1,p_2,k^-)= 
x_-^2 \bra{\bf 2}^2 \bigg\{1+ 
i\bigg(\frac{k_\mu \ve_\nu^- J_1^{\mu\nu}}{p_1\cdot\ve^-}\bigg)
- \frac{1}{2}\bigg(\frac{k_\m\ve_\n^- J_1^{\m\n}}{p_1\cdot\ve^-}\bigg)^{\!2}\bigg\} \ket{\bf1}^2 \,.       
\end{equation}
Written in this form, we can clearly see how to interpret the terms in the amplitude \eq{eq:spin1amp3pt3} as the multipole contributions with respect to the chiral spinor basis, despite the fact that they do not equal the multipoles in \eq{eq:spin1amp3pt2} individually.

As a final remark, note that since the operator $(k_\m \ve_\n^-J^{\m\n})^j$ annihilates the spin-1 state for $j \geq 3$, the preceding expression can be expressed compactly in an exponential form:
\begin{equation}\label{eq:m3m} 
{\cal M}_3(p_1,p_2,k^-) = x_-^2 \bra{\bf2}^2 \exp\!{\bigg(i\frac{k_\m \ve^-_\n J^{\m\n}}{p\cdot\ve^-}} \bigg) \ket{\bf 1}^2 \,.   
\end{equation}

It can be checked explicitly that acting with the operator on the
state $\bra{\bf 2}^2$ yields the same result.
On the other hand, choosing the other helicity of the graviton will yield the parity conjugated version of \eq{eq:m3m}:
\begin{equation}\label{eq:m3p} 
{\cal M}_3(p_1,p_2,k^+) = {x_+^2} [\bf2|^2 \exp\!\bigg(i\frac{k_\mu \varepsilon^+_\nu J^{\mu\nu}}{p\cdot\varepsilon^+} \bigg) |\bf1]^2 .
\end{equation}

\subsubsection{Exponential form of three-particle Amplitude}
\label{sec:exponentiation3pt}

In this section I will generalize the previous discussion to arbitrary spin-$s$.
I will focus only on integer spin to ignore the factors $(-1)^{2s}$ that appear when dealing with also half-integer spin.

We start by considering the three-point amplitudes for massive matter minimally coupled to gravity in \eq{eq:Minimal_Graviton_3pt}:
\begin{equation}
{\cal M}_3^{(s)}(p_1,p_2,k^+)
= \frac{\k}{2}\frac{\wa{\bf1}{\bf2}^{2s} x^{2}}{m^{2s-2}} , \qquad \quad
{\cal M}_3^{(s)}(p_1,p_2,k^-)
= \frac{\k}{2}\frac{\wb{\bf1}{\bf2}^{2s} x^{-2}}{m^{2s-2}} .    
\end{equation}
We focus once again on the minus-helicity amplitude.\\
In such a compact form all the dependence on the spin tensor is completely hidden. 
In order to restore it, we need to rewrite the minus-helicity amplitude in the chiral basis 
\begin{align}\label{eq:bin}
{\cal M}_3^{(s)}(p_1,p_2,k^-)
& =\frac{\k}{2}\, \frac{x_-^2}{m^{2s-2}}
\bigg( \wa{\bf2}{\bf1} + \frac{\wa{\bf2}{k}\wa{k}{\bf1}}{m x_-}
\bigg)^{\!\odot 2s} \,= \nn \\
& \quad = \frac{\k}{2}\,\frac{x_-^2}{m^{2s-2}} \bra{\bf 2}^{2s}
\Bigg[ \sum_{j=0}^{2s} \binom{2s}{j}
\bigg( \frac{\ket{k}\bra{k}}{m x_-}\bigg)^j \Bigg] \ket{\bf1}^{2s} \,.
\end{align}
We now recast the operator in square brackets into exponential form using the differential angular momentum operator in \eq{eq:angular_2}. 
Indeed, the formulae in \eq{eq:raising} can be easily generalized
to the product states of spin-$s$, namely
\begin{equation}\label{eq:raisinggeneral}
\bigg(i\frac{k_\mu \ve^-_\nu J^{\mu\nu}}{p\cdot\ve^-}\bigg)^{\!j} \ket{\bf p}^{2s} =
\left\{
\begin{array}{ll}
\dfrac{(2s)!}{(2s-j)!} \ket{\bf p}^{2s-j}
\bigg( \dfrac{\ket{k} \wa{k}{\bf p}}{mx_-}
\bigg)^{\!j} &, \quad j \leq 2s \,, \\
\hspace{2cm}0 &, \quad j > 2s \,.
\end{array}
\right.
\end{equation}
From this we can derive the formal relations
\begin{equation}\label{eq:raisinggeneral2}
\bigg(i\frac{k_\mu \varepsilon^-_\nu J^{\mu\nu}}{p\cdot\varepsilon^-}\bigg)^{\!\odot j}  =
\left\{
\begin{array}{ll}
\dfrac{(2s)!}{(2s-j)!}
\bigg( \dfrac{\ket{k} \bra{k}}{mx_-}
\bigg)^{\!\otimes j}\!\odot \iden^{\otimes 2s-j} &, \quad j \leq 2s \,,\\
\hspace{2cm}0 &, \quad j > 2s \,.
\end{array}
\right.
\end{equation}
Hence, 
\begin{align}\label{eq:bin2exp}
\bra{\bf 2}^{2s} \Bigg[ \sum_{j=0}^{2s} \binom{2s}{j}
\bigg( \frac{\ket{k}\bra{k}}{m x_-}\bigg)^j \Bigg] \ket{\bf 1}^{2s}\, & = \nn \\
& \hspace{-3cm}= \bra{\bf 2}^{2s}\!\sum_{j=0}^{\infty}\frac{1}{j!} \bigg( i\frac{k_\mu \varepsilon^-_\nu J^{\mu\nu}}{p\cdot\varepsilon^-}\bigg)^{\!j} \ket{\bf 1}^{2s}\!
= \bra{\bf 2}^{2s}\! \exp\!\bigg(i\frac{k_\mu \varepsilon^-_\nu J^{\mu\nu}}{p\cdot\varepsilon^-}\bigg) \ket{\bf 1}^{2s} \,.
\end{align}    
Therefore, for finite (or infinite) spins we have:
\begin{equation}\label{eq:exp3m}
\hat{{\cal M}}_3^{(s)}(p_1,p_2,k^-) = {\cal M}_3^{(0)} \exp\!\bigg(i\frac{k_\mu \varepsilon^-_\nu J^{\mu\nu}}{p\cdot\varepsilon^-}\bigg) \,, \qquad \quad
{\cal M}_3^{(s)} = \frac{1}{m^{2s}}\bra{\bf 2}^{2s}\hat{{\cal M}}_3^{(s)}\ket{\bf 1}^{2s} \,,
\end{equation}
For the plus-helicity amplitude we can follow the same procedure, rewriting the amplitude in the anti-chiral basis, and obtain
\begin{equation}\label{eq:exp3p}
\hat{{\cal M}}_3^{(s)}(p_1,p_2,k^+) = {\cal M}_3^{(0)}
   \exp\!\bigg(i\frac{k_\mu \varepsilon^+_\nu J^{\mu\nu}}
                     {p\cdot\varepsilon^+}\bigg) ,
\qquad \quad
{\cal M}_3^{(s)} = \frac{1}{m^{2s}}
    [\bf2|^{2s}\hat{{\cal M}}_3^{(s)}|\bf1]^{2s} \,.
\end{equation}
In the above formulae $\MC^{(0)}_3(p_1,p_2,k^\pm)\equiv\frac{\k}{2}(mx_\pm)^2$ corresponds to the scalar piece, \ie the Schwarsczhild case.
These formulae match precisely the Kerr energy-momentum tensor, see \Refe{Vines_2018, Guevara_2019}.

%-------------------
\subsection{Classical infinite-spin limit}\label{sec:classical_inf_spin}
In \app{app:spin_tensor} we motivated the definition of a generalized expectation value of an operator $\mathcal{O}$ acting on two massive states,
represented by their polarization tensors,
\begin{equation}
\la{{\cal O}}\ra
= \frac{ \ve_{2,\mu_1\dots\mu_s}{\cal O}^{\mu_1\dots\mu_s,\nu_1\dots\nu_s}\ve_{1,\nu_1\ldots\nu_s} }{ \ve_{2,\mu_1\ldots\mu_s} \ve_1^{\mu_1\ldots\mu_s} } \,.
\end{equation}
It is possible to obtain the correct result just using the above formula, \eg see section 3 of \Refe{Guevara_2019}. 
Nonetheless, for the rest of this paper I will follow the machinery presented in \Refe{Guevara_2019_b}, as it is completely equivalent but I find it more convenient.
In this approach, we recover the entire spin information from the amplitudes by combining the spinor-helicity formalism with the covariant approach to multipoles introduced in \Refe{bautista2019}, while the GEV will only serve to fix the normalization.

\begin{minipage}[l]{0.55\linewidth}
We begin our discussion by considering the three-point minimal-coupling amplitudes, \eq{eq:Minimal_Graviton_3pt}:
\begin{subequations}\label{eq:GravityMatter3pt}
\begin{align}
\label{eq:GravityMatter3ptPlus}
{\cal M}_3(p_1^{\{a\}}\!,-p_2^{\{b\}}\!,k^{+}) &= -\frac{\kappa}{2}
\frac{\wa{1^a}{2^b}^{\odot 2s}}{m^{2s-2}} x^2 , \\*
\label{eq:GravityMatter3ptMinus}
{\cal M}_3(p_1^{\{a\}}\!,-p_2^{\{b\}}\!,k^{-}) &= -\frac{\kappa}{2}
\frac{[1^a 2^b]^{\odot 2s}}{m^{2s-2}} x^{-2} .
\end{align} 
\end{subequations}
\end{minipage} 
\begin{minipage}[c]{0.4\linewidth}
\begin{figure}[H]
    \centering
    \includegraphics[width=0.5\textwidth]{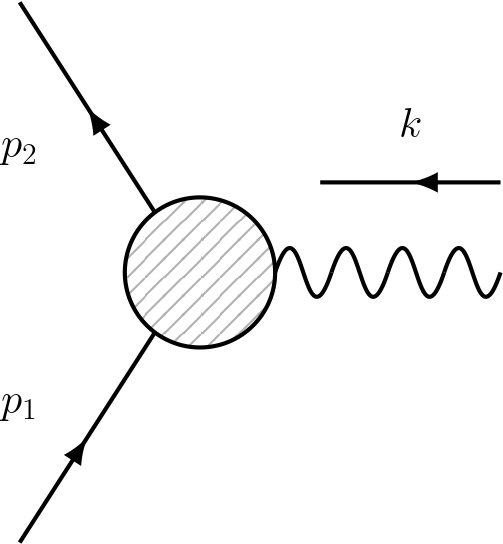}
    \caption{3-pt amplitude. Figure reproduced from \Refe{Guevara_2019_b}}
\end{figure}
\end{minipage}

\noindent As we know from \sect{sec:multipole}, it is possible to recast the spin structure of the amplitudes in \eq{eq:GravityMatter3pt} in an exponential form:
\begin{subequations}\label{eq:GravityMatter3ptExp}
\begin{align}\label{eq:GravityMatter3ptExpPlus}
\wa{2^b}{1^a}^{\odot 2s} &
= [2^b|^{\odot 2s}\! \exp\!\bigg(\!{-i}\frac{k_\mu \ve_\nu^+ \bar{\sigma}^{\mu\nu}}{p_1\cdot\ve^+}\bigg) |1^a]^{\odot 2s} \,, \\
\label{eq:GravityMatter3ptExpMinus}
[2^b 1^a]^{\odot 2s} &
        = \bra{2^b}^{\odot 2s}\!
          \exp\!\bigg(\!{-i}\frac{k_\mu \varepsilon_\nu^- \sigma^{\mu\nu}}
                        {p_1\cdot\varepsilon^-}\bigg) \ket{1^a}^{\odot 2s} \,,
    \end{align} 
\end{subequations}
where we have used as the algebraic realizations of $J^{\m\n}$ the tensor-product \\
$- (\s^{\m\n}\otimes\iden + \iden\otimes\ \s^{\m\n})\equiv -\s^{\m\n}$ for the chiral basis and $ - (\bar{\s}^{\m\n}\otimes\iden + \iden\otimes\ \bar{\s}^{\m\n})\equiv -\bar{\s}^{\m\n}$ for the anti-chiral basis.

Note that a consistent picture of spin-induced multipoles of a pointlike particle must be formulated in the particle's rest frame.
However, the spin operator is defined in \app{app:spin_tensor} for momentum $p_{\rm a}$, with $a=1,2$, but acts, in the rest frame, on the states with momenta $p_{\rm a} \pm k/2$.
The cure for that is to take into account an additional Lorentz boost,
needed to bridge the gap between two states with different momenta,
\Refe{bautista2019}.\\
To start, we note that any two four-vectors $p_1$ and $p_2$ of equal mass $m$ may be related by the spinorial transformations:
\begin{equation}\label{eq:Lorentz1to2Spinor} 
\ket{2^b}  = U_{12~a}^{~~b}\exp\!\bigg( \frac{i}{m^2} p_1^\mu k^\nu \sigma_{\mu\nu}\!\bigg)\ket{1^a} , \qquad 
|2^b]  = U_{12~a}^{~~b}\exp\!\bigg( \frac{i}{m^2} p_1^\mu k^\nu \bar{\sigma}_{\mu\nu}\!\bigg)|1^a] ,  
\end{equation}
where $U_{12} \in SU(2)$ is a little-group transformation
that depends on the specifics of the massive-spinor realization.\\
The self-duality property of the spinorial generators, $\s^{\m\n}=\frac{i}{2}\e^{\m\n\r\s}\s_{\r\s}$, allow us to easily rewrite the above exponents as
\begin{equation}\label{eq:Tensor2Vector}
   \frac{i}{m^2}\, p_1^\mu k^\nu \sigma_{\mu\nu,\alpha}^{~~~~\;\beta}
    = k \cdot a_{\alpha}^{~\beta} , \qquad
   \frac{i}{m^2} \,p_1^\mu k^\nu \bar{\sigma}^{~~~\dot{\alpha}}_{\mu\nu,~\dot{\beta}}
    =-k \cdot a^{\dot{\alpha}}_{~\,\dot{\beta}} ,
\end{equation}
where we have defined chiral representations for the Pauli-Lubanski operators, \Refe{Guevara_2019_b},
\begin{equation}\label{eq:PauliLubanskiSpinor}
a_{~~\a}^{\m,~\,\b} = \frac{1}{2m^2} \e^{\m\n\r\s} p_{a\,\n}\s_{\r\s,\a}^{~~~~\;\b} \,, 
\qquad 
a^{\m,\dot{\a}}_{~~~\,\dot{\b}} = \frac{1}{2m^2} \e^{\m\n\r\s} p_{a\,\n}\bar{\s}^{~~~\dot{\a}}_{\r\s,~\dot{\b}} .
\end{equation}
Notice that, already at this level, we are free to pick $p_{\rm a}=p_1+k/2$ instead of $p_1$, as the product $k \cdot a$ remains the same.

The natural extension to higher-spin states of the Pauli-Lubanski operators in the chiral representations, is simply the analogous of \eq{eq:Lorentz_Generator}:
\begin{equation}\label{eq:PauliLubanski2s}
(a^{\m})_{\a_1\dots\a_{2s}}^{~~~~~~~~\b_1\dots\b_{2s}}
= a_{~~\a_1}^{\m,~~\b_1}\d_{\a_2}^{\b_2}\!\dots \d_{\a_{2s}}^{\b_{2s}} 
+ \ldots + \d_{\a_1}^{\b_1}\!\dots \d_{\a_{2s-1}}^{\b_{2s-1}}a_{~~\a_{2s}}^{\m,~~~\b_{2s}} \,.
\end{equation}
Therefore, we have
\begin{equation}\label{eq:Lorentz1to2SpinorPL}
\begin{aligned}
\ket{\bf 2}^{\odot 2s}\!& = e^{k \cdot a}
      \big\{ U_{12} \ket{\bf 1} \big\}^{\!\odot 2s} \,, \qquad
   |\bf2]^{\odot 2s}\!= e^{-k \cdot a} \big\{ U_{12} |\bf1] \big\}^{\!\odot 2s} \,, \\
   \bra{\bf2}^{\odot 2s}\!& = \big\{ U_{12} \bra{\bf1} \big\}^{\!\odot 2s}
      e^{-k \cdot a} \,,\:\,\quad
   [\bf2|^{\odot 2s}\!= \big\{ U_{12} [\bf1| \big\}^{\!\odot 2s} e^{k \cdot a} \,,    
\end{aligned}
\end{equation}
where the second line follows from the antisymmetry of $\s^{\m\n}$ and $\bar{\s}^{\m\n}$ in the sense of $\epsilon^{\alpha\beta} \sigma_{~~~\beta}^{\mu\nu,~\gamma} \epsilon_{\gamma\delta} = -\sigma_{~~~\delta}^{\mu\nu,~\alpha}$.

We are finally ready to inspect the spin dependence of the three-point amplitudes. Using \eq{eq:Lorentz1to2SpinorPL}, we can rewrite \eq{eq:GravityMatter3pt} as
\begin{subequations}\label{eq:GravityMatter3pt2}
\begin{align}
\label{eq:GravityMatter3ptPlus2}
{\cal M}_3^{(s)}(k^+) & = -\frac{\k}{2}\frac{x^2}{m^{2s-2}} [\bf2|^{\odot 2s} e^{-2k \cdot a} |\bf1]^{\odot 2s} \,, \\
\label{eq:GravityMatter3ptMinus2}
{\cal M}_3^{(s)}(k^-) & = -\frac{\k}{2}\frac{x^{-2}}{m^{2s-2}}\bra{\bf2}^{\odot 2s} e^{2k \cdot a}  \ket{\bf1}^{\odot 2s} \,.
\end{align} 
\end{subequations}
The apparent spin dependence in the amplitude formulae above is of the form $e^{\mp 2k \cdot a}$, while no such dependence seems to exist in the original formulae \eq{eq:GravityMatter3pt}.
Still, the true angular-momentum dependence inherent to the minimal-coupling amplitudes should not depend on the choice of the spinorial basis.
This apparent contradiction is resolved by taking into account the transformations \eq{eq:Lorentz1to2Spinor}. \\
To fully grasp this statement, it is helpful to examine the following example:
consider $\wa{\bf 2}{\bf 1}^{\odot 2s}$, in the chiral representation we have 
\begin{equation}
\wa{\bf 2}{\bf 1}^{\odot 2s} = \big\{U_{12}\qla{\bf1}\big\}^{\odot 2s}e^{-k\cdot a}\qra{\bf1}^{\odot 2s}\,,
\end{equation}
whereas in the anti-chiral representation we have
\begin{equation}
\wa{\bf 2}{\bf 1}^{\odot 2s} = 
\qlb{\bf 2}^{\odot 2s}e^{-2k\cdot a}\qrb{\bf 1}^{\odot 2s}=
\big\{U_{12}\qlb{\bf1}\big\}^{\odot 2s}e^{-k\cdot a}\qrb{\bf1}^{\odot 2s}\,.
\end{equation}
We can now clearly see how in both cases we have the same spin dependence.

In light of these considerations, we strip the spin-states from the amplitudes in \eq{eq:GravityMatter3pt2} to cleanly obtain the spin dependence (see \app{app:boosts}). 
Furthermore, as the the total angular momentum of a Kerr Black Hole is large, it is a good approximation to take the infinite-spin limit. \\
For example, if we were to consider once again $\wa{\bf 2}{\bf 1}^{\odot 2s}$ after this manipulation we would get:
\begin{equation}
   \lim_{s \to \infty} m^{2s} (U_{12})^{\odot 2s} e^{-k \cdot a} .
\end{equation}
Note that the factor of $m^{2s}$, that apparently diverges, cancels in the actual amplitudes. Thus
\begin{subequations}\label{eq:GravityMatter3pt3}
\begin{align}
\label{eq:GravityMatter3ptPlus3}
   {\cal M}_3^{(\infty)}(k^+) & \approx -\frac{\kappa}{2} m^2 x^2
      e^{-k \cdot a}\!\lim_{s \to \infty} (U_{12})^{\odot 2s} , \\*
\label{eq:GravityMatter3ptMinus3}
   {\cal M}_3^{(\infty)}(k^-) & \approx -\frac{\kappa}{2} m^2 x^{-2}
      e^{k \cdot a}\!\lim_{s \to \infty} (U_{12})^{\odot 2s} .
\end{align} 
\end{subequations}
The remaining unitary factor of $(U_{12})^{\odot 2s}$ parametrizes an arbitrary little-group transformation that corresponds to the choice of the spin quantization axis (see \Refe{Guevara_2019_b}).
As such, it is inherently quantum-mechanical and therefore should be removed in the classical limit.
Indeed, it also appears in the simple product of polarization tensors
\begin{align}\label{eq:SpinTransitionNorm}
\lim_{s \to \infty} \ve_2 \cdot \ve_1 
&= \lim_{s \to \infty} \frac{1}{m^{2s}}\wa{\bf2}{\bf1}^{\odot s} \odot [\bf2 \bf1]^{\odot s}= \\
& = \lim_{s \to \infty} \frac{1}{m^{2s}}\big\{ U_{12} \bra{\bf1} \big\}^{\!\odot s} e^{-k \cdot a}\ket{\bf1}^{\odot s} \big\{ U_{12} [\bf1| \big\}^{\!\odot s} e^{k \cdot a} |\bf1]^{\odot s}
= \lim_{s \to \infty} (U_{12})^{\odot 2s} \,. \nn
\end{align}
So we can interpret the factor of $(U_{12})^{\odot 2s}$ as the state normalization in accord with the notion of GEV, and as such it cancels out whenever we compute a physical quantity.

%%%%%%%%%%%%%%%%%%%%%%%%%%%%%%%%%%%%%%%%%%%%%%%%
%----------------------------------------------%
%%%%%%%%%%%%%%%%%%%%%%%%%%%%%%%%%%%%%%%%%%%%%%%%

\clearpage

\section{Computing higher-point Amplitudes from the Soft Theorems}\label{sec:cap4}

In this section, I will discuss how to use the soft theorems, in the spinor-helicity formalism, to obtain amplitudes with an increasing number of gravitons.\\
Soft theorems describe general properties of scattering amplitudes in the infrared (IR) regime where the four-momentum of one  massless particle approaches zero, $p^\mu \rightarrow \epsilon p^\mu $ with $\epsilon \to 0$. 
In this limit it is possible to write down a recursion relation connecting the amplitude to a lower-point one with the soft particle removed.    
For soft gravitons with $h= \pm 2$ the recursions take the form
\begin{align}\label{eq:soft}
\MC_{n+1}^{\pm 2} = \left( \frac{S^{(0)}_{\pm 2}}{\epsilon^3} +\frac{S^{(1)}_{\pm 2}}{\epsilon^2}   +\frac{S^{(2)}_{\pm 2}}{\epsilon}  \right)\MC_{n}\,, 
\end{align}
where $S^{(i)}_{\pm h}$ are called the soft factors.
The soft factors are {\em universal}, in the sense that they do not depend on what other ``hard" particles are involved in the scattering. \\
The soft graviton theorem was first formulated by Steven Weinberg in 1965 \Refe{Weinberg_1965}. However, after some initial progress \Refe{Low_1958}, \Refe{Low_Burnett_Kroll}, this approach saw little use for nearly twenty years.\\
A new life into the study of soft theorems was brought by the emergence of on-shell methods for calculating scattering processes.
An important step in this development was made by Cachazo and Strominger \Refe{cachazo_strominger}. Using the BCFW recursion relations \Refe{Britto_2005}, they were able to extend the graviton soft theorems up to sub-subleading order at tree level.
Despite these rapid developments, all of the results derived with on-shell techniques were valid only for massless theories.\\
Recently, using the spinor-helicity formalism for {\em massive} particles introduced in \Refe{arkanihamed2021} and discussed in \sect{sec:cap1}, Falkowski and Machado established soft recursion relations valid for general theories with massive particles of any spin. 

%-------------
\subsection{BCFW Recursion Relations}\label{sec:BCFW}

In this section I will discuss the BCFW recursion relations, \Refe{Britto_2005}. Using these relations we depart from the traditional QFT workflow: the on-shell recursion relations allow us to compute amplitudes without making any reference to a Lagrangian or to Feynman diagrams. \\
Note that the `original' BCFW relations apply exclusively to amplitudes involving only massless particles. In this section, I will focus solely on the massless case, while the massive case will be addressed in subsequent sections.

The idea behind the Britto-Cachazo-Feng-Witten (BCFW) recursion relations is to consider the amplitude $\MC_n(p_1, \dots,p_n)$ as an analytic function of its complex momenta $p_1, \dots ,p_n$. 
The momenta are complexified by introducing a shift, which preserves \emph{on-shellness} and \emph{momentum conservation}, and which is linear in a complex variable $z$.

Let us shift one of the momenta, say $p_j^\m$ with $j=1,\dots n$, by a vector $q^\m$, \ie \\ 
$\hat{p}_j(z)=p_j+zq$, with $z \in \IC$.
Furthermore, in order to preserve the total momentum conservation, let us shift another momentum $\hat{p}_i(z)=p_i-zq$.
On-shellness requires that $\hat{p}^2_j(z)=\hat{p}^2_i(z)=0$, which in turn implies the orthogonality conditions $q^2=p_j\cdot q=p_i \cdot q=0$.
The amplitude $\MC_n(p_1,\dots ,p_n)$ then becomes an analytic function $\MC(z)$ of this shift parameter $z$.

It is a well known fact that unitarity, in the sense of the optical theorem, implies that the poles of an amplitude correspond to the exchange of on-shell intermediate states. 
This means that the original amplitude factorizes on its poles into two smaller amplitudes connected by a vanishing propagator.
Diagrammatically,
\vspace{-0.5 cm}
\begin{figure}[H]
    \centering
    \includegraphics[width=0.9\textwidth]{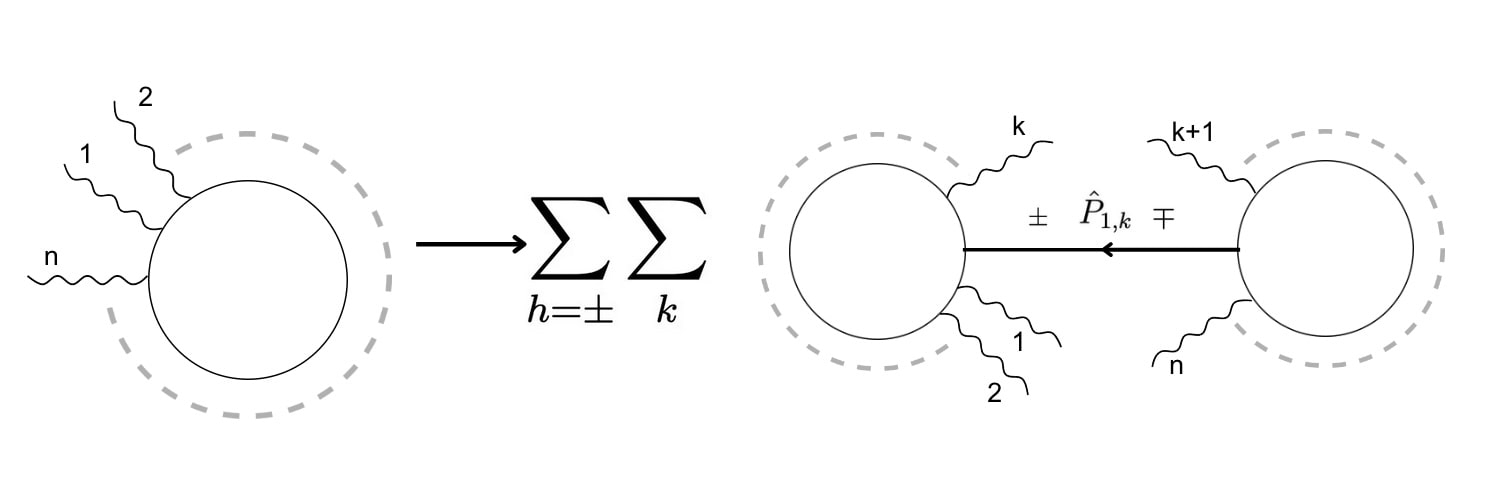}
\end{figure}
\vspace{-0.7 cm}
\noindent where we have defined the momentum flowing through the propagator as 
\begin{equation}
\hat{P}^\m_{1,k}\equiv p_1^\m+\dots+\hat{p}^\m_j+\dots+p_k^\m  \,.  
\end{equation}
Notice that if $i$ and $j$ are on the same side of the diagram, $P_{1,k}^\m$ is independent of $z$ and there is no pole.\\
Now, $\hat{P}_{1,k}^2=(P_{1,k}+zq)^2=P_{1,k}^2+2zP_{1,k}\cdot q$. 
The factorization poles occur when $\hat{P}_{1,k}(z)$ goes on-shell for some $z_k$, \ie
$0=\hat{P}_{1,k}^2=(P_{1,k}+z_k q)^2=P_{1,k}^2+2z_k P_{1,k}\cdot q$,
therefore the poles are located at 
\begin{equation}
z_k=-\frac{P^2_{1,k}}{2 P_{1,k}\cdot q}\,, \qquad  
\mathrm{with}\quad z_k \neq 0\,.
\end{equation}
Note that $\hat{P}^2_{1,k}(z)=2 P_{1,k}\cdot q \,(z-z_k)=-\dfrac{P^2_{1,k}}{z_k}(z-z_k)$. \\
Now, $\MC_n(z=0)$ corresponds to the original amplitude, which is analytic around the origin, as the poles occur at $z_k\neq0$.
Cauchy's theorem states that on a contour $\mathcal{C}$ very close to the origin
\begin{equation}
\frac1{2\pi i}\oint_\mathcal{C} dz \frac{\MC_n(z)}{z}=\Res{z\rightarrow 0}\frac{\MC_n(z)}{z}=\MC_n(0)\,.
\end{equation}
Furthermore, the global residue theorem states that over contour large enough to encompass all its poles,
\begin{equation}
\MC_n(0)+\sum_k \Res{z\rightarrow z_k}\frac{\MC_n(z)}{z}=\Res{z \rightarrow \infty}\frac{\MC_n(z)}{z}\,,
\end{equation}
where $z_k$ are the locations of the factorization poles.\\
If $\MC_n(z)\rightarrow0$ as $z\rightarrow\infty$, then the residue at infinity vanishes and we are left with
\begin{equation}
\MC_n(0)+\sum_k \Res{z\rightarrow z_k}\frac{\MC_n(z)}{z}=0
\end{equation}
Using then the factorization into two lower-point amplitudes connected by a vanishing propagator,
\begin{align}
i\MC_n(0) &=
-\sum_k \Res{z\rightarrow z_k}\frac{i\MC_n(z)}{z} \,=  \\
& \hspace{-0.9cm} =-\sum_{h=\pm}\sum_{k=2}^{n-2}\Res{z\rightarrow z_k}\biggl[
i\MC_{k+1}(1,2,\dots,\hat{j},\dots,-\hat{P}^{-h}_{1,k})\frac{i}{z \hat{P}^2_{1,k}}\, i\MC_{n-k+1}(\hat{P}_{1,k}^h,k+1,\dots,\hat{i},\dots,n)\biggr] \,, \nn
\end{align}
where $\MC_{k+1}$ and $\MC_{n-k+1}$ are the amplitudes on either side of the pole. \\
Recalling now that 
\begin{equation}
\Res{z\rightarrow z_k}\frac{\MC_n(z)}{z}=\lim_{z\rightarrow z_k}(z-z_k)\frac{\MC_n(z)}{z}\,, \qquad \mathrm{and} \quad z-z_k=-\frac{z_k \hat{P}^2_{1,k}}{P^2_{1,k}}\,,
\end{equation}
we finally obtain the BCFW recursion relations:
\begin{align}
 i\MC(p_1 \dots p_n) &= \\
& \hspace{-0.8cm} =\sum_{h=\pm}\sum_{k=2}^{n-2}\,\,i\MC_{k+1}(1,\dots,\hat{j},\dots,-\hat{P}^{-h}_{1,k})\frac{i}{P^2_{1,k}}\, i\MC_{n-k+1}(\hat{P}_{1,k}^h,\dots,\hat{i},\dots,n)\,. \nn
\end{align}
Notice that the sum is over the $(n-3)$ partitions of the $n$ momenta into two sets, with at least three momenta in both sets (even when considering complex momenta, the smallest possible on-shell amplitudes are the three-point ones).

To complete the proof of the on-shell recursion relations, it is necessary to show that \(\MC_n(z)\rightarrow0\) as \(z\rightarrow\infty\). This point is rather delicate, as for a generic amplitude, this behavior is not always assured; however, in the cases I will discuss, the residue at infinity will consistently be zero. A detailed discussion of all possible scenarios can be found in \Refe{Arkani_Hamed_2008}.

%-------------
\subsection{Soft Recursions}\label{sec:soft}

In this section, I will derive a formula that connects an \(n+1\)-point amplitude with at least one massless particle to an \(n\)-point amplitude. 
This formula captures the leading terms in the limit where the momentum of a massless particle is continuously taken to zero. 
This massless particle is referred to as {\em soft}, and the formula is referred to as the {\em soft recursion}.
I will mainly follow \Refe{Falkowski_2021} and \Refe{Elvang_2017}.\\
I will employ a similar BCFW approach, but with generalizations to include massive particles. Additionally, the soft limits will be taken in a manifestly on-shell manner using an extra complex deformation of the external massless momenta.\\
In addition to the new soft theorems, the resulting master formula yields consistency conditions such as the conservation of electric charge, the Einstein equivalence principle, supergravity Ward identities, and the Weinberg-Witten theorem.

\subsubsection{Momentum shift}

As seen in \sect{sec:BCFW}, a crucial step for any amplitude to factorize is performing a shift of the complex momenta.
I will now examine how to implement such a shift in the massive case and, as an additional step, deform the external massless momenta. \\
Consider a set of $n+1$ particles with momenta $p_0 \dots p_n$ satisfying $\sum_{i=0}^n p_i = 0$ and the on-shell conditions $p_i^2 = m_i^2$.
The particle labeled as $0$ is always massless, $p_0^2=0$, and it will be our soft particle. 
We perform a two-parameter deformation of this kinematics.  
One parameter $z$ controls complex deformation of the momenta of the particle $0$ and of two other particles, below labeled as $j$ and $k$. The other parameter $\epsilon$ controls the soft limit of the particle 0. \\
A convenient choice of the shifts is the following \footnote{One advantage of the two-parameter shift in \eq{eq:ST_ss} compared to simpler BCFW-like one-parameter shifts is that it will directly lead to soft expansion of recursion relations
in powers of $1/\e$. As a bonus, it will allow for more control over the recursions’ boundary terms,
as the latter will start at higher orders in $1/\e$.}:
\begin{equation}\label{eq:ST_ss}
\hat p_0 = \epsilon p_0 - z q_0\,, \quad 
\hat p_j = p_j - 
\frac{(\epsilon-1) (p_0 p_k)  - z (q_0 p_k)}{ q_j p_k }  q_j \,, \quad
\hat p_k = p_k - 
\frac{ (\epsilon-1)(p_0 p_j) -  z (q_0 p_j)}{ q_k p_j}  q_k \,, 
\end{equation}
and the remaining particles remain un-shifted.\\
We now must fix the four-vectors $q_0$, $q_j$, $q_k$ in such a way that on-shellness and momentum conservation are preserved also on the shifted kinematics.\\
In order to preserve on-shellness, we impose
\begin{equation}
\begin{cases}
\hat{p}_0^2= (\e p_0 - zq_0)^2 =0 \\
\hat{p}_j=\biggl(p_j - 
\dfrac{(\epsilon-1) (p_0 p_k) - z (q_0 p_k)}{ q_j p_k }  q_j\biggr)^2 =m_j^2 \\
\hat p_k^2= p_k - \biggl(\dfrac{ (\epsilon-1)(p_0 p_j) -  z (q_0 p_j)}{ q_k p_j}  q_k \biggr)^2=m_k^2
\end{cases}   
\rightarrow
\begin{cases}
p_0\cdot q_0 = 0 \,,\,\,\, q_0^2=0 \\
p_j\cdot q_j = 0  \,,\,\,\, q_j^2=0 \\
p_k\cdot q_k = 0  \,,\,\,\, q_k^2=0 \\
\end{cases}\,.
\end{equation}
Furthermore, in order to preserve momentum conservation, we impose
\begin{align}\label{eq:conservazione_impulso_2}
\sum_{i=0}^n \hat{p}_i &=
\sum_{i=0}^n p_i+ \\
& \, +\biggl[(\e-1)p_0 -zq_0 -\dfrac{(\epsilon-1) (p_0 p_k) - z (q_0 p_k)}{ q_j p_k }  q_j
- \dfrac{ (\epsilon-1)(p_0 p_j) -  z (q_0 p_j)}{ q_k p_j}  q_k 
\biggr]=0  \,,\nn
\end{align}
thus the term in square bracket must vanish. 
Notice that as $\e$ and $z$ are completely unrelated, in order for the second line of \eq{eq:conservazione_impulso_2} to be null, each contribution must individually vanish, so
\begin{equation}
\begin{cases}
p_0-\dfrac{p_0p_k}{q_jp_k}q_j-\dfrac{p_0p_j}{q_kp_j}q_k=0 \\
q_0+\dfrac{q_0p_k}{q_jp_k}q_j+\dfrac{q_0p_j}{q_kp_j}q_k=0 
\end{cases}\,.
\end{equation}
A possible solution for this system is simply obtained by squaring both equations
\begin{equation}
\begin{cases}
\biggl(p_0-\dfrac{p_0p_k}{q_jp_k}q_j-\dfrac{p_0p_j}{q_kp_j}q_k\biggr)^2=0 \\
\biggl(q_0+\dfrac{q_0p_k}{q_jp_k}q_j+\dfrac{q_0p_j}{q_kp_j}q_k\biggr)^2=0
\end{cases}
\Rightarrow
\begin{cases}
p_0 \cdot q_j =0 \,,\,\,\,p_0 \cdot q_k =0 \,,\,\,\,q_j \cdot q_k =0 \\
q_0 \cdot q_j =0 \,,\,\,\,q_0 \cdot q_k =0 \,,\,\,\,q_j \cdot q_k =0 \\
\end{cases}\,.
\end{equation}
Altogether, the four-vectors $q_0, q_j, q_k$ satisfy 
\begin{equation} \label{eq:ST_ss_qcondition}
q_0^2 = q_j^2 = q_k^2 =  q_0 p_0 = q_j p_j = q_k p_k = 0\,, 
\qquad 
 q_j p_0 = q_k p_0  = q_j q_0  = q_k q_0 = q_j q_k  = 0\,. 
\end{equation} 
 
\subsubsection{Recursion}

Consider an $n+1$ point amplitude $\MC_{n+1} \equiv \MC(1 \dots n 0)$. 
We denote $\hat \MC^{z,\epsilon}_{n+1}$ its shifted version, which is  $\MC_{n+1}$ evaluated for the ``hatted" kinematics in \eq{eq:ST_ss}. 
Similarly, we denote $\MC_{n} \equiv \MC(1 \dots n)$, and  its shifted version by $\hat \MC_n^{z,\epsilon}$. \\
Note that the trivial shift is $\hat \MC^{0,1} = \MC$.

At tree level $\hat \MC^{z,\epsilon}$ is a meromorphic function of $z$, \ie a function that is holomorphic on an open subset of the complex plane, except for a set of isolated points, which are poles of the function.
Thus, all the singularities of $\hat \MC^{z,\epsilon}$ are given by a finite set of simple poles at $z = z_i$.
As in \sect{sec:BCFW}, we may now apply the Cauchy formula, 
\begin{equation}\label{eq:ST_cauchy}
\hat \MC^{0,\epsilon}_{n+1} = -\sum_i \Res{z \to z_i}\dfrac{\hat \MC^{z,\epsilon}_{n+1}}{z}  + B_\infty \,,    
\end{equation}
where the boundary term $B_\infty$ is the residue at infinity. Once again, if $\hat \MC^{z,\epsilon}_{n+1}$ goes to zero as $z \to \infty$ then  $B_\infty = 0$. \\
Assuming the residue at infinity to be null or at least finite in the limit $\e \to 0$, the only poles of the integrand are at the solutions of
\begin{equation}
(\hat{p}_0(z)+p_{a_1}+\dots +p_{a_m})^2=0\,,
\end{equation}
for any non-empty subset $\{a_1, a_2, \dots,a_m \}$ of $\{1,2,\dots,\dots,n\}$. \\
As seen in \sect{sec:BCFW}, the residue at each of these poles is determined by unitarity to be the product of two lower
point amplitudes.
For a pole at say $z=z_m$, the vector $\hat{P}_{0,m}-m^2_{0,m}=\hat{p}_0+p_{a_1}+\dots + p_{a_m}$ is null by construction, and the residue is given by
\begin{equation}
\hat \MC^{0,\epsilon}_{n+1}=-\sum_{\mathcal I}\,\dfrac{\MC_{m+1}(\hat{p}_0,p_{a_1},\dots,p_{a_m},-\hat{P}_{0,m})\, \MC_{n-m+1}(\hat{P}_{0,m},\overline{p_{a_1},\dots,p_{a_m}})}{P^2_{0,m}-m^2_{0,m}}\,,
\end{equation}
where $\overline{a_1,\dots,a_m}$ is the complement of $a_1,\dots,a_m$, and we sum over all possible non-empty subsets $\mathcal{I}=\{a_1,\dots,a_m\}$ of $\{1,\dots,n\}$.\\
Let us now separate the terms in the sum into two groups. The first group is those where the
set $\mathcal{I}$ consists of a single element and the second group contains all other sets. The reason for this separation is that, as proven in Appendix A of \Refe{cachazo_strominger}, elements of the former becomes singular when $\l_0 \to 0$ while elements of the latter remains finite.\\
In conclusion, we only focus on a subset of the poles related to the emission of the particle-0 from an external leg of the $n$-point amplitude  $\MC_n$, as $1/\epsilon^m$ terms in $\hat \MC^{z,\epsilon}_{n+1}$ can only come from their residues. \\
There are $n$ such poles $z_l$ related to the solutions of the equations $\hat P_{0l}^2|_{z = z_l}= m_l^2$, where $\hat P_{0l} \equiv \hat p_0 + \hat p_l$  and  it is understood that $ \hat p_l = p_l$ for $l \neq j,k$.
The poles are located at:
\begin{align}
0=\hat{P}_{0l}^2\big|_{z=z_l}-m_l^2 &=(\hat{p}_0+\hat{p}_l)^2|_{z = z_l}-m_l^2=(\e p_0 -z_l q_0+\hat{p}_l)^2-m_l^2\,= \nn \\
&= \e^2p_0^2+z_l^2 q_0^2+\hat{p}_l^2+2\e p_0 \cdot \hat{p}_l-2\e z_l p_0 \cdot q_0 -2 z_l q_0 \cdot \hat{p}_l-m_l^2 \,
\end{align}
using now on-shellness and \eq{eq:ST_ss_qcondition}, we find
\begin{equation}\label{eq:ST_zl}   
z_l = \epsilon \frac{p_0 p_l }{ q_0 p_l} , 
\qquad l=1 \dots n\,,  
\end{equation} 
notice that this result holds true also for the shifted particles.
Furthermore, we have $\hat P_{0l}^2 -  m_l^2 = -2 q_0 p_l (z - z_l)$, \ie poles at $\hat P_{0l}^2 \to  m_l^2$ are in one-to-one correspondence with the poles at $z \to z_l$.

To sum up, we have:
\begin{subequations}
\begin{align}
&\hat \MC^{z,\epsilon}(1 \dots n 0)|_{\hat P_{0l}^2 \to m_l^2}  =
- \frac{\hat \MC^{z_l,\epsilon}(1 \dots P_{0l} \dots  n) \hat \MC^{z_l,\epsilon}((-P_{0l}) l 0 )}{\hat P_{0l}^2  -  m_l^2 }\, = \nn\\
&\hspace{4cm}=\frac{ \hat \MC^{z_l,\epsilon}(1 \dots P_{0l} \dots  n) \hat \MC^{z_l,\epsilon}((-P_{0l}) l 0)}{2 q_0 p_l (z - z_l) }\,,\\
&\dfrac{\Res{z \to z_l} \hat \MC^{z,\epsilon}(0 \dots n)  }{ z_l} 
 =  \frac{1 }{ \epsilon}\frac{ \hat \MC^{z_l,\epsilon}(1 \dots P_{0l} \dots  n) \hat \MC^{z_l,\epsilon}((-P_{0l}) l 0 )  }{ 
2 p_0 p_l } \,. 
\end{align} 
\end{subequations}
 
Plugging this back into the Cauchy formula in \eq{eq:ST_cauchy} one obtains the soft recursion: 
\begin{equation}\label{eq:ST_recursion}
\hat \MC^{0,\epsilon}(1 \dots n 0) =  - \frac{1 }{ \epsilon } \sum_{l = 1}^n 
\frac{ \hat \MC^{z_l,\epsilon}(1 \dots P_{0l} \dots  n) \hat \MC^{z_l,\epsilon}( (-P_{0l}) l  0 )  }{ 2 p_0 p_l }  + \ \mathcal{O}(\epsilon^0). 
\end{equation}
As a final remark, notice that the boundary term in \eq{eq:ST_cauchy} is absorbed in $\mathcal{O}(\epsilon^0)$, as it cannot produce singular terms in the $\epsilon \to 0$ limit.

\subsubsection{Spinor shift}
To compute the desired amplitudes, we find it convenient to use the spinor-helicity formalism. Accordingly, we will now focus on implementing the framework discussed in the previous sections at the spinorial level.

We begin by discussing spinor shifts that realize the momentum shift in \eq{eq:ST_ss}. 
The discussion varies depending on whether the shifted particles $j$ and $k$ are massive or massless.
In our case $m_j > 0$ and  $m_k > 0$; the other possible cases are discusssed in \Refe{Falkowski_2021}.\\
One possible choice of the shift vectors $q$ in \eq{eq:ST_ss} is 
\footnote{In the following, the spinor Lorentz indices will be often left implicit to keep the notation concise.}:
\begin{equation}\label{eq:ST_0mmbar_qsigma}
q_0 \s = y \tl_0\,, \qquad 
q_j \s = p_j \s \tl_0 \tl_0\,, \qquad 
q_k \s = p_k \s \tl_0 \tl_0\,,   
\end{equation}
where $y$ is an arbitrary spinor satisfying $\wa{y}{0}\neq 0$.
It is easy to check that the constraints in \eq{eq:ST_ss_qcondition} are satisfied by this choice. \\
We may now decompose also the shifted momenta into spinors: 
\begin{equation}
\hat p_0 \sigma = \lambda_0^z \tilde \lambda_0\,, \qquad   
\hat p_j \sigma =\chi_j^J \tilde \chi_{j \, J}^z \,, \qquad
\hat p_k \sigma =\chi_k^J \tilde \chi_{k \, J}^z \,.
\end{equation}
At the spinor level, a momentum shift can be achieved by shifting only the holomorphic (or anti-holomorphic) spinor while keeping the other spinor fixed. \\
In particular I will use the $\{0\bar{m}\bar{m}\}$ shift, defined as:
\begin{align}\label{eq:ST_0mmbar_spinor}
{\rm \bf \{ 0 \bar m \bar m \} }: \qquad  
\l_{0}^z & = \e  \l_0 - z  y \,, \nn \\
\tx_j^z & = \tx_j + \frac{(\e -1)\wab{0}{p_k}{0}- z \wab{y}{p_k}{0}}{\wbb{0}{p_j}{p_k}{0}} \wb{\bf j}{0} \tl_0 \,,  \\
\tx_k^z & = \tx_k - \frac{(\e -1)\wab{0}{p_j}{0}- z \wab{y}{p_j}{0}}{\wbb{0}{p_j}{p_k}{0}} \wb{\bf k}{0} \tl_0 \,, \nn
\end{align}
with all other spinors kept unshifted. 

A crucial point that will become more relevant in the following sections, is that this shift is non-trivial for $z = \epsilon = 0$: 
in this limit the spinors $\tx_{j/k}$ must shift to absorb the original momentum of the particle $0$. 
On the other hand, the shift is trivial for  $z =0$ and  $\e = 1$, but this is not the soft limit $\e \to 0$. \\

For the remainder of this section, I will derive some identities that will be useful in later computations.\\
For the $\{ 0 \bar m \bar m \}$ shift,  the location of the poles in \eq{eq:ST_zl} can be recast as:  
\begin{equation}
z_l =  \e \frac{ \wab{0}{p_l}{0}}{ \wab{y}{p_l}{0}} \, ; \qquad 
\text {which, for $m_l = 0$, reduces to:} \quad 
z_l = \e \frac{ \wa{0}{l}}{\wa{y}{l}} \,. 
\end{equation}
At the poles, the shifted spinors can be expressed as 
\begin{align}\label{eq:ST_0mmbar_uf}
\l_0^{z_l} & = 
\e \frac{\wa{y}{0}}{\wab{y}{p_l}{0}} p_l \s \tl_0 \,,  \nn\\ 
\tx_j^{z_l} & = 
\tx_j -\frac{2 p_0 p_k }{ \wbb{0}{p_j}{p_k}{0}} \wb{\bf j}{0}\tl_0 
+ \e \frac{\wa{y}{0} \wbb{0}{p_k}{p_l}{0}}{ \wab{y}{p_l}{0} \wbb{0}{p_j}{p_k}{0}} \wb{\bf j}{0}\tl_0 \, \\
\tx_k^{z_l} & =
\tx_k +\frac{2 p_0 p_j}{\wbb{0}{p_j}{p_k}{0}}\wb{\bf k}{0} \tl_0
- \e \frac{ \wa{y}{0} \wbb{0}{p_j}{p_l}{0}}{\wab{y}{p_l}{0}\wbb{0}{p_j}{p_k}{0}} \wb{\bf k}{0} \tl_0 \,. \nn
\end{align} 
These identities allow us to rewrite $\hat P_{0l}$ at the poles as:
\begin{equation}
\text{for $l \neq j,k$} \quad \rightarrow \quad 
\hat P_{0l} \s |_{z_l} = \hat p_0 \s |_{z_l} + p_l \s |_{z_l} = \l_0^{z_l} \tl_0 + p_l \s=
\e \frac{\wa{y}{0}}{\wab{y}{p_l}{0}}p_l \s \tl_0 \tl_0 + p_l\s \,.
\end{equation}
Furthermore, since $\hat P_{0l}|_{z=z_l}$ is on shell by definition, we may also factorize it into spinors: 
\begin{equation}
l\neq j,k\,:
\begin{cases}
\hat P_{0l} \s|_{z_l} = \x_{0l}^L \tx_{0l \, L}, \quad \x_{0l} = \x_l \,, \quad \tx_{0l} = \tx_l +\e \dfrac{\wa{y}{0}}{\wab{y}{p_l}{0}}\wb{\bf l}{0} \tl_0 \,, \quad \text{if} \,\, m_l>0  \\
\hat P_{0l} \s|_{z_l} = \l_{0l} \tl_{0l}\,, \quad \,\,\,\l_{0l} = \l_l\,, \quad \tl_{0l} = \tl_l + \e \dfrac{\wa{y}{0}}{\wa{y}{l}} \tl_0 
\,, \qquad \quad \,\text{if} \,\, m_l=0 
\end{cases}
\end{equation}
Similarly at $z_j$ and $z_k$ we have the decomposition 
\begin{equation}\label{eq:no_epsilon}
\begin{aligned}
\hat P_{0j} \s|_{z_j} &= 
p_j \s - \frac{ 2 p_0 p_k}{\wbb{0}{p_j}{p_k}{0}} p_j \s \tl_0 \tl_0 \, \Rightarrow \, \x_{0j}  =  \x_j \,, \quad \tx_{0j} =\tx_j - \frac{2 p_0 p_k}{\wbb{0}{p_j}{p_k}{0}} \wb{\bf j}{0} \tl_0\,,\\
\hat P_{0k} \s|_{z_k} &= 
p_k \s + \frac{2 p_0 p_j}{\wbb{0}{p_j}{p_k}{0}} p_k \s \tl_0 \tl_0 
\, \Rightarrow \,\x_{0k} = \x_k \,, \quad \tx_{0k} = \tx_k + \frac{2 p_0 p_j}{\wbb{0}{p_j}{p_k}{0}} \wb{\bf k}{0} \tl_0 \,. 
\end{aligned} 
\end{equation}
Note that $\hat P_{0j} \s|_{z_j}$ and $\hat P_{0k} \s|_{z_k}$, and similarly the shifts $\tx_j^{z_k}$ and $\tx_k^{z_j}$, are independent of $\e$.

%-------------
\subsection{Soft Theorems}\label{sec:st}
We are now ready to apply the recursion formula in \eq{eq:ST_recursion} to derive the soft factors for amplitudes involving particles of any mass and spin, with the graviton as the soft particle.
I will restrict the analysis to the case of single soft particle emission.

Consider a gravity theory where the massless spin-2 graviton is minimally coupled to matter and to itself.\\
In the on-shell formalism, the minimal coupling of the graviton to two matter particle of same mass $m>0$ and spin $s$ corresponds to the on-shell 3-point amplitudes in \eq{eq:Minimal_Graviton_3pt}:
\begin{equation} \label{eq:GR_Mxxh_spinS_minimal}
\MC(p_1,-p_2,k_3^-)= -\frac{\k}{2} \frac{\wab{3}{p_1}{\z}^2}{\wb{3}{\z}^2} \frac{\wb{\bf 2}{\bf 1}^{2s}}{m^{2s}}\,, \qquad
\MC(p_1,-p_2,k_3^+) = -\frac{\k}{2} \frac{\wab{\z}{p_1}{3}^2}{\wb{3}{\z}^2} \frac{\wa{\bf 2}{\bf 1}^{2s}}{m^{2s}}\,, 
\end{equation}
whereas the pure gravity on-shell 3-point amplitudes correspond to \eq{eq:ampiezze_3_gravitoni}:
\begin{equation}\label{eq:GR_Mhhh}
\MC(k_1^-, -k_2^-, k_3^+) = 
-\frac{\k}{2}\frac{\wa{1}{2}^6}{\wa{1}{3}^2 \wa{2}{3}^2 }\,, \qquad 
\MC(k_1^+, -k_2^+, k_3^-) =   
-\frac{\k}{2}\frac{\wb{1}{2}^6}{\wb{1}{3}^2 \wb{2}{3}^2}\,. 
\end{equation} 
Without loss of generality, we may focus on the emission of a soft graviton with plus-helicity 
\footnote{The minus-helicity and plus-helicity cases do not interfere with each other. Moreover, the minus-helicity case can be easily obtained from the plus-helicity one by applying a parity transformation to the amplitude}. 
Thus, the shifted three-point amplitude computed at the $z_l$-poles is:  
\begin{equation}
\label{eq:ST_hatM_minimal}
\hat \MC^{z_l,\e} ( (-\hat{P}_{0l}) l 0^+_h )=  
-\frac{\k}{2} \,\frac{\wab{y}{p_l}{0}^2}{\e^2 \wa{0}{y}^2}\,. 
\end{equation}
Plugging this back into \eq{eq:ST_recursion} we obtain:   
\begin{equation}\label{eq:ST_softGravity}
\hat \MC^{0,\e}(1 \dots n 0^+_h) = \frac{\k}{2}\,\frac{1}{\e^3} \sum_{l=1}^n \frac{\wab{y}{p_l}{0}^2}{ (2 p_0 p_l) \wa{0}{y}^2}  \hat \MC^{z_l,\e}(1 \dots \hat{P}_{0l} \dots  n)+ \mathcal{O}(\e^0)\,. 
\end{equation} 

To proceed we need to expand the $n$-point amplitude above in powers of $\e$. \\
As shown in \app{app:taylor_expansion}, we have
\begin{equation}\label{eq:ST_epsilonExpansion}
\hat\MC^{z_l,\e}(1 \dots \hat{P}_{0l} \dots n)=
\left\{ 1+ \e \dfrac{\wa{0}{y}}{\wab{y}{p_l}{0}} \tD_l+
\frac{\e^2}{2}\dfrac{\wa{0}{y}^2}{\wab{y}{p_l}{0}^2} \tD_l^2
\right\}\hat\MC^{0,0}(1\dots n) +\mathcal{O}(\e^3) \,,
\end{equation}
where the differential operator $\tilde {\cal D}_l$ is 
\begin{equation}\label{eq:ST_calDl}
\tD_l = \wb{0}{\bf l}\wb{0}{\partial_l} 
+\dfrac{\wbb{0}{p_k}{p_l}{0}}{\wbb{0}{p_j}{p_k}{0}}
\wb{0}{\bf j}\wb{0}{\partial_j} 
-\dfrac{\wbb{0}{p_j}{p_l}{0}}{\wbb{0}{p_j}{p_k}{0}}
\wb{0}{\bf k}\wb{0}{\partial_k}\,,
 \qquad  {\rm for}\,\,l \neq j,k, 
\end{equation} 
and $\tilde {\cal D }_l = 0$ for $l = j,k$ 
\footnote{The operator $\tD_l$ can be related to the angular momentum operator $J^{\mu \nu}_l$}.

Plugging the expanded $\hat \MC^{z_l,\epsilon}$  back into \eq{eq:ST_softGravity} we obtain the soft theorem in the standard form:  
\begin{equation}\label{eq:ST_softTheoremGravity}
\hat \MC^{0,\e}(1 \dots n 0_h^+ )  = \bigg \{ \frac{1}{\e^3} S_{+2}^{(0)} + \frac{1}{\e^2} S_{+2}^{(1)}+ \frac{1}{\e} S_{+2}^{(0)}\bigg \} \hat \MC^{0,0}(1 \dots  n) + \mathcal{O}(\e^0) \,,  
\end{equation} 
where we have defined
\begin{equation}\label{eq:fattori_soffici}
S^{(0)}_{+2}= \frac{\k}{2}\sum_{l=1}^n 
\dfrac{\wab{y}{p_l}{0}^2}{(2 p_0\cdot p_l) \wa{0}{y}^2  }, \quad 
S^{(1)}_{+2}= \frac{\k}{2}\sum_{l=1}^n 
\dfrac{\wab{y}{p_l}{0}}{ (2 p_0\cdot p_l) \wa{0}{y}} \tD_l, \quad
S^{(2)}_{+2}= \frac{\k}{4}\sum_{l=1}^n 
\frac{1}{ 4 p_0\cdot p_l} \tD_l \tD_l \,. 
\end{equation}
It is important to remark that the soft factors do not depend on $y$ (either as a consequence of momentum conservation or of angular momentum conservation), see \Refe{Falkowski_2021}. \\
The soft factors above are the natural extension to the massive case of the ones presented in \Refe{cachazo_strominger} and \Refe{Elvang_2017}, moreover $S^{(0)}_{+2}$ is nothing but the Weinberg soft factor.

Finally, notice that for the emission of a minus-helicity graviton the computation is the same, and the final result is related to the former by a parity transformation, see \Refe{Falkowski_2021}.

\subsubsection{Soft exponentiation}\label{sec:exp}

I now discuss the exponential representation of the soft theorems.\\
\Refe{He_2014} showed that the soft theorem for gravitons and gluons could be written in terms of an exponential operator acting on lower point amplitudes. 
Furthermore, in \Refe{Falkowski_2021}, it was shown that the exponential operator can be extended for amplitudes with
massive particles of any spin and with a soft particle of any integer helicity $|h|\leq 2$. 
   
The Taylor expansion in \eq{eq:ST_epsilonExpansion} has been trucated at second order, nonetheless we may consider the complete series. 
It is straightforward to check that it can be recast in the exponential form: 
\begin{equation}
\hat \MC^{z_l,\epsilon}(1 \dots \hat{P}_{0l} \dots  n)   = \exp\left(\epsilon   \frac{ \wa{0}{y}}{\wab{y}{p_l}{0}} \tD_l  \right)
 \hat \MC^{0,0}(1 \dots  n) +  \mathcal{O}(\epsilon^3) .
\end{equation} 
Substituting now in \eq{eq:ST_softTheoremGravity}, we obtain:
\begin{align}\label{eq:soft_theorem_falkowski}
\hat \MC^{0,\e}(1 \dots n 0^+_h) 
&= \frac{\k}{2}\,\frac{1}{\e^3} \sum_{l=1}^n \frac{\wab{y}{p_l}{0}^2}{ (2 p_0 p_l) \wa{0}{y}^2}  \hat \MC^{z_l,\e}(1 \dots \hat{P}_{0l} \dots  n)+ \mathcal{O}(\e^0) = \nn\\
&= \frac{\k}{2}\,\frac{1}{\e^3} \sum_{l=1}^n \frac{\wab{y}{p_l}{0}^2}{ (2 p_0 p_l) \wa{0}{y}^2} \exp\left(\epsilon   \frac{ \wa{0}{y}}{\wab{y}{p_l}{0}} \tD_l  \right)
 \hat \MC^{0,0}(1 \dots  n) + \mathcal{O}(\e^0) \,, 
\end{align}
where the differential operator $\tD_l$ is given in \eq{eq:ST_calDl}.

\subsubsection{Consistency checks}\label{sec:consistency_checks}

I conclude this discussion on the soft theorems by verifying that the formulae derived in the previous sections using the spinor-helicity formalism are consistent with those found in the literature, which were obtained through different approaches.

As seen in \Refe{Falkowski_2021}, the Taylor expansion around $\e \to 0$ of the minimally coupled gravity amplitude in the soft limit can be recast in an exponential form \eq{eq:soft_theorem_falkowski}:
\begin{flalign}\label{eq:forma_esp1}
\hat \MC^{0,\e}(1 \dots n 0^+_h)& = \frac{\k}{2}\,\frac{1}{\e^3} \sum_{l=1}^n \frac{\wab{y}{p_l}{0}^2}{ (2 p_0 p_l) \wa{0}{y}^2} \exp\left(\epsilon   \frac{ \wa{0}{y}}{\wab{y}{p_l}{0}} \tD_l  \right) \hat \MC^{0,0}(1 \dots  n) + \mathcal{O}(\e^0) = \nn\\
& \hspace{-1.5cm}= \frac{\k}{2}\,\frac{1}{\e^3} \sum_{l=1}^n \frac{\wab{y}{p_l}{0}^2}{(2 p_0 p_l) \wa{0}{y}^2}\exp\Biggl\{ \epsilon   \frac{ \wa{0}{y}}{\wab{y}{p_l}{0}} \,\times  \\
&\times \left( \wb{0}{\bf l}\wb{0}{\partial_l} 
+\dfrac{\wbb{0}{p_k}{p_l}{0}}{\wbb{0}{p_j}{p_k}{0}}
\wb{0}{\bf j}\wb{0}{\partial_j} 
-\dfrac{\wbb{0}{p_j}{p_l}{0}}{\wbb{0}{p_j}{p_k}{0}}
\wb{0}{\bf k}\wb{0}{\partial_k}
\right) \Biggr\}\hat \MC^{0,0}(1 \dots  n) + \mathcal{O}(\e^0) \,.\nn
\end{flalign} 

We are now interested in comparing this result with the one originally obtained by Cachazo and Strominger in \Refe{cachazo_strominger} and analyzed comprehensively by Guevara et all in \Refe{Guevara_2019}:
\begin{equation}\label{eq:cachazostrominger2}
   {\cal M}_{n+1} = \frac{\k}{2}\sum_{i=1}^n
   \left[ \frac{(p_i\cdot\varepsilon)^2}{p_i\cdot k}
        +i\frac{(p_i\cdot\varepsilon)
                (k_\mu \varepsilon_\nu J_i^{\mu\nu})}{p_i\cdot k}
        - \frac{1}{2}\frac{(k_\mu \varepsilon_\nu J_i^{\mu\nu})^2}
                          {p_i\cdot k}
   \right] {\cal M}_n + {\cal O}(k^2) \,.
\end{equation}
Here the soft momentum $k$ corresponds to the external soft
graviton, and we have constructed its polarization tensor as 
$\ve_{\m\n}=\ve_\m \ve_\n$. \\
In \Refe{Guevara_2019} it was shown that for the three- and four- point amplitudes we are able to recast the term in square bracket in an exponential form. Following this intuition, we may write
\begin{flalign}\label{eq:soft_theorem_2}
{\cal M}_{n+1}(1\dots n, 0^+)=\frac{\k}{2}
\sum_{i=1}^n\frac{\left(p_i\cdot \varepsilon \right)^2}{p_i\cdot k}
\exp\left( i\frac{k_\mu\varepsilon_\nu J^{\mu\nu}_i}{p_i\cdot\varepsilon}\right)
{\cal M}_n(1\dots n) \,.
\end{flalign}

We now recall that $\hat{\cal M}=\hat{\cal M}^{(0,1)}$. 
Thus, to make contact with \eq{eq:forma_esp1} we generalize to the case of arbitrary $\e$ shift. 
To this end, we perform the $\{ 0, \bar{m}, \bar{m}\}$ shift in \eq{eq:ST_0mmbar_spinor}, where we choose once again the shifted spinors to be the one associated with the soft graviton ``0", and the massive spinors $\textbf{j}$, $\textbf{k}$, 
and expand around $\epsilon \rightarrow 0$:
\begin{align}
&\MC^{(0,\e)}_n(1\dots n)=
\MC^{(0, \e)}_n(1\dots\hat{\bf j}\dots \hat{\bf k} \dots n)\,, 
\qquad 
\MC^{(0,\e)}_n=\sum_{m=0}^\infty \dfrac{\e^m}{m!}\bigg(\de{\e}\bigg)^{\!\!m}\MC^{(0,\e)}_n\Big|_{\e=0}\,, \\
& \de{\e}\MC_n^{(0,\e)}=
\left(\dee{\tx_j}{\e}\de{\tx_j}+\dee{\tx_k}{\e}\de{\tx_k}\right)\MC_n^{(0,\e)}= \nn\\
& \qquad\qquad = \left(\frac{\wa{0}{y}\wbb{0}{p_k}{p_l}{0}}{\wab{y}{p_l}{0}\wbb{0}{p_j}{p_k}{0}}\wb{0}{\bf j}\wb{0}{\de{\tx_j}}-
\frac{\wa{0}{y}\wbb{0}{p_j}{p_l}{0}}{\wab{y}{p_l}{0}\wbb{0}{p_j}{p_k}{0}}\wb{0}{\bf k}\wb{0}{\de{\tx_k}}\right)\MC_n^{(0,\e)} \,.
\end{align}
Altogether
\begin{align}\label{eq:consistenza}
&\MC_{n+1}^{(0,\e)}(1\dots n, 0^+)=
\frac{\k}{2}\sum_{i=1}^n\frac{1}{\e^3}\frac{\left(p_i\cdot \varepsilon \right)^2}{p_i\cdot k}\exp\left( i\frac{k_\mu\varepsilon_\nu J^{\mu\nu}_i}{p_i\cdot\ve}\right) 
\left(\sum_{m=0}^\infty \dfrac{\e^m}{m!}\bigg(\de{\e}\bigg)^{\!\!m}\MC^{(0,\e)}_n\Big|_{\e=0}\right)= \nn \\
&\qquad =\frac{\k}{2}\sum_{i=1}^n\frac{1}{\e^3}\frac{\left(p_i\cdot \varepsilon \right)^2}{p_i\cdot k}\exp\left( i\frac{k_\mu\ve_\nu J^{\mu\nu}_i}{p_i\cdot\varepsilon}\right)\,\times \nn\\
& \qquad \quad \times \left(\sum_{m=0}^\infty \dfrac{\e^m}{m!}\left(\frac{\wa{0}{y}\wbb{0}{p_k}{p_l}{0}}{\wab{y}{p_l}{0}\wbb{0}{p_j}{p_k}{0}}\wb{0}{\bf j}\wb{0}{\de{\tx_j}}-
\frac{\wa{0}{y}\wbb{0}{p_j}{p_l}{0}}{\wab{y}{p_l}{0}\wbb{0}{p_j}{p_k}{0}}\wb{0}{\bf k}\wb{0}{\de{\tx_k}}\right)^{\!\!m}\MC^{(0,\e)}_n\Big|_{\e=0}\right)= \nn\\
&\qquad =\frac{\k}{2}\sum_{i=1}^n\frac{1}{\e^3}\frac{\left(p_i\cdot \varepsilon \right)^2}{p_i\cdot k}\exp\left(\e \frac{\wa{0}{y}}{\wab{y}{p_l}{0}}\wb{0}{\bf l}\wb{0}{\de{\tx_l}}\right) \, \times \nn\\
& \qquad \quad \times \exp\left(\frac{\wa{0}{y}\wbb{0}{p_k}{p_l}{0}}{\wab{y}{p_l}{0}\wbb{0}{p_j}{p_k}{0}}\wb{0}{\bf j}\wb{0}{\de{\tx_j}}-
\frac{\wa{0}{y}\wbb{0}{p_j}{p_l}{0}}{\wab{y}{p_l}{0}\wbb{0}{p_j}{p_k}{0}}\wb{0}{\bf k}\wb{0}{\de{\tx_k}}\right)\MC^{(0,\e)}_n\Big|_{\e=0}=\nn\\
&\qquad =\frac{\k}{2}\sum_{i=1}^n\frac{1}{\e^3}\frac{\left(p_i\cdot \varepsilon \right)^2}{p_i\cdot k}
\exp\left(\e \frac{\wa{0}{y}}{\wab{y}{p_l}{0}}\tD_l \right)\MC^{(0,0)}_n\,,
\end{align}
where we have used the parity reversed form of \eq{eq:angular_2} to go from the second to the third line.\\
From \eq{eq:consistenza} it is clear that the two formulas match perfectly, confirming the consistency of our procedure. Moreover, by recalling that the soft expansion was performed by defining the parameter $\e$ via the rescaling $p^\m=\e p^\m$, or in spinor variables $\l_0=\e\l_0$, we can fully absorb the $\e$ in this formula. As a result, the two outcomes are exactly the same, further validating our consistency check.

%%%%%%%%%%%%%%%%%%%%%%%%%%%%%%%%%%%%%%%%%%%%%%%%
%----------------------------------------------%
%%%%%%%%%%%%%%%%%%%%%%%%%%%%%%%%%%%%%%%%%%%%%%%%

\clearpage

\section{Gravitational Compton Amplitude from the Soft Theorems}\label{sec:cap5}

In this section I will apply the soft theorems discussed in \sect{sec:cap4} to the three-point amplitudes obtained in \sect{sec:cap3} to obtain the four-point Compton gravitational amplitude.\\
As seen in \sect{sec:consistency_checks}, our soft theorems can be formulated using \eq{eq:forma_esp1} or \eq{eq:soft_theorem_2} equivalently. 
In this section, I will use the latter formulation as most of the results found in literature are expressed in this way.
Furthermore as I'm not interested in this section to explore the soft limit of the four-point amplitude, but rather to use the soft theorems to obtain this higher-point amplitude, I find it more convenient to use directly \eq{eq:soft_theorem_2}, rather than computing \eq{eq:forma_esp1} and taking the $\e \to 1$ limit.\\
I will start by discussing the spinless Schwarzschild case and later generalize also to the spinning Kerr case.

\subsection{Compton Amplitude for Schwarzschild Black Holes}\label{sec:compton_sc}

Our starting point are the three-point amplitudes \eq{eq:exp3p}, that I rewrite below for convenience,
\begin{equation}
{\cal M}_3^{(s)}(p_1,p_2,k^+_3)
= \frac{\k}{2}\frac{\wa{\bf1}{\bf2}^{2s} x_3^{2}}{m^{2s-2}} , \qquad \quad
{\cal M}_3^{(s)}(p_1,p_2,k^-_3)
= \frac{\k}{2}\frac{\wb{\bf1}{\bf2}^{2s} x_3^{-2}}{m^{2s-2}} \,.    
\end{equation}
As I'm now interested the Schwarzscild case, $s=0$; hence, after expliciting the $x$-factors, these amplitudes reduce to: 
\begin{equation}
{\cal M}_3^{(0)}(p_1,p_2,k^+_3)
= \k (p_1 \cdot \ve_3^+)^2 \, , \qquad \quad
{\cal M}_3^{(0)}(p_1,p_2,k^-_3)
= \k (p_1 \cdot \ve_3^-)^2  \,.    
\end{equation}

Let us start by considering the plus-helicity case.\\
To get the corresponding four-point amplitude we may now use \eq{eq:soft_theorem_2}:
\begin{flalign}\label{eq:M_3_0_a}
{\cal M}_{4}^{(0)}(p_1, p_2, k_3^+, k_4^+)=\frac{\k}{2}
\sum_{i=1}^3\frac{\left(p_i\cdot \ve_4^+ \right)^2}{p_i\cdot k_4}
\exp\left( i\frac{k_{4,\mu}\ve_{4,\nu}^+ J^{\mu\nu}_i}{p_i\cdot\ve_4^+}\right)
{\cal M}_3^{(0)}(p_1, p_2, k_3^+) \,.
\end{flalign}
This result is gauge-independent, therefore we can make a convenient gauge choice to simplify the computations. Nonetheless, to make the final result general, we must express it too in a gauge-independent way. \\
A convenient gauge choice is the one in which we fix the reference vectors for the massless particles labeled $3$ and $4$ to be $\ve_3=\ve(3,4)$, $\ve_4=\ve(4,3)$, this way all products $p_i\cdot \ve_j=0$, for $i,j=3,4$,
and the sum in \eq{eq:M_3_0_a} reduces to:
\begin{flalign}\label{eq:M_3_0_b}
&{\cal M}_{4}^{(0)}(p_1, p_2, k_3^+, k_4^+)= \\
& \qquad =\frac{\k}{2}\left[
\frac{\left(p_1\cdot \ve_4^+ \right)^2}{p_1\cdot k_4}
\exp\left( i\frac{k_{4,\mu}\ve_{4,\nu}^+ J^{\mu\nu}_1}{p_1\cdot\ve_4^+}\right)+\frac{\left(p_2\cdot \ve_4^+ \right)^2}{p_2\cdot k_4}
\exp\left( i\frac{k_{4,\mu}\ve_{4,\nu}^+ J^{\mu\nu}_2}{p_2\cdot\ve_4^+}\right)\right]
{\cal M}_3^{(0)}(p_1, p_2, k_3^+) \,. \nn
\end{flalign}
Notice now that, as we are studying the scalar case, whenever the arguments of the exponential operators in \eq{eq:M_3_0_b} act on the three-point scalar amplitudes $\MC_3^{(0)}$ they vanish. Thus:
\begin{flalign}\label{eq:M_3_0_c}
{\cal M}_{4}^{(0)}(p_1, p_2, k_3^+, k_4^+) & = \frac{\k}{2}\left[
\frac{\left(p_1\cdot \ve_4^+ \right)^2}{p_1\cdot k_4}
+\frac{\left(p_2\cdot \ve_4^+ \right)^2}{p_2\cdot k_4}\right]
{\cal M}_3^{(0)}(p_1, p_2, k_3^+)= \nn \\
& = \left(\dfrac{\k}{2}\right)^{\!2}\left[
\frac{\left(p_1\cdot \ve_4^+ \right)^2}{p_1\cdot k_4}
+\frac{\left(p_2\cdot \ve_4^+ \right)^2}{p_2\cdot k_4}\right]
2(p_1 \cdot \ve_3^+)^2 \,.  
\end{flalign}
Furthermore, from the three-point amplitude, we inherit the relation $p_1\cdot k_3=0$ 
\footnote{Note that this relation, in the context of BCFW, naturally arises when considering the $t$-channel residue within the on-shell formalism. Furthermore, it is important to emphasize that a consistent description of spin-induced multipoles for a pointlike particle must be formulated in the particle’s rest frame. This relation aligns perfectly with that requirement.}. 
Using total momentum conservation, and the relations previously discussed we are able to rewrite \eq{eq:M_3_0_c} as:
\begin{align}
{\cal M}_{4}^{(0)}(p_1, p_2, k_3^+, k_4^+) & = 
\left(\dfrac{\k}{2}\right)^{\!2}\left[
\frac{\left(p_1\cdot \ve_4^+ \right)^2}{p_1\cdot k_4}
+\frac{\left((p_1+k_3+k_4)\cdot \ve_4^+ \right)^2}{p_2\cdot k_4}\right]
2(p_1 \cdot \ve_3^+)^2 = \nn \\
&= \dfrac{\k^2}{2}\left(p_1\cdot \ve_3^+ \right)^2\left(p_1\cdot \ve_4^+ \right)^2 \left[
\dfrac{1}{p_1\cdot k_4}+\dfrac{1}{p_2 \cdot k_4}\right] = \\
&= -\dfrac{\k^2}{2}\dfrac{\left(p_1\cdot \ve_3^+ \right)^2\left(p_1\cdot \ve_4^+ \right)^2(k_3 \cdot k_4)^2}{(p_1\cdot k_4)(p_2 \cdot k_4)(k_3 \cdot k_4)} \,. \nn  
\end{align}
We remark that this expression is still gauge-invariant, and it is possible to rewrite it in a manifestly gauge invariant way if we notice that
\footnote{This relation is easily proven using the fact that in our specific gauge choice, $p_1 \cdot k_3=0$ and $p_i\cdot \ve_j=0$, for $i,j=3,4$.}:
\begin{equation}
p_1 \cdot F_3 \cdot F_4 \cdot p_1= -\left(p_1\cdot \ve_3^+ \right)\left(p_1\cdot \ve_4^+ \right)(k_3 \cdot k_4) \,,
\end{equation}
where $F_{\m\n}$ is the field strength $F_{\m\n}=k_\m\ve_\n-k_\n\ve_\m$. 
Therefore,
\begin{align}\label{eq:M_3_0_fin}
{\cal M}_{4}^{(0)}(p_1, p_2, k_3^+, k_4^+) = 
 \dfrac{-\k^2\omega_{(0)}^2}{(2p_1\cdot k_4)(2p_2 \cdot k_4)(2k_3 \cdot k_4)} \,,
\end{align}
where we have defined the quantity $\omega_{(0)}\equiv -2 \, p_1 \cdot F_3 \cdot F_4 \cdot p_1 $. \\

The computation for the minus-helicity case is exactly the same, the only difference being the helicity of graviton-$3$ in $\omega_{(0)}$. \\
The remaining helicity configurations, \ie when the graviton-$4$ has negative helicity, are easily obtained from \eq{eq:M_3_0_fin} using parity and charge conjugation transformations.

This result is in perfect agreement with the ones found in the literature,
\eg see \Refe{bautista2019}, \Refe{bautista2023scatteringblackholebackgrounds}.

\subsection{Compton Amplitude for Kerr Black Holes}\label{sec:aa}
In this section I generalize the discussion in \sect{sec:compton_sc} to the case of spinning black holes.\\
Once again our starting point are the amplitudes
\begin{equation}
{\cal M}_3^{(s)}(p_1,p_2,k^+_3)
= \frac{\k}{2}\frac{\wa{\bf1}{\bf2}^{2s} x_3^{2}}{m^{2s-2}} , \qquad \quad
{\cal M}_3^{(s)}(p_1,p_2,k^-_3)
= \frac{\k}{2}\frac{\wb{\bf1}{\bf2}^{2s} x_3^{-2}}{m^{2s-2}} \,.    
\end{equation}
In this form the angular momentum dependence of the amplitude is hidden inside the spinors, however it is convenient to make it manifest:
\begin{subequations}
\begin{align}
{\cal M}_3^{(s)}(p_1,p_2,k^+_3) &=
\dfrac{\k}{m^{2s}} (p_1 \cdot \ve_3^+)^2 \qlb{\bf 1}^{2s} \exp\left(i\dfrac{k_{3,\m}\ve_{3,\n}^+J_1^{\m\n}}{p_1\cdot\ve_3^+}\right)\qrb{\bf 2}^{2s} \,, \\    
{\cal M}_3^{(s)}(p_1,p_2,k^-_3) &=
\dfrac{\k}{m^{2s}} (p_1 \cdot \ve_3^-)^2 \qla{\bf 1}^{2s} \exp\left(i\dfrac{k_{3,\m}\ve_{3,\n}^-J_1^{\m\n}}{p_1\cdot\ve_3^-}\right)\qra{\bf 2}^{2s} \,,
\end{align}   
\end{subequations}
where I have used the fact that the action of the operator: $\exp\left(i\frac{k_{3,\m}\ve_{3,\n}^+J^{\m\n}}{p_1\cdot\ve_3^+}\right)$, on either the spinors $\qla{\bf 1}^{2s}$ ($\qlb{\bf 1}^{2s}$)
or $\qra{\bf 2}^{2s}$ ($\qrb{\bf 2}^{2s}$) gives the same result, to pick without loss of generality $J\to J_1$ 
\footnote{For a rigorous proof of the above statement see section 2.3 of \Refe{Guevara_2019}}.

Let us start by considering the plus-helicity case.\\
To get the corresponding four-point amplitude we may use once again \eq{eq:soft_theorem_2}:
\begin{flalign}\label{eq:M_3_s_a}
{\cal M}_{4}^{(s)}(p_1, p_2, k_3^+, k_4^+)=\frac{\k}{2}
\sum_{i=1}^3\frac{\left(p_i\cdot \ve_4^+ \right)^2}{p_i\cdot k_4}
\exp\left( i\frac{k_{4,\mu}\ve_{4,\nu}^+ J^{\mu\nu}_i}{p_i\cdot\ve_4^+}\right)
{\cal M}_3^{(s)}(p_1, p_2, k_3^+) \,.
\end{flalign}
Also this time, it is convenient to pick $\ve_3=\ve(3,4)$ and $\ve_4=\ve(4,3)$, this way all products $p_i\cdot \ve_j=0$, for $i,j=3,4$ and the last term of the sum drops out. Hence,
\begin{flalign}\label{eq:M_3_s_b}
&{\cal M}_{4}^{(s)}(p_1, p_2, k_3^+, k_4^+)= \\
& \qquad =\frac{\k}{2}\left[
\frac{\left(p_1\cdot \ve_4^+ \right)^2}{p_1\cdot k_4}
\exp\left( i\frac{k_{4,\mu}\ve_{4,\nu}^+ J^{\mu\nu}_1}{p_1\cdot\ve_4^+}\right)+\frac{\left(p_2\cdot \ve_4^+ \right)^2}{p_2\cdot k_4}
\exp\left( i\frac{k_{4,\mu}\ve_{4,\nu}^+ J^{\mu\nu}_2}{p_2\cdot\ve_4^+}\right)\right]
{\cal M}_3^{(s)}(p_1, p_2, k_3^+) \,. \nn
\end{flalign}
Recalling now that the action of the operator $\exp\left(i\frac{k_{4,\m}\ve_{4,\n}^+J^{\m\n}}{p_1\cdot\ve_4^+}\right)$ gives the same result whether it acts on the spinors-$1$ or -$2$, the two exponential operators in \eq{eq:M_3_s_b} are completely equivalent and we can write:
\begin{flalign}\label{eq:M_3_s_c}
{\cal M}_{4}^{(s)}(p_1, p_2, k_3^+, k_4^+) & = \frac{\k}{2}\left[
\frac{\left(p_1\cdot \ve_4^+ \right)^2}{p_1\cdot k_4}
+\frac{\left(p_2\cdot \ve_4^+ \right)^2}{p_2\cdot k_4}\right]\exp\left(i\frac{k_{4,\m}\ve_{4,\n}^+J_1^{\m\n}}{p_1\cdot\ve_4^+}\right) {\cal M}_3^{(s)}(p_1, p_2, k_3^+)= \nn \\
& \hspace{-1.5 cm}= \left(\dfrac{\k}{2}\right)^{\!2}\left[
\frac{\left(p_1\cdot \ve_4^+ \right)^2}{p_1\cdot k_4}
+\frac{\left(p_2\cdot \ve_4^+ \right)^2}{p_2\cdot k_4}\right]
2(p_1 \cdot \ve_3^+)^2 \, \times \\
& \times \dfrac{1}{m^{2s}} \qlb{\bf 1}^{2s} \exp\left(i\frac{k_{4,\m}\ve_{4,\n}^+J_1^{\m\n}}{p_1\cdot\ve_4^+}\right) \exp\left(i\dfrac{k_{3,\m}\ve_{3,\n}^+J_1^{\m\n}}{p_1\cdot\ve_3^+}\right)\qrb{\bf 2}^{2s} \,. \nn
\end{flalign}
Notice that the first line of \eq{eq:M_3_s_c} is simply \eq{eq:M_3_0_c}. This is to be expected, as the scalar contribution relative to the scattering of Kerr Black Holes is simply the Schwarzschild Compton scattering.

Focus now on the the second line. I define the operators:
\begin{equation}
J_3\equiv i\dfrac{k_{3,\m}\ve_{3,\n}^+J_1^{\m\n}}{p_1\cdot\ve_3^+}\,,\qquad \qquad
J_4\equiv i\dfrac{k_{4,\m}\ve_{4,\n}^+J_1^{\m\n}}{p_1\cdot\ve_4^+}\,.
\end{equation}
In general these operators do not commute, therefore the product of the two exponentials is not just the exponential of the sum. Nonetheless using the Baker-Campbell-Hausdorf formula:
\begin{equation}
e^Xe^Y=e^Z\,, \quad Z=X+Y+\dfrac12[X,Y]+\dfrac{1}{12}[X,[X,Y]]+\dfrac{1}{12}[[X,Y],Y]+\dots
\end{equation}
we can write 
\begin{equation}
\qlb{\bf 1}^{2s} e^{J_4} e^{J_3} \qrb{\bf 2}^{2s}=
\qlb{\bf 1}^{2s} \exp\left\{J_3+J_4-\dfrac12[J_3,J_4]+\dots \right\} \qrb{\bf 2}^{2s} \,.
\end{equation}
The sum is then guaranteed to truncate due to the Cayley-Hamilton theorem.
From the Lorentz algebra, $[J,J]\sim J$, thus even higher commutators 
can be put in terms of lower multipoles. It is important however to notice that starting from the third commutator we begin developing unphysical poles in the argument of the exponential, which make us question the validity of this approximation. \\
The amplitude before the tensorization only has the singlet, the dipole and the quadrupole contribution, that we grouped in an exponential form, furthermore also the soft theorems have been proven to be universal (in the sense that they are related to a physical symmetry) only up to second order in the series expansion.
Thus when considering the multipole expansion order by order for miniamal gravity, the only universal terms are the ones up to the first commutator and all higher contributions are negligible.
Indeed up to order $J^3$, we have:
\begin{equation}\label{eq:esponenziale}
e^{J_4}e^{J_3}=e^{J_3+J_4-\frac12[J_3,J_4]}\,.
\end{equation}
Let us then study in detail the argument of the exponential in \eq{eq:esponenziale}:
\begin{align}\label{eq:somma_esp}
J_3+J_4-\frac12[J_3,J_4]=
i\dfrac{k_{3,\m}\ve_{3,\n}^+J_1^{\m\n}}{p_1\cdot\ve_3^+}+
i\dfrac{k_{4,\m}\ve_{4,\n}^+J_1^{\m\n}}{p_1\cdot\ve_4^+}
-\dfrac12\left[i\dfrac{k_{3,\m}\ve_{3,\n}^+J_1^{\m\n}}{p_1\cdot\ve_3^+}, 
i\dfrac{k_{4,\r}\ve_{4,\s}^+J_1^{\r\s}}{p_1\cdot\ve_4^+}
\right]   \,.
\end{align}
Using the antisymmetry of $J^{\m\n}$, and recalling that the field strength $F_{\m\n}$ is defined as $F_{\m\n}=2\,k_{[\m}\ve_{\n]}$, we can rewrite \eq{eq:somma_esp} as
\begin{equation}
i\dfrac{F_{3,\m\n}J_1^{\m\n}}{2\,p_1\cdot\ve_3^+}+
i\dfrac{F_{4,\m\n}J_1^{\m\n}}{2\,p_1\cdot\ve_4^+}
+\dfrac12\left[\dfrac{F_{3,\m\n}J_1^{\m\n}}{2\,p_1\cdot\ve_3^+}, 
\dfrac{F_{4,\r\s}J_1^{\r\s}}{2\, p_1\cdot\ve_4^+}
\right]  \,.     
\end{equation}
Let us now focus on this commutator:
\begin{equation}\label{eq:commutatore}
\dfrac12\left[\dfrac{F_{3,\m\n}J_1^{\m\n}}{2\,p_1\cdot\ve_3^+}, 
\dfrac{F_{4,\r\s}J_1^{\r\s}}{2\, p_1\cdot\ve_4^+}
\right]=
\dfrac{F_{3,\m\n}J_1^{\m\n}F_{4,\r\s}J_1^{\r\s}-F_{4,\r\s}J_1^{\r\s}F_{3,\m\n}J_1^{\m\n}}{8\, (p_1\cdot\ve_3^+)(p_1\cdot\ve_4^+)}\,,
\end{equation}
notice that the numerator in \eq{eq:commutatore} can be rewritten as
\begin{equation}
F_{3,\m\n}F_{4,\r\s}J_1^{\m\n}J_1^{\r\s}-F_{3,\m\n}F_{4,\r\s}J_1^{\r\s}J_1^{\m\n}=F_{3,\m\n}F_{4,\r\s}\left[J_1^{\m\n},J_1^{\r\s}\right] \,,
\end{equation}
and the remaining commutator is fixed by the Lorentz algebra, 
\begin{equation}\label{eq:aaa}
\left[ J^{\mu\nu} , J^{\r\s} \right] = -i \left( \eta^{\mu\r} J^{\n\s} + \eta^{\n\s} J^{\m\r} - \eta^{\m\s} J^{\n\r} - \eta^{\n\r} J^{\m\s} \right) \,.   
\end{equation}
Thus 
\begin{equation}
F_{3,\m\n}F_{4,\r\s}\left[J_1^{\m\n},J_1^{\r\s}\right]=2i\left[F_3,F_4\right]_{\m\n}J_1^{\m\n} \,,
\end{equation}
and substituting in \eq{eq:commutatore}, we find
\begin{equation}
\dfrac12\left[\dfrac{F_{3,\m\n}J_1^{\m\n}}{2\,p_1\cdot\ve_3^+}, 
\dfrac{F_{4,\r\s}J_1^{\r\s}}{2\, p_1\cdot\ve_4^+}
\right]=
i\dfrac{\left[F_3,F_4\right]_{\m\n}J_1^{\m\n}}{4\, (p_1\cdot\ve_3^+)(p_1\cdot\ve_4^+)}\,.
\end{equation}

Let us now return to \eq{eq:somma_esp}, which we can now rewrite as
\begin{equation}
J_3+J_4-\frac12[J_3,J_4]=
i\dfrac{F_{3,\m\n}J_1^{\m\n}}{2\,p_1\cdot\ve_3^+}+
i\dfrac{F_{4,\m\n}J_1^{\m\n}}{2\,p_1\cdot\ve_4^+}-
i\dfrac{\left[F_3,F_4\right]_{\m\n}J_1^{\m\n}}{4\, (p_1\cdot\ve_3^+)(p_1\cdot\ve_4^+)}\,.
\end{equation}
Once again, I stress the fact that this expression is gauge-invariant, and it is possible to perform the following manipulation to obtain a manifestly gauge invariant result:
\begin{align}\label{eq:somma_espon}
& i\dfrac{F_{3,\m\n}J_1^{\m\n}}{2\,p_1\cdot\ve_3^+}+
i\dfrac{F_{4,\m\n}J_1^{\m\n}}{2\,p_1\cdot\ve_4^+}-
i\dfrac{\left[F_3,F_4\right]_{\m\n}J_1^{\m\n}}{4\, (p_1\cdot\ve_3^+)(p_1\cdot\ve_4^+)}= \\
& \qquad =i\dfrac{F_{3,\m\n}J_1^{\m\n} (p_1\cdot\ve_4^+)(k_4\cdot k_3)
+F_{4,\m\n}J_1^{\m\n}(p_1\cdot\ve_3^+)(k_3\cdot k_4)-\left[F_3,F_4\right]_{\m\n}J_1^{\m\n}(k_3\cdot k_4)/2}{2\,(p_1\cdot\ve_3^+)(p_1\cdot\ve_4^+)(k_4\cdot k_3)}\,. \nn
\end{align}
We now recall that
\begin{equation}
\omega_{(0)}\equiv 2\left(p_1\cdot \ve_3^+ \right)\left(p_1\cdot \ve_4^+ \right)(k_3 \cdot k_4)= -2 p_1 \cdot F_3 \cdot F_4 \cdot p_1\,, 
\end{equation}
furthermore, using $p_i\cdot \ve_j=0$, for $i,j=3,4$:
\begin{equation}
k_3 \cdot F_4 \cdot p_1= (p_1\cdot\ve_4^+)(k_4\cdot k_3)\,, \qquad 
k_4 \cdot F_3 \cdot p_1= (p_1\cdot\ve_3^+)(k_3\cdot k_4)\,,
\end{equation}
and thanks to the conservation of the total momentum and the on-shellness of the amplitude, we have
\begin{equation}
\begin{aligned}
& \hspace{2.5cm} p_1+p_2+k_3+k_4=0 \,\,\Rightarrow \,\,\left(p_1+(k_3+k_4)\right)^2=p_2^2 \,\,\Rightarrow \\
&\Rightarrow \,\, 2p_1 \cdot (k_3+k_4)= -(k_3+k_4)^2=-2(k_3\cdot k_4) \,\,\Rightarrow\,\, -(k_3 \cdot k_4)=-p_1 \cdot (k_3+k_4) 
\end{aligned}  
\end{equation}
Altogether, we have 
\begin{align}
& J_3+J_4-\frac12[J_3,J_4]= \dfrac{\omega_{(1)}}{\omega_{(0)}}=\\
& \qquad = -i\dfrac{F_{3,\m\n} (p_1\cdot F_4 \cdot k_3)+F_{4,\m\n}(p_1\cdot F_3\cdot k_4)+\left[F_3,F_4\right]_{\m\n}p_1 \cdot (k_3+k_4)/2}{\omega_{(0)}} J_1^{\m\n} \nn \,, 
\end{align}
where I have defined the quantity:
\begin{equation}
i\omega_{(1)} \equiv F_{3,\m\n} J^{\m\n}_1(p_1\cdot F_4 \cdot k_3)+F_{4,\m\n}J^{\m\n}_1(p_1\cdot F_3\cdot k_4)+\left[F_3,F_4\right]_{\m\n}J^{\m\n}_1p_1 \cdot \frac{(k_3+k_4)}{2}
\end{equation}
Therefore, 
\begin{equation}\label{eq:Compton_fin_a}
{\cal M}_{4}^{(s)}(p_1, p_2, k_3^+, k_4^+)=\dfrac{-\k^2\omega_{(0)}^2}{(2p_1\cdot k_4)(2p_2 \cdot k_4)(2k_3 \cdot k_4)}
\dfrac{1}{m^{2s}} \qlb{\bf 1}^{2s} \exp\left(\frac{\omega_{(1)}}{\omega_{(0)}}\right)\qrb{\bf 2}^{2s} \,.
\end{equation}
It is often convenient to consider the kinematic configuration in which the particles $1$ and $3$ are incoming and particle $2$ and $4$ are outgoing, in this case \eq{eq:Compton_fin_a} becomes:
\begin{equation}\label{eq:Compton_fin_b}
{\cal M}_{4}^{(s)}(p_1, -p_2, k_3^+, -k_4^+)=\dfrac{\k^2\omega_{(0)}^2}{(2p_1\cdot k_4)(2p_2 \cdot k_4)(2k_3 \cdot k_4)}
\dfrac{1}{m^{2s}} \qlb{\bf 1}^{2s} \exp\left(\frac{\tilde{\omega}_{(1)}}{\tilde{\omega}_{(0)}}\right)\qrb{\bf 2}^{2s} \,,
\end{equation}
where
\begin{equation}
\begin{gathered}
\tilde{\omega}_{(0)}=\omega_{(0)}\,,\\
i\tilde{\omega}_{(1)} \equiv F_{3,\m\n}J^{\m\n}_1 (p_1\cdot F_4 \cdot k_3)+F_{4,\m\n}J^{\m\n}_1(p_1\cdot F_3\cdot k_4)-\left[F_3,F_4\right]_{\m\n}J^{\m\n}_1p_1 \cdot \frac{(k_3-k_4)}{2} 
\end{gathered}
\end{equation}
\\
The computation for the minus-helicity case is exactly the same, the only difference being the helicity of graviton-3 in the $\omega$-s.\\

We can now compute the classical infinite-spin limit. This is achieved in a straightforward manner by replacing all instances of \(J_{\mu\nu}\) with \(-S_{\mu\nu}\), as the entire contribution from the orbital angular momentum is utilized to boost the spinors-2 to match spinors-1. Moreover, the contraction of these spinors cancels out the \(1/m^{2s}\) term.
As a final remark, the boost needed to map $1$ to $2$ is the same one that aligns $p_1$ and $p_2$.
Thus
\begin{equation}\label{eq:Compton_fin_cl}
{\cal M}_{4}^{(\infty)}(p_1, -p_2, k_3, -k_4)=\dfrac{\k^2\omega_{(0)}^2}{(2p_1\cdot k_4)^2(2k_3 \cdot k_4)}
\exp\left(\frac{\Omega_{(1)}}{\Omega_{(0)}}\right)\,,
\end{equation}
where
\begin{equation}
\begin{gathered}
\Omega_{(0)}=\omega_{(0)}\,, \\
-i\Omega_{(1)} \equiv F_{3,\m\n}S^{\m\n}_1 (p_1\cdot F_4 \cdot k_3)+F_{4,\m\n}S^{\m\n}_1(p_1\cdot F_3\cdot k_4)-\left[F_3,F_4\right]_{\m\n}S^{\m\n}_1 p_1 \cdot \frac{(k_3-k_4)}{2}
\end{gathered} 
\end{equation}
This result is in perfect agreement with the one presented in \Refe{art_4}.

\subsection{Classical formulation of the Soft Theorems}

In the previous sections, we obtained the classical three-point amplitude and subsequently the classical four-point amplitude.
To obtain the classical four-point amplitude we started from the quantum version of the three-point amplitude, used the soft theorems discussed in \sect{sec:cap4} to obtain the quantum version of the four-point amplitude and only then computed the classical limit.\\
A natural question arises: is it possible to obtain directly the classical four point amplitude starting from the classical three-point amplitude in a completely classical formulation? \\
In this section I will discuss this procedure.\\

\noindent Let us begin by considering the classical three- and four- point amplitudes
\begin{subequations}\label{eq:clas_1}
\begin{gather}
{\cal M}_3^{(\infty)}(p_1, -p_2, k) = 
-\k (p_1 \cdot \ve_3)^2 \exp\left(i\dfrac{\ve \cdot S \cdot k}{\ve \cdot p}  \right) \, ,\\
{\cal M}_{4}^{(\infty)}(p_1, -p_2, k_3, -k_4)=\dfrac{\k^2\omega_{(0)}^2}{(2p_1\cdot k_4)^2(2k_3 \cdot k_4)}
\exp\left(\frac{\Omega_{(1)}}{\Omega_{(0)}}\right)\,,
\end{gather} 
\end{subequations}
As we are considering a classical amplitude, it is useful to express this amplitude using the rules presented in \sect{sec:cap2}, that can be summarized as follows: we restore all the $\hbar$-s, replace all massless momenta with their corresponding wavenumber, and associate all massive momenta with their classical four-velocity, \ie
\begin{equation}
\k \to \dfrac{\k}{\sqrt\hbar}\,, \qquad J \to \dfrac{J}{\hbar} \,,\qquad F^{\m\n}\to \hbar \bar{F}^{\m\n} \,, \qquad
k \to \hbar \kb \,, \qquad p \to m v\,.
\end{equation}
In light of such substitutions, it is convenient to define the mass-normalized spin tensor $\tS^{\m\n}$:
\begin{equation}
S^{\m\n}=\e^{\m\n\r\s}p_\r a_\s \,\,\Rightarrow \,\,
\tS^{\m\n}=\e^{\m\n\r\s}v_\r a_\s
\end{equation}
Using these definitions we rewrite \eq{eq:clas_1} as
\footnote{In the following equations the subscript-0 is not related to the spin of the massive particles involved in the scattering, but rather it indicates that the amplitude considered is the tree-level contribution}
\begin{subequations}
\begin{gather}
\label{eq:clas_2_a}
\mathcal{M}^{(0)}_{3,\rm cl}(v,k) = - \dfrac{\k}{\sqrt{\hbar}}\, m^2 \, (\varepsilon \cdot v)^2 \, \exp\left( i \frac{ \, \varepsilon \cdot \tS \cdot \kb}{\varepsilon \cdot v} \right)\, , \\
\label{eq:clas_2_b}
\mathcal{M}^{(0)}_{4,\rm cl} (v, \kb_3, \kb_4) = \frac{\kappa^2 m^2 \bar{\omega}_{(0)}^2}{8\hbar (\kb_3 \cdot \kb_4) (\kb_4 \cdot v)^2} \exp \left( \frac{\bar{\omega}_{(1)}}{\bar{\omega}_{(0)}} \right)\, ,
\end{gather}
\end{subequations}
where we have defined
\begin{equation}
\begin{gathered}
\bar{\omega}_{(0)}= -2\, v \cdot \bar{F}_3 \cdot \bar{F}_4 \cdot v\,, \\
-i\bar{\omega}_{(1)} \equiv \bar{F}_{3,\m\n}\tS^{\m\n}_1 (v\cdot \bar{F}_4 \cdot \bar{k}_3)+\bar{F}_{4,\m\n}\tS^{\m\n}_1(v\cdot \bar{F}_3\cdot \bar{k}_4)-\left[\bar{F}_3,\bar{F}_4\right]_{\m\n}\tS^{\m\n}_1 v \cdot \frac{(\bar{k}_3-\bar{k}_4)}{2}
\end{gathered}
\end{equation}
Notice that in \eq{eq:clas_2_a} the argument of the exponential scale as $\mathcal{O}(\hbar^0)$, therefore the result is completely classical at all orders in the multipole expansion.
Meanwhile in \eq{eq:clas_2_b}, even if the argument of the exponential apparently scales as $\mathcal{O}(\hbar^0)$, this result is valid only up to cubic order in the spin (actually up to fourth order as pointed out in \Refe{art_4}):
this is due to the fact that starting from this order we should start considering also the higher-order commutators, that scale as $\mathcal{O}(\hbar^n)$, with $n\geq 1$ and are related to the quantum contributions to the scattering. \\
Nonetheless, these contributions are strongly suppressed as $\hbar \ll 1$ and we can reasonably assume \eq{eq:clas_2_a} to be exact.

The goal is now to construct some purely classical soft recursion relations, that relate the four-point classical amplitude to the three point one. \\
Let us consider the quantum soft recursion relations:
\begin{flalign}
{\cal M}_{4}^{(s)}(p_1, p_2, k_3, k_4)=\frac{\k}{2}
\sum_{i=1}^3\frac{\left(p_i\cdot \ve_4 \right)^2}{p_i\cdot k_4}
\exp\left( i\frac{k_{4,\mu}\ve_{4,\nu} J^{\mu\nu}_i}{p_i\cdot\ve_4}\right)
{\cal M}_3^{(s)}(p_1, p_2, k_3) \,
\end{flalign}
an educated guess consists in replacing them with their classical counterpart
\begin{flalign}\label{eq:classi_a}
\mathcal{M}^{(0)}_{4,\rm cl}(v_1, v_2, \kb_3, \kb_4)=\frac{\k}{2 \hbar^{3/2}} \left[ \sum_{i=1}^2 m\frac{\left(v_i\cdot \ve_4 \right)^2}{v_i \cdot \kb_4} \right]
\exp\left( -i\frac{\bar{F}_{4,\m\nu} \tS^{\mu\nu}}{v \cdot\ve_4} \right)
\mathcal{M}^{(0)}_{3,\rm cl}(v_1, \kb_3) \,
\end{flalign}
where once again the gauge freedom is used to fix $\bar{k}_i \cdot \ve_j=0$ for $i,j=3,4$. \\
Substituting now $\mathcal{M}^{(0)}_{3,\rm cl}(v_1, \kb_3)$ in \eq{eq:classi_a} we obtain
\begin{flalign}\label{eq:classi_b}
\mathcal{M}^{(0)}_{4,\rm cl}(v_1, v_2, \kb_3, \kb_4) & =
-\frac{\k^2}{2 \hbar^{2}} \left[ \sum_{i=1}^2 m\frac{\left(v_i\cdot \ve_4 \right)^2}{v_i \cdot \kb_4} \right]m^2\left(v_1\cdot \ve_3 \right)^2 \times \\
& \qquad \qquad \times
\exp\left( -i\frac{\bar{F}_{4,\m\nu} \tS^{\mu\nu}}{v \cdot\ve_4} \right)
\exp\left( -i\frac{\bar{F}_{3,\m\nu} \tS^{\mu\nu}}{v \cdot\ve_3} \right) \,. \nn
\end{flalign}

Let us focus on the first line of \eq{eq:classi_b}
\begin{align}
& -\frac{\k^2}{2 \hbar^{2}} \left[ \sum_{i=1}^2 m\frac{\left(v_i\cdot \ve_4 \right)^2}{v_i \cdot \kb_4} \right]m^2\left(v_1\cdot \ve_3 \right)^2 = 
-\frac{\k^2}{2\hbar^2} \left[m\frac{\left(v_1\cdot \ve_4 \right)^2}{v_1 \cdot \kb_4} + m\frac{\left(v_2\cdot \ve_4 \right)^2}{v_2 \cdot \kb_4} \right]m^2\left(v_1\cdot \ve_3 \right)^2 = \nn \\
& \hspace{2cm} =\frac{\k^2 m^2}{2 \hbar} \dfrac{\left(v_1\cdot \ve_4 \right)^2\left(v_1\cdot \ve_3 \right)^2 (\kb_3 \cdot \kb_4)^2}{(v_1 \cdot \kb_4)(v_2 \cdot \kb_4)(\kb_3 \cdot \kb_4)} =
\frac1\hbar \frac{\k^2 m^2 \bar{\omega}_{(0)}^2}{(2\,v_1 \cdot \kb_4)(2\,v_2 \cdot \kb_4)(2\,\kb_3 \cdot \kb_4)} \,.
\end{align}
Recalling now that we the momenta $p_1$ and $p_2$ are aligned, we can write
\begin{equation}
-\frac{\k^2}{2 \hbar^{2}} \left[ \sum_{i=1}^2 m\frac{\left(v_i\cdot \ve_4 \right)^2}{v_i \cdot \kb_4} \right]m^2\left(v_1\cdot \ve_3 \right)^2 = 
\frac{\k^2 m^2 \bar{\omega}_{(0)}^2}{8\hbar(v \cdot \kb_4)^2(\kb_3 \cdot \kb_4)} \,,
\end{equation}
which matches precisely the scalar contribution of $\MC_{4 \rm cl}^{(0)}(v,\kb_3,\kb_4)$.

Let us now consider the exponentials. Notice that on the three-point kinematics we can trade the classical spin operator with the classical angular momentum operator, \ie
\begin{equation}
\dfrac{F_{\m\n}J^{\m\n}}{2\, p\cdot \ve}=-
\dfrac{F_{\m\n}S^{\m\n}}{p\cdot \ve} \,.
\end{equation}
We can then perform the same procedure as in \sect{sec:aa}. 
Notice however that after expliciting all $\hbar$-s, \eq{eq:aaa} becomes:
\begin{equation}
\left[ J^{\mu\nu} , J^{\r\s} \right] = -i \hbar \left( \eta^{\mu\r} J^{\n\s} + \eta^{\n\s} J^{\m\r} - \eta^{\m\s} J^{\n\r} - \eta^{\n\r} J^{\m\s} \right) \,,   
\end{equation}
therefore, apparently, another power of $\hbar$ arises when taking the commutator.
However in a purely classical setting the commutator should be replaced by the corresponding Poisson brackets. 
Let us discuss in detail in what they differ.

Both classical mechanics and quantum mechanics use bi-linear brackets of variables with similar algebraic properties. In classical mechanics the variables are functions of the canonical
coordinates and momenta, and the Poisson bracket of two such variables $A(q, p)$ and $B(q, p)$ are defined as
\begin{equation}
[A,B]_P = \sum_i \left(\dee{A}{q_i}\dee{B}{p_i}-\dee{A}{p_i}\dee{B}{q_i}\right) \,.
\end{equation}
In quantum mechanics the variables are linear operators in some Hilbert space, and the commutator bracket of two operators is
\begin{equation}
[A,B]_C=AB-BA \,.
\end{equation}
Both types of brackets share similar algebraic properties: they are linear, antisymmetric, satisfy the Leibniz rule, and possess the Jacobi identity.\\
Also, both types of brackets involving the Hamiltonian can be used to describe the time dependence of the classical/quantum variables. In classical mechanics,
\begin{equation}
\frac{d}{dt}A(q,p)=[A,H]_P \,,
\end{equation}
where $H$ is the Hamiltonian operator and I have used the Hamilton equations. \\
Meanwhile in quantum-mechanics, and in particular in the Heisenberg picture, we have
\begin{equation}
i\hbar \frac{d}{dt}\hat{A}=[\hat{A},\hat{H}]_C \,.
\end{equation}
Furthermore, it is well known that, once we generalize the Poisson brackets to the non-commuting variables of quantum mechanics, they become proportional to the commutator brackets, \ie
\begin{equation}
[\hat{A},\hat{B}]_P=\frac{[\hat{A},\hat{B}]_C}{i\hbar}
\end{equation}
Thus the correct $\hbar$ scaling is restored also in the classical computation and with the same procedure it is possible to check that also the exponentials match.

%%%%%%%%%%%%%%%%%%%%%%%%%%%%%%%%%%%%%%%%%%%%%%%%
%----------------------------------------------%
%%%%%%%%%%%%%%%%%%%%%%%%%%%%%%%%%%%%%%%%%%%%%%%%

\clearpage

\section{The tree-level four- and five- point Amplitude in the Soft Expansion}\label{sec:cap6}

In this section, I start by computing the four-point amplitude corresponding to the scattering of two Kerr black holes with different masses and spins. I then apply the soft theorems discussed in \sect{sec:cap4} to derive the corresponding five-point amplitude. \\
This amplitude is crucial as it relates to the emission of radiation, allowing us to compute a waveform using the KMOC formalism explained in \sect{sec:cap2}. Moreover, within the KMOC approach, this amplitude is the only one relevant at leading order.

\subsection{Four-point Amplitude}\label{sec:4pt}

Let us begin by computing the four-point amplitude. This amplitude is of particular interest because it contains the complete information about the classical 1PM scattering of two spinning black holes without radiation emission.

\begin{figure}[H] 
\centering
\includegraphics[width=0.4\textwidth]{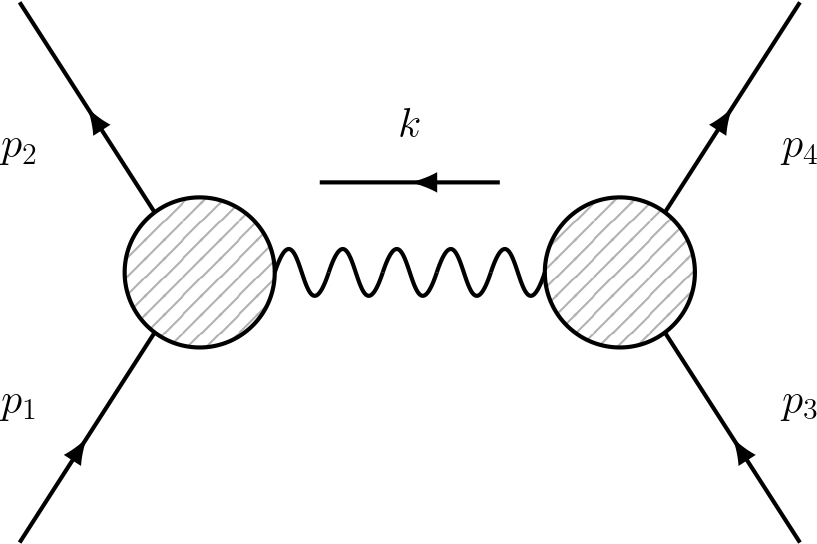}
\caption{4-pt amplitude. Figure reproduced from \Refe{Guevara_2019}}
\label{fig:4-pt}
\end{figure}

Consider two Kerr black holes with spins $s_A$ and $s_B$ and masses $m_A$ and $m_B$, respectively.\\
To compute the four-point amplitude, we would typically consider contributions from the $s$, $t$, and $u$ channels. However, the only contribution relevant to classical scattering arises from the $t$-channel.
Therefore, to compute this amplitude, we can utilize the BCFW recursion relations discussed in \sect{sec:BCFW}, calculating the residue of the scattering amplitude in \fig{fig:4-pt} just at the pole $t\equiv k^2=0$ on finite complex kinematics and then analytically continuing the result to real kinematics at a later stage.
\vspace{-1cm}
\begin{align}
&\MC_4^{(s_A,s_B)}(p_1,-p_2,p_3,-p_4) =
\sum_{h=\pm}\MC_3^{(s_A)}(p_1,-p_2,k^h)\dfrac{-1}{t}\MC_3^{(s_B)}(p_3,-p_4,k^{-h})+\mathcal{O}(t^0)= \nn \\
& \qquad \quad= \frac{-(\k/2)^2}{m_A^{2s_A-2}m_B^{2s_B-2}t}
\Bigl(x_A^2 x_B^{-2}\wa{\bf 2}{\bf 1}^{\odot 2s_A} \wb{\bf 4}{\bf 3}^{\odot 2s_B}+
x_A^{-2} x_B^{2}\wb{\bf 2}{\bf 1}^{\odot 2s_A} \wa{\bf 4}{\bf 3}^{\odot 2s_B}\Bigr) +\mathcal{O}(t^0) \,.
\end{align}
It is now convenient to introduce the following kinematic variables
\begin{equation}\label{eq:Momentum_Variables}
x_A/x_B=\g(1-v)\,, \quad x_B/x_A=\g(1+v)\,, \quad 
\text{with} \quad 
\gamma=\frac{1}{\sqrt{1-v^2}}=\dfrac{p_A\cdot p_B}{m_A m_B}=u_A\cdot u_B \,.
\end{equation}
Following the procedure outlined in \sect{sec:classical_inf_spin}, we can thus write:
\begin{align}\label{eq:Spin_Amplitude_2}
\MC_4^{(s_A,s_B)} &= \frac{-(\k/2)^2 \g^2}{m_A^{2s_A-2}m_B^{2s_B-2}t} \,\times \nn\\
& \times \Bigl((1-v)^2\{U_{12}\qla{\bf 1}\}^{\odot 2s_A} e^{-k\cdot a_A}\qra{\bf 1}^{\odot 2s_A} \{U_{34}\qlb{\bf 3}\}^{\odot 2s_B} e^{-k\cdot a_B}\qrb{\bf 3}^{\odot 2s_B}+ \\
& \qquad + (1+v)^2\{U_{12}\qlb{\bf 1}\}^{\odot 2s_A} e^{k\cdot a_A}\qrb{\bf 1}^{\odot 2s_A} \{U_{34}\qla{\bf 3}\}^{\odot 2s_B} e^{k\cdot a_B}\qra{\bf 3}^{\odot 2s_B} \Bigr) +\mathcal{O}(t^0)\,.\nn
\end{align}
It is now useful to perform a manipulation of these exponentials to obtain a more convenient expression.
We recall that on the pole kinematics, \(k^2 = 0\); furthermore, due to the transversality of the exchange momentum, \(p_A \cdot k = p_B \cdot k = 0\), and by the spin supplementary condition (SSC), \(p_A \cdot a_A = p_B \cdot a_B = 0\).\\
Therefore the following relations hold true
\footnote{These equalities can be derived by squaring the left-hand
sides and computing the resulting Gram determinants, using the fact that $\e^{\a_1\dots\a_n}\e_{\b_1\dots\b_n}=\d^{\a_1\dots\a_n}_{\b_1\dots\b_n}$.}:
\begin{equation}\label{eq:spin1pole}
   i\e_{\m\n\r\s} p_A^\m p_B^\n k^\r a_A^\s
     = m_A m_B \g v (k \cdot a_A) \,, \quad
   i\e_{\m\n\r\s} p_A^\m p_B^\n k^\r a_B^\s
     = m_A m_B \g v (k \cdot a_B) \,.
\end{equation}
Hence, if we define
\begin{equation}\label{eq:wInForm}
w^{\m\n}=\frac{2p_A^{[\m} p_B^{\n]}}{m_A m_B \g v} \,, \qquad
[w*a]_\mu = \frac{1}{2} \e_{\m\n\a\b} w^{\a\b} a^\n \,,
\end{equation}
and strip the unitary transition factors
$U_{12}^{\odot 2s_A}$ and $U_{34}^{\odot 2s_B}$ via the GEV,
we obtain the classical limit of the scattering amplitude \eq{eq:Spin_Amplitude_2} as
\begin{equation}
\la \MC_4(k) \ra=
-\Big(\frac{\k}{2}\Big)^{\!2} \frac{m_A^2 m_B^2}{k^2}\g^2 \sum_\pm (1 \pm v)^2 \exp[{\pm i} \bigl(k \cdot (w*a_0)\bigr)] \,,
\label{eq:SpinDeflectionAmplitude3}
\end{equation}
where $a_0^\mu = a_A^\mu + a_B^\mu$ is the total spin pseudo-vector.

%----------------
\subsection{Five-point Amplitude}

In this section, I will compute the five-point amplitude
\footnote{This section has been inspired by the notes on the spinning waveforms based on results of Riccardo Gonzo and inspired by discussions with Marc Canaj, David Kosower, Pavel Novichkov and Arshia Momeni.
}. \\
Before delving into the full calculations, I will first review and summarize key results from the previous sections.

\subsubsection{Preliminaries}

\begin{figure}[H]
\centering
\begin{tikzpicture}
\begin{feynman}
\begin{scope}
	\vertex (ip1) ;
	\vertex [below=2.0 of ip1] (ip2);
    
    \vertex [above left=0.12 and 1.70 of ip1] (q1) {$ p_1$};
	\vertex [below left=0.12 and 1.70 of ip2] (qp1) {$ p_2$};
	\vertex [above right=0.12 and 1.70 of ip1] (q2) {$ p_3$};
	\vertex [below right=0.12 and 1.70 of ip2] (qp2) {$ p_4$};

    \diagram* {
		(ip1) -- [photon, momentum'=$p_1+p_3$]  (ip2)
	};
    
    \draw[postaction={decorate}, decoration={markings, mark = at position 0.5 with {\arrow{Stealth}}}] (q1) -- (ip1);
	\draw[postaction={decorate}, decoration={markings, mark = at position 0.5 with {\arrow{Stealth}}}] (q2) -- (ip1);
	\draw[postaction={decorate}, decoration={markings, mark = at position 0.5 with {\arrow{Stealth}}}] (qp1) -- (ip2);
	\draw[postaction={decorate}, decoration={markings, mark = at position 0.5 with {\arrow{Stealth}}}] (qp2) -- (ip2);
 \end{scope}
 	\filldraw [color=white] ($ (ip1)$) circle [radius=8pt];
	\filldraw  [fill=allOrderBlue] ($ (ip1) $) circle [radius=8pt];
	
	\filldraw [color=white] ($ (ip2) $) circle [radius=8pt];
	\filldraw  [fill=allOrderBlue] ($ (ip2) $) circle [radius=8pt];
 \end{feynman};
 \end{tikzpicture}

\caption{Four-point amplitude with all incoming external momenta}\label{fig:4pt}
\end{figure}
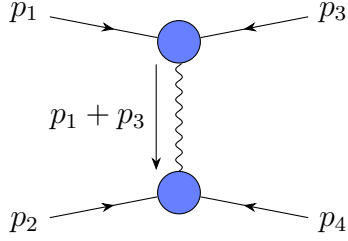

Our starting point is the 4-point amplitude in \fig{fig:4pt}:
\begin{equation}
\begin{aligned}
\MC_4^{s_A, s_B}(p_1,\,p_2,-p_3,-p_4) &= \frac{8\pi G}{m_A^{2s_A-2}m_B^{2s_B-2}(p_1+p_3)^2}\, \times  \\
&\hspace{-1cm} \times \big(\g^2(1-v)^2 \wa{\bf 3}{\bf 1}^{2s_A} 
\wb{\bf 4}{\bf 2}^{2s_B}+
\g^2(1+v)^2 \wb{\bf 3}{\bf 1}^{2s_A} \wa{\bf 4}{\bf 2}^{2s_B}\big)\,, \nn
\end{aligned}
\end{equation}
where we work with all external momenta incoming and we have defined
\[\gamma = \dfrac{1}{\sqrt{1-v^2}}= \dfrac{p_1\!\cdot p_2}{m_1 m_2}\].

To simplify the calculations, I will restrict to the case where the particle A is spinless, \ie $s_A=0$.
The amplitude then becomes:
\begin{equation}
\MC_4^{0, s_B}(p_1,\,p_2,\,p_3,\,p_4)= \frac{8\pi G m_A^2}{m_B^{2s_B-2}(p_1+p_3)^2} \big(\g^2(1-v)^2 \wb{\bf 4}{\bf 2}^{2s_B}+
\g^2(1+v)^2 \wa{\bf 4}{\bf 2}^{2s_B}\big) 
\end{equation}
It is also convenient to define the function $I_\pm(p_1\cdot p_2) \equiv \gamma^2(1\pm v)^2$. So,
\begin{equation}\label{eq:4pt_non_shiftata}
\MC_4^{0, s_B}(p_1,\,p_2,\,p_3,\,p_4)= \frac{8\pi G m_A^2}{m_B^{2s_B-2}(p_1+p_3)^2} \big(I_-(p_1\cdot p_2) \wb{\bf4}{\bf2}^{2s_B}+ I_+(p_1\cdot p_2) \wa{\bf4}{\bf2}^{2s_B}\big)   
\end{equation}

As outlined earlier, our aim is to apply the soft theorems discussed in \sect{sec:cap4} to derive the five-point amplitude with an emitted graviton starting from the four-point amplitude in \eq{eq:4pt_non_shiftata}.\\
Our strategy consists of shifting the four-point amplitude in the soft limit, \(\e \to 0\); then, multiplying it by the appropriate soft factors, we are able to reconstruct the five-point amplitude in the soft limit, where the additional leg corresponds to the soft graviton, labeled as particle $5$.\\
This approach is somewhat unconventional, as the soft limit is typically used to study the infrared behavior of an amplitude. In such cases, one generally starts with a higher-point amplitude and expands it to express it as a lower-point amplitude multiplied by soft factors.
However, by utilizing \eq{eq:ST_softTheoremGravity}, we can reverse this process, provided we remain consistent with our definition of the shifts.

The natural BCFW shifts we want to use are the ones in \eq{eq:ST_ss}:
\begin{equation}
\label{eq:ST_sss}
\hat p_5 = \e p_5 - z q_5\,, \quad 
\hat p_1 = p_1 - 
\frac{(\e-1)(p_5 p_3)-z (q_5 p_3)}{q_1 p_3} q_1 \,, \quad
\hat p_3 = p_3 - 
\frac{ (\e-1)(p_5 p_1) - z (q_5 p_1)}{q_3 p_1}  q_3\,, 
\end{equation}
where $q_5$, $q_1$, $q_3$ are four-vectors satisfying 
\begin{equation}\label{eq:ST_sss_qcondition}
q_5^2 = q_1^2 = q_3^2 =  q_5 p_5 = q_1 p_1 = q_3 p_3 = 0\,, 
\qquad 
 q_1 p_5 = q_3 p_5  = q_1 q_5  = q_3 q_5 = q_1 q_3  = 0\,. 
\end{equation}
Once again, we choose the the shift vectors $q$ as  
\begin{equation}\label{eq:ST_0mmbar_qsigma_2}
q_5 \s = y \tl_5\,, \qquad 
q_1 \s = p_1 \s \tl_5 \tl_5\,, \qquad 
q_3 \s = p_3 \s \tl_5 \tl_5\,.  
\end{equation}
We recall that at the spinor level, this shift is achieved by promoting the momenta to be complex, thereby making the holomorphic and antiholomorphic spinors as independent. The shift is then implemented as follows:
\begin{align}\label{eq:ST_0mmbar_spinor_2}
\{ 5 \bar{\bf1} \bar{\bf3}\} : \qquad  
\l_5^z & = \e  \l_5 - z  y\,, \nn \\
\tx_1^z & = \tx_1 + \frac{(\e-1)\wab{5}{p_3}{5} - z \wab{y}{p_3}{5}}{\wbb{5}{p_1}{p_3}{5}}\wb{\bf 1}{5}\tl_5 \,, \nn \\
\tx_3^z & = \tx_3 - \frac{(\e-1)\wab{5}{p_1}{5} - z \wab{y}{p_1}{5}}{\wbb{5}{p_1}{p_3}{5}}\wb{\bf 3}{5}\tl_5 \,, 
\end{align}
while keeping all other spinors fixed.\\
The poles $z_l$ are located at $\hat{P}_{5l}^2\big|_{z=z_l}-m_l^2=0$, so
\begin{equation}
z_l=\e\dfrac{p_5 \cdot p_l}{q_5 \cdot p_l}=
\e \dfrac{\wab{5}{p_l}{5}}{\wab{y}{p_l}{5}} \,,
\end{equation}
and the shifted spinors evaluated at the poles location are:
\begin{align}\label{eq:shift_res}
\l_5^{z_l}&=\e\frac{\wa{y}{5}}{\wab{y}{p_l}{5}}p_l \s \tl_5 \,,\nn\\ 
\tx_1^{z_l}&=\tx_1 -\frac{2 p_5 p_3}{\wbb{5}{p_1}{p_3}{5}}\wb{\bf 1}{5} \tl_5 +\e \frac{\wa{y}{5} \wbb{5}{p_3}{p_l}{5}}{\wab{y}{p_l}{5}\wbb{5}{p_1}{p_3}{5}}\wb{\bf 1}{5}\tl_5 \,,  \\ 
\tx_3^{z_l}&=\tx_3
+\frac{2 p_5 p_1}{\wbb{5}{p_1}{p_3}{5}}\wb{\bf 3}{5}\tl_5
-\e \frac{\wa{y}{5} \wbb{5}{p_1}{p_l}{5}}{\wab{y}{p_l}{5}\wbb{5}{p_1}{p_3}{5}}\wb{\bf 3}{5}\tl_5 \,. \nn
\end{align}
Finally we can rewrite $\hat P_{5l}\s\big|_{z_l}$, that being evaluated at the poles is on-shell, as
\begin{align}
&\hat P_{5l}\s|_{z_l}=p_l\s +\e\dfrac{\wa{y}{5}}{\wab{y}{p_l}{5}}p_l\s\tl_5\tl_5\,, \,\,\,\Rightarrow\,\,\,
\x_{5l}=\x_l\,, \quad 
\tx_{5l}=\tx_l + \e \frac{\wa{y}{5}}{\wab{y}{p_l}{5}}\wb{\bf l}{5}\tl_5\,, \quad l\neq 1,3, 
\nn \\
& \hat P_{51}\s|_{z_1}=p_1\s -\frac{2 p_5 p_3}{\wbb{5}{p_1}{p_3}{5}} p_1\s\tl_5\tl_5 \,, \, \Rightarrow \, 
\x_{51}=\x_1\,, \quad 
\tx_{51}=\tx_1 - \frac{2 p_5 p_3}{\wbb{5}{p_1}{p_3}{5}}\wb{\bf 1}{5}\tl_5  \,, \\
& \hat P_{53}\s|_{z_3}=p_3\s -\frac{2 p_5 p_1}{\wbb{5}{p_1}{p_3}{5}} p_3\s\tl_5\tl_5 \,, \, \Rightarrow \, 
\x_{53}=\x_3\,, \quad 
\tx_{53}=\tx_1 - \frac{2 p_5 p_1}{\wbb{5}{p_1}{p_3}{5}}\wb{\bf 3}{5}\tl_5  \,. \nn
\end{align}
Using now the soft theorems discussed in \sect{sec:cap4} we have:
\begin{equation}\label{eq:ST_softTheoremGravity_2}
\hat \MC^{0,\epsilon}_5(\bf{1},\bf{2},\bf{3},\bf{4},5_h^+ )=    \bigg\{\frac{1}{\e^3}S_{+2}^{(0)}+\frac{1}{\e^2}S_{+2}^{(1)}+ \frac{1}{\e} S_{+2}^{(0)}\bigg\} 
\hat \MC^{0,0}_4(\hat{\bf 1}, \bf{2},\hat{\bf 3},\bf{4}) 
+ \mathcal{O}(\e^0)\,,  
\end{equation}
where
\begin{equation}
S^{(0)}_{+2}= \frac{\k}{2}\sum_{l=1}^4 \frac{\wab{y}{p_l}{5}^2}{(2 p_5\cdot p_l) \wa{5}{y}^2}\,, \quad 
S^{(1)}_{+2}= \frac{\k}{2}\sum_{l=1}^4 \frac{\wab{y}{p_l}{5}}{(2 p_5\cdot p_l) \wa{5}{y}} \tD_l \,, \quad
S^{(2)}_{+2}= \frac{\k}{4}\sum_{l=1}^4 \frac{1}{4 p_5\cdot p_l} \tD_l \tD_l \,, 
\end{equation}
and we have defined the quantities 
\begin{equation}\label{eq:D_tilde}
\tD_l= \tilde{d}_l+\frac{\wbb{5}{p_3}{p_l}{5}}{\wbb{5}{p_1}{p_3}{5}}\wb{5}{\bf 1} \wb{5}{\de{\tx_1}} -\frac{\wbb{5}{p_1}{p_l}{5}}{\wbb{5}{p_1}{p_3}{5}}\wb{5}{\bf 3} \wb{5}{\de{\tx_3}}\,, \qquad 
\tilde{d}_l= \wb{5}{\bf l}\wb{5}{\de{\tx_l}}\,,
\end{equation}
with $\tilde{\mathcal{D}}_l=0$ for l = 1, 3.

Notice that for $\e=0$, 
\begin{equation}\label{eq:eps=0}
\hat{P}_{5l}\s\big|_{z_l, \e=0}=p_l \s \big|_{\e=0}\,, \qquad
\tx_1^{z_l}\big|_{\e=0}=\tx_1 -\frac{2 p_5 p_3}{\wbb{5}{p_1}{p_3}{5}}\wb{\bf 1}{5} \tl_5 \,,  \qquad 
\tx_3^{z_l}\big|_{\e=0}=\tx_3 +\frac{2 p_5 p_1}{\wbb{5}{p_1}{p_3}{5}}\wb{\bf 3}{5}\tl_5 \,.    
\end{equation}

\subsubsection{Explicit soft factors}

\newcommand{\yy}{p_1\cdot p_2}
\newcommand{\xx}{\hat{p}_1\cdot p_2}
\newcommand{\aaa}{m_1 m_2}
\newcommand{\aaaa}{m_1^2 m_2^2}

I will now compute \eq{eq:ST_softTheoremGravity_2} explicitly. \\
First, I will shift the four-point amplitude \eq{eq:4pt_non_shiftata} to obtain its shifted version, \(\hat{\MC}^{0,0}\). Note that the trivial shift corresponds to \((0, 1)\), not \((0, 0)\). Therefore, to transition from \eq{eq:4pt_non_shiftata} to its shifted version, I will apply the shift to the momenta \(p_1 \to \hat{p}_1\) and \(p_3 \to \hat{p}_3\):
\begin{equation}\label{eq:4pt_shiftata}
\hat{\MC}_4^{0, s_B}(\hat{p}_1,\,p_2,\,\hat{p}_3,\,p_4)\Big|_{\e=0}= \frac{8\pi G m_A^2}{m_B^{2s_B-2}(p_1+p_3)^2} \big(\hat{I}_-(\hat{p}_1\cdot p_2) \wb{\bf4}{\bf2}^{2s_B}+ \hat{I}_+(\hat{p}_1\cdot p_2) \wa{\bf4}{\bf2}^{2s_B}\big)   
\end{equation}
Let us now focus on the functions $\hat{I}_\pm(\hat{p}_1\cdot p_2)$ and see how they are related to their un-shifted version $I_\pm(p_1\cdot p_2)$. 
To this end we find convenient to write $I_\pm(p_1\cdot p_2)$ explicitly as:
\begin{align}
I_{\pm}(p_1\cdot p_2) & =\g^2 (1\pm v)^2 = \\ 
&=\left(\frac{\yy}{\aaa}\left(1\pm\sqrt{1-\frac{\aaaa}{(\yy)^2}}\right)\right)^2=
\left(\frac{\yy \pm \sqrt{(\yy)^2-\aaaa}}{\aaa}\right)^2 \,.\nn
\end{align}
Thus $\hat{I}_{\pm}(\hat{p}_1\cdot p_2)$ is nothing but:
\begin{equation}
\label{eq:I_pm_tilde}
I_{\pm}(\xx)=\left(\frac{\xx \pm \sqrt{(\xx)^2-\aaaa}}{\aaa}\right)^2\,.
\end{equation}
We can now easily express the product $\xx$ in terms of the un-shifted spinors using \eq{eq:eps=0} and the identity: 
\begin{equation}
\xx=\frac12\wa{\hat{\bf 1}}{\bf 2}\wb{\bf 2}{\hat{\bf 1}}\,.
\end{equation}
Therefore,
\begin{align}
\xx &=\frac12\wa{\hat{\bf 1}}{\bf 2}\wb{\bf 2}{\hat{\bf 1}} = \\
&= \frac12\wa{{\bf 1}}{\bf 2}\wb{\bf 2}{{\bf 1}}-\frac{p_5 p_3}{\wbb{5}{p_1}{p_3}{5}}\wa{\bf 1}{\bf 2}\wb{\bf 1}{5}\wb{\bf 2}{5}=\yy-\frac{p_5 p_3}{\wbb{5}{p_1}{p_3}{5}}\wb{\bf 1}{5}\wa{\bf 1}{\bf 2}\wb{\bf 2}{5}\,. \nn
\end{align}

We are now ready to compute the desired five-point amplitude.\\
For simplicity, I will work out the terms in \eq{eq:ST_softTheoremGravity_2} one by one, starting with $S^{(0)}_{+2}$. 
Remembering that $\k^2=32 \pi G$, we have:
\begin{flalign}\label{eq:termine_s0}
\dfrac{1}{\e^3} S_{+2}^{(0)} \hat{\MC}^{0,0}_4(\hat{\bf 1},\bf 2, \hat{\bf 3},\bf{4}) & =  \nn\\
& \hspace{-2cm}= \dfrac{1}{\e^3}\left(\dfrac{\k}{2}\right)^{\!3} \dfrac{m_A^2}{m_B^{2s_B-2}(p_1+p_3)^2}\,\times \\
& \times \sum_{l=1}^4 \frac{\wab{y}{p_l}{5}^2}{ (2 p_5\cdot p_l) \wa{5}{y}^2} 
\left(\hat{I}_-(\xx) \wb{\bf4}{\bf2}^{2s_B}+
\hat{I}_+(\xx) \wa{\bf 4}{\bf 2}^{2s_B}\right) \,. \nn &
\end{flalign}

I then compute the term $S^{(1)}_{+2}$:
\begin{flalign}\label{eq:termine_s1}
\dfrac{1}{\e^2} S_{+2}^{(1)} \hat{\MC}^{0,0}_4(\hat{\bf 1},\bf 2, \hat{\bf 3},\bf{4}) & = \nn \\
& \hspace{-2cm} = \dfrac{1}{\e^2} \left(\frac{\k}{2}\right)^{\!3} \dfrac{m_A^2}{m_B^{2s_B-2}(p_1+p_3)^2}\,\times \\
& \times \sum_{l=1}^4 \dfrac{\wab{y}{p_l}{5}}{ (2 p_5\cdot p_l) \wa{5}{y}} \tD_l \left(\hat{I}_-(\xx) \wb{\bf 4}{\bf 2}^{2s_B}+ \hat{I}_+(\xx) \wa{\bf 4}{\bf 2}^{2s_B}\right) \,.\nn
\end{flalign}
Notice now that the operator $\tD_l$ is simply the realization of the derivative $\de{\e}$, hence it behaves as the usual derivative, in the sense that it satisfies the usual Leibniz product rule and the chain rule. Hence,
\begin{flalign}\label{eq:derivata_prima}
&\tD_l \left(\hat{I}_-(\xx) \wb{\bf 4}{\bf 2}^{2s_B}+\hat{I}_+(\xx) \wa{\bf 4}{\bf 2}^{2s_B}\right)=\\
&\qquad \qquad = \left(\tD_l (\hat{I}_-(\xx) \right)\wb{\bf4}{\bf2}^{2s_B}+ \hat{I}_-(\xx) \left(\tD_l \wb{\bf 4}{\bf 2}^{2s_B}\right) +\left(\tD_l \hat{I}_+(\xx)\right) \wa{\bf 4}{\bf 2}^{2s_B}\,+ \nn \\
& \hspace{3cm} + \hat{I}_+(\xx) \left(\tD_l \wa{\bf 4}{\bf 2}^{2s_B}\right) \nn &
\end{flalign}
Let us examine each of these contributions individually. Consider the first and the third terms in \eq{eq:derivata_prima}:
\begin{equation}\label{eq:derivata_1}
\tD_l (\hat{I}_{\pm}(\xx))=\dee{\big(\hat{I}_\pm(\xx)\big)}{(\xx)}\tD_l(\xx)
\equiv\hat{I}'_{\pm}(\xx)\tD_l(\xx)\,,
\end{equation}
where we have defined the function
\begin{equation}\label{eq:I_pm_tilde_primo}
\hat{I}'_{\pm}(\xx)\equiv \dee{\big(\hat{I}_\pm(\xx)\big)}{(\xx)}=
\pm \frac{2}{\aaaa}
\dfrac{\left(\xx \pm \sqrt{(\xx)^2- \aaaa}\right)^2}{\sqrt{(\xx)^2- \aaaa}}    
\end{equation}
We recall that $\tD_l=0$ for $l = 1, 3$; so we are left to compute just $\tD_2(\xx)$ and $\tD_4(\xx)$.
To compute such derivatives, I will use the parity conjugate version of the identity in \eq{eq:angular_2}, that I rewrite below for convenience:
\begin{equation}
\wb{k}{\de{\tx_p^b}}\qrb{p^a}=\d^a_b \qrb{k}\,.
\end{equation}
So, 
\begin{subequations}
\begin{flalign}
\tD_2(\xx) & =
\left(\wb{5}{\bf 2}\wb{5}{\de{\tx_2}} +\dfrac{\wbb{5}{p_3}{p_2}{3}}{\wbb{5}{p_1}{p_3}{5}}\wb{5}{\bf 1}  \wb{5}{\de{\tx_1}}-\dfrac{\wbb{5}{p_1}{p_2}{5}}{\wbb{5}{p_1}{p_3}{5}}\wb{5}{\bf 3} \wb{5}{\de{\tx_3}}\right) \left(\frac12 \wa{\bf 1}{\bf 2}\wb{\bf 2}{\hat{\bf1}} \right)=  \nn \\
& = \dfrac12 \wb{5}{\bf 1}\wa{\bf 1}{\bf 2}\wb{\bf 2}{5}+ \dfrac12 \dfrac{\wbb{5}{p_3}{p_2}{5}}{\wbb{5}{p_1}{p_3}{5}}\wb{5}{\bf 1} \wa{\bf 1}{\bf 2}\wb{\bf 2}{5}= \\
& = \frac12 \wb{5}{\bf 1}\wa{\bf 1}{\bf 2}\wb{\bf 2}{5}\left(1+\dfrac{\wbb{5}{p_3}{p_2}{5}}{\wbb{5}{p_1}{p_3}{5}}\right) \,,\nn \\
\tD_4(\xx)  & =
\left(\wb{5}{\bf 4}\wb{5}{\de{\tx_4}} +\dfrac{\wbb{5}{p_3}{p_4}{3}}{\wbb{5}{p_1}{p_3}{5}}\wb{5}{\bf 1}  \wb{5}{\de{\tx_1}}-\dfrac{\wbb{5}{p_1}{p_4}{5}}{\wbb{5}{p_1}{p_3}{5}}\wb{5}{\bf 3} \wb{5}{\de{\tx_3}}\right) \left(\frac12 \wa{\bf 1}{\bf 2}\wb{\bf 2}{\hat{\bf1}} \right)=  \nn \\
& = \dfrac12\wb{5}{\bf 1}\wa{\bf 1}{\bf 2}\wb{\bf2}{5}\dfrac{\wbb{5}{p_3}{p_4}{5}}{\wbb{5}{p_1}{p_3}{5}} \,.
\end{flalign}
\end{subequations}
Consider now the second term in \eq{eq:derivata_prima}. In this case the derivative acts only on the un-shifted spinors, so we simply have:
\begin{subequations}\label{eq:derivata_2}
\begin{flalign}
\tD_2 \wb{\bf 4}{\bf 2}^{2s_B} & =\left(\wb{5}{\bf 2}\wb{5}{\de{\tx_2}}\right) \wb{\bf 4}{\bf 2}^{2s_B}=\tilde{d}_2 \wb{\bf 4}{\bf 2}^{2s_B} \\
\tD_4 \wb{\bf 4}{\bf 2}^{2s_B} & =\left(\wb{5}{\bf 4}\wb{5}{\de{\tx_4}}\right) \wb{\bf 4}{\bf 2}^{2s_B}=\tilde{d}_4 \wb{\bf 4}{\bf 2}^{2s_B} 
\end{flalign} 
\end{subequations}
Furthermore, it is straightforward to check that: $\tilde{d}_2 \wb{\bf{4}}{\bf{2}}^{2s_B}=-\tilde{d}_4 \wb{\bf{4}}{\bf{2}}^{2s_B}$.

At last, consider the fourth term in. in \eq{eq:derivata_prima}.
As we are working with complex momenta, holomorphic and anti-holomorphic spinors are completely independent, thus
\begin{equation}\label{eq:derivata_3}
\tD_2\wa{\bf{4}}{\bf{2}}^{2s_B}=\tD_4\wa{\bf{4}}{\bf{2}}^{2s_B}=0\,.
\end{equation}

We can now finally substitute \eq{eq:derivata_prima} in \eq{eq:termine_s1} to get:
\begin{equation}\label{eq:sUno}
\begin{aligned}
\dfrac{1}{\e^2} S_{+2}^{(1)} \hat{\MC}^{0,0}_4(\hat{\bf 1},\bf 2, \hat{\bf 3},\bf{4}) & = \\
& \hspace{-2.5cm}= \dfrac{1}{\e^2} \left(\dfrac{\k}{2}\right)^{\!3} \dfrac{m_A^2}{m_B^{2s_B-2}(p_1+p_3)^2} \, \times  \\
& \hspace{-1.5cm}\times \Biggl\{\dfrac{\wab{y}{p_2}{5}}{(2 p_5\cdot p_2) \wa{5}{y}} \biggl(\hat{I}'_-(\xx)\tD_2(\xx)\wb{\bf 4}{\bf 2}^{2s_B}+\hat{I}_-(\xx)\big(\tilde{d}_2 \wb{\bf 4}{\bf 2}^{2s_B}\big)\,+ \\
& \hspace{2.5cm} +\hat{I}'_+(\xx)\tD_2(\xx)\wa{\bf{4}}{\bf{2}}^{2s_B}\biggr)\,+  \\
& \hspace{-0.5cm}+ \dfrac{\wab{y}{p_4}{5}}{(2 p_5\cdot p_4)\wa{5}{y}}
\biggl(\hat{I}'_-(\xx)\tD_4(\xx)\wb{\bf{4}}{\bf{2}}^{2s_B}+\hat{I}_-(\xx)\big(\tilde{d}_4 \wb{\bf{4}}{\bf{2}}^{2s_B}\big)\,+ \\
& \hspace{3cm} +\hat{I}'_+(\xx)\tD_4(\xx)\wa{\bf{4}}{\bf{2}}^{2s_B}\biggr)\,\Biggr\}= \\
& \hspace{-2.5cm}= \dfrac{1}{\e^2} \left(\dfrac{\k}{2}\right)^{\!3} \dfrac{m_A^2}{m_B^{2s_B-2}(p_1+p_3)^2} \, \times  \\
&\hspace{-1.5cm} \times \Biggl\{ \left[
\dfrac{ \wab{y}{p_2}{5}}{(4 p_5\cdot p_2)\wa{5}{y}}
\left(1+ \dfrac{\wbb{5}{p_3}{p_2}{5}}{\wbb{5}{p_1}{p_3}{5}}\right)+\dfrac{\wab{y}{p_4}{5}}{(4 p_5\cdot p_4)\wa{5}{y}}
\dfrac{\wbb{5}{p_3}{p_4}{5}}{\wbb{5}{p_1}{p_3}{5}}\right]\wb{5}{\bf{1}}\wa{\bf{1}}{\bf{2}}\wb{\bf{2}}{5}\, \times \\
& \times \left(\hat{I}'_-(\xx)\wb{\bf{4}}{\bf{2}}^{2s_B}+\hat{I}'_+(\xx)\wa{\bf{4}}{\bf{2}}^{2s_B}\right) \,+  \\
&+\left[\dfrac{ \wab{y}{p_2}{5}}{(2 p_5\cdot p_2)\wa{5}{y}}-\dfrac{ \wab{y}{p_4}{5}}{(2 p_5\cdot p_4)\wa{5}{y}} \right]\hat{I}_-(\xx) \, \big(\tilde{d}_2 \wb{\bf{4}}{\bf{2}}^{2s_B}\big)\Biggr\} \,.
\end{aligned}
\end{equation}

I then compute the term $S^{(2)}_{+2}$:
\begin{flalign}\label{eq:termine_s2}
\dfrac{1}{\e} S_{+2}^{(2)} \hat{\MC}^{0,0}_4(\hat{\bf 1},\bf 2, \hat{\bf 3},\bf{4}) & = \\
& \hspace{-2.5cm} =\dfrac{1}{\e}\left(\dfrac{\k}{2}\right)^{\!3} \dfrac{m_A^2}{m_B^{2s_B-2}(p_1+p_3)^2}
\sum_{l=1}^4 \frac{1}{4 p_5\cdot p_l}  \tD_l \tD_l\left(\hat{I}_-(\xx) \wb{\bf 4}{\bf 2}^{2s_B}+ \hat{I}_+(\xx) \wa{\bf 4}{\bf 2}^{2s_B}\right) \,.\nn
\end{flalign}
Let us now focus on this second derivative:
\begin{equation}\label{eq:derivata_seconda}
\begin{aligned}
&\tD_l\left(\tD_l \left(\hat{I}_-(\xx) \wb{\bf 4}{\bf 2}^{2s_B}+\hat{I}_+(\xx) \wa{\bf 4}{\bf 2}^{2s_B}\right)\right)=\\
&\hspace{2cm}=\tD_l \biggl(\hat{I}'_-(\xx)\tD_l(\xx)\wb{\bf4}{\bf2}^{2s_B}+\hat{I}_-(\xx)\left(\tilde{d}_l \wb{\bf4}{\bf2}^{2s_B}\right)\,+ \\
&\hspace{5cm}+\hat{I}'_+(\xx)\tD_l(\xx)\wa{\bf4}{\bf2}^{2s_B}\biggr) =  \\
&\hspace{2cm} =\hat{I}''_-(\xx)\left(\tD_l(\xx)\right)^{\!2} \wb{\bf4}{\bf2}^{2s_B}+\hat{I}'_-(\xx)\,\tD_l\left(\tD_l(\xx)\right)\wb{\bf4}{\bf2}^{2s_B} \,+ \\
& \hspace{3cm} +2\,\hat{I}'_-(\xx)\tD_l(\xx)\tilde{d}_l \wb{\bf4}{\bf2}^{2s_B}+\hat{I}_-(\xx)\,\tD_l\left(\tilde{d}_l \wb{\bf4}{\bf2}^{2s_B}\right)\,+ \\
& \hspace{3cm} +\hat{I}''_+(\xx)\left(\tD_l(\xx)\right)^{\!2} \wa{\bf4}{\bf2}^{2s_B}+\hat{I}'_+(\xx)\,\tD_l\left(\tD_l(\xx)\right)\wa{\bf4}{\bf2}^{2s_B}\,, 
\end{aligned}
\end{equation}
where we have defined the function
\begin{flalign}\label{eq:I_pm_tilde_secondo}
\hat{I}''_{\pm}(\xx) &\equiv
\dee{{}^{\,2} \big(\hat{I}_\pm(\xx)\big)}{(\xx)^2}= \\
& =\dfrac{2\left(\xx \pm \sqrt{(\xx)^2-\aaaa}\right)^2}{\aaaa\left[(\xx)^2-\aaaa\right]^{3/2}}\left(2\sqrt{(\xx)^2-\aaaa}\mp \xx\right) \,.\nn
\end{flalign}

Let’s examine, once again, each of the new contributions individually. 
Consider the second and and sixth terms in \eq{eq:derivata_seconda}:
\begin{align}
& \tD_2\left(\tD_2(\xx)\right)=\tD_2 \left(\frac12\wb{5}{\bf1} \wa{\bf1}{\bf2}\wb{\bf2}{5}\left(1+\dfrac{\wbb{5}{p_3}{p_2}{5}}{\wbb{5}{p_1}{p_3}{5}}\right) \right) \propto \wb{5}{5}=0 \,, \\
& \tD_4\left(\tD_4(\xx)\right)=\tD_4 \left(\dfrac12\wb{5}{\bf 1} \wa{\bf1}{\bf2}\wb{\bf2}{5}\dfrac{\wbb{5}{p_3}{p_4}{5}}{\wbb{5}{p_1}{p_3}{5}}\right)\propto \wb{5}{5}=0 \nn\,.
\end{align}
It is worth noting that this outcome is expected: although the amplitude itself has no specific $\e$-dependence, the momentum shifts are linear in $\e$. Given now that $\tD_l$ is merely the realization via the chain rule of $\de{\e}$, it follows that the second derivative vanishes when acting on (the linear $\e$-dependent term) $\hat{p_1}\cdot p_2$.

Consider now the fourth term in \eq{eq:derivata_seconda}. As for \eq{eq:derivata_2} it is straightforward to check that:
\begin{equation}
\begin{gathered}
\tD_2\left(\tilde{d}_2 \wb{\bf4}{\bf2}^{2s_B}\right)=
\tilde{d}_2\left(\tilde{d}_2 \wb{\bf4}{\bf2}^{2s_B}\right)\,, \qquad
\tD_4\left(\tilde{d}_4 \wb{\bf4}{\bf2}^{2s_B}\right)=
\tilde{d}_4\left(\tilde{d}_4 \wb{\bf4}{\bf2}^{2s_B}\right)\,, \\
\tilde{d}_2\left(\tilde{d}_2 \wb{\bf4}{\bf2}^{2s_B}\right)=
\tilde{d}_4\left(\tilde{d}_4 \wb{\bf4}{\bf2}^{2s_B}\right)\,.
\end{gathered}
\end{equation}
\\

We can now finally substitute \eq{eq:derivata_seconda} in \eq{eq:termine_s2} to get:
\begin{equation}\label{eq:sDue}
\begin{aligned}
\dfrac{1}{\e} S_{+2}^{(2)} \hat{\MC}^{0,0}_4(\hat{\bf 1},\bf 2, \hat{\bf 3},\bf{4}) & = \\
& \hspace{-3cm} =\dfrac{1}{\e}\left(\dfrac{\k}{2}\right)^{\!3} \dfrac{m_A^2}{m_B^{2s_B-2}(p_1+p_3)^2} \, \times \\
& \hspace{-2.5cm}\times \Biggl\{ \dfrac{1}{4 p_5\cdot p_2}
\biggl(\hat{I}''_-(\xx)\left(\tD_2(\xx)\right)^{\!2} \wb{\bf4}{\bf2}^{2s_B}+2\,\hat{I}'_-(\xx)\tD_2(\xx)\tilde{d}_2\wb{\bf4}{\bf2}^{2s_B}\,+\\
& + \hat{I}_-(\xx)\,\tilde{d}_2\left(\tilde{d}_2 \wb{\bf4}{\bf2}^{2s_B}\right) +\hat{I}''_+(\xx)\left(\tD_2(\xx)\right)^{\!2} \wa{\bf4}{\bf2}^{2s_B}\biggr)\,+\\
& \hspace{-1.5cm} +\frac{1}{4 p_5\cdot p_4}
\biggl(\hat{I}''_-(\xx)\left(\tD_4(\xx)\right)^{\!2} \wb{\bf4}{\bf2}^{2s_B}+2\,\hat{I}'_-(\xx)\tD_4(\xx)\tilde{d}_4\wb{\bf4}{\bf2}^{2s_B}\,+\\
& \quad +\hat{I}_-(\xx)\,\tilde{d}_4\left(\tilde{d}_4\wb{\bf4}{\bf2}^{2s_B}\right)+\hat{I}''_+(\xx)\left(\tD_4(\xx)\right)^{\!2} \wa{\bf4}{\bf2}^{2s_B}\biggr)\Biggr\}\,= 
\end{aligned}
\end{equation}
\begin{equation}
\begin{aligned}
& \hspace{-1.5cm} =\dfrac{1}{\e}\left(\dfrac{\k}{2}\right)^{\!3} \dfrac{m_A^2}{m_B^{2s_B-2}(p_1+p_3)^2} \, \times \nn\\
& \hspace{-0.5cm}\times\Biggl\{\Biggl[\frac{1}{16 p_5\cdot p_2} \left(1+\dfrac{\wbb{5}{p_3}{p_2}{5}}{\wbb{5}{p_1}{p_3}{5}}\right)^{\!\!2}+\dfrac{1}{16 p_5\cdot p_4}\left(\dfrac{\wbb{5}{p_3}{p_4}{5}}{\wbb{5}{p_1}{p_3}{5}}\right)^{\!\!2}\Biggr]\left(\wb{5}{\bf1} \wa{\bf1}{\bf2}\wb{\bf2}{5}\right)^{2} \times \\
& \qquad \times \left(\hat{I}''_-(\xx)\wb{\bf4}{\bf2}^{2s_B}+\hat{I}''_+(\xx)\wa{\bf4}{\bf2}^{2s_B}\right) \,+ \\
&  +\Biggl[\dfrac{1}{4 p_5\cdot p_2}\left(1+\dfrac{\wbb{5}{p_3}{p_2}{5}}{\wbb{5}{p_1}{p_3}{5}}\right)-\dfrac{1}{4 p_5\cdot p_4} \dfrac{\wbb{5}{p_3}{p_4}{5}}{\wbb{5}{p_1}{p_3}{5}}\Biggr]\wb{5}{\bf1} \wa{\bf1}{\bf2}\wb{\bf2}{5} \, \times \\ 
&\qquad \times \hat{I}'_-(\xx)\left(\tilde{d}_2\wb{\bf4}{\bf2}^{2s_B}\right) \, + \\
&+\left[\dfrac{1}{4 p_5\cdot p_2}+\dfrac{1}{4 p_5\cdot p_4}\right]\hat{I}_-(\xx)\left(\tilde{d}_2(\tilde{d}_2 \wb{\bf4}{\bf2}^{2s_B})\right) \Biggr\}
\end{aligned}   
\end{equation}
\\

Substituting now \eq{eq:termine_s0}, \eq{eq:sUno} and \eq{eq:sDue} in \eq{eq:ST_softTheoremGravity_2}, we finally obtain the five-point amplitude in the soft expansion up to sub-sub-leading order:
\begin{equation}\label{eq:5pti_soft}
\begin{aligned}
\hat \MC^{0,\epsilon}_5(\bf{1},\bf{2},\bf{3},\bf{4},5_h^+ ) &=
\left(\dfrac{\k}{2}\right)^{\!3} \dfrac{m_A^2}{m_B^{2s_B-2}(p_1+p_3)^2} \, \times \\
& \hspace{-3cm}\times \Biggl\{ \dfrac{1}{\e^3}\sum_{l=1}^4 \frac{\wab{y}{p_l}{5}^2}{ (2 p_5\cdot p_l) \wa{5}{y}^2} 
\left(\hat{I}_-(\xx) \wb{\bf4}{\bf2}^{2s_B}+
\hat{I}_+(\xx) \wa{\bf 4}{\bf 2}^{2s_B}\right)  \,+ \\
& \hspace{-2cm}+ \dfrac{1}{\e^2} \Biggl\{ \left[
\dfrac{ \wab{y}{p_2}{5}}{(4 p_5\cdot p_2)\wa{5}{y}}
\left(1+ \dfrac{\wbb{5}{p_3}{p_2}{5}}{\wbb{5}{p_1}{p_3}{5}}\right)+\dfrac{\wab{y}{p_4}{5}}{(4 p_5\cdot p_4)\wa{5}{y}}
\dfrac{\wbb{5}{p_3}{p_4}{5}}{\wbb{5}{p_1}{p_3}{5}}\right]\wb{5}{\bf{1}}\wa{\bf{1}}{\bf{2}}\wb{\bf{2}}{5}\, \times \\
& \times \left(\hat{I}'_-(\xx)\wb{\bf{4}}{\bf{2}}^{2s_B}+\hat{I}'_+(\xx)\wa{\bf{4}}{\bf{2}}^{2s_B}\right) \,+  \\
&\hspace{-0.5cm}+\left[\dfrac{ \wab{y}{p_2}{5}}{(2 p_5\cdot p_2)\wa{5}{y}}-\dfrac{ \wab{y}{p_4}{5}}{(2 p_5\cdot p_4)\wa{5}{y}} \right]\hat{I}_-(\xx) \, \big(\tilde{d}_2 \wb{\bf{4}}{\bf{2}}^{2s_B}\big)\Biggr\} \, +\\
& \hspace{-2cm} +\dfrac{1}{\e}\Biggl\{\Biggl[\frac{1}{16 p_5\cdot p_2} \left(1+\dfrac{\wbb{5}{p_3}{p_2}{5}}{\wbb{5}{p_1}{p_3}{5}}\right)^{\!\!2}+\dfrac{1}{16 p_5\cdot p_4}\left(\dfrac{\wbb{5}{p_3}{p_4}{5}}{\wbb{5}{p_1}{p_3}{5}}\right)^{\!\!2}\Biggr]\left(\wb{5}{\bf1} \wa{\bf1}{\bf2}\wb{\bf2}{5}\right)^{2} \times \\
& \qquad \times \left(\hat{I}''_-(\xx)\wb{\bf4}{\bf2}^{2s_B}+\hat{I}''_+(\xx)\wa{\bf4}{\bf2}^{2s_B}\right) \,+ \\
& \hspace{-0.5cm} +\Biggl[\dfrac{1}{4 p_5\cdot p_2}\left(1+\dfrac{\wbb{5}{p_3}{p_2}{5}}{\wbb{5}{p_1}{p_3}{5}}\right)-\dfrac{1}{4 p_5\cdot p_4} \dfrac{\wbb{5}{p_3}{p_4}{5}}{\wbb{5}{p_1}{p_3}{5}}\Biggr]\wb{5}{\bf1} \wa{\bf1}{\bf2}\wb{\bf2}{5} \, \times \\ 
&\qquad \times \hat{I}'_-(\xx)\left(\tilde{d}_2\wb{\bf4}{\bf2}^{2s_B}\right) \, + \\
&\hspace{-0.5cm}+\left[\dfrac{1}{4 p_5\cdot p_2}+\dfrac{1}{4 p_5\cdot p_4}\right]\hat{I}_-(\xx)\left(\tilde{d}_2(\tilde{d}_2 \wb{\bf4}{\bf2}^{2s_B})\right) \Biggr\}\Biggr\} \,.
\end{aligned}
\end{equation}

\subsubsection{Classical large spin limit}

We are now interested in taking the large classical spin limit of the
amplitude \eq{eq:5pti_soft} in order to compute a sensible amplitude.
I will follow the works presented in \Refe{Guevara_2019, Guevara_2019_b} and discussed in \sect{sec:classical_inf_spin}, and extend it to the five-point amplitude.

We recall that a consistent picture of spin-induced multipoles of a pointlike particle must be formulated in the particle’s rest frame and that
all spin multipoles of the amplitude can be extracted through a finite Lorentz boost which is needed to bridge the gap between two states with different momenta.
To start, we note that any two four-vectors $p_4$ and $p_2$ of equal mass $m$ may be related by the spinorial transformations:
\begin{align}
\qra{\bf{4}^b} & = U_{24~a}^{~~b}\exp\!\left(\dfrac{i}{m^2} p_{2}^\m q^\n \s_{\m\n}\!\right)\qra{\bf{2}^a} \,, \qquad
\qrb{\bf{4}^b} & = U_{24~a}^{~~b} \exp\!\left(\dfrac{i}{m^2} p_{2}^\m q^\n \bar{\s}_{\m\n}\!\right) \qrb{\bf{2}^a} \,,
\end{align}
where $U_{24} \in SU(2)$ is a little-group transformation
that depends on the specifics of the massive-spinor realization, $q$ is the boost $q\equiv p_2+p_4$ needed to connect the two spinors.
Note that $q^2=(p_2+p_4)^2=0$ as we are working in the on-shell framework and we are in the center of mass frame of the interactions.\\
Furthermore, using \eq{eq:Tensor2Vector}, we write the above
exponents as:
\begin{equation}
\frac{i}{m^2} p_2^\mu q^\nu \sigma_{\mu\nu,\alpha}^{~~~~\;\beta} = q \cdot (a_B)_{\alpha}^{~\beta} , \qquad
\frac{i}{m^2} p_2^\mu q^\nu \bar{\s}^{~~~\dot{\a}}_{\mu\nu,~\dot{\beta}}
=-q \cdot (a_B)^{\dot{\alpha}}_{~\,\dot{\beta}} \,
\end{equation}
and using \eq{eq:Lorentz1to2SpinorPL}, we write
\begin{equation}
\begin{aligned}
&\qra{\bf 4}^{\odot 2s_B}\! = e^{q \cdot a_B}\big\{U_{24} \qra{\bf 2}\big\}^{\!\odot 2s_B} \,, \qquad
\qrb{\bf 4}^{\odot 2s_B}\!= e^{-q \cdot a_B}\big\{ U_{24} \qrb{\bf 2}\big\}^{\!\odot 2s_B} \,, \\
&\qla{\bf 4}^{\odot 2s_B}\! = \big\{ U_{24} \qla{\bf 2} \big\}^{\!\odot 2s_B} e^{-q \cdot a_B} \,, \qquad \!\!\!
\qlb{\bf 4}^{\odot 2s_B}\!= \big\{ U_{24} \qla{\bf 4} \big\}^{\!\odot 2s_B} e^{q \cdot a_B} \,.
\end{aligned} 
\end{equation}

With this brief recap complete, we now have all the necessary tools to compute the classical limit of the amplitude in \eq{eq:5pti_soft}. 

I begin by considering the terms
\begin{subequations}\label{eq:inf_a}
\begin{flalign}
\lim_{s_B\to \infty} \wa{\bf4}{\bf2}^{2s_B} & =
\lim_{s_B\to \infty} \big\{U_{24}\qla{\bf2}\big\}^{\odot 2s_B}e^{-q\cdot a_B}\qra{\bf2}^{\odot 2s_B} = \nn \\
\label{eq:inf_uno_a}
& \qquad =e^{-q\cdot a_B} \lim_{s_B\to \infty} m_B^{2s_B}(U_{24})^{\odot 2s_B} \,,\\
\lim_{s_B\to \infty} \wb{\bf4}{\bf2}^{2s_B} & =
\lim_{s_B\to \infty} \big\{U_{24}\qlb{\bf 2}\big\}^{\odot 2s_B}e^{q\cdot a_B}\qrb{\bf 2}^{\odot 2s_B}= \nn \\
\label{eq:inf_uno_b}
& \qquad =e^{q\cdot a_B} \lim_{s_B\to \infty} m_B^{2s_B}(U_{24})^{\odot 2s_B} \,.
\end{flalign}
\end{subequations}
We remark that, as expected, the mass-infinities ($m^{2s_B}$) in \eq{eq:inf_uno_a}, \eq{eq:inf_uno_b} cancel out the terms ($m^{-2s_B}$) that appear in the amplitude \eq{eq:5pti_soft}; and that the remaining unitary factor of $(U_{24})^{\odot 2s_B}$ parametrizes an arbitrary little-group transformation, and it cancels out in a GEV sense as in \sect{sec:classical_inf_spin}

The discussion about the terms $\tilde{d}_l \wb{\bf4}{\bf2}^{2s_B}$, and $\tilde{d}_l\left(\tilde{d}_l \wb{\bf4}{\bf2}^{2s_B}\right)$ is more delicate. Consider
\begin{equation}
\lim_{s_B\to \infty} \tilde{d}_2 \wb{\bf4}{\bf2}^{2s_B}= \lim_{s_B\to\infty}\wb{5}{\bf 2}\wb{5}{\de{\tx_2}}\wb{\bf4}{\bf2}^{2s_B}\,,
\end{equation}
we aim to express $\tilde{d}_2$ in a more familiar way, extract it from the limit, and subsequently evaluate the limit. \\
I will use the relation 
\begin{equation}
k_\mu \ve_{k,\nu}^+ J_{\bf p}^{\m\n}=-\frac{i}{\sqrt{2}}\wb{k}{\bf p}\wb{k}{\de{\tx_p}} \,. 
\end{equation}
I will focus on the $\tilde{d}_2$ derivatives, the computation for the $\tilde{d}_4$ is perfectly analogous.
\begin{subequations}\label{eq:inf_b}
\begin{align}
& \lim_{s_B\to \infty} \tilde{d}_2 \wb{\bf4}{\bf2}^{2s_B} = 
\wb{5}{\bf 2}\wb{5}{\de{\tx_2}}\lim_{s_B\to\infty}\wb{\bf4}{\bf2}^{2s_B}= i\sqrt{2} \big( p_{5,\mu} \ve_{5,\nu}^+  J_2^{\mu\nu} \big)
\lim_{s_B\to \infty}\wb{\bf4}{\bf2}^{2s_B}= \nn \\
\label{eq:inf_due}
&\hspace{2.8cm} = i\sqrt{2} \big( p_{5,\mu} \ve_{5,\nu}^+  J_2^{\mu\nu} \big)
e^{q\cdot a_B} \lim_{s_B\to \infty} m_B^{2s_B}(U_{24})^{\odot 2s_B}\\ 
\label{eq:inf_tre}
& \lim_{s_B\to \infty} \tilde{d}_2\left(\tilde{d}_2 \wb{\bf4}{\bf2}^{2s_B}\right) = -2 \big( p_{5,\mu} \ve_{5,\nu}^+  J_2^{\mu\nu} \big)^2
e^{q\cdot a_B} \lim_{s_B\to \infty} m_B^{2s_B}(U_{24})^{\odot 2s_B} 
\end{align}
\end{subequations}
Using now \eq{eq:GOV}, we can write:
\begin{equation}
k_\m\ve_{k,\n}^+J^{\m\n}_{\bf p}= 2i\, (a_B \cdot k)(p\cdot \ve^+)\,,
\end{equation}
thus
\begin{subequations}\label{eq:inf_c}
\begin{align}
& \lim_{s_B\to \infty} \tilde{d}_2 \wb{\bf4}{\bf2}^{2s_B} = 
-2\sqrt{2}(a_B\cdot k)(p_2\cdot \ve_5^+) e^{q\cdot a_B} \lim_{s_B\to \infty} m_B^{2s_B}(U_{24})^{\odot 2s_B}=\nn \\
&\hspace{2.8cm}=2\dfrac{\wab{y}{p_2}{5}}{\wa{5}{y}}(a_B\cdot k)e^{q\cdot a_B} \lim_{s_B\to \infty} m_B^{2s_B}(U_{24})^{\odot 2s_B} \\
& \lim_{s_B\to \infty} \tilde{d}_2\left(\tilde{d}_2 \wb{\bf4}{\bf2}^{2s_B}\right) = 
8 (a_B\cdot k)^2(p_2\cdot \ve_5^+)^2 e^{q\cdot a_B} \lim_{s_B\to \infty} m_B^{2s_B}(U_{24})^{\odot 2s_B} = \nn \\
&\hspace{3.7cm}=4\dfrac{\wab{y}{p_2}{5}^2}{\wa{5}{y}^2}(a_B\cdot k)^2e^{q\cdot a_B} \lim_{s_B\to \infty} m_B^{2s_B}(U_{24})^{\odot 2s_B} 
\end{align}
\end{subequations}

Substituting now \eq{eq:inf_a} and \eq{eq:inf_c} in \eq{eq:5pti_soft} we finally obtain the classical 5-point amplitude in the in infinite spin limit, normalized in the GEV sense:\\

\begin{equation}\label{eq:5pti_soft_classica}
\begin{aligned}
\Big\la \hat \MC^{0,\e}_5(\bf{1},\bf{2},\bf{3},\bf{4},5_h^+ ) \Big\ra_{\!\text{cl}} &=
\left(\dfrac{\k}{2}\right)^{\!3} \dfrac{m_A^2 m_B^2}{(p_1+p_3)^2} \, \times \\
& \hspace{-3cm}\times \Biggl\{ \dfrac{1}{\e^3}\sum_{l=1}^4 \frac{\wab{y}{p_l}{5}^2}{ (2 p_5\cdot p_l) \wa{5}{y}^2} 
\left(\hat{I}_-(\xx) e^{q\cdot a_B}+
\hat{I}_+(\xx) e^{-q\cdot a_B}\right)  \,+ \\
& \hspace{-2cm}+ \dfrac{1}{\e^2} \Biggl\{ \left[
\dfrac{ \wab{y}{p_2}{5}}{(4 p_5\cdot p_2)\wa{5}{y}}
\left(1+ \dfrac{\wbb{5}{p_3}{p_2}{5}}{\wbb{5}{p_1}{p_3}{5}}\right)+\dfrac{\wab{y}{p_4}{5}}{(4 p_5\cdot p_4)\wa{5}{y}}
\dfrac{\wbb{5}{p_3}{p_4}{5}}{\wbb{5}{p_1}{p_3}{5}}\right]\wb{5}{\bf{1}}\wa{\bf{1}}{\bf{2}}\wb{\bf{2}}{5}\, \times \\
& \times \left(\hat{I}'_-(\xx)e^{q\cdot a_B}+\hat{I}'_+(\xx)e^{-q\cdot a_B}\right) \,+  \\
&\hspace{-0.5cm}+ 2\dfrac{\wab{y}{p_2}{5}}{\wa{5}{y}} \left[\dfrac{ \wab{y}{p_2}{5}}{(2 p_5\cdot p_2)\wa{5}{y}}-\dfrac{ \wab{y}{p_4}{5}}{(2 p_5\cdot p_4)\wa{5}{y}} \right]\hat{I}_-(\xx) \, (a_B\cdot k)e^{q\cdot a_B} \Biggr\} \, +\\
& \hspace{-2cm} +\dfrac{1}{\e}\Biggl\{\Biggl[\frac{1}{16 p_5\cdot p_2} \left(1+\dfrac{\wbb{5}{p_3}{p_2}{5}}{\wbb{5}{p_1}{p_3}{5}}\right)^{\!\!2}+\dfrac{1}{16 p_5\cdot p_4}\left(\dfrac{\wbb{5}{p_3}{p_4}{5}}{\wbb{5}{p_1}{p_3}{5}}\right)^{\!\!2}\Biggr]\left(\wb{5}{\bf1} \wa{\bf1}{\bf2}\wb{\bf2}{5}\right)^{2} \times \\
& \qquad \times \left(\hat{I}''_-(\xx)e^{q\cdot a_B}+\hat{I}''_+(\xx)e^{-q\cdot a_B}\right) \,+ \\
& \hspace{-0.5cm} + 2\dfrac{\wab{y}{p_2}{5}}{\wa{5}{y}} \Biggl[\dfrac{1}{4 p_5\cdot p_2}\left(1+\dfrac{\wbb{5}{p_3}{p_2}{5}}{\wbb{5}{p_1}{p_3}{5}}\right)-\dfrac{1}{4 p_5\cdot p_4} \dfrac{\wbb{5}{p_3}{p_4}{5}}{\wbb{5}{p_1}{p_3}{5}}\Biggr]\wb{5}{\bf1} \wa{\bf1}{\bf2}\wb{\bf2}{5} \, \times \\ 
&\qquad \times \hat{I}'_-(\xx)(a_B\cdot k)e^{q\cdot a_B} \, + \\
&\hspace{-0.5cm}+4\dfrac{\wab{y}{p_2}{5}^2}{\wa{5}{y}^2}\left[\dfrac{1}{4 p_5\cdot p_2}+\dfrac{1}{4 p_5\cdot p_4}\right]\hat{I}_-(\xx)(a_B\cdot k)^2 e^{q\cdot a_B} \Biggr\}\Biggr\} \,.
\end{aligned}
\end{equation}

This result is consistent with the one presented in section 4.4 of \Refe{Britto_2022} for the scattering of two classical black holes, found via the standard BCFW approach.

%%%%%%%%%%%%%%%%%%%%%%%%%%%%%%%%%%%%%%%%%%%%%%%%
%----------------------------------------------%
%%%%%%%%%%%%%%%%%%%%%%%%%%%%%%%%%%%%%%%%%%%%%%%%

\clearpage

\section{Spinning waveforms from KMOC at leading order}\label{sec:cap7}

In this section, I employ the methods discussed in \sect{sec:cap2} to compute the strain at future null infinity. At leading order, the only relevant amplitude is the five-point amplitude. While it may seem straightforward to simply substitute this amplitude and compute the observable, this naive approach proves difficult to apply in practice. A more efficient strategy is the one used in \Refe{art_4}:
we factorize the five-point amplitude into the three- and four-point amplitudes obtained in \sect{sec:cap3} and \sect{sec:cap5}, respectively. \\

Let us start with a brief recap of the results obtained in \sect{sec:cap2}.\\
The strain at future null infinity is:
\begin{align}\label{eq:strain}
&h(x)=\frac{\k}{8 \pi |\vec{x}\;\!|} 
\int_{0}^{\infty}\! \hat{\rm{d}} \omega \,\Big[W(b; k^-) e^{-i \omega u }  +\left[ W(b; k^+) \right]^*  e^{i \omega u} \Big]\,,
\end{align}
where $k^\mu = \omega n^\mu = \omega(1, \hat{x})$ with \(\hat{x} = \vec{x} / \lvert \vec{x} \rvert\), \(u = x^0 - \lvert \vec{x} \rvert\) is the retarded time, $W(b; k^\pm)$ is the helicity-dependent \textit{spectral waveform} of the emitted gravitational wave and $b$ is the impact parameter.  \\
We recall that, for the classical scattering of two massive particles, the spectral waveform can be computed from the S-matrix
\begin{equation}\label{eq:KMOCwaveform}
i W(b,k^h)= \Lexp \int \mathrm{d}\m \, e^{i(q_1\cdot b_1 + q_2 \cdot b_2)}\ \mathcal{I}_{a_h} \Rexp \, ,
\end{equation}
where the double-angle brackets enforce the
classical limit of the expression inside, and we have defined
\begin{subequations}
\begin{gather}
\hat{\d}^{(D)} (q_1 + q_2 - k)\, \mathcal{I}_{a_h} = \la p_1' p_2'  |  S^{\dagger}a_h(k) S | p_1 p_2 \ra \, , \\ 
\label{eq:measureinnn}
\mathrm{d}\m = \left[\prod_{i=1,2} \hat{\mathrm{d}}^D q_i\ \hat{\d}(-2{p}_i\cdot q_i +q_i^2)\right]\hat{\d}^{D} (q_1 + q_2 - k) \, ,
\end{gather}
\end{subequations}
and $q_i = p_i - p_i'$. \\
At leading order in perturbation theory, we have
\begin{equation}\label{eq:KMOCtree}
W^{(0)}(b, k^h)= \int \mathrm{d}\m \, e^{i(q_1\cdot b_1 + q_2 \cdot b_2)}\, \MC_{5,\rm cl}^{(0)}(q_1, q_2, k^h) \, ,	
\end{equation}
where $\mathcal{M}_{5,\rm cl}^{(0)}(q_1, q_2, k^h)$ denotes the classical tree-level scattering amplitude. 

It is now useful to perform the following manipulation to set $b_1=b$ and $b_2=0$. We are interested in the strain at future null infinity so we will eventually consider the waveform in the time domain; notice now that 
\begin{equation}
\begin{aligned}
&\int \hat{\mathrm{d}}\omega \, e^{-i\omega u}e^{i(q_1\cdot b_1+q_2\cdot b_2)} \,\hat{\d}^{D} (q_1 + q_2 - k)=\int \hat{\mathrm{d}}\omega \, e^{-i\omega u}e^{i(q_1\cdot b_1+k\cdot b_2-q_1 \cdot b_2)} \,\hat{\d}^{D} (q_1 + q_2 - k)=\\
&= \int \hat{\mathrm{d}}\omega \, e^{-i\omega u}e^{iq_1\cdot(b_1-b_2)}e^{i\omega n\cdot b_2} \,\hat{\d}^{D} (q_1 + q_2 - k)=\int \hat{\mathrm{d}}\omega \, e^{-i\omega (u-n\cdot b_2)}e^{iq_1\cdot(b_1-b_2)}\,\hat{\d}^{D} (q_1 + q_2 - k) \,,
\end{aligned}
\end{equation}
We can now perform the translation $u \to u+n\cdot b_2$ and define $b=b_1-b_2$ to obtain
\begin{equation}
\int \hat{\mathrm{d}}\omega \, e^{-i\omega u}e^{iq_1\cdot b}\,\hat{\d}^{D} (q_1 + q_2 - k) \,,    
\end{equation}
From this discussion, we conclude that for the purpose of the computation, we can set \( b_1 = b \) and \( b_2 = 0 \); the symmetric result can be recovered later by applying the appropriate translation at the level of the time-domain waveform.\\

\noindent We are now ready to compute \eq{eq:KMOCtree} $D=4$ dimensions. \\
Let us start by integrating out $q_2$ using the momentum-conserving delta function.
Recalling that in the classical limit, at leading order, we can neglect the shift \(q_i^2\) in the delta functions of \eq{eq:measureinnn}, we have
\begin{equation}
\begin{aligned}
W^{(0)}(b, k^h) & =\int \hat{\mathrm{d}}^4 q_1 \hat{\mathrm{d}}^4 q_2\, \hat{\d}(-2{p}_1\cdot q_1) \hat{\d}(-2{p}_2\cdot q_2)\hat{\d}^{4} (q_1 + q_2 - k)  \, e^{i(q_1\cdot b_1 + q_2 \cdot b_2)}\, \MC_{5,\rm cl}^{(0)}(q_1, q_2, k^h) = \\
& = \int \hat{\mathrm{d}}^4 q_1 \, \hat{\d}(-2{p}_1\cdot q_1) \hat{\d}\big(-2{p}_2\cdot (k-q_1)\big) e^{iq_1\cdot b}\, \MC_{5,\rm cl}^{(0)}(q_1, q_2, k^h)
\end{aligned}    
\end{equation}
It is now useful to parameterize the remaining integration variable $q_1$ as
\begin{align}\label{eq:qparam}
q_1= z_1 v_1 + z_2 v_2 + z_v \tilde{v} + z_b \tilde{b}\, ,
\end{align}
with
\begin{align}
v_1^\mu= \frac{p_1^\mu}{m_1}\, ,\quad 
v_2^\mu= \frac{p_2^\mu}{m_2}\, , \quad
\tilde{v}^\mu = \frac{v^\mu}{\sqrt{-v^2}}\, , 
\quad\tilde{b}^\mu = \frac{b^\mu}{\sqrt{-b^2}}\, ,
\end{align}
where $p_i$ are the incoming momenta of the scattered objects,
and
\begin{align}
 v^\mu = \epsilon^{\mu \nu \rho \sigma} v_{1 \nu}  v_{2 \rho}  \tilde{b}_{\sigma}\, , \quad v^2 = - (\gamma^2-1)\,, \quad  \gamma = v_1 \cdot v_2 \,.
 \end{align}
Furthermore, if we identify $b$ to be the asymptotic impact parameter, we also have $b\cdot v_1 = b\cdot v_2 =0$.

Let us now focus on the measure $d\m$ and see how it transforms under this manipulation. 
Under the change of variables \eq{eq:qparam}, the Jacobian of the transformation is 
\begin{equation}
\det J= |\e_{\m\n\r\s} \tilde{v}^\m v_1^\n v_2^\r \tilde{b}^\s|=\sqrt{\g^2-1} \,,
\end{equation}
thus 
\begin{equation}
\mathrm{d}^4q_1=\sqrt{\g^2-1} \,\mathrm{d}z_1\mathrm{d}z_2\mathrm{d}z_v\mathrm{d}z_b \,.
\end{equation}
Let us now see how the $\d$-s transform after the change of variables of \eq{eq:qparam}:
\begin{subequations}
\begin{align}
&-2p_1\cdot q_1  = -2z_1 (p_1\cdot v_1)-2z_2(p_1\cdot v_2)-2z_v (p_1\cdot \tilde{v})-2z_b (p_1\cdot \tilde{b}) =  \nn\\
& \hspace{1.75cm}= -2m_1(z_1+z_2\g) \\
&-2p_2\cdot (k-q_1)  = -2m_2(v_2\cdot k) +2m_2(\g z_1+z_2)=
-2m_2\omega_2 +2m_2(\g z_1+z_2)\,,
\end{align}
\end{subequations}
where we have defined
\begin{align}
w_1 = v_1 \cdot k\,,  \qquad w_2 = v_2 \cdot k \,. 
\end{align}
Using now the property of the delta function: $\d(ax)=\frac{1}{|a|}\d(x)$; the measure becomes 
\begin{align}
\mathrm{d}\mu &= \frac{\sqrt{\gamma^2-1} \, \mathrm{d}z_1\mathrm{d}z_2\mathrm{d}z_v\mathrm{d}z_b}{(4\pi)^2 m_1 m_2} \,
\d\left(z_1 + \g z_2\right) \d\left(\omega_2-\g z_1-z_2\right) =\nn\\
&= \frac{\sqrt{\gamma^2-1} \, \mathrm{d}z_1\mathrm{d}z_2\mathrm{d}z_v\mathrm{d}z_b}{(4\pi)^2 m_1 m_2} \,
\d\left(z_1 + \g z_2\right) \d\left((\g^2-1)z_2+\omega_2\right)\,,
\end{align}

Altogether, \eq{eq:KMOCtree} becomes
\begin{align}
W^{(0)}(b,k^h) &= \int \mathrm{d} z_1 \mathrm{d} z_2 \mathrm{d} z_v \mathrm{d} z_b \,e^{ - i z_b \sqrt{-b^2} } \d\left(z_1 + \g z_2\right) \d\left((\g^2-1)z_2+\omega_2\right) \times  \nn \\
& \qquad \qquad\times \sqrt{\gamma^2-1}\,\frac{{\mathcal{M}}_{5,\rm cl}^{(0)}(q_1,q_2.k^h)}{(4\pi)^2 m_1 m_2} \,.
\end{align}
We may now use the two delta functions to integrate out $z_1$ and $z_2$, and obtain
\begin{align}\label{eq:z_integrals}
W^{(0)}(b,k^h) &= \int \mathrm{d} z_v \mathrm{d} z_b \,  e^{ - i z_b \sqrt{-b^2} } \frac{\hat{\mathcal{M}}_{5,\rm cl}^{(0)}}{(4\pi)^2 m_1 m_2 \sqrt{\gamma^2-1}} \ ,
\end{align}
where
\begin{equation}
	\hat{\mathcal{M}}_{5,\rm cl}^{(0)} = \left.\mathcal{M}_{5,\rm cl}^{(0)}(q_1,q_2,k^h) \right|_{ z_1 =  \frac{\gamma}{\gamma^2-1} w_2\, , \ z_2 = -  \frac{1}{\gamma^2-1} w_2 }\, .
\end{equation}
\\

\begin{figure}[H]
    \centering
	\includegraphics[scale=0.7]{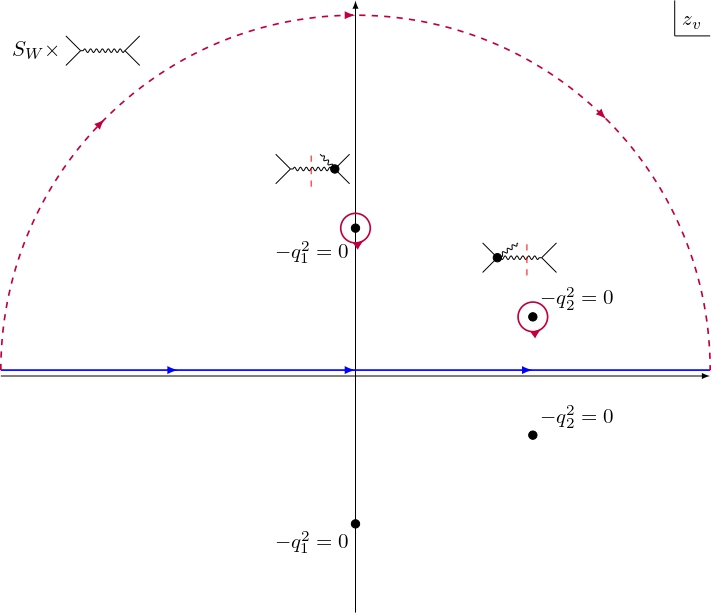}
	\caption{The deformation of the contour $\color{red}{\mathcal{C}^{(+)}}$ in the complex $z_v$ plane allows to evaluate the integral of the five-point tree-level amplitude directly in terms of the factorization channels. Figure reproduced from \Refe{art_4}.}
	\label{fig:analytic_continuation}
\end{figure}

\noindent We are now left to evaluate this integral. 
Let us start by evaluating the $z_v$ integral. 
To accomplish this we deform the contour of integration from the real axis to infinity through the upper half-plane (UHP) (or equivalently the LHP) as in \fig{fig:analytic_continuation}. 
Thanks to the Riemann-Lebesgue's lemma it is possible to show that the principal value
\footnote{Notice that we can consider the principal value of the integral without loss of generality, as, by the Sokhotski–Plemelj theorem, the integral and its principal value differ by a term proportional to a delta function. This delta function is proportional to non-negative powers of \( z_b \), which, when substituted into \eq{eq:strain}, result in terms that integrate to short-range interactions, \( \delta^{(n)}(\sqrt{-b^2}) \), and are thus irrelevant in the classical approximation.}
of the integral over the dotted half-circumference vanishes in the limit of infinite radius;
hence, using the Cauchy theorem, we have
\begin{equation}
\begin{aligned}
& \int_{-\infty}^{+\infty}\mathrm{d} z_v\, \hat{\MC}_{5,\rm cl}^{(0)}(q_1,q_2,k^\l)= \int_{\color{red}{\mathcal{C}^{(+)}}} \! \mathrm{d} z_v \, \hat{\mathcal{M}}_{5,\rm cl}^{(0)} (q_1,q_2,k^{\lambda}) \equiv I^{\lambda}_{\text{UHP}} \\
& \hspace{2.5cm} \Rightarrow I^{\lambda}_{\text{UHP}}= 2\pi i\sum_i \Res{z=z_i} \hat{\MC}_{5,\rm cl}^{(0)}(q_1,q_2,k^\l)    
\end{aligned}
\end{equation}
Let us now study the poles of this integrand. 

We have simple poles of the tree-level five-point amplitude in the complex $z_v$-plane corresponding to $q_1^2=0$ and $q_2^2=0$ which correspond to factorization channels:
\begin{equation}
\begin{split}
\frac{I_{\mathcal{C}^{(+)}_{q_1}}^{\lambda}}{2 \pi i} &= \sum_{h} \mathcal{M}_{4, \mathrm{cl}}^{(0)}(p_2,k^{\lambda},-q_1^{h}) \mathcal{M}_{3, \mathrm{cl}}^{(0)}(p_1,q_1^{h}) \operatornamewithlimits{Res}_{z = \hat{z}_{1}} \frac{-1}{q_1^2} \,,\\
\frac{I_{\mathcal{C}^{(+)}_{q_2}}^{\lambda}}{2 \pi i} &= \sum_{h} \mathcal{M}_{4, \mathrm{cl}}^{(0)}(p_1,k^{\lambda},-q_2^{h}) \mathcal{M}_{3, \mathrm{cl}}^{(0)}(p_2,q_2^{h}) \operatornamewithlimits{Res}_{z = \hat{z}_{2}} \frac{-1}{q_2^2}\,,
\label{eq:Cq1q2_zv}
\end{split}
\end{equation}
where $\hat{z}_{1}$ and $\hat{z}_{2}$ are the solutions in the UHP of the pole constraints $-q_i^2=0$.\\
The pole(s) $\hat{z}_1$ are given by the solutions of 
\begin{align}
-q_1^2=-(z_1^2+z_2^2+2\g z_1z_2-z_v^2-z_b^2)=0 \,,
\end{align}
recalling now that in the hatted-kinematics
\begin{equation}\label{eq:constr_z}
z_1=\dfrac{\g}{\g^2-1}\omega_2 \,,\qquad
z_2=-\dfrac{1}{\g^2-1}\omega_2 \,,
\end{equation}
we find
\begin{align}
-q_1^2=z_v^2+z_b^2+\omega_{2,r}^2=(z_v - \hat{z}_1)(z_v - \hat{z}_1^*)=0 \,,
\end{align}
with
\begin{equation}\label{eq:z1}
\hat{z}_1 = i \sqrt{z_b^2+w_{2,r}^2} \,.    
\end{equation}
Meanwhile, the pole(s) $\hat{z}_2$ are given by the solutions of 
\begin{align}
-q_2^2&=-(k-q_1)^2=-q_1^2+2k\cdot q_1= \\
&=-(z_1^2+z_2^2+2\g z_1z_2-z_v^2-z_b^2)+2z_1(k\cdot v_1)+2z_2(k\cdot v_2)+2z_v(k\cdot \tilde{v})+2z_b(k\cdot \tilde{b})=0 \,, \nn
\end{align}
which, after imposing the constraints in \eq{eq:constr_z}, becomes
\begin{align}\label{eq:qu2}
-q_2^2&=z_v^2+z_b^2+\omega_{2,r}^2+2\g\omega_{1,r}\omega_{2,r}-
2\omega_{2,r}^2+2z_v(k\cdot \tilde{v})+2z_b(k\cdot \tilde{b})= \nn\\
&= (z_v+k\cdot\tilde{v})^2 +(z_b+k\cdot\tilde{b})^2-(k\cdot \tilde{v})^2-(k\cdot \tilde{b})^2 -\omega_{2,r}^2+2\g\omega_{1,r}\omega_{2,r}\,.
\end{align}
Notice now that 
\begin{equation}
-(k\cdot \tilde{v})^2=(k\cdot \tilde{b})^2+\omega_{1,r}^2+\omega_{2,r}^2 -2\g\omega_{1,r}\omega_{2,r}\,,
\end{equation}
henceforth \eq{eq:qu2} reduces to
\begin{align}
-q_2^2&=(z_v+k\cdot\tilde{v})^2+(z_b+k\cdot\tilde{b})^2+\omega_{1,r}^2= (z_v - \hat{z}_2)(z_v - \hat{z}_2^*)=0 \,,
\end{align}
with
\begin{equation}\label{eq:z2}
\hat{z}_2 = -\tilde{v} \cdot k +i \sqrt{(z_b + \tilde{b}\cdot k)^2+w_{1,r}^2} \,,
\end{equation}

Other than the residues related to these simple poles, we should also take into account (half) the residue at infinity of the five-point amplitude, which we label as $I_{\mathcal{C}^{(+)}_{\infty}}^{\lambda}$
\footnote{It is interesting to notice that the residue at infinity 
is completely determined by the leading soft theorem in the classical limit. For a more detailed discussion see App. A of \Refe{art_4}.}.\\

\noindent Summarizing, we have the following equality
\begin{equation}
	I_{\text{UHP}}^{\lambda}  = I^{\lambda}_{\mathcal{C}^{(+)}_{q_1}} + I^{\lambda}_{\mathcal{C}^{(+)}_{q_2}} + I^{\lambda}_{\mathcal{C}^{(+)}_{\infty}}\,, 
 \label{eq:IUHP}
\end{equation}
The treatment of the residue at infinity is quite delicate, as it can have spurious pole that cancel only at later stages.
To avoid having to deal with this problem we can use the following trick: we split the $z_v$-integration into two equal pieces and deform the integration contour in the UHP and LHP. As a consequence, the contributions from the arcs at infinity cancel each other ($I^{\lambda}_{\mathcal{C}^{(+)}_{\infty}} = I^{\lambda}_{\mathcal{C}^{(-)}_{\infty}}$) and only the finite poles contribute:
\begin{align}
	\label{eq:zv_integral_toresidues}
	\hspace{-10pt}	\int_{-\infty}^{+\infty} \mathrm{d} z_v\ \hat{\mathcal{M}}_{5,\rm cl}^{(0)} &= \frac{1}{2} \left(I_{\text{UHP}}^{\lambda} - I_{\text{LHP}}^{\lambda}\right) \nonumber = \frac{1}{2} \left(I^{\lambda}_{\mathcal{C}^{(+)}_{q_1}} + I^{\lambda}_{\mathcal{C}^{(+)}_{q_2}}  - I^{\lambda}_{\mathcal{C}^{(-)}_{q_1}} - I^{\lambda}_{\mathcal{C}^{(-)}_{q_2}} \right) \nonumber \\
 \hspace{-10pt} &= i \pi \sum_{i=1,2}\left(\underset{z_v=z_i}{Res} \hat{\mathcal{M}}_{5,\rm cl}^{(0)} - \underset{z_v=z_i^*}{Res} \hat{\mathcal{M}}_{5,\rm cl}^{(0)}\right) \,.
\end{align}

The direct computation of the $z_b$ integral is rather involved and evaluates to Bessel functions, as shown in \Refe{Jakobsen_2021} and \Refe{Brandhuber_2023}; however, since we are interested in the strain at future null infinity we need to compute the waveform in the time domain.
By switching the order of integration, the computation greatly simplifies. \\
To compute the $\omega$ integral, it is useful to extract 
all $\omega$ dependencies and define the dimensionless variables
\begin{equation}\label{eq:scalingwz}
w_i \to \omega\, \bar{w}_i\ , \,\,\, \bar{w}_i = n \cdot v_i \,, \quad z_v \to \omega\, z_v \,, \quad z_b \to \omega\, z_b \,. 
\end{equation} 
To determine the scaling of $\hat{\MC}^{(0)}_{5, \rm cl}$ with $\omega$, notice that the five-point amplitude has dimension $4 - 5 = -1$, however the gravitational coupling $\k$ has dimension $-1$ and in the perturbative expansion we have $\hat{\MC}^{(L)}_{5, \rm cl}\sim \k^{3+2L}$. 
Notice that if we express the Mandelstam variables of $\hat{\MC}^{(0)}_{5, \rm cl}(q_1,q_2,k)$ using \eq{eq:scalingwz}, the only dimensionful parameter is $\omega$. Indeed
\begin{equation}
\begin{gathered}
-q_1^2=\omega^2(z_v^2+z_b^2+\bar{\omega}^2_{2,r}) \,, \qquad -q_2^2=\omega^2\left[(z_v+n\cdot \tilde{v})^2+(z_b+n\cdot\tilde{b})^2+\bar{\omega}^2_{1,r}\right] \,,\\
k^2=0\,, \qquad 
k\cdot q_1=\omega \bar{\omega}_1\,, \qquad 
k\cdot q_2=\omega \bar{\omega}_2\,.
\end{gathered}
\end{equation}
If we now express $\hat{\MC}^{(0)}_{5, \rm cl}(q_1,q_2,k)$ as 
\begin{equation}
\hat{\MC}^{(0)}_{5, \rm cl}(q_1,q_2,k)=\k^{3}\omega^{\a}f(\bar{\omega}_i, z_v, z_b)\,.
\end{equation}
Since $\omega$ has dimension $1$, by dimensional analysis we find the $\omega$ dependence of this amplitude to be fixed by 
\begin{equation}
[\mathcal{M}_{5, {\rm cl}}^{(0)}]=[\k^{3}][\omega^{\a}] \quad
\Rightarrow \quad -1=-3+\a \,.
\end{equation}
Since $\omega$ has dimension $1$, we finally obtain $\a=2$.
Therefore to extract all $\omega$ dependencies out of $\mathcal{M}_{5,\rm cl}^{(0)}$ we rescale it as 
\footnote{If we where to consider higher order contributions the formula would be modified to \\$\mathcal{M}_{5,\rm cl}^{(L)}\to \omega^{-2+L}\mathcal{M}_{5,\rm cl}^{(L)}$, see \Refe{art_4}}
\begin{equation}\label{eq:omega_scaling_scalar}
\mathcal{M}_{5,\rm cl}^{(0)} \to \omega^{-2}  \mathcal{M}_{5, {\rm cl}}^{(0)}\,.
\end{equation}
Using now the fact that, at tree level,
\begin{equation}
W^{(0)}(b,k^{\pm})=[W^{(0)}(-b,k^{\mp})]^*\,,
\end{equation}
we are finally able to rewrite \eq{eq:strain} as:
\begin{align}\label{eq:strain_fin}
h^{(0)} ( x)  =&  \frac{\kappa}{8 \pi |\vec{x}\, | } \nn 
\int_{-\infty}^{\infty}\!\hat{\mathrm{d}} \omega
\ \Big[   W^{(0)}(b; k^-)  e^{-i \omega u }\Big]=\\
 & =\frac{\kappa}{8 \pi |\vec{x}\, | } \nonumber 
\int_{-\infty}^{\infty}\!
\hat{\mathrm{d}} \omega
\ \left[   \int\! \mathrm{d} z_v \mathrm{d} z_b \,  e^{ - i z_b \sqrt{-b^2} } \frac{\hat{\mathcal{M}}_{5,\rm cl}^{(0)}}{(4\pi)^2 m_1 m_2 \sqrt{\gamma^2-1}}  e^{-i \omega u} \right]=\\
& =\frac{\kappa}{8 \pi |\vec{x}\, | } \nonumber 
\int_{-\infty}^{\infty}\!
\hat{\mathrm{d}} \omega
\ \left[   \int\! \mathrm{d}  z_v \mathrm{d} z_b \,\omega^2\,  e^{ - i \omega\,z_b \sqrt{-b^2} } \frac{\omega^{-2}\,\hat{\mathcal{M}}_{5,\rm cl}^{(0)}}{(4\pi)^2 m_1 m_2 \sqrt{\gamma^2-1}}  e^{-i \omega u} \right]=\\
 & =\frac{\kappa}{8 \pi |\vec{x}\, | } \nonumber 
\int_{-\infty}^{\infty}\!
\hat{\mathrm{d}} \omega
\ \left[ \int\!\mathrm{d} z_b \,  e^{ - i \omega\,z_b \sqrt{-b^2} } \frac{\left(I_{\text{UHP}}^{\lambda} - I_{\text{LHP}}^{\lambda}\right) \nonumber}{2(4\pi)^2 m_1 m_2 \sqrt{\gamma^2-1}}  e^{-i \omega u} \right]=\\
& =\frac{\kappa}{8 \pi |\vec{x}\, | } \nonumber 
\int\!\mathrm{d} z_b \frac{\left(I_{\text{UHP}}^{\lambda} - I_{\text{LHP}}^{\lambda}\right) \nonumber}{2(4\pi)^2 m_1 m_2 \sqrt{\gamma^2-1}}\int_{-\infty}^{\infty} \!\hat{\mathrm{d}} \omega \,e^{ - i \omega\,(z_b \sqrt{-b^2}+u) }=\\
& =\frac{\kappa}{8 \pi |\vec{x}\, | } 
\int\!\mathrm{d} z_b \frac{\left(I_{\text{UHP}}^{\lambda} - I_{\text{LHP}}^{\lambda}\right) }{2(4\pi)^2 m_1 m_2 \sqrt{\gamma^2-1}}\d(z_b \sqrt{-b^2}+u) 
\end{align}
Therefore, we have
\begin{equation}\label{eq:tree-waveform}
h^{(0)}(x) = \frac{\kappa}{(4 \pi)^3 |\vec{x}| \sqrt{-b^2} } \frac{(I_{\text{UHP}}^{\lambda} - I_{\text{LHP}}^{\lambda})}{ 4 m_1 m_2 \sqrt{\gamma^2 - 1}} \Bigg|_{z_b = - u/\sqrt{-b^2}}\,,
\end{equation}
From this point onwards a direct computation becomes rather involved, therefore I will resort to \textit{Mathematica} to obtain the final result.\\

\subsection{Tree-level waveform for Schwarzschild black holes}

We now compute the tree-level scattering waveform for spinless particles by combining \eq{eq:tree-waveform}, \eq{eq:Cq1q2_zv} and \eq{eq:zv_integral_toresidues} with the three-point and four-point amplitudes previously obtained:
\begin{align}
	\mathcal{M}_{3,\rm cl}^{(0)}(p, k) &= - \kappa \, m^2 (\varepsilon \cdot v)^2 \,, \nonumber \\
 \mathcal{M}_{4,\text{cl}}^{(0)}(p,k_1,k_2) &= \frac{\kappa^2 m^2}{q^2}\left(\frac{v \cdot F_1 \cdot F_2 \cdot v}{v \cdot k_1}\right)^2 \,,
\end{align}
with $q=(q_1-q_2)/2$.

Substituting in \eq{eq:tree-waveform}, we find
\begin{equation}
\begin{split}
&h^{(0)} ( x) = \frac{G^2 m_1 m_2}{ 
|\vec{x}\, | \sqrt{-b^2}} \frac{1}{\bar{w}_{1}^2 \bar{w}_{2}^2 \sqrt{1+T_{2}^2} \left(\gamma +\sqrt{\left(1+T_{1}^2\right) \left(1+T_{2}^2\right)}+T_{1} T_{2}\right)}\\
&\quad \times \Bigg( \frac{3 \bar{w}_{1} + 2\gamma \left(2 T_1 T_2 \bar{w}_{1}-T_2^2 \bar{w}_{2}+\bar{w}_{2}\right)-\left(2 \gamma ^2-1\right) \bar{w}_{1}}{\gamma ^2-1}f_{1,2}^2 \\
&\quad-\frac{4\gamma T_2 \bar{w}_{2} f_{1}+2\left(2 \gamma ^2-1\right) \left[T_1 \left(1+T_2^2\right) \bar{w}_{2} f_{1}+T_2 (T_1 T_2 \bar{w}_{1}+\bar{w}_{2}) f_{2}\right]}{ \sqrt{\gamma ^2-1}} f_{1,2} \\
&\quad+4 \left(1+T_2^2\right) \bar{w}_{2}f_{1} f_{2}-4 \gamma  \left(1+T_2^2\right) \bar{w}_{2} \left(f_{1}^2+f_{2}^2\right)+2 \left(2 \gamma ^2-1\right) \left(1+2 T_2^2\right) \bar{w}_{2} f_{1} f_{2} \Bigg) + \left(1\leftrightarrow 2\right)\ ,
\end{split}
\end{equation}
where we have defined
\begin{align}
T_i &\coloneq \frac{\sqrt{\gamma^2-1} \, (u-b_i\cdot n)}{\sqrt{-(b_1-b_2)^2} \, \bar{w}_i}\ ,\\
	S^2 &\coloneq \frac{T_1^2-2 \gamma  T_1 T_2+T_2^2}{\gamma ^2-1} - 1\ , \\
f_{1,2} &= v_1 \cdot \varepsilon_k\, v_2 \cdot n - v_1 \cdot n\, v_2 \cdot \varepsilon_k\ ,\\
f_{i} &= \tilde{b} \cdot \varepsilon_k\, v_i \cdot n - \tilde{b} \cdot n\, v_i \cdot \varepsilon_k\ .
\end{align}

\subsection{Tree-level waveform for Kerr black holes}
\label{sec:Kerr}
Let us now compute tree-level scattering waveform Kerr black holes, that we identify with spinning point particles. 
In this case the three-point and the four-point Compton amplitudes are:
\begin{equation}
\begin{gathered}
\mathcal{M}^{(0)}_{3,\rm cl} = - \kappa \, m^2 \, (\varepsilon \cdot v)^2 \, \exp\left( \frac{i \, \varepsilon \cdot S \cdot k}{\varepsilon \cdot v} \right)\ ,\\
\mathcal{M}^{(0)}_{4,\rm cl} = \frac{\kappa^2 m^2 \omega_0^2}{8 (k_1 \cdot k_2) (k_1 \cdot v)^2} \exp \left( \frac{\omega_1}{\omega_0} \right) + \mathcal{O}(S^5)\, ,   
\end{gathered}
\end{equation}
with 
\begin{equation}
\begin{gathered}
\omega_0 = -2\, v \cdot F_1 \cdot F_2 \cdot v\, ,  \\
i\, \omega_1 = k_1 \cdot F_2 \cdot v \, S \cdot F_1 - \frac{(k_1 - k_2) \cdot v}{2} S \cdot F_1 \cdot F_2 + \left( 1 \leftrightarrow 2\right)\, ,  
\end{gathered}
\end{equation}
where $S$ is the (mass-normalized) spin tensor of the Kerr black holes and we have defined $A\cdot B\cdot C=A^{\m\n}B_{\n\r}C^{\r\s}\eta_{\s\m}$ and $A\cdot B= A^{\m\n}B_{\n\m}$. \\
Note that 
\begin{equation}
\dfrac{\omega_2}{\omega_0}=\dfrac12\left(\dfrac{\omega_1}{\omega_0}\right)^2 \,,
\end{equation}
where
\begin{align}
\omega_2 &= (k_1\cdot k_2) \bigg[ S \cdot S \left( \frac{F_1 \cdot F_2}{2} - v \cdot F_1 \cdot F_2 \cdot v \right) - 2\, S \cdot S \cdot F_1 \cdot F_2 \bigg] \nonumber \\
& \qquad \qquad  - \frac{\omega_0}{2} (k_1 + k_2) \cdot S \cdot S \cdot (k_1 + k_2)\ ,
\end{align}
and $A\cdot B\cdot C\cdot D=A^{\m\n}B_{\n\r}C^{\r\s}D_{\s\m}$.
Hence, we can write, as anticipated in \sect{sec:cap5}, the Compton amplitude up to fifth order in the spin tensor without any spurious pole as
\begin{equation}\label{eq:Kerr_S2}
\mathcal{M}^{(0)}_{4,\rm cl} = \k^2 m^2
\frac{\omega_0^2+\omega_0\omega_1+\omega_0\omega_2+\frac13\omega_1\omega_2+\frac16\omega_2^2}{8 (k_1 \cdot k_2) (k_1 \cdot v)^2} + \mathcal{O}(S^5) \, .
\end{equation}
We may now be tempted to directly substitute in \eq{eq:tree-waveform} in order to compute the waveform up to the fourth order in spin for both black holes, i.e. $\mathcal{O}(S_1^4,S_2^4)$. 
However in this case the discussion is more delicate.
Indeed, after applying the rescaling \eq{eq:scalingwz}, the degree of homogeneity in $\omega$ is not the one predicted in \eq{eq:omega_scaling_scalar}, as the mass-dimension of $S_i^{\mu \nu}$ is $-1$.
If we write the five-point amplitude in the spin-multipole expansion 
\begin{align}
  \mathcal{M}^{(L)}_{5,{\rm cl}} &= \sum_{s_1,s_2=1}^{\infty} \left[ S_1^{\mu_1 \nu_1} \dots S_1^{\mu_{s_1} \nu_{s_1}} \right] \left[S_2^{\rho_1 \sigma_1} \dots S_2^{\rho_{s_2} \sigma_{s_2}} \right] \mathcal{M}^{(L),(s_1,s_2)}_{\mu_1 \nu_1, \dots \mu_{s_1} \nu_{s_1}, \rho_1 \sigma_1, \dots \rho_{s_2} \sigma_{s_2}}\ ,
 \end{align}
the scaling will be
\begin{equation}
  \mathcal{M}^{(L),(s_1,s_2)} \rightarrow \omega^{-2+L+s_1+s_2} \mathcal{M}^{(L),(s_1,s_2)} \, .
\end{equation}
Therefore it is convenient to compute the tree-level five-point spinning waveform order by order in its spin-multipole expansion,\ie
\begin{equation}
h^{(0)}(x) = \sum_{s_1,s_2=0}^\infty h^{(s_1,s_2)}(x)\, .
\end{equation}
Let us now study how the $z_v$ integral gets modified. 
\begin{equation}
\begin{aligned}
h^{(s_1,s_2)} ( x)  &=  \frac{\k}{8 \pi |\vec{x}\, |}
\int_{-\infty}^{\infty}\!
\hat{\mathrm{d}} \omega
\ \left[   \int\! \mathrm{d} z_v \mathrm{d} z_b \,  e^{ - i z_b \sqrt{-b^2} } \frac{\hat{\mathcal{M}}_{5,\rm cl}^{(0),(s_1,s_2)}}{(4\pi)^2 m_1 m_2 \sqrt{\gamma^2-1}}  e^{-i \omega u} \right]=  \\
& =\frac{\k}{8 \pi |\vec{x}\, | } \int_{-\infty}^{\infty}\!
\hat{\mathrm{d}} \omega \, \left[\int\! \mathrm{d}  z_v \mathrm{d} z_b \,\omega^2\,  e^{ - i \omega\,z_b \sqrt{-b^2} } \frac{{\omega^{-2+s_1+s_2}\,\hat{\mathcal{M}}}_{5,\rm cl}^{(0)(s_1,s_2)}}{(4\pi)^2 m_1 m_2 \sqrt{\gamma^2-1}}  e^{-i \omega u} \right]=\\
&=\frac{\k}{8 \pi |\vec{x}\, | } \int\! \mathrm{d}  z_v \mathrm{d} z_b \, \dfrac{\hat{\mathcal{M}}_{5,\rm cl}^{(0)(s_1,s_2)}}{(4\pi)^2 m_1 m_2 \sqrt{\g^2-1}} \int_{-\infty}^{\infty}\!\hat{\mathrm{d}} \omega \, \omega^{s_1+s_2} e^{-i \omega (u+z_b\sqrt{-b^2}})= \\
&=\frac{\k}{8 \pi |\vec{x}\, | } \int\! \mathrm{d}  z_v \mathrm{d} z_b \, \dfrac{\hat{\mathcal{M}}_{5,\rm cl}^{(0)(s_1,s_2)}}{(4\pi)^2 m_1 m_2 \sqrt{\g^2-1}} \times \\
& \hspace{3cm}\times \left(\dfrac{i}{\sqrt{-b^2}}\right)^{s_1+s_2}\dee{^{(s_1+s_2)}}{z_b^{s_1+s_2}}
\int_{-\infty}^{\infty}\!\hat{\mathrm{d}} \omega \, e^{-i \omega (u+z_b\sqrt{-b^2}})= \\
&=\frac{\k}{8 \pi |\vec{x}\, | }\left(\dfrac{i}{\sqrt{-b^2}}\right)^{s_1+s_2}\!\! \int\! \mathrm{d}  z_v \mathrm{d} z_b \, \dfrac{\hat{\mathcal{M}}_{5,\rm cl}^{(0)(s_1,s_2)}}{(4\pi)^2 m_1 m_2 \sqrt{\g^2-1}} \left(\dee{^{(s_1+s_2)}}{z_b^{s_1+s_2}} \d(u+z_b\sqrt{-b^2})\right)= \\
&=\frac{\k}{2|\vec{x}\, | (4\pi)^3 m_1 m_2 \sqrt{\g^2-1}}\dfrac{(-i)^{s_1+s_2}}{(\sqrt{-b^2})^{s_1+s_2}}\times \\
&\hspace{3cm}\times \int\! \mathrm{d}  z_v \mathrm{d} z_b \,  \d(u+z_b\sqrt{-b^2})\left(\dee{^{(s_1+s_2)}}{z_b^{s_1+s_2}}\hat{\MC}_{5,\rm cl}^{(0)(s_1,s_2)}\right) = \\
&=\frac{\k}{2|\vec{x}\, | (4\pi)^3 m_1 m_2 \sqrt{\g^2-1}}\dfrac{(-i)^{s_1+s_2}}{(\sqrt{-b^2})^{s_1+s_2+1}} \dee{^{(s_1+s_2)}}{z_b^{s_1+s_2}}\int\! \mathrm{d}  z_v \, \hat{\MC}_{5,\rm cl}^{(0)(s_1,s_2)}\Bigg|_{z_b=-u/\sqrt{-b^2}} 
\end{aligned}   
\end{equation}
Consequently, the spinning tree-level waveform can be computed order by order, isolating the $s_1^{\rm th}$ and $s_2^{\rm th}$ multipoles in the factorized amplitudes, as
\begin{equation}\label{eq:tree-waveform-spin}
\hspace{-2pt}h^{(s_1,s_2)}(x) = \frac{\k}{(4 \pi)^3 \, |\vec{x}| \, (\sqrt{-b^2})^{s_1+s_2+1}} \frac{(-i)^{s_1+s_2}}{4 m_1 m_2 \sqrt{\gamma^2 - 1}} \frac{\partial^{(s_1+s_2)}}{\partial z_b^{s_1+s_2}} (I_{\text{UHP}}^{(s_1,s_2)} - I_{\text{LHP}}^{(s_1,s_2)}) \Bigg|_{z_b = - u/\sqrt{-b^2}}\,.
\end{equation}
In conclusion, we have
\begin{equation}
h^{(0)}(x)=\sum_{s_1,s_2=0}^4\dfrac{G^2m_1m_2}{|\vec{x}|(\sqrt{-b^2})^{s_1+s_2+1}}\mathfrak{h}_{s_1,s_2}(x)+\mathcal{O}(S_1^5,S_2^5) \,,
\end{equation}
where the $\mathfrak{h}_{s_1,s_2}$’s are provided in the \href{https://bitbucket.org/spinning-gravitational-observables/tree-level-waveform/}{ancillary files} obtained in \Refe{art_4}.

%%%%%%%%%%%%%%%%%%%%%%%%%%%%%%%%%%%%%%%%%%%%%%%%
%----------------------------------------------%
%%%%%%%%%%%%%%%%%%%%%%%%%%%%%%%%%%%%%%%%%%%%%%%%

\clearpage
\section{Conclusions}\label{sec:concl}
In this thesis, I have studied the classical scattering problem of two Kerr black holes in general relativity using novel quantum field theory techniques within the Post-Minkowskian expansion, extending the subleading soft theorem to account for spinning particles. \\
I began by describing the spinor-helicity formalism, which served as a foundational tool for computing basic amplitudes. I then introduced a general method for obtaining higher-point amplitudes, utilizing the BCFW recursion relations to compute them in a general context. This was later refined to the soft recursion relations, which I applied to compute \((n+1)\)-point amplitudes involving one additional external graviton, starting from the corresponding \(n\)-point amplitudes.\\
With these tools in place, I computed the gravitational Compton amplitude for both spinless and spinning black holes, and the elastic 2-to-2 black hole scattering amplitudes. I also derived the five-point amplitude, which represents the first original result in this thesis. While the five-point amplitude exists in the literature, to my knowledge, no prior work has computed it using soft recursion on the four-point amplitude.

In parallel, I emphasized the necessity of a classical formulation for scattering amplitudes and introduced the KMOC formalism, detailing its complexities and offering a general method for computing observables. In this context, I re-evaluated all quantum amplitudes to derive their classical counterparts, presenting a straightforward method for calculating the classical (infinite-spin) limit.\\
Additionally, I introduced a classical version of the soft recursion relations to derive higher-point classical amplitudes from lower-point ones in a purely classical setting—an approach that, to my knowledge, is presented here for the first time. 

Finally, I applied the KMOC formalism to compute the leading-order waveform for black hole scattering with radiation emission, covering both the spinning and spinless cases.\\

There are several compelling future directions that can be explored:
\begin{itemize}
\item Investigating the structure of spurious poles starting from the \(\mathcal{O}(S^4)\) term in the Compton amplitude for spinning particles could provide new insights into the dynamics of spinning black holes.
\item A deeper understanding of the higher orders in the five-point amplitude, especially in connection with effects like the Spin Memory Effect, could pave the way for more accurate calculations of spin-related phenomena in classical black hole scattering.
\item Further exploration of the relationship between soft factors and asymptotic symmetries is another promising avenue. 
\item It would also be valuable to investigate the double copy structure of soft theorems for massive particles.
\item Finally, it wold be interesting to extend these results to higher-loop or PM orders, including radiative corrections and finite-size effects.
\end{itemize}

%%%%%%%%%%%%%%%%%%%%%%%%%%%%%%%%%%%%%%%%%%%%%%%%
%----------------------------------------------%
%%%%%%%%%%%%%%%%%%%%%%%%%%%%%%%%%%%%%%%%%%%%%%%%

\vspace{4cm}

\section{Acknowledgements}
I would like to express my heartfelt gratitude to Prof. Vittorio Del Duca for guiding and supporting me during this journey.
A sincere thank you to Dr. Riccardo Gonzo, my thesis co-advisor, for his immense patience, invaluable advice, and the knowledge they imparted throughout the entire process of writing this thesis.\\
I am also deeply grateful to the professors at Sapienza, whose teachings sparked my passion for theoretical physics, without whom I would not have reached this milestone.\\
To my family, friends, and university colleagues, thank you for making this achievement truly special.

%%%%%%%%%%%%%%%%%%%%%%%%%%%%%%%%%%%%%%%%%%%%%%%%
%----------------------------------------------%
%%%%%%%%%%%%%%%%%%%%%%%%%%%%%%%%%%%%%%%%%%%%%%%%

\clearpage
\appendix

\section{Conventions}\label{Conventions} 

\subsection{Lorentz algebra}
I follow the work discussed in chapter 27 of \Refe{Schwartz_2013}, and in particular I will mostly follow the notation presented in \Refe{Chung_2019}. \\
We work with the metric $\eta_{\mu\nu} = \text{diag}(+1, -1, -1, -1)$, so that $p^2 = p^\mu p_\mu = (p_0)^2 - (\vec{p})^2$. 
Our convention for the Lorentz generators are fixed by the algebra, 
\begin{equation}\label{eq:LorentzGenDef}
\left[ J^{\mu\nu} , J^{\l\s} \right] = -i \left( \eta^{\mu\l} J^{\n\s} + \eta^{\n\s} J^{\m\l} - \eta^{\m\s} J^{\n\l} - \eta^{\n\l} J^{\m\s} \right) \,.   
\end{equation}

To relate the (connected part) of the Lorentz group $SO(1,3)$ and its double cover $SL(2,\mathbb{C})$, we follow a widely adopted convention for spinors and gamma matrices, 
\begin{align}
\gamma^\mu 
=
\begin{pmatrix}
0 & (\sigma^\mu)_{\a\dot{\b}}
\\
(\bar{\sigma}^\mu)^{\dot{\a}\b} & 0 
\end{pmatrix}
\,, 
\quad
\sigma^\mu = (\iden, \vec{\sigma}) \,, 
\quad
\bar{\sigma}^\mu = (\iden, -\vec{\sigma}) \,, 
\end{align}
where $\vec{\sigma}$ denote Pauli matrices. 
Complex conjugation exchanges undotted and dotted indices. 
Spinor indices are raised and lowered by the invariant tensor of $SL(2,\mathbb{C})$ satisfying 
\begin{equation}\label{SL2C-epsilon} 
\epsilon_{\alpha\beta} = - \epsilon_{\beta\alpha} \,,
\quad 
\epsilon_{\alpha\beta} \epsilon^{\beta\gamma} = \delta_\alpha{}^\gamma \,,
\quad 
\epsilon^{12} = + 1\,, 
\quad
\epsilon_{\dot{\alpha}\dot{\beta}} = (\e_{\alpha\beta})^*  \,. 
\end{equation}
For example, $\lambda^\alpha = \epsilon^{\alpha\beta} \lambda_\beta$ and $\tilde{\lambda}_{\dot{\alpha}} = \epsilon_{\dot{\alpha} \dot{\beta}} \tilde{\lambda}^{\dot{\beta}}$.

For any (momentum) 4-vector, the bi-spinor notation is defined by 
\begin{equation}
p_{\alpha\dot{\alpha}} = p_\mu (\sigma^\mu)_{\alpha\dot{\alpha}} 
\,,
\quad 
p^2 = \det(p_{\alpha\dot{\alpha}}) 
= \frac{1}{2} \epsilon^{\alpha\beta} \e^{\dot{\alpha}\dot{\beta}} p_{\alpha\dot{\alpha}} p_{\beta\dot{\beta}} \,.    
\end{equation}

\subsection{Massless momenta} \label{sec:masslessmom}

For massless momenta, $p_{\a\dot{\a}}$ as a $(2\times 2)$ matrix has rank 1, so it can be written as 
\begin{equation}
p_{\a\dot{\a}} = \l_\a \tilde{\l}_{\dot{\a}} \,.   
\end{equation}
For a real momentum, the spinors satisfy the reality condition, 
\begin{equation}
(\l_\a)^* = \text{sign}(p_0) \tilde{\l}_{\dot{\a}} \,.    
\end{equation}
The Little group $U(1)$ acts on the spinors as 
\begin{equation}
\l \rightarrow e^{-i \frac{\theta}{2}} \l 
\,, \quad 
\tilde{\l} \rightarrow e^{i \frac{\theta}{2}} \tilde{\l} \,.     
\end{equation}
The spinors for $p$ and those for $(-p)$ must be proportional. 
We fix the relation by setting 
\begin{equation}
\l(-p) = i\l(p) \,,
\quad 
\tilde{\l}(-p) =  i\tilde{\l}(p) \,.
\end{equation}
It is customary to introduce a bra-ket notation, 
\begin{align}
|p \rangle \leftrightarrow \l_\a \,,
\quad 
\langle p | \leftrightarrow \l^\a \,,
\quad
|p ] \leftrightarrow \tilde{\l}^{\dot{\a}} \,,
\quad 
[p | \leftrightarrow \tilde{\l}_{\dot{\a}} \,, 
\end{align}
which leads to the Lorentz invariant, Little group covariant brackets, 
\begin{equation}
\langle i j \rangle = \l_i^\a \l_j^\b \e_{\a\b} = \l_i^\a \l_{j\b}
\,, \quad
[i j] = \tilde{\l}_{i \dot{\a}} \tilde{\l}_j^{\dot{\a}} \,.
\end{equation}
The massless Mandelstam variables, which are both Lorentz invariant and Little group invariant, 
can be expressed as 
\begin{align}
2 p_i \cdot p_j =  \e^{\a\b} \e^{\dot{\a}\dot{\b}} 
(p_i)_{\a\dot{\a}} (p_j)_{\b\dot{\b}} = \langle i j \rangle [j i] \,.
\end{align}

\subsection{Massive momenta} \label{sec:massivemom}

For massive momenta, the on-shell condition in the bi-spinor notation is given by
\begin{equation}
\det(p_{\a\dot{\a}}) =  m^2 \,.
\end{equation}
The massive helicity spinor variables are defined by
\begin{equation}
p_{\a\dot{\a}} = \l_\a{}^I \tilde{\l}_{I\dot{\a}} \,, 
\quad 
\det(\l_\a{}^I) = m = \det(\tilde{\l}_{I\dot{\a}})\,. 
\end{equation}
The index $I$ indicates a doublet of the $SU(2)$ Little group. 
The reality condition reads 
\begin{align}
(\l_\a{}^I)^* &= \text{sign}(p_0) \tilde{\l}_{I\dot{\a}} 
\\ (\l_\a{}_I)^* &= -\text{sign}(p_0) \tilde{\l}^{I}_{\dot{\a}} \,.     
\end{align}
\paragraph{SU(2)-invariant tensor} 
Given a matrix representation of the doublet of $SU(2)$, 
\begin{equation}
\psi^I \rightarrow U^I{}_J \psi^J \,, 
\end{equation}
the two defining properties of $SU(2)$ can be written as
\begin{equation}
\e_{IK} U^I{}_J U^K{}_L = \e_{JL} \,,
\quad 
U^I{}_J (U^\dagger)^J{}_K %= U^I{}_J (U^K{}_J)^* 
= \delta^I{}_K \,, 
\label{SU2-def}
\end{equation}
where the $SU(2)$-invariant tensor $\e_{IJ}$ shares, by convention, the first three properties in eq.\eqref{SL2C-epsilon}. Just like spinor indices, the Little group indices are raised and lowered by $\e_{IJ}$ and $\e^{IJ}$.
It follows from eq.\eqref{SU2-def} that the two variables below transform in the same way. 
\begin{equation}
\psi_I := \e_{IJ} \psi^J \quad \mbox{and} \quad 
\bar{\psi}_I := (\psi^I)^* \,. 
\end{equation}
Then, 
\begin{equation}
p_{\a\dot{\a}} 
= \l_\a{}^I \tilde{\l}_{I\dot{\a}} 
= - \l_{\a I} \tilde{\l}^I{}_{\dot{\a}} \,,
\quad 
\bar{p}^{\dot{\a}\a} 
= p_\m (\bar{\s}^\m)^{\dot{\a}\a} 
= \e^{\dot{\a}\dot{\b}} \e^{\a\b} p_{\b \dot{\b}} 
= \l^{\a I} \tilde{\l}_I{}^{\dot{\a}} 
= - \l^\a{}_I \tilde{\l}^{I\dot{\a}} \,.
\end{equation}
It is also useful to note that
\begin{equation}
\e^{\a\b} \l_\a{}^I \l_\b{}^J = \det(\l) \e^{IJ} = m \e^{IJ} \,,
\quad 
\e^{\dot{\a}\dot{\b}} \tilde{\l}_{I\dot{\a}} \tilde{\l}_{J\dot{\b}} = -\det(\tilde{\l}) {\e}_{IJ} = -m {\e}_{IJ} \,.
\end{equation}

\paragraph{Dirac spinors}

By definition, the massive spinor helicity variables satisfy the Dirac equation
\begin{equation}
p_{\a\dot{\a}} \tilde{\l}^{\dot{\a} I} = m \lambda_\a{}^I \,, 
\quad 
p^{\dot{\a}\a} \l_\a{}^I = m \tilde{\l}^{\dot{\a} I}\,.
\label{Dirac-eq}
\end{equation}
For a fixed massive particle, the $SU(2)$ Little group is always completely symmetrized. For convenience, we use the \textbf{BOLD} notation to suppresses the $SU(2)$ little group indices.

The Dirac equation eq.\eqref{Dirac-eq} can be written in the BOLD bra-ket notation as
\begin{equation}
p \qrb{\bf{p}}= m \qra{\bf{p}} \,, 
\quad 
\qla{\bf{p}} p = - m \qlb{\bf{p}} \,.
\label{Dirac-bold}
\end{equation}

As a final consistency check, it is important to see how these massive momenta behave in the high energy limit. This discussion is explained in detail in the appendix A.4 of \Refe{Chung_2019}: massive momenta behave as massless ones, with corrections proportional to the quantity $m/E$. Note that those arguments become crucial when we study the UV behaviour of our theory.

%%%%%%%%%%%%%%%%%%%%%%%%%%%%%%%%%%%%%%%%%%%%%%%%
%----------------------------------------------%
%%%%%%%%%%%%%%%%%%%%%%%%%%%%%%%%%%%%%%%%%%%%%%%%

\section{Spin tensor for spin-1 matter}\label{app:spin_tensor}
I will now discuss the spin tensor for a massive spin-1 particle.
The starting point is the one-particle expectation value
of the angular-momentum operator in the quantum-\\mechanical sense:
\begin{equation}\label{eq:spin1particle}
 \hspace{-0.3cm}   S_p^{\mu\nu}
     = \frac{\bra{p} J^{\mu\nu}\ket{p}}{\braket{p}{p}}
     = \frac{ \ve_{p,\sigma}^*
              J^{\mu\nu,\sigma}_{~~~~\;\tau} \e_p^\tau }
            { \varepsilon_p^*\cdot\varepsilon_p }
     = 2i \ve_p^{*[\mu} \ve_p^{\nu]}\,, \quad \text{where} \quad
    J^{\mu\nu,\sigma}_{~~~~~\tau}
     = i[\eta^{\mu\sigma} \delta^\nu_\tau - \eta^{\nu\sigma} \delta^\mu_\tau] \,,    
\end{equation}
where $J^{\mu\nu}$ are the Lorentz generators in the vector representation. Notica that, due to the transversality of the massive polarization vectors, we have $p_\mu S^{\mu\nu} = 0$.\\
We are now interested in generalizing \eq{eq:spin1particle} to the case in which the two external states are different, in particular we will denote the incoming momentum as $p_1$ and the outgoing with $p_2$.

We may now introduce the notion of \emph{generalized expectation value} (GEV) in the following way:
\begin{equation}
S_{12}^{\mu\nu} = \frac{\bra{\bf 2}J^{\mu\nu}\ket{\bf 1}}{\braket{\bf 2}{\bf 1}}
= \frac{ \ve_{2\sigma}^* J^{\mu\nu,\sigma}_{~~~~\;\tau} \ve_1^\tau}{ \ve_2^*\cdot \ve_1 }
= \frac{ 2i \varepsilon_2^{*[\mu} \varepsilon_1^{\nu]} }
{ \varepsilon_2^*\cdot \varepsilon_1 } .
\end{equation}
Notice that, when computing amplitudes using the spinor-helicity methods, we often prefer to work with all momenta incoming. 
To compare results obtained using these different choices, we analytically continue our bi-spinors to negative energies using the conventions explained in \app{Conventions} and suppress the conjugation sign, thus
\begin{equation}\label{eq:spin1tensornaive}
S_{12}^{\mu\nu} = -2i\ve_1^{[\mu} \ve_2^{\nu]}/(\ve_1\cdot\ve_2) ,
\end{equation}

We end this discussion with the following consideration:
in a classical computation it is desirable to consider a spin tensor that satisfies the \emph{spin supplementary condition} (SSC), \ie $p_\mu S^{\mu\nu} = 0$. However,
\begin{equation}
p_\mu S_{12}^{\mu\nu} = -\frac{i}{2}
\frac{(k\cdot\ve_2) \ve_1^\nu + (k\cdot\ve_1) \ve_2^\nu}
    {\ve_1\cdot \ve_2} \neq 0 , \quad \text{where $k=-p_1-p_2$ }
\end{equation}
Yet, the spin tensor is intrinsically ambiguous,
as the separation between the orbital and intrinsic pieces
of the total angular momentum is relativistically frame-dependent.
This ambiguity allows the spin tensor to be transformed
as $S^{\mu\nu}\rightarrow S^{\mu\nu}+p^{[\mu}r^{\nu]}$, where the difference $p^{[\m}r^{\n]}$ for some vector $r^\n$ accounts for the relative shift between $S^{\m\n}$ and $ L^{\m\n} \sim p^{[\m}\de{p^{\n]}}$. Adjusting $r^\nu$ to accommodate for the SSC, we finally obtain
\begin{equation}\label{eq:spin_operator}
   S^{\mu\nu} = S_{12}^{\mu\nu}
    + \frac{2}{m^{2}} \,p_\lambda S_{12}^{\lambda[\mu} p^{\nu]}
    =-\frac{i}{\varepsilon_1\cdot \varepsilon_2} \bigg\{
      2\varepsilon_1^{[\mu} \varepsilon_2^{\nu]}
    - \frac{1}{m^2} p^{[\mu}
      \big( (k\cdot\varepsilon_2) \varepsilon_1
          + (k\cdot\varepsilon_1) \varepsilon_2 \big)^{\nu]} \bigg\} \,.
\end{equation}

%%%%%%%%%%%%%%%%%%%%%%%%%%%%%%%%%%%%%%%%%%%%%%%%
%----------------------------------------------%
%%%%%%%%%%%%%%%%%%%%%%%%%%%%%%%%%%%%%%%%%%%%%%%%

\section{Angular-momentum operator}\label{app:angularmomentum}
Let us delve into the details of the angular momentum operator in the spinor-helicity formalism.
Consider the total angular momentum $J_{\mu\nu}= L_{\mu\nu} + S_{\mu\nu}$.
The orbital-piece in momentum-space is 
\begin{equation}\label{eq:orbitalmomentum}  
L_{\m\n}=2i\,p_{[\m}\de{p^{\n]}}=p_\s \Sigma_{\m\n,~\t}^{~~~\s}
\de{p_\t} \,, 
\end{equation}
where $\Sigma^{\mu\nu}$ are the Lorentz generators (in vector representation they are simply $J^{\mu\nu}$).

I will focus on the massive case, for a more general treatment see the appendix C of \Refe{Guevara_2019}. Recall that the massive momenta is $p^\mu=\frac12\qla{p^a}\sigma^\mu\qrb{p_a}$. The angular-momentum operator in the space of massive spinors $\{\l_\a^a, \tl_{\dot{\b}}^b\}$ is given by (see \Refe{Witten_2004}):
\begin{equation}\label{eq:orbitalspinormassive_1}
J^{\mu\nu}
= \bigg[ \l^{\alpha a} \sigma_{~~~\alpha}^{\mu\nu,~\,\beta}
         \frac{\partial~}{\partial \l^{\beta a}}
       + \tl_{\dot{\alpha}}^{\,a}
         \bar{\sigma}^{\mu\nu,\dot{\alpha}}_{~~~~\,\dot{\beta}}
         \frac{\partial~}{\partial \tl_{\dot{\beta}}^a} \bigg] \,,     
\end{equation}
where the matrices
\begin{equation}
\s^{\m\n,~\b}_{~~~\a} =\frac{i}{4}(\s^\m_{\a\dot\g}\bar{\s}^{\n,\dot\g \b}-\s^\n_{\a\dot\g}\bar{\s}^{\m,\dot\g \b})\,,  \qquad
\bar\s^{\m\n,\dot\a}_{~~~~\dot\b}=\frac{i}{4}(\bar\s^{\m,\dot\a \g}\s^\n_{\g\dot\b}-\bar\s^{\n,\dot\a \g}\s^\m_{\g\dot\b})\,,
\end{equation}
are the left-handed and right-handed representations of the Lorentz-group algebra.
Equivalently, we may express \eq{eq:orbitalspinormassive_1} in spinor indices as
\begin{equation}\label{eq:orbitalspinormassive_2}
J_{\a\dot{\a},\b\dot{\b}}=\s^\m_{\a\dot{\a}} \s^\n_{\b\dot{\b}} J_{\mu\nu}
= 2i\bigg[ \l_{(\a}^{~a} \de{\l^{\b)a}}\e_{\dot{\a}\dot{\b}}
+ \e_{\a\b} \tl_{(\dot{\a}}^{~a} \de{\tl^{\dot{\b})a}} \bigg]\, .
\end{equation}
Notice that this operator is by construction invariant under the little group ${\rm SU}(2)$.

We are now ready to discuss the action of the aforementioned angular momentum operator on massive spin-$s$ tensors.
The (integer) spin-$s$ tensors are parametrized in terms of massive spinor-helicity variables as
\begin{equation}
\ve_{\a_1\dot{\a}_1\dots\a_s\dot{\a}_s}^{a_1\dots a_{2s}}
= \frac{2^{s/2}}{m^s} \l_{\a_1}^{(a_1} \l_{\dot{\a}_1}^{a_2}
  \cdots \l_{\a_s}^{a_{2s-1}} \l_{\dot{\a}_s}^{a_{2s})} .
\label{eq:poltensors}
\end{equation}\label{eq:Lorentz_Generator}
Since $J^{\mu\nu}$ is a first-order differential operator,
it distributes when acting on $\varepsilon^{a_{1}\cdots a_{2s}}$
according to the standard product rule for derivatives, \ie
\begin{align}
J^{\mu\nu} \e_{\a_1\dot{\a}_1\dots \a_s\dot{\a}_s}^{a_1\dots a_{2s}} =
\frac{2^{s/2}}{m^s} \bigg\{ \Big[\e_{\a_1 \b} \big( \l^{\a(a_1} \s_{~~~\a}^{\m\n,~\,\b} \big) \Big]\l_{\dot{\a}_1}^{a_2} & \cdots \l_{\a_s}^{a_{2s-1}} \tl_{\dot{\a}_s}^{a_{2s})} \,+ \nn\\
+ \,\l_{\a_1}^{(a_1} \Big[ \tl_{\dot{\a}}^{a_2} \bar{\s}^{\m\n,\dot{\a}}_{~~~~\,\dot{\a_2}}\Big] & \cdots \l_{\a_s}^{a_{2s-1}} \tl_{\dot{\a}_s}^{a_{2s})}+\dots \bigg\} \,.   
\end{align}
Finally, we remark that the action of the differential operator \eq{eq:orbitalspinormassive_2} is precisely that of the algebraic generator $\Sigma_{\mu\nu}$, indeed:
\begin{equation}
L_{\m\n}p^\r=-\Sigma_{\m\n,~\s}^{~~~\r}p^\s\,, \qquad
(\ve S^{\m\n})_\t=\ve_\s\Sigma^{\m\n,\s}_{~~~~\t} \,.
\end{equation}
Therefore, for on-shell amplitudes,the spinorial differential operator \eq{eq:orbitalspinormassive_1} incorporates both the orbital and intrinsic contributions, so it is the total angular-momentum operator.

%%%%%%%%%%%%%%%%%%%%%%%%%%%%%%%%%%%%%%%%%%%%%%%%
%----------------------------------------------%
%%%%%%%%%%%%%%%%%%%%%%%%%%%%%%%%%%%%%%%%%%%%%%%%

\section{Spin multipoles from boosts}\label{app:boosts}

Let us closely examine some crucial points discussed in \sect{sec:classical_inf_spin}. Consider the three-point amplitude
\begin{equation}
   {\cal M}_3^{(s)} = {\cal M}_3^{(0)}\:\!
      \varepsilon_2 \cdot
      \exp\!\bigg(\!{-i}\frac{k_\mu \varepsilon_\nu \Sigma^{\mu\nu}}
                             {p_1\cdot\varepsilon}\bigg)
      \cdot \varepsilon_1 \,,
\label{eq:Exp}
\end{equation}
where $\varepsilon_1$ and $\varepsilon_2$ are spin-$s$ polarization tensors and $-\Sigma^{\mu\nu}$ are the Lorentz generators, which in the vector representation reduce to $J^{\m\n}$.\\
On the three-point kinematics, the polarization states for $p_1$ and $p_2$ are related via
\begin{equation}
   \varepsilon_2
    = \exp\!\bigg( \frac{i}{m^2} p_1^\mu k^\nu \Sigma_{\mu\nu}\!\bigg)
      \tilde{\varepsilon}_1 , \qquad
   \tilde{\varepsilon}_1 = U_{12}^{(s)} \varepsilon_1 ,
\label{eq:Boost1}
\end{equation}
where $U_{12}^{(s)}$ is the tensor representation of an $SU(2)$ little-group transformation.\\
The two exponents commute on the three-point kinematics, so
\begin{align}
\label{eq:Arg}
{\cal M}_3^{(s)}\! &= {\cal M}_3^{(0)}
\tilde{\varepsilon}_1 \exp\!\bigg(\!{-}\frac{i}{m^2} p_1^\mu k^\nu \Sigma_{\mu\nu}\!\bigg)\exp\!\bigg(\!{-i}\frac{k_\mu \varepsilon_\nu \Sigma^{\mu\nu}}{p_1\cdot\varepsilon}\bigg)\varepsilon_1 \,=\nn \\
&=  {\cal M}_3^{(0)}\tilde{\varepsilon}_1\exp\!\bigg(\!{-i}\frac{k_\mu \varepsilon_\nu \Sigma_\perp^{\mu\nu}}{p_1\cdot\varepsilon}\bigg)\varepsilon_1\,, 
\end{align}
where we have defined
\begin{equation}
\Sigma_\perp^{\mu\nu} = \Sigma^{\mu\nu} + \frac{2}{m^2} p_1^{[\mu}\Sigma^{\nu]\rho} p_{1\:\!\rho} \,, \qquad \textrm{with} \quad p_{1\:\!\mu} \Sigma_\perp^{\mu\nu} = 0    
\end{equation}
Notice that, being a transverse tensor, it can be used to construct representations of the little group, in fact this is nothing but the spin operator as defined in \eq{eq:spin_operator}.\\
Let us now apply the spinor-helicity formalism to the
above argument. Picking the negative helicity, we may write \eq{eq:Exp} as:
\begin{equation}
   {\cal M}_3^{(s)} = \frac{{\cal M}_3^{(0)}}{m^{2s}} \bra{\bf 2}^{\odot 2s}
      \exp\!\bigg(\!{-i}\frac{k_\mu \varepsilon_\nu^- \sigma^{\mu\nu}}
                        {p_1\cdot\varepsilon^-}\bigg)
      \ket{\bf1}^{\odot 2s} \,.
\end{equation}
Furthermore, in section 3.1 of \Refe{Guevara_2019} the following relation is proven on the three-particle kinematics:
\begin{equation}
   {-i}\frac{k_\mu \varepsilon_\nu^- \sigma^{\mu\nu}}{p_1\cdot\varepsilon^-}
    = -2i\frac{k_\mu \varepsilon_\nu^- \sigma_\perp^{\mu\nu}}{p_1\cdot\varepsilon^-}
    = 2 k \cdot a \,.
\label{eq:GOV}
\end{equation}
After rewriting \eq{eq:Boost1} in this formalism, \ie
\begin{equation}
   \frac{i}{m^2} p_1^\mu k^\nu \sigma_{\mu\nu} = k \cdot a \quad \Rightarrow \quad
   \ket{\bf2}^{\odot 2s}\!= e^{k \cdot a} \big\{ U_{12} \ket{\bf1} \big\}^{\!\odot 2s}\!\!
\label{eq:Boost}
\end{equation}
We finally obtain:
\begin{align}
   {\cal M}_3^{(s)} = \frac{{\cal M}_3^{(0)}}{m^{2s}}
      \big\{ U_{12} \bra{\bf1} \big\}^{\!\odot 2s}
      e^{-k \cdot a} e^{2k \cdot a} \ket{\bf1}^{\odot 2s} 
    = \frac{{\cal M}_3^{(0)}}{m^{2s}}
      \big\{ U_{12} \bra{\bf1} \big\}^{\!\odot 2s} e^{k \cdot a} \ket{\bf1}^{\odot 2s} .
\end{align}

As a final remark, note that, despite the appearance of the factor $p\cdot\varepsilon$ in the denominator of the exponential operator in 
\eq{eq:Exp}, the three-point amplitude does not actually have such `apparent' pole, as we can clearly see from \eq{eq:GOV}.

%%%%%%%%%%%%%%%%%%%%%%%%%%%%%%%%%%%%%%%%%%%%%%%%
%----------------------------------------------%
%%%%%%%%%%%%%%%%%%%%%%%%%%%%%%%%%%%%%%%%%%%%%%%%

\section{Second-order Taylor expansion}\label{app:taylor_expansion}

In this appendix we complete the argument in \sect{sec:st} by expanding the `near-soft' amplitude to second order in $\e$.

Consider a generic function of $\e$. Its Taylor expansion around zero up to second order in $\e$~is given by
\begin{equation}
G(\e)=G(\e)\big|_{\e=0}+\e \de{\e} G(\e)\big|_{\e=0}+\frac{\e^2}{2} \dee{^{\,2}}{\e^2}G(\e)\big|_{\e=0} \,.
\end{equation}

We are now interested in performing a similar expansion for the amplitude \\$\hat\MC^{z_l,\e}(1 \dots P_{0l} \dots n)$. The only quantities that depend on $\e$ in this amplitude are the shifted spinors and $z_l$, that are rewritten below for convenience:
\begin{align}
& \hspace{3cm} z_l =  \e \dfrac{ \wab{0}{p_l}{0}}{ \wab{y}{p_l}{0}} \\
& \begin{cases}
\tx_j^{z_l} = 
\tx_j -\dfrac{2 p_0 p_k }{ \wbb{0}{p_j}{p_k}{0}} \wb{\bf j}{0}\tl_0 
+ \e \dfrac{\wa{y}{0} \wbb{0}{p_k}{p_l}{0}}{ \wab{y}{p_l}{0} \wbb{0}{p_j}{p_k}{0}} \wb{\bf j}{0}\tl_0 \, \\
\tx_k^{z_l} =
\tx_k +\dfrac{2 p_0 p_j}{\wbb{0}{p_j}{p_k}{0}}\wb{\bf k}{0} \tl_0
- \e \dfrac{ \wa{y}{0} \wbb{0}{p_j}{p_l}{0}}{\wab{y}{p_l}{0}\wbb{0}{p_j}{p_k}{0}} \wb{\bf k}{0} \tl_0 \,, \\
\tx_{0l}^{z_l} = \tx_l +\e \dfrac{\wa{y}{0}}{\wab{y}{p_l}{0}}\wb{\bf l}{0} \tl_0  
\end{cases}
\end{align}
Notice that, as $z_l$ is linear in $\e$, $\hat\MC^{z_l,\e}\big|_{\e=0}=\hat\MC^{0,0}$. Furthermore, for $l\neq j,k$, $\hat{P}_{0l}=p_l$ and $\tx_{0l}=\tx_l$ for $\e=0$. \\
Using the chain rule we trade the derivatives on $\e$ with derivatives on the above spinors:
\begin{equation}
\begin{aligned}
\de{\e} & =
\dee{\tx_{0l}^{z_l}}{\e}\de{\tx_{0l}^{z_l}}+
\dee{\tx_{j}^{z_l}}{\e}\de{\tx_{j}^{z_l}}+
\dee{\tx_{k}^{z_l}}{\e}\de{\tx_{k}^{z_l}} = \\
&= \dfrac{\wa{y}{0}}{\wab{y}{p_l}{0}}\wb{\bf l}{0} \tl_0 \de{\tx_{0l}^{z_l}}+
\dfrac{\wa{y}{0} \wbb{0}{p_k}{p_l}{0}}{ \wab{y}{p_l}{0} \wbb{0}{p_j}{p_k}{0}} \wb{\bf j}{0}\tl_0 \de{\tx_{j}^{z_l}}-
\dfrac{ \wa{y}{0} \wbb{0}{p_j}{p_l}{0}}{\wab{y}{p_l}{0}\wbb{0}{p_j}{p_k}{0}} \wb{\bf k}{0} \tl_0 \de{\tx_{k}^{z_l}} = \\
& = \dfrac{\wa{y}{0}}{\wab{y}{p_l}{0}} \left(
\wb{\bf l}{0} \tl_0 \de{\tx_{0l}^{z_l}}+
\dfrac{ \wbb{0}{p_k}{p_l}{0}}{\wbb{0}{p_j}{p_k}{0}} \wb{\bf j}{0}\tl_0 \de{\tx_{j}^{z_l}}-
\dfrac{\wbb{0}{p_j}{p_l}{0}}{\wbb{0}{p_j}{p_k}{0}} \wb{\bf k}{0} \tl_0 \de{\tx_{k}^{z_l}} \right) = \\
&=\dfrac{\wa{y}{0}}{\wab{y}{p_l}{0}}\left(
\wb{\bf l}{0}\wb{0}{\de{\tx_{0l}^{z_l}}}+
\dfrac{ \wbb{0}{p_k}{p_l}{0}}{\wbb{0}{p_j}{p_k}{0}} \wb{\bf j}{0} \wb{0}{\de{\tx_{j}^{z_l}}}-
\dfrac{\wbb{0}{p_j}{p_l}{0}}{\wbb{0}{p_j}{p_k}{0}} \wb{\bf k}{0} \wb{0}{\de{\tx_{k}^{z_l}}} \right) \,.
\end{aligned}
\end{equation}
Thus, in the limit $\e \to 0$, we have:
\begin{equation}
\de{\e}=\dfrac{\wa{0}{y}}{\wab{y}{p_l}{0}} \tD_l
\end{equation}
where we have defined the quantities 
\begin{align}\label{eq:D_tilde_1}
\tD_l=\tilde{d}_l+
\dfrac{ \wbb{0}{p_k}{p_l}{0}}{\wbb{0}{p_j}{p_k}{0}} \wb{0}{\bf j}\wb{0}{\de{\tx_{j}^{z_l}}}-
\dfrac{\wbb{0}{p_j}{p_l}{0}}{\wbb{0}{p_j}{p_k}{0}} \wb{0}{\bf k} \wb{0}{\de{\tx_{k}^{z_l}}} \,, \qquad
\tilde{d}_l=\wb{0}{\bf l}\wb{0}{\de{\tx_l^{z_l}}}\,.
\end{align}
Notice that $\tD_l=0$ for $l = j, k$. This is to be expected, indeed \eq{eq:no_epsilon} clearly shows that in this case $\hat P_{0j} \s|_{z_j}$, $\hat P_{0k} \s|_{z_k}$, $\tx_j^{z_k}$ and $\tx_k^{z_j}$ are independent of $\e$.

In conclusion, we have:
\begin{equation}
\hat\MC^{z_l,\e}(1 \dots \hat{P}_{0l} \dots n)=
\left\{ 1+ \e \dfrac{\wa{0}{y}}{\wab{y}{p_l}{0}} \tD_l+
\frac{\e^2}{2}\dfrac{\wa{0}{y}^2}{\wab{y}{p_l}{0}^2} \tD_l^2
\right\}\hat\MC^{0,0}(1\dots  n) +\mathcal{O}(\e^3)\,.
\end{equation}

%%%%%%%%%%%%%%%%%%%%%%%%%%%%%%%%%%%%%%%%%%%%%%%%
%----------------------------------------------%
%%%%%%%%%%%%%%%%%%%%%%%%%%%%%%%%%%%%%%%%%%%%%%%%

%
\bibliographystyle{unsrt}
\bibliography{ref}

\end{document}